\documentclass[twoside,12pt]{article}
\usepackage{graphicx}
\usepackage{epsfig}
\usepackage{amssymb,amsmath,cite}
\usepackage{bm}       

\newcommand{\ba}{\begin{eqnarray}}
\newcommand{\ea}{\end{eqnarray}}
\newcommand{\ban}{\begin{eqnarray*}}
\newcommand{\ean}{\end{eqnarray*}}
\newcommand{\bsub}{\begin{subequations}}
\newcommand{\esub}{\end{subequations}}
\newcommand{\hbo}{\hbar \omega}
\newcommand{\om}{\omega}
\newcommand{\la}{\lambda}
\newcommand{\s}{\sigma}
\newcommand{\lm}{(\lambda,\mu)}
\newcommand{\lms}{(\lambda_{\sigma},\mu_{\sigma})}

\newcommand{\be}{\begin{equation}}
\newcommand{\ee}{\end{equation}}
\newcommand{\bea}{\begin{eqnarray}}
\newcommand{\eea}{\end{eqnarray}}

\begin{document}

\title{\vspace{1cm}
Partial Dynamical Symmetries}

\author{A. Leviatan\\
Racah Institute of Physics, The Hebrew University, 
Jerusalem 91904, Israel}
\maketitle

\begin{abstract}
This overview focuses on the notion of partial dynamical 
symmetry (PDS), for which a prescribed symmetry is obeyed by 
a subset of solvable eigenstates, but is not shared by the Hamiltonian. 
General algorithms are presented to 
identify interactions, of a given order, with such intermediate-symmetry 
structure. Explicit bosonic and fermionic Hamiltonians with PDS 
are constructed in the framework of models based on spectrum generating 
algebras. PDSs of various types are shown to be relevant to
nuclear spectroscopy, quantum phase transitions
and systems with mixed chaotic and regular dynamics.
\end{abstract}



\newpage

\tableofcontents

\newpage

\section{Introduction}
\label{sec:intro}

Symmetries play an important role in dynamical systems. 
Constants of motion associated with a symmetry govern the integrability of a 
given classical system. At the quantum level, symmetries 
provide quantum numbers for the classification of states, 
determine spectral degeneracies and selection rules, and facilitate 
the calculation of matrix elements.  
An exact symmetry occurs when the Hamiltonian of the system commutes 
with all the generators ($g_i$) of the symmetry-group $G$,  
$[\, \hat{H} \, , \, g_i\,] = 0$. 
In this case, all states have good symmetry and are labeled by the 
irreducible representations (irreps) of $G$. 
The Hamiltonian admits a block structure so that 
inequivalent irreps do not mix and all eigenstates 
in the same irrep are degenerate. In a dynamical symmetry 
the Hamiltonian commutes with the Casimir operator of $G$, 
$[\, \hat{H} \, , \, \hat{C}_{G}\,] = 0$, 
the block structure of $\hat{H}$ is retained, the states preserve 
the good symmetry but, in general, are no longer degenerate. 
When the symmetry is completely broken 
then $[\, \hat{H} \, , \, g_i\,] \neq 0$, and none 
of the states have good symmetry. In-between these limiting cases there 
may exist intermediate symmetry structures, called partial (dynamical) 
symmetries, for which the symmetry is neither exact nor completely broken. 
This novel concept of symmetry and its implications for 
dynamical systems, in particular nuclei, 
are the focus of the present review. 

Models based on spectrum generating algebras form a convenient framework
to examine underlying symmetries in many-body systems and
have been used extensively in diverse areas of physics~\cite{BNB}.
Notable examples in nuclear physics are 
Wigner's spin-isospin SU(4) supermultiplets~\cite{WIG}, 
SU(2) single-$j$ pairing~\cite{Kerman61}, 
Elliott's SU(3) model~\cite{Elliott58}, symplectic model~\cite{Rowe85}, 
pseudo SU(3) model~\cite{ps_su3},  
Ginocchio's monopole and quadrupole pairing models~\cite{GIN}, 
interacting boson models (IBM) for
even-even nuclei~\cite{ibm} and boson-fermion models (IBFM) for 
odd-mass nuclei~\cite{ibfm}. 
Similar algebraic techniques have proven to be useful in the 
structure of molecules~\cite{vibron,Frank94} and of hadrons~\cite{BIL}.
In such models the Hamiltonian is expanded in elements 
of a Lie algebra, ($G_0$), 
called the spectrum generating algebra. 
A dynamical symmetry occurs if the Hamiltonian
can be written in terms of the Casimir operators 
of a chain of nested algebras, 
$G_0\supset G_1 \supset \ldots \supset G_n$~\cite{Iachello06}. 
The following properties are then observed. 
(i)~All states are solvable and analytic expressions
are available for energies and other observables. 
(ii)~All states are classified by quantum numbers, 
$\vert\alpha_0,\alpha_1,\ldots,\alpha_n\rangle$, 
which are the labels of the irreps of the algebras in the chain. 
(iii)~The structure of wave functions is completely dictated by symmetry
and is independent of the Hamiltonian's parameters. 
\begin{figure}[t]
\begin{center}
\includegraphics[height=15cm]{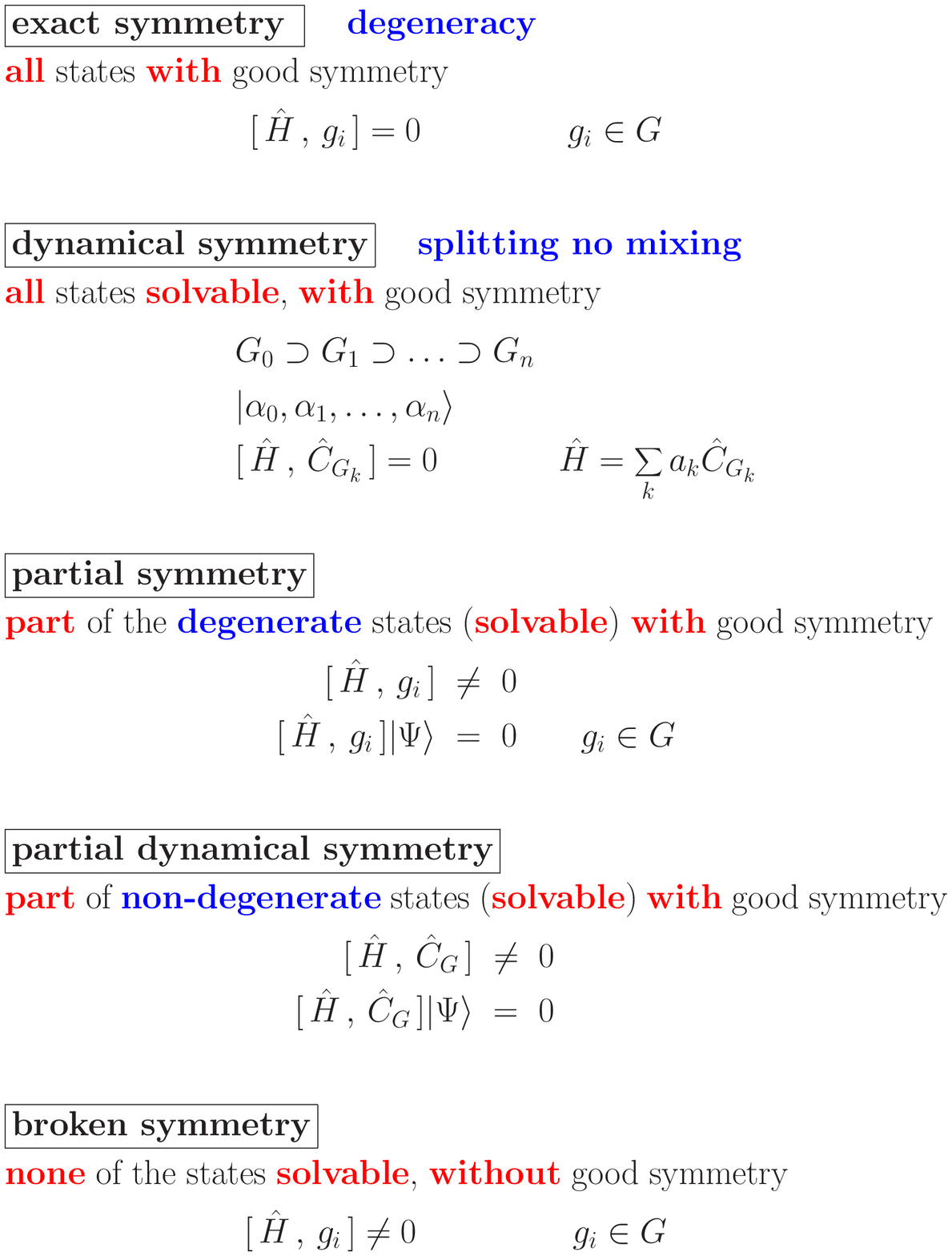}
\caption{
\protect\small 
Hierarchy of symmetries. 
\label{figSymmetry}}
\end{center}
\end{figure}

A dynamical symmetry provides clarifying insights 
into complex dynamics and its 
merits are self-evident. 
However, in most applications to realistic systems,
the predictions of an exact dynamical symmetry are rarely fulfilled 
and one is compelled to break it. 
The breaking of the symmetry is required for a number of reasons. 
First, one often finds that the assumed symmetry is 
not obeyed uniformly, {\it i.e.}, 
is fulfilled 
by only some of the states but not by others. Certain degeneracies
implied by the assumed symmetry are not always realized, 
({\it e.g.}, axially deformed nuclei rarely fulfill the IBM SU(3) 
requirement of degenerate $\beta$ and $\gamma$ 
bands~\cite{ibm}). Secondly, forcing the Hamiltonian to be 
invariant under a symmetry group may impose constraints which are too severe
and incompatible with well-known features of nuclear dynamics 
({\it e.g.}, the models
of~\cite{GIN} require degenerate single-nucleon energies). 
Thirdly, in describing 
transitional nuclei in-between two different structural phases,  
{\it e.g.}, spherical and deformed, the Hamiltonian 
by necessity mixes terms with different symmetry character. 
In the models mentioned above, the required symmetry breaking is achieved
by including in the Hamiltonian terms associated with (two or more) 
different sub-algebra chains of the parent spectrum generating algebra. 
In general, under such circumstances, solvability is lost,
there are no remaining non-trivial conserved quantum numbers and all
eigenstates are expected to be mixed.
A partial dynamical symmetry (PDS) corresponds to
a particular symmetry breaking for which some (but not all) of the 
virtues of a dynamical symmetry are retained. 
The essential idea is to relax the stringent conditions
of {\em complete} solvability
so that the properties (i)--(iii)
are only partially satisfied.
It is then possible to identify the following types of 
partial dynamical symmetries 
\begin{itemize}
\item {\em PDS type I:} 
$\qquad\;\;\;$ {\bf part} of the states have {\bf all} the 
dynamical symmetry
\item{\em PDS type II:}
$\qquad\;\,$ {\bf all} the states have {\bf part} of the 
dynamical symmetry
\item{\em PDS type III:}
$\qquad$ {\bf part} of the states have {\bf part} of the dynamical 
symmetry.
\end{itemize}

In PDS of type~I, only part of the eigenspectrum is analytically 
solvable and retains all the dynamical symmetry (DS) quantum numbers. 
In PDS of type~II, the 
entire eigenspectrum retains some of the DS quantum numbers. 
PDS of type~III has a hybrid character, in the sense that 
some (solvable) eigenstates keep some of the quantum numbers.  

The notion of partial dynamical symmetry generalizes the concepts of exact and 
dynamical symmetries. In making the transition from an exact to a 
dynamical symmetry, states which 
are degenerate in the former scheme are split but not mixed in the latter,
and the block structure of the Hamiltonian is retained.
Proceeding further to partial symmetry, some blocks or selected states in a 
block remain pure, while other states mix and lose the symmetry character. 
A partial dynamical symmetry lifts 
the remaining degeneracies, but preserves the symmetry-purity of the 
selected states. The hierarchy of broken symmetries is depicted in Fig.~1. 

The existence of Hamiltonians with partial symmetry or partial 
dynamical symmetry is by no means obvious. An Hamiltonian with 
the above property is not invariant under the group $G$ 
nor does it commute with the Casimir invariants of $G$, 
so that various irreps are in general mixed in its eigenstates. 
However, it posses a subset of solvable states, denoted by 
$\vert\Psi\rangle$ in Fig.~1, which respect the symmetry. 
The commutator $[\, \hat{H} \, , \, g_i\,]$ 
or $[\, \hat{H} \, , \, \hat{C}_{G}\,]$ vanishes only when it acts on 
these `special' states with good $G$-symmetry.

In this review, we survey the various types of partial dynamical symmetries 
(PDS) and discuss algorithms for their realization in bosonic and fermionic 
systems. Hamiltonians with PDS are explicitly constructed, 
including higher-order terms. We present empirical examples 
of the PDS notion and demonstrate its relevance to nuclear spectroscopy, 
to quantum phase transitions and 
to mixed systems with coexisting regularity and chaos.

\subsection{The interacting boson model}
\label{subsec:ibm}

In order to illustrate  the various notions of symmetries and 
consider their implications, it is beneficial 
to have a framework that has a rich algebraic structure 
and allows tractable yet detailed calculations of observables. 
Such a framework is provided by the interacting boson model 
(IBM)~\cite{ibm,arimaiac76,arimaiac78,arimaiac79}, 
widely used in the description of low-lying quadrupole collective states 
in nuclei. The degrees of freedom of the model are one monopole boson 
($s^{\dag}$) and five quadrupole bosons ($d^{\dag}_{\mu}$). 
The bilinear combinations 
$\{s^{\dag}s,\,s^{\dag}d_{\mu},\, d^{\dag}_{\mu}s,\, 
d^{\dag}_{\mu}d_{\mu'}\}$ span a U(6) algebra. 
These generators can be transcribed in spherical tensor form as
\ba
\hat{n}_{s} &=& s^{\dag}s \;\; , \;\;
U^{(L)}_{\mu} = (d^{\dag} \tilde{d})^{(L)}_{\mu}
\quad  
L=0,1,2,3,4
\nonumber\\
\Pi^{(2)}_{\mu} &=& d^{\dag}_{\mu}s + s^{\dag}\tilde{d}_{\mu} \;\; , \;\;
\bar{\Pi}^{(2)}_{\mu} = i(d^{\dag}_{\mu}s - s^{\dag}\tilde{d}_{\mu}) ~,
\label{u6gen}
\ea
where $\tilde{d}_{\mu} = (-1)^{\mu}d_{-\mu}$, and standard notation of 
angular momentum coupling is used. 
U(6) serves as the spectrum generating algebra 
and the invariant (symmetry) algebra is O(3), 
with generators $L^{(1)}_{\mu} = \sqrt{10}\,U^{(1)}_{\mu}$. 
The IBM Hamiltonian is expanded in terms of 
the operators~(\ref{u6gen}) and consists of Hermitian, 
rotational-scalar interactions which conserve the total number of $s$- 
and $d$- bosons, $\hat N = \hat{n}_s + \hat{n}_d = 
s^{\dagger}s + \sum_{\mu}d^{\dagger}_{\mu}d_{\mu}$.
Microscopic interpretation of the model suggests that for a given even-even 
nucleus the total number of bosons, $N$, is fixed and is taken as the 
sum of valence neutron and proton particle and hole pairs counted from 
the nearest closed shell~\cite{Talmi93}. 

The three dynamical symmetries of the IBM are 
\ba
\begin{array}{lllllll}
{\rm U(6)} & \supset & {\rm U(5)}  & \supset {\rm O(5)} & 
\supset & {\rm O(3)}\qquad & 
\;{\rm anharmonic\; spherical\; 
vibrator}\\
{\rm U(6)} & \supset & {\rm SU(3)} & \supset {\rm O(3)} &         &  & 
\;{\rm axially}{\rm -deformed \; rotovibrator} \\
{\rm U(6)} & \supset & {\rm O(6)}  & \supset {\rm O(5)} & 
\supset & {\rm O(3)} & 
\;\gamma{\rm -unstable\; deformed\; rotovibrator} \\
\end{array}
\label{IBMds}
\ea
The associated analytic solutions 
resemble known limits of the geometric model of nuclei~\cite{bohr75}, 
as indicated above. Each chain provides a complete basis, 
classified by the irreps of the corresponding algebras, which can 
be used for a numerical diagonalization of the Hamiltonian in the general 
case. In the Appendix we collect the relevant information concerning the 
generators and Casimir operators of the algebras in Eq.~(\ref{IBMds}). 
Electromagnetic moments and rates can be calculated in the IBM with 
transition operators of appropriate rank. For example, the most 
general one-body E2 operator reads
\ba
T(E2) = e_{B}\left [ \, \Pi^{(2)} + \chi\,U^{(2)} \, \right ] ~.
\label{Te2}
\ea
A geometric visualization of the model is obtained by 
an energy surface 
\ba
E_{N}(\beta,\gamma) &=& 
\langle \beta,\gamma; N\vert \hat{H} \vert \beta,\gamma ; N\rangle ~, 
\label{enesurf}
\ea
defined by the expectation value of the Hamiltonian in the coherent 
(intrinsic) state~\cite{gino80,diep80}
\bsub
\ba
\vert\beta,\gamma ; N \rangle &=&
(N!)^{-1/2}(b^{\dagger}_{c})^N\,\vert 0\,\rangle ~,\\
b^{\dagger}_{c} &=& (1+\beta^2)^{-1/2}[\beta\cos\gamma 
d^{\dagger}_{0} + \beta\sin{\gamma} 
( d^{\dagger}_{2} + d^{\dagger}_{-2})/\sqrt{2} + s^{\dagger}] ~. 
\ea
\label{condgen}
\esub
Here $(\beta,\gamma)$ are
quadrupole shape parameters whose values, $(\beta_0,\gamma_0)$, 
at the global minimum of $E_{N}(\beta,\gamma)$ define the equilibrium 
shape for a given Hamiltonian. 
The shape can be spherical $(\beta =0)$ or deformed $(\beta >0)$ 
with $\gamma =0$ (prolate), $\gamma =\pi/3$ (oblate), 
$\gamma$-independent, or triaxial $(0 < \gamma < \pi/3)$. 
The latter shape requires terms of order higher than two-body 
in the boson Hamiltonian~\cite{isachen81,levshao90}. 
The equilibrium deformations associated with the Casimir operators of the 
leading subalgebras in the dynamical symmetry chains~(\ref{IBMds}) are, 
$\beta_0=0$ for U(5), $(\beta_0=\sqrt{2},\gamma_0=0)$ for SU(3) and 
$(\beta_0=1,\gamma_0\,{\rm arbitrary})$ for O(6).

\section{PDS type I}
\label{sec:PDStypeI}

PDS of type I corresponds to a situation for which the defining 
properties of a dynamical symmetry (DS), namely, solvability, 
good quantum numbers, and symmetry-dictated structure are fulfilled exactly, 
but by only a subset of states. An algorithm for constructing Hamiltonians 
with this property has been developed in~\cite{AL92} and further elaborated 
in~\cite{RamLevVan09}. The analysis starts from the chain of nested algebras
\begin{equation}
\begin{array}{ccccccc}
G_{\rm dyn}&\supset&G&\supset&\cdots&\supset&G_{\rm sym}\\
\downarrow&&\downarrow&&&&\downarrow\\[0mm]
[h]&&\langle\Sigma\rangle&&&&\Lambda
\end{array}
\label{chain}
\end{equation}
where, below each algebra,
its associated labels of irreps are given. Eq.~(\ref{chain}) implies 
that $G_{\rm dyn}$ is the dynamical (spectrum generating) 
algebra of the 
system such that operators of all physical observables 
can be written in terms of its generators~\cite{Frank94,Iachello06}; 
a single irrep of $G_{\rm dyn}$
contains all states of relevance in the problem.
In contrast, $G_{\rm sym}$ is the symmetry algebra
and a single of its irreps contains states that are degenerate in energy.
A frequently encountered example is $G_{\rm sym}={\rm O}(3)$,
the algebra of rotations in 3 dimensions,
with its associated quantum number
of total angular momentum $L$.
Other examples of conserved quantum numbers
can be the total spin $S$ in atoms
or total isospin $T$ in atomic nuclei.

The classification~(\ref{chain}) is generally valid
and does not require conservation of particle number.
Although the extension from DS to PDS
can be formulated under such general conditions,
let us for simplicity assume in the following
that particle number is conserved.
All states, and hence the representation $[h]$,
can then be assigned a definite particle number~$N$.
For $N$ identical particles
the representation $[h]$
of the dynamical algebra $G_{\rm dyn}$
is either symmetric $[N]$ (bosons)
or antisymmetric $[1^N]$ (fermions)
and will be denoted, in both cases, as $[h_N]$.
For particles that are non-identical
under a given dynamical algebra $G_{\rm dyn}$,
a larger algebra can be chosen
such that they become identical under this larger algebra
(generalized Pauli principle). 
The occurrence of a DS of the type~(\ref{chain})
signifies that the Hamiltonian is written in terms of the Casimir 
operators of the algebras in the chain
\ba
\hat{H}_{DS} &=& \sum_{G} a_{G}\,\hat{C}_{G} 
\label{hDS}
\ea 
and its eigenstates can be labeled as
$|[h_N]\langle\Sigma\rangle\dots\Lambda\rangle$;
additional labels (indicated by $\dots$)
are suppressed in the following.
Likewise, operators can be classified
according to their tensor character under~(\ref{chain})
as $\hat T_{[h_n]\langle\sigma\rangle\lambda}$.

Of specific interest in the construction of a PDS
associated with the reduction~(\ref{chain}),
are the $n$-particle annihilation operators $\hat T$
which satisfy the property
\begin{equation}
\hat T_{[h_n]\langle\sigma\rangle\lambda}
|[h_N]\langle\Sigma_0\rangle\Lambda\rangle=0,
\label{anni}
\end{equation}
for all possible values of $\Lambda$
contained in a given irrep~$\langle\Sigma_0\rangle$ of $G$. 
Any $n$-body, number-conserving normal-ordered interaction
written in terms of these annihilation operators 
and their Hermitian conjugates (which transform as the
corresponding conjugate irreps)
\ba
\hat{H}' &=& \sum_{\alpha,\beta} 
A_{\alpha\beta}\, \hat{T}^{\dag}_{\alpha}\hat{T}_{\beta}
\label{PS}
\ea
has a partial G-symmetry. This comes about since for 
arbitrary coefficients, $A_{\alpha\beta}$, $\hat{H}'$ 
is not a G-scalar, hence most of its eigenstates will be a 
mixture of irreps of G, yet relation~(\ref{anni}) ensures that a subset of 
its eigenstates $\vert [h_N]\langle\Sigma_0\rangle\Lambda\rangle$, 
are solvable and have good quantum numbers under the chain~(\ref{chain}). 
An Hamiltonian with partial dynamical symmetry is obtained by adding 
to $\hat{H}'$ the dynamical symmetry Hamiltonian, 
$\hat{H}_{DS}$~(\ref{hDS}), still preserving the solvability
of states with $\langle\Sigma\rangle=\langle\Sigma_0\rangle$
\ba
\hat{H}_{PDS} &=& \hat{H}_{DS} + \hat{H}' ~.
\label{hPDS}
\ea  

If the operators $\hat{T}_{\beta}\equiv
\hat T_{[h_n]\langle\sigma\rangle\lambda}$ 
span the entire irrep $\langle\sigma\rangle$ of G, 
then the annihilation condition~(\ref{anni}) is satisfied
for all $\Lambda$-states in $\langle\Sigma_0\rangle$, 
if none of the $G$ irreps $\langle\Sigma\rangle$
contained in the $G_{\rm dyn}$ irrep $[h_{N-n}]$
belongs to the $G$ Kronecker product
$\langle\sigma\rangle\times\langle\Sigma_0\rangle$. 
So the problem of finding interactions
that preserve solvability
for part of the states~(\ref{chain})
is reduced to carrying out a Kronecker product.
In this case, although 
the generators $g_i$ of $G$ do not commute with $\hat{H}'$, 
their commutator does vanish when it acts on the 
solvable states~(\ref{anni})
\bsub
\ba
\left [\, g_i\, , \hat{H}'\, \right ] &\neq& 0 ~,\\ 
\left [\, g_i\, , \hat{H}'\, \right ]
|[h_N]\langle\Sigma_0\rangle\Lambda\rangle &=& 0 ~, \quad g_i \in G ~.
\label{gicommub}
\ea
\esub 
Eq.~(\ref{gicommub}) follows from 
$\left [\, g_i\, , \, \hat{T}^{\dag}_{\alpha}\hat{T}_{\beta}\, \right ] = 
\hat{T}^{\dag}_{\alpha}\, \left [\, g_i \, , \hat{T}_{\beta}\,\right ]
+ \left [ \, g_i \, , \hat{T}^{\dag}_{\alpha}\, \right ] \hat{T}_{\beta}$
and the fact that $\left [\, g_i \, , \hat{T}_{\beta}\,\right ]$ 
involves a linear combination 
of $G$-tensor operators which satisfy Eq.~(\ref{anni}). 
The arguments for choosing the special irrep 
$\langle\Sigma\rangle=\langle\Sigma_0\rangle$ in Eq.~(\ref{anni}), 
which contains the solvable states, are based on 
physical grounds. A~frequently encountered choice is the irrep which 
contains the ground state of the system. 

The above algorithm for constructing Hamiltonians with PDS of type I 
is applicable to any semisimple group. It can also address more general 
scenarios, in which relation~(\ref{anni}) holds only for some states 
$\Lambda$ in the irrep $\langle\Sigma_0\rangle$ and/or some 
components $\lambda $  of the tensor 
$\hat T_{[h_n]\langle\sigma\rangle\lambda}$. 
In this case, the Kronecker product rule does not apply, yet the 
PDS Hamiltonian is still of the form as in Eqs.~(\ref{PS})-(\ref{hPDS}), 
but now the solvable states span only part of the corresponding 
$G$-irrep. This is not the case 
in the quasi-exactly solvable Hamiltonians, introduced 
in~\cite{Turbiner88}, where the solvable states form complete 
representations. The coexistence of solvable and unsolvable states, 
together with the availability of an algorithm, distinguish the notion 
of PDS from the notion of accidental degeneracy~\cite{mosh83}, 
where all levels are arranged in degenerate multiplets. 

An Hamiltonian with PDS of type I does not have good symmetry but 
some of its eigenstates do. The symmetry of the latter states 
does not follow from invariance properties of the Hamiltonian. 
This situation is opposite to that encountered in a spontaneous symmetry 
breaking, where the Hamiltonian respects the symmetry 
but its ground state breaks it. 
The notion of PDS differs also from the notion of 
quasi-dynamical symmetry~\cite{rowe0405}. 
The latter corresponds to a situation in 
which selected states in a system continue to exhibit characteristic 
properties ({\it e.g.}, energy and B(E2) ratios) of a dynamical symmetry, 
in the face of strong symmetry-breaking interactions. Such an ``apparent'' 
persistence of symmetry is due to the coherent nature of the mixing 
in the wave functions of these states. In contrast, in a PDS of 
type~I, although the symmetry is broken (even strongly) in most states, 
the subset of solvable states preserve it exactly. In this sense, 
the symmetry is partial but exact!. 

In what follows we present concrete constructions of Hamiltonians with PDS 
associated with the three DS chains~(\ref{IBMds}) of the IBM. 
The partial symmetries in question involve 
continuous Lie groups. PDS can, however, be associated also 
with discrete groups which are relevant, {\it e.g.}, to molecular physics. 
An example of a partial symmetry which involves 
point groups was presented in~\cite{pinchen97}, employing a model of 
coupled anharmonic oscillators to describe the
molecule $XY_6$. The partial symmetry of the Hamiltonian allowed 
a derivation of analytic expressions
for the energies and eigenstates of a set of unique levels. 
Furthermore, the numerical calculations required to obtain the energies of the
remaining (non-unique) levels were greatly simplified since the Hamiltonian
could be diagonalized in a much smaller space.

\subsection{U(5) PDS (type I)}
\label{subsec:u5PDStypeI}
 
The U(5) DS chain of the IBM and related quantum numbers are given 
by~\cite{arimaiac76}
\ba
\begin{array}{ccccccc}
{\rm U}(6)&\supset&{\rm U}(5)&\supset&{\rm O}(5)&
\supset&{\rm O}(3)\\
\downarrow&&\downarrow&&\downarrow&&\downarrow\\[0mm]
[N]&&\langle n_d \rangle&&(\tau)&n_\Delta& L
\end{array} ~,
\label{chainu5}
\ea 
where the generators of the above groups are defined in 
Table~\ref{TabIBMcas} of the Appendix. 
For a given U(6) irrep~$[N]$, the allowed U(5) and O(5) irreps 
are $n_d=0,1,2,\ldots, N$ and  $\tau=n_d,\,n_d-2,\dots 0$ 
or $1$, respectively. The values of $L$ contained in the O(5) 
irrep $(\tau)$, are obtained by partitioning 
$\tau=3n_{\Delta}~+~\lambda$, with $n_{\Delta},\,\lambda\geq 0$ 
integers, and $L=2\lambda,\,2\lambda-2,2\lambda-3,\ldots, \lambda$. 
The multiplicity label $n_\Delta$ in the
${\rm O(5)}\supset {\rm O(3)}$ reduction, 
counts the maximum number of $d$-boson 
triplets coupled to $L=0$~\cite{Szpik}. The eigenstates 
$\vert[N]\langle n_d \rangle(\tau)n_\Delta LM\rangle$ 
are obtained with a Hamiltonian
with U(5) DS which, for one- and two-body interactions, can be
transcribed in the form
\ba
\hat{H}_{\rm DS} &=& \epsilon\,\hat{n}_d + A\,\hat{n}_d(\hat{n}_d+4)
+ B\, \hat{C}_{{\rm O(5)}} + C\,\hat{C}_{{\rm O(3)}} ~.
\label{hDSu5}
\ea 
Here $\hat{n}_d$ and $\hat{n}_d(\hat{n}_d+4)$  are the linear and quadratic 
Casimir operators of U(5), respectively, and    
$\hat{C}_{G}$ denotes the quadratic Casimir operator of $G$, as 
defined in the Appendix. 
The Casimir operators of U(6) are omitted from Eq.~(\ref{hDSu5}) 
since they are functions of the 
total boson number operator, $\hat{N}=\hat{n}_s + \hat{n}_d$, which 
is a constant for all $N$-boson states. The spectrum of 
$\hat{H}_{\rm DS}$ is completely solvable with eigenenergies
\ba
E_{\rm DS} &=& \epsilon\, n_d + A\, n_d ( n_d+4) 
+ B\,\tau(\tau+3)
+\, C\,L(L+1).
\label{eDSu5}
\ea
The U(5)-DS spectrum of Eq.~(\ref{eDSu5})
resembles that of an anharmonic spherical vibrator, 
describing quadrupole excitations of a spherical shape. 
The splitting of states in a given U(5) multiplet, $\langle n_d \rangle$, 
is governed by the O(5) and O(3) terms in 
$\hat{H}_{\rm DS}$~(\ref{hDSu5}).
The lowest U(5) multiplets involve states with quantum numbers 
$(n_d=0,\,\tau=0,\, L=0)$, 
$(n_d=1,\,\tau=1,\, L=2)$, and $(n_d=2,\,\tau=0,\,L=0)$, 
$(n_d=2,\,\tau=2,\,L=2,4)$.
\begin{table}
\begin{center}
\caption{\label{Tabu5tens}
\protect\small
Normalized one- and two-boson U(5) tensors.}
\vspace{1mm}
\begin{tabular}{cccccl}
\hline
& & & & &\\[-3mm]
$n$&$n_d$&$\tau$& $n_{\Delta}$ &$\ell$&
$\hat B^\dag_{[n]\langle n_d\rangle(\tau)n_{\Delta}\ell m}$\\
& & & & &\\[-3mm]
\hline
& & & & &\\[-2mm]
1& 0& 0& 0& 0& $s^{\dag}$\\[2pt]
1& 1& 1& 0& 2& $d^{\dag}_{m}$\\[2pt]
2& 0& 0& 0& 0& $\sqrt{\frac{1}{2}}(s^{\dag})^2$\\[2pt]
2& 1& 1& 0& 2& $s^{\dag} d^{\dag}_{m}$\\[2pt]
2& 2& 0& 0& 0& $\sqrt{\frac{1}{2}}(d^{\dag} d^{\dag})^{(0)}_0$\\[2pt]
2& 2& 2& 0& 2& $\sqrt{\frac{1}{2}}(d^\dag d^\dag)^{(2)}_m$\\[2pt]
2& 2& 2& 0& 4& $\sqrt{\frac{1}{2}}(d^\dag d^\dag)^{(4)}_m$\\
& & & & &\\[-3mm]
\hline
\end{tabular}
\end{center}
\end{table}

The construction of Hamiltonians with U(5)-PDS is based on identification 
of $n$-boson operators which annihilate all states in a given 
U(5) irrep~$\langle n_d\rangle$. A physically relevant choice is the
irrep $n_d=0$ which consists of the ground state, with $\tau=L=0$, 
built of $N$ $s$-bosons
\ba
\vert [N],n_d=\tau=L=0\rangle &=& (N!)^{-1/2}\,(s^{\dag})^N
\vert 0 \rangle ~.
\label{condu5}
\ea
Considering U(5) tensors, 
$\hat B^\dag_{[n]\langle n_d\rangle(\tau)n_{\Delta}\ell m}$, 
composed of $n$ bosons of which $n_d$ are $d$-bosons then, 
clearly, the Hermitian conjugate of such operators 
with $n_d\neq 0$ will annihilate the $n_d=0$ state of Eq.~(\ref{condu5}).  
Explicit expressions for $n$-boson U(5) tensors, 
with $n=1,2$ are shown in Table~\ref{Tabu5tens}. From them one can 
construct the following one- and two-body Hamiltonian with U(5)-PDS 
\ba
\hat{H}_{PDS} &=&
\epsilon_{d}\,d^{\dag}\cdot \tilde{d}
+ u_{2}\,s^{\dag}d^{\dag}\cdot\tilde{d}s 
+ v_{2}\,\left [\, 
s^{\dag}d^{\dag}\cdot (\tilde{d} \tilde{d})^{(2)} + H.c. \, \right ] 
\nonumber\\
&&
+ \sum_{L=0,2,4}c_{L}\,(d^{\dag}d^{\dag})^{(L)}\cdot
(\tilde{d}\tilde{d})^{(L)} ~,
\label{u5hPDS}
\ea
where $H.c.$ means Hermitian conjugate. By construction, 
\ba
\hat{H}_{PDS}\vert [N],n_d=\tau=L=0\rangle &=& 0 ~.
\label{nd0}
\ea
Using Eq.~(\ref{hIBMcas}), we can rewrite 
$\hat{H}_{PDS}$ in the form
\ba
\hat{H}_{PDS} &=& \hat{H}_{DS} + \hat{V}_2 ~, 
\label{hPDSu5}
\ea
where $\hat{H}_{DS}$ is the U(5) dynamical symmetry Hamiltonian, 
Eq.~(\ref{hDSu5}), and $\hat{V}_2$ is given by 
\ba
\hat{V}_{2} &=&   
 v_{2}\left [\, 
s^{\dag}d^{\dag}\cdot (\tilde{d} \tilde{d})^{(2)} + H.c. \, \right ] 
=  v_{2}\,\Pi^{(2)}\cdot U^{(2)}
=  v_{2}\,U^{(2)}\cdot \Pi^{(2)} ~.
\label{V2}
\ea
The operators $\Pi^{(2)}_{\mu}$ and $U^{(2)}_{\mu}$ are defined in 
Eq.~(\ref{u6gen}). The $\hat{V}_2$ term breaks the U(5) DS, however, 
it still has the U(5) basis states with $(n_d=\tau=L=0)$, Eq.~(\ref{nd0}), 
and $(n_d=\tau=L=3)$ as zero-energy eigenstates 
\ba
\hat{V}_{2}\,\vert [N],n_d=\tau=L=3\rangle &=& 0 ~.
\label{V2nd3}
\ea
The last property follows from the U(5) selection rules of 
$\hat{V}_2$, $\Delta n_d = \pm 1$, and the fact that the irreps 
$(n_d=4,\tau=0,2,4)$ and $(n_d=2,\tau=0,2)$ do not contain an $L=3$ state. 
Altogether, $\hat{H}_{PDS}$~(\ref{hPDSu5}) 
is not diagonal in the U(5) chain, but retains the following solvable 
U(5) basis states with known eigenvalues
\bsub
\ba
\vert [N], n_d=\tau=L=0\rangle \;\;\;\; 
&&E_{PDS} = 0 ~,\\
\vert [N], n_d=\tau=L=3\rangle \;\;\;\; 
&&E_{PDS} = 3\epsilon + 21A + 18B + 12C ~. \qquad\qquad
\ea
\label{ePDSu5}
\esub
As will be discussed in Section~\ref{sec:PDSQPT}, 
this class of Hamiltonians with U(5)-PDS of type I is relevant to 
first-order quantum shape-phase transitions in nuclei. 

A second class of Hamiltonians with U(5)-PDS can be obtained by 
considering the interaction
\ba
\hat{V}_{0} &=&   
v_{0}\left [\, 
(s^{\dag})^2\tilde{d}\cdot \tilde{d} + H.c. \, \right ] ~. 
\label{V0}
\ea
This interaction breaks the U(5) DS, however, it still has selected 
U(5) basis states as zero-energy eigenstates
\bsub
\ba
\hat{V}_{0}\,\vert [N],n_d=\tau=N,L\,\rangle &=& 0 ~,\\
\hat{V}_{0}\,\vert [N],n_d=\tau=N-1,L\,\rangle &=& 0 ~,
\ea
\label{V0ndN}
\esub
where $L$ takes the values compatible with the 
${\rm O(5)}\supset{\rm O(3)}$ reduction. 
These properties follow from the fact that $s^2$ annihilates 
states with $n_s=0,\,1$ ($n_d=N,\, N-1$) and $\tilde{d}\cdot\tilde{d}$ 
annihilates states with $n_d=\tau$~\cite{arimaiac76}. 
Adding the interaction $\hat{V}_0$ 
to the U(5) dynamical symmetry Hamiltonian $\hat{H}_{DS}$~(\ref{hDSu5}), 
we obtain the following Hamiltonian with U(5)-PDS
\ba
\hat{H}_{PDS}' &=& \hat{H}_{DS} + \hat{V}_0 ~. 
\label{hPDSu5b}
\ea
$\hat{H}_{PDS}'$ is not diagonal in the U(5) chain, but retains the 
following solvable U(5) basis states with known eigenvalues
\bsub
\ba
&&\vert [N], n_d=\tau=N,L\,\rangle \;\;\qquad\quad 
E_{PDS}' = N[\,\epsilon_{d} + A(N+4) + B(N+3)\,] + CL(L+1) ~,\qquad\qquad\\
&&\vert [N], n_d=\tau=N-1,L\,\rangle \;\;\;\quad 
E_{PDS}' = (N-1)[\,\epsilon_{d} + A(N+3) 
+ B(N+2)\,] + CL(L+1) ~. \qquad\qquad
\ea
\label{ePDSu5b}
\esub
The Hamiltonian $\hat{H}_{PDS}'$~(\ref{hPDSu5b}) with U(5)-PDS of type I, 
contains terms from both the U(5) and O(6) chains~(\ref{IBMds}), 
hence preserves the common segment of subalgebras, 
${\rm O(5)}\supset {\rm O(3)}$. 
As such, it exhibits also an O(5)-PDS of type II, to be discussed 
in Section~\ref{sec:PDStypeII}. 
As will be shown in Section~\ref{sec:PDSQPT}, 
this class of Hamiltonians is relevant to second-order quantum 
shape-phase transitions in nuclei.

\subsection{SU(3) PDS (type I)}
\label{subsec:su3PDStypeI}

The SU(3) DS chain of the IBM and related quantum numbers are given 
by~\cite{arimaiac78}
\ba
\begin{array}{ccccc}
{\rm U}(6)&\supset&{\rm SU}(3)&\supset&{\rm O}(3)\\
\downarrow&&\downarrow&&\downarrow\\[0mm]
[N]&&\left (\lambda,\mu\right )& K & L
\end{array} ~,
\label{chainsu3}
\ea
where the generators of the above groups $G$ are defined in 
Table~\ref{TabIBMcas} of the Appendix. 
For a given U(6) irrep $[N]$, the allowed SU(3) irreps are 
$(\lambda,\mu)=(2N-4k-6m,2k)$ 
with $k,m$ non-negative integers, such that, $\lambda,\mu\geq 0$. 
The multiplicity label $K$ 
is needed for complete classification and corresponds geometrically to the
projection of the angular momentum on the symmetry axis. 
The values of $L$ contained in the above SU(3) irreps 
are $L=K,K+1,K+2,\ldots,K+{\rm max}\{\lambda,\mu\}$, where 
$K=0,\, 2,\ldots, {\rm min}\{\lambda,\mu\}$;
with the exception of $K=0$ for which 
$L=0,\, 2,\ldots, {\rm max}\{\lambda,\mu\}$. 
The states $\vert [N](\lambda,\mu)KLM\rangle$ 
form the (non-orthogonal) Elliott basis~\cite{Elliott58} 
and the Vergados basis 
$\vert [N](\lambda,\mu)\tilde{\chi}LM\rangle$~\cite{VER} is obtained 
from it by a standard orthogonalization procedure. 
The two bases coincide in the large-N limit and both are eigenstates 
of a Hamiltonian with SU(3) DS. The latter, 
for one- and two-body interactions, can be transcribed in the form 
\ba
\hat{H}_{DS} &=& h_{2}\left [-\hat C_{{\rm SU(3)}} 
+ 2\hat N (2\hat N+3)\right ]
+ C\, \hat C_{{\rm O(3)}} \quad ~,
\label{hDSsu3}
\ea
where $\hat{C}_{G}$ is  
the quadratic Casimir operator 
of $G$, as defined in the Appendix. 
The spectrum of $\hat{H}_{DS}$ is completely solvable with eigenenergies
\ba
E_{\rm DS} &=& h_{2}\,
\left [\, -f_{2}(\lambda,\mu) + 2N(2N+3) \, \right ] 
+ C\, L(L+1)
\nonumber\\
&=& h_{2}\,
6\left [2N(k+2m) - k(2k-1)-3m(2m-1) -6km\right ]
+ CL(L+1) ~,\qquad
\label{eDSsu3}
\ea
where 
$f_{2}(\lambda,\mu) = \lambda^2 + \mu^2 + \lambda\mu + 3\lambda + 3\mu$ 
and $(\lambda,\mu)=(2N-4k-6m,2k)$. The spectrum 
resembles that of an axially-deformed rotovibrator and the corresponding  
eigenstates are arranged in SU(3) multiplets. 
In a given SU(3) irrep $(\lambda,\mu)$, each $K$-value is associated 
with a rotational band and states 
with the same L, in different $K$-bands, are degenerate. 
The lowest SU(3) irrep is $(2N,0)$, which describes the ground band 
$g(K=0)$ of a prolate deformed nucleus. 
The first excited SU(3) irrep
$(2N-4,2)$ contains both the $\beta(K=0)$ and $\gamma(K=2)$ bands. 
States in these bands with the same angular momentum are degenerate.
This $\beta$-$\gamma$ degeneracy is a characteristic feature of the SU(3) 
limit of the IBM which, however, is not commonly observed~\cite{CW}. 
In most deformed nuclei the $\beta$ band lies above the $\gamma$ band. 
In the IBM framework, with at most two-body interactions, one
is therefore compelled to break SU(3) 
in order to conform with the experimental data. 
To do so, the usual approach has been to include in the Hamiltonian
terms from other chains
so as to lift the undesired $\beta$-$\gamma$ degeneracy.
Such an approach was taken by Warner Casten and Davidson (WCD) 
in~\cite{WCD}, where an O(6) term was added to the SU(3) Hamiltonian. 
However, in this procedure,
the SU(3) symmetry is completely broken,
all eigenstates are mixed and no analytic solutions are retained. 
Similar statements apply to the description in the consistent Q formalism 
(CQF)~\cite{CQF}, where the Hamiltonian involves a non-SU(3) quadrupole 
operator. In contrast, partial SU(3) symmetry, to be discussed below, 
corresponds to breaking SU(3), but in a very particular way
so that {\it part} of the states (but not all) will still be solvable with
good symmetry. As such, the virtues of a dynamical symmetry 
({\it e.g.}, solvability) are fulfilled but by only a subset of states. 

The construction of Hamiltonians with SU(3)-PDS is based on identification 
of $n$-boson operators which annihilate all states in a given 
SU(3) irrep $(\lambda,\mu)$, 
chosen here to be the ground band irrep $(2N,0)$. 
For that purpose, we consider the following boson-pair operators with 
angular momentum $L =0,\,2$
\bsub
\ba
P^{\dagger}_{0} &=& d^{\dagger}\cdot d^{\dagger} - 2(s^{\dagger})^2 ~,\\
P^{\dagger}_{2\mu} &=& 2d^{\dagger}_{\mu}s^{\dagger} + 
\sqrt{7}\, (d^{\dagger}\,d^{\dagger})^{(2)}_{\mu} ~.
\ea
\label{PL}
\esub
\begin{table}
\begin{center}
\caption{\label{Tabsu3tensII}
\protect\small
Normalized one- and two-boson SU(3) tensors. 
For the indicated irreps the labels of the Vergados basis 
($\tilde{\chi}$) and Elliott basis ($K$) are identical, $\tilde{\chi}=K$.} 
\vspace{1mm}
\begin{tabular}{ccccl}
\hline
& & & & \\[-3mm]
$n$&$(\lambda,\mu)$&$\tilde{\chi}$&$\ell$&
$\qquad\hat B^\dag_{[n](\lambda,\mu)\tilde{\chi};\ell m}$\\
& & & & \\[-3mm]
\hline
& & & & \\[-2mm]
1& (2,0) & 0& 0&
$s^{\dag}$ \\[2pt]
1& (2,0) & 0& 2&
$d^{\dag}_{m}$ \\[2pt]
2& (4,0) & 0& 0& 
$\sqrt{\frac{5}{18}}(s^{\dag})^2 
+ \sqrt{\frac{2}{9}}(d^{\dag}d^{\dag})^{(0)}_{0} $ \\[2pt]
2& (4,0) & 0& 2& 
$\sqrt{\frac{7}{9}}s^{\dag}d^{\dag}_{m} 
- \frac{1}{3}(d^{\dag}d^{\dag})^{(2)}_{m} $ \\[2pt]
2& (4,0) & 0& 4& 
$\frac{1}{\sqrt{2}}(d^{\dag}d^{\dag})^{(4)}_{m} $ \\[2pt]
2& (0,2) & 0& 0& 
$-\sqrt{\frac{2}{9}}(s^{\dag})^2 
+ \sqrt{\frac{5}{18}}(d^{\dag}d^{\dag})^{(0)}_{0} \qquad$ \\[4pt]
2& (0,2) & 0& 2& 
$\sqrt{\frac{2}{9}}s^{\dag}d^{\dag}_{m} 
+ \sqrt{\frac{7}{18}}(d^{\dag}d^{\dag})^{(2)}_{m} $ \\[2pt]
& & & & \\[-3mm]
\hline
\end{tabular} 
\end{center}
\end{table}
As seen from Table~\ref{Tabsu3tensII}, these operators 
are proportional to two-boson SU(3) tensors, 
$B^{\dagger}_{[n](\lambda,\mu)\tilde{\chi};\ell m}$, with $n=2$ and 
$(\lambda,\mu)=(0,2)$
\bsub
\ba
B^{\dagger}_{[2](0,2)0;00} &=& \frac{1}{3\sqrt{2}}\,P^{\dagger}_{0} ~,\\
B^{\dagger}_{[2](0,2)0;2\mu} &=& \frac{1}{3\sqrt{2}}\,P^{\dagger}_{2\mu} ~.
\ea
\esub
The corresponding Hermitian conjugate boson-pair annihilation operators,  
$P_0$ and $P_{2\mu}$, transform 
as $(2,0)$ under SU(3), and satisfy
\ba
P_{0}\,\vert [N](2N,0)K=0, LM\rangle &=& 0 ~,
\nonumber\\
P_{2\mu}\,\vert [N](2N,0)K=0, LM\rangle &=& 0 ~.
\label{P0P2}
\ea
Equivalently, these operators satisfy
\ba
P_{0}\vert c;\, N\rangle &=& 0
\nonumber\\
P_{2\mu}\vert c;\, N\rangle &=& 0
\label{PLcond}
\ea
where 
\ba 
\vert c; \, N\rangle &=& 
(N!)^{-1/2} (b^{\dag}_{c})^{N}\vert 0\rangle \;\;\; , \;\;\;
b^{\dag}_{c}= (\sqrt{2}\,d^{\dag}_{0} + s^{\dag})/\sqrt{3} ~. 
\label{condsu3}
\ea
The state $\vert c; \, N\rangle$ is a condensate of $N$ bosons and 
is obtained by substituting the SU(3) equilibrium deformations in the 
coherent state of Eq.~(\ref{condgen}), 
$\vert c; \, N\rangle = \vert\beta=\sqrt{2},\gamma=0 ; N \rangle$.  
It is the lowest-weight state in the SU(3) irrep $(\lambda,\mu)=(2N,0)$ 
and serves as an intrinsic state for the SU(3) ground band~\cite{CA83}. 
The rotational members of the band 
$\vert [N](2N,0)K=0, LM\rangle$, Eq.~(\ref{P0P2}), 
are obtained by angular momentum projection from $\vert c;\, N\rangle$. 
The relations in Eqs.~(\ref{P0P2})-(\ref{PLcond}) follow from the 
fact that the action of the operators $P_{L\mu}$ leads to a state with 
$N-2$ bosons in the U(6) irrep $[N-2]$, 
which does not contain the SU(3) irreps obtained from the product 
$(2,0)\times (2N,0)= (2N+2,0)\oplus (2N,1)\oplus (2N-2,2)$. 

Following the general algorithm, a two-body Hamiltonian with partial 
SU(3) symmetry can now be constructed as~\cite{AL92,lev96}
\ba
\hat{H}(h_0,h_2) &=& h_{0}\, P^{\dagger}_{0}P_{0} 
+ h_{2}\,P^{\dagger}_{2}\cdot \tilde{P}_{2} ~,
\label{HPSsu3}
\ea
where $\tilde P_{2,\mu} = (-)^{\mu}P_{2,-\mu}$. 
For $h_{2}=h_{0}$, the Hamiltonian is an SU(3) scalar, related to the 
quadratic Casimir operator of SU(3) 
\ba
\hat{H}(h_0=h_2) &=& 
h_{2}\,\left [P^{\dagger}_{0}P_{0} 
+ P^{\dagger}_{2}\cdot \tilde{P}_{2}\right ] 
= h_{2}\left [-\hat C_{{\rm SU(3)}} + 2\hat N (2\hat N+3)\right ]
~.\qquad
\label{Hsu3}
\ea
For $h_0=-5h_2$, the Hamiltonian transforms as a $(2,2)$ SU(3) 
tensor component. For arbitrary $h_{0}, h_{2}$ coefficients, 
$\hat{H}(h_0,h_2)$ is not an SU(3) scalar, 
nevertheless, the relations in Eqs.~(\ref{P0P2})-(\ref{PLcond}) ensure 
that it has a solvable zero-energy band with good SU(3) quantum numbers  
$(\lambda,\mu)=(2N,0)$. When the coefficients 
$h_{0},\,h_{2}$ are positive, $\hat{H}(h_0,h_2)$ 
becomes positive definite by construction, 
and the solvable states form its SU(3) ground band.

$\hat{H}(h_0,h_2)$ of Eq.~(\ref{HPSsu3}) 
has additional solvable eigenstates with good SU(3) character. 
This comes about from Eq.~(\ref{PLcond}) and the following properties 
\ba
&&P_{L,\mu}\vert c;N\rangle \;=\; 0
\;\;\; , \;\;\;
\left [P_{L,\mu}\, ,\, P^{\dagger}_{2,2}\right ]\vert c;N\rangle\;= \;
\delta_{L,2}\delta_{\mu,2}\,6(2N+3)\vert c;N\rangle \;\;\; , \qquad
\nonumber\\
&&\left [\left [P_{L,\mu}\, , \,P^{\dagger}_{2,2}\right ]\, , \,
P^{\dagger}_{2,2}\right ]\; = \;
\delta_{L,2}\delta_{\mu,2}\,24P^{\dagger}_{2,2}\;\;\; ,
\qquad\quad L=0,2
\quad ~.
\label{PLprop}
\ea
These relations imply that the sequence of states
\ba
\vert k\rangle \; \propto \; \left (P^{\dagger}_{2,2}\right )^{k}
\vert c; N-2k\rangle ~,
\label{k}
\ea
are eigenstates of $\hat{H}(h_0,h_2)$ 
with eigenvalues $E_{k}\;=\; 6h_{2}\left (2N+1 -2k \right )k $.
A comparison with Eq.~(\ref{eDSsu3}) shows that
these energies are the SU(3) eigenvalues of $\hat{H}(h_0=h_2)$, 
Eq.~(\ref{Hsu3}), and identify the states $\vert k\rangle$ to be 
in the SU(3) irreps $(2N-4k,2k)$. 
They can be further shown to be the 
lowest-weight states in these irreps. Furthermore, $P_0$ satisfies
\ba
P_{0}\,\vert k\rangle &=& 0 ~,
\label{P0k}
\ea
or equivalently, 
\ba
P_{0}\,\vert [N](2N-4k,2k)K=2k, LM \rangle &=& 0 ~.
\label{P0}
\ea
The states $\vert k\rangle$~(\ref{k}) 
are deformed and serve as intrinsic states 
representing $\gamma^{k}$ bands with angular momentum projection ($K=2k$) 
along the symmetry axis~\cite{CA83}. In particular, 
as noted earlier, 
$\vert k=0\rangle = \vert c; N\rangle$
represents the ground band ($K=0$) and $\vert k=1\rangle$
is the $\gamma$-band ($K=2$). The rotational members of these bands, 
$\vert [N](2N-4k,2k)K=2k, LM \rangle 
\propto \hat{\cal P}_{LM}\vert k\rangle$, 
Eq.~(\ref{P0}), can be obtained by O(3) projection from the corresponding 
intrinsic states $\vert k\rangle$. 
Relations~(\ref{P0k}) and~(\ref{P0}) are equivalent, since 
the angular momentum projection operator, $\hat{\cal P}_{LM}$, 
is composed of O(3) generators, hence commutes with $P_0$. 
The intrinsic states $\vert k \rangle$ break the O(3) 
symmetry, but since the Hamiltonian in Eq.~(\ref{HPSsu3}) is 
O(3) invariant, the projected states with good~$L$ 
are also eigenstates of $\hat{H}(h_0,h_2)$ with energy 
$E_k$ and with good SU(3) symmetry. 
For the ground band $(k=0)$ the projected states span the
entire SU(3) irrep $(2N,0)$. For excited bands $(k\neq 0)$, the projected 
states span only part of the corresponding 
SU(3) irreps. There are other states originally in these irreps 
(as well as in other irreps) which do
not preserve the SU(3) symmetry and therefore get mixed.
In particular, the ground $g(K=0)$, and 
$\gamma(K=2)$ bands retain their SU(3) character $(2N,0)$ and $(2N-4,2)$ 
respectively, but the first excited $K=0$ band is mixed. 
This situation corresponds precisely to that of partial SU(3) symmetry.
An Hamiltonian $\hat{H}(h_0,h_2)$ which is not an SU(3) scalar has a 
subset of {\it solvable} eigenstates which continue to have 
good SU(3) symmetry. 

All of the above discussion is applicable also to the case when
we add to the Hamiltonian of Eq.~(\ref{HPSsu3}) 
the Casimir operator of O(3), and by doing so, convert the partial 
SU(3) symmetry into partial dynamical SU(3) symmetry. 
The additional rotational term contributes just an $L(L+1)$ splitting 
but does not affect the wave functions.
The most general one- and two-body Hamiltonian with SU(3)-PDS 
can thus be written in the form
\ba
\hat{H}_{PDS} &=& \hat{H}(h_0,h_2) + C\, \hat{C}_{{\rm O(3)}}
=  \hat{H}_{DS} + (h_0 - h_2)\, P^{\dagger}_{0}P_{0} ~.
\label{hPDSsu3}
\ea
Here $\hat{H}_{DS}$ is the SU(3) dynamical symmetry Hamiltonian, 
Eq.~(\ref{hDSsu3}), with parameters $h_2$ and $C$.
The $P^{\dag}_{0}P_0$ term in Eq.~(\ref{hPDSsu3}) 
is not diagonal in the SU(3) chain~(\ref{chainsu3}), 
but Eqs.~(\ref{P0P2}) and~(\ref{P0}) ensure that it annihilates 
a subset of states with good SU(3) quantum numbers. 
Consequently, $\hat{H}_{PDS}$ retains selected solvable bands with good 
SU(3) symmetry. Specifically, the solvable states are members of the 
ground $g(K=0)$ band
\bsub
\ba
&&\vert N,(2N,0)K=0,L\rangle \;\;\;\; L=0,2,4,\ldots, 2N\\
&& E_{PDS}= CL(L+1)
\ea
\label{gband}
\esub
and $\gamma^{k}(K=2k)$ bands
\bsub
\ba
&&\vert N,(2N-4k,2k)K=2k,L\rangle
\;\;\;\;\;\; 
L=K,K+1,\ldots, (2N-2k) \qquad\qquad\\
&&
E_{PDS} =  h_{2}\,6k \left (2N - 2k+1 \right ) + CL(L+1)
\qquad k>0 ~.
\ea
\label{gamband}
\esub 
The solvable states~(\ref{gband})-(\ref{gamband}) are those projected 
from the intrinsic states $\vert k\rangle$ of Eq.~(\ref{k}), and are 
simply selected members of the Elliott basis 
$\phi_{E}((\lambda,\mu)KLM)$~\cite{Elliott58}. In particular, 
the states belonging to the ground and $\gamma$ bands are the Elliott 
states $\phi_{E}((2N,0)K=0,LM)$ and $\phi_{E}((2N-4,2)K=2,LM)$ 
respectively. Their wave functions 
can be expressed in terms of the orthonormal Vergados basis, 
$\Psi_{V}((\lambda,\mu)\tilde{\chi} LM)$~\cite{VER}. 
For the ground band and for members of the $\gamma$ band with $L$ odd, the
Vergados and Elliott bases are identical. 
The Elliott states in the $\gamma (K=2)$ band with $L$ even 
are mixtures of Vergados states 
in the SU(3) irrep $(2N-4,2)$ 
\ba
&&\phi_{E}((2N-4,2)K=2,LM) =
\nonumber\\
&&\qquad\quad
\left [\,
\Psi_{V}((2N-4,2)\tilde{\chi}=2,LM)  - x^{(L)}_{20}\,
\Psi_{V}((2N-4,2)\tilde{\chi}=0,LM)
\,\right ]/  
x^{(L)}_{22} ~,\qquad\quad
\label{phigam}
\ea
where $x^{(L)}_{20},\, x^{(L)}_{22}$ 
are known coefficients which appear in the
transformation between the two bases~\cite{VER}.

Since the wave functions of the solvable states are known, it is 
possible to obtain {\it analytic} expressions for matrix 
elements of observables between them. For calculating E2 rates, 
it is convenient to rewrite the relevant E2 operator, Eq.~(\ref{Te2}), 
in the form
\ba
T(E2) \; = \; \alpha\, Q^{(2)} + \theta\, \Pi^{(2)} ~,
\label{Te2su3}
\ea
where $Q^{(2)}$ is the quadrupole SU(3) generator 
[$\,Q^{(2)} =\Pi^{(2)} -(\sqrt{7}/2)\,U^{(2)}\,$] 
and $\Pi^{(2)}$ is a $(2,2)$ tensor under SU(3). 
The B(E2) values for $g\to g$ and $g\to \gamma$ transitions 
\bsub
\ba
B(E2;g,L\rightarrow g,L') =
\qquad\qquad\qquad\qquad\qquad\qquad
\qquad\qquad\qquad\qquad\qquad
\nonumber\\
\frac{\vert\langle\phi_{E}((2N,0)K=0,L')||
\alpha\, Q^{(2)} + \theta\, \Pi^{(2)}||
\phi_{E}((2N,0)K=0,L)\rangle\vert^{2}}{(2L+1)} ~,
\label{gtog}
\\
B(E2;\gamma,L\rightarrow g,L') =
\qquad\qquad\qquad\qquad\qquad\qquad
\qquad\qquad\qquad\qquad\qquad
\nonumber\\
\theta^2\,
\frac{\vert\langle\phi_{E}((2N,0)K=0,L')||\Pi^{(2)}||
\phi_{E}((2N-4,2)K=2,L)\rangle\vert^{2}}{(2L+1)} ~.
\label{gamtog}
\ea
\label{E2tran}
\esub
can be expressed in terms of E2 matrix elements in the Vergados 
basis, for which analytic expressions are available~\cite{arimaiac78,ISA}. 

\begin{figure}[t]
\begin{center}
\includegraphics[height=8cm]{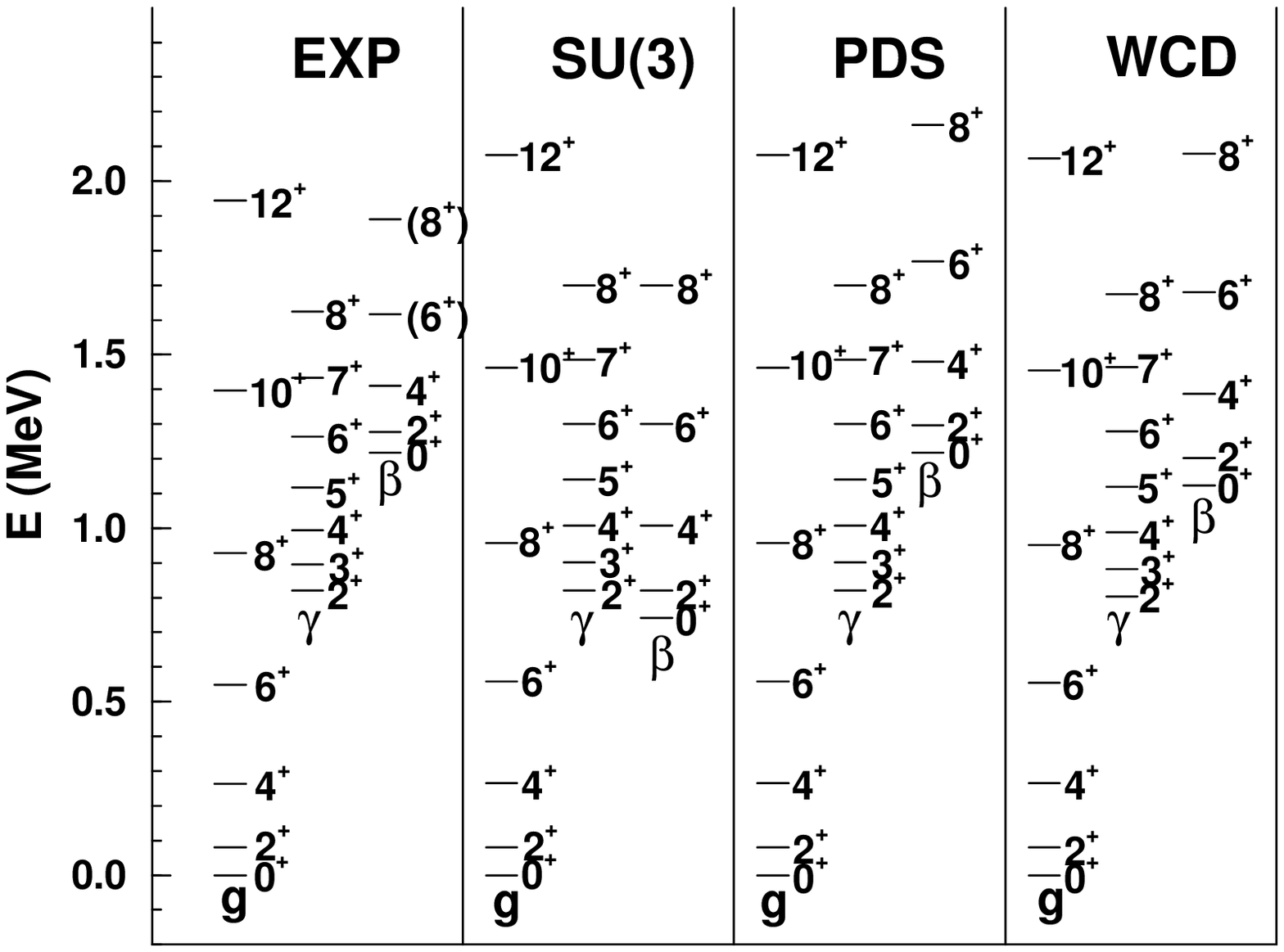}
\caption{
\small
Spectra of $^{168}$Er ($N=16$). Experimental energies
(EXP) are compared with IBM calculations in an exact SU(3) dynamical 
symmetry [SU(3)], in a broken SU(3) symmetry (WCD)~{\protect\cite{WCD}} and 
in a partial dynamical SU(3) symmetry (PDS). 
The latter employs the Hamiltonian of Eq.~(\ref{hPDSsu3}) with 
$h_0=8,\,h_2=4,\,C=13$ keV. Adapted from~\cite{lev96}. 
\label{figEr168}}
\end{center}
\end{figure}

The Hamiltonian of Eq.~(\ref{hPDSsu3}), with SU(3)-PDS,  
was used in~\cite{lev96} to describe spectroscopic data of $^{168}$Er. 
The experimental spectra~\cite{WCD} of the 
ground $g(K=0^{+}_1)$, 
$\gamma(K=2^{+}_1)$ and $K=0^{+}_2$  bands in this nucleus 
is shown in Fig.~\ref{figEr168}, 
and compared with an exact DS, PDS 
and broken SU(3) calculations. 
\begin{figure}[t]
\begin{center}
\includegraphics[height=14cm]{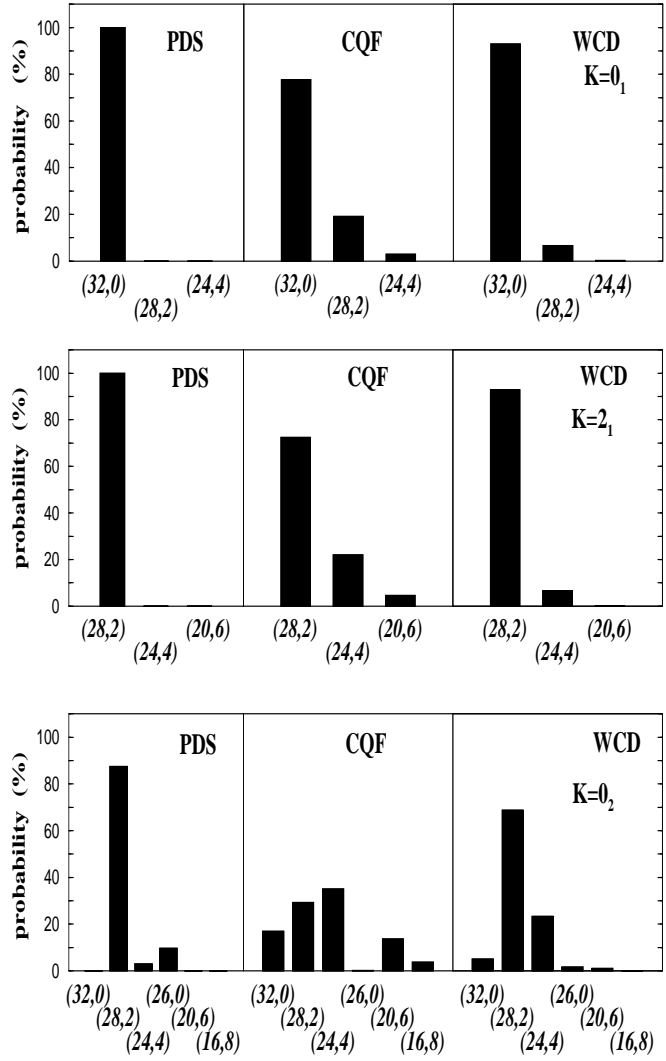}
\caption{
\small
SU(3) decomposition of wave functions of the ground ($K=0_1$), 
$\gamma$ ($K=2_1$),
and $K=0_2$ bands of $^{168}$Er ($N=16$) in a SU(3)-PDS 
calculation
and in broken-SU(3) calculations: 
CQF~{\protect\cite{CQF}} and WCD~{\protect\cite{WCD}}. 
Adapted from~{\protect\cite{LevSin99}}. 
\label{figsu3decomp}}
\end{center}
\end{figure}
According to the previous discussion, the SU(3)-PDS spectrum
of the solvable ground and $\gamma$ bands is given by 
\ba
E_{g}(L) &=& C L(L+1) ~,
\nonumber\\
E_{\gamma}(L) &=& 6h_{2}(2N-1) + C L(L+1) ~.
\ea 
$\hat{H}_{PDS}$~(\ref{hPDSsu3}) 
is specified by three parameters (N=16 for 
$^{168}$Er according to the usual boson counting). 
The values of $C$ and $h_{2}$ were extracted from the 
experimental energy differences
$[E(2^{+}_{g})-E(0^{+}_{g})]$ and $[E(2^{+}_{\gamma})-E(2^{+}_{g})]$ 
respectively. The spectrum of an exact SU(3) DS, Eq.~(\ref{eDSsu3}), 
obtained for $h_0=h_2$, deviates considerably from the experimental data 
since empirically the $K=0^{+}_2$ and $\gamma(K=2^{+}_1)$ 
bands are not degenerate. In the SU(3)-PDS calculation, 
the parameter $h_{0}$ was varied so as to reproduce the 
bandhead energy of the $K=0^{+}_2$ band. The prediction 
for other rotational members of the $K=0^{+}_1,\,0^{+}_2,\,2^{+}_1$ 
bands is shown in Fig.~\ref{figEr168}. 
Clearly, the SU(3) PDS spectrum is an
improvement over the schematic, exact SU(3) dynamical symmetry 
description, since the $\beta$-$\gamma$ degeneracy is lifted. 
The ground and $\gamma$ bands are solvable, still, 
the quality of the calculated PDS spectrum is similar to that obtained
in the broken-SU(3) calculation (WCD)~\cite{WCD}. 

The resulting SU(3) decomposition of the lowest bands 
for the SU(3)-PDS calculation 
is shown in Fig.~\ref{figsu3decomp}, and compared~\cite{LevSin99} 
to the conventional broken-SU(3) calculations 
WCD~\cite{WCD} and CQF~\cite{CQF}. 
In the latter calculations all states are mixed with 
respect to SU(3). In contrast, in the PDS calculation, 
the good SU(3) character, $(32,0)$ for
the ground band and $(28,2)$ for $\gamma$ band, is retained, 
while the $K=0^{+}_2$ band contains about $13\%$ 
admixtures into the dominant $(28,2)$ irrep. 
Thus, the breaking of the SU(3) symmetry induced by 
$\hat{H}_{PDS}$~(\ref{hPDSsu3}) is not small, and can lead to an 
appreciable SU(3) mixing in the remaining non-solvable states. 
Nevertheless, the special 
solvable states carry good SU(3) labels. The SU(3) symmetry is 
therefore partial but exact.
\begin{table}[t]
\begin{center}
\caption{\label{Tab3Er168}
\protect\small
B(E2) branching ratios from states in the $\gamma$ band in
$^{168}$Er. The column EXP lists the experimental ratios~\cite{WCD}, 
PDS is the SU(3) partial dynamical symmetry 
calculation~{\protect\cite{lev96}} and 
WCD is the broken SU(3) calculation~{\protect\cite{WCD}}. 
Adapted from~{\protect\cite{lev96}}.} 
\vspace{1mm}
\begin{tabular}{lcccc|rlcccc}
\hline
& & & & & & & & & &\\[-3mm]
$L^{\pi}_{i}$ & $L^{\pi}_{f}$ &  EXP &  PDS &  WCD &    &
$L^{\pi}_{i}$ & $L^{\pi}_{f}$ &  EXP &  PDS &  WCD \\
& & & & & & & & & &\\[-3mm]
\hline
& & & & & & & & & &\\[-2mm]
$2^{+}_{\gamma}$ & $0^{+}_{g}$      & $54.0$   &  $64.27$ &  $66.0$  &    &
$6^{+}_{\gamma}$ & $4^{+}_{g}$      &   $0.44$ &   $0.89$ &   $0.97$ \\[2pt]
                 & $2^{+}_{g}$      & $100.0$  & $100.0$  & $100.0$  &    &
                 & $6^{+}_{g}$      &   $3.8$  &   $4.38$ &   $4.3$  \\[2pt]
                 & $4^{+}_{g}$      &   $6.8$  &   $6.26$ &   $6.0$  &    &
                 & $8^{+}_{g}$      &   $1.4$  &   $0.79$ &   $0.73$ \\[2pt]
$3^{+}_{\gamma}$ & $2^{+}_{g}$      &   $2.6$  &   $2.70$ &   $2.7$  &    &
                 & $4^{+}_{\gamma}$ & $100.0$  & $100.0$  & $100.0$  \\[2pt]
                 & $4^{+}_{g}$      &   $1.7$  &   $1.33$ &   $1.3$  &    &
                 & $5^{+}_{\gamma}$ &  $69.0$  &  $58.61$ &  $59.0$  \\[2pt]
                 & $2^{+}_{\gamma}$ & $100.0$  & $100.0$  & $100.0$  &    &
$7^{+}_{\gamma}$ & $6^{+}_{g}$      &   $0.74$ &   $2.62$ &   $2.7$  \\[2pt]
$4^{+}_{\gamma}$ & $2^{+}_{g}$      &   $1.6$  &   $2.39$ &   $2.5$  &    &
                 & $5^{+}_{\gamma}$ & $100.0$  & $100.0$  & $100.0$  \\[2pt]
                 & $4^{+}_{g}$      &   $8.1$  &   $8.52$ &   $8.3$  &    &
                 & $6^{+}_{\gamma}$ &  $59.0$  &  $39.22$ &  $39.0$  \\[2pt]
                 & $6^{+}_{g}$      &   $1.1$  &   $1.07$ &   $1.0$  &    &
$8^{+}_{\gamma}$ & $6^{+}_{g}$      &   $1.8$  &   $0.59$ &   $0.67$ \\[2pt]
                 & $2^{+}_{\gamma}$ & $100.0$  & $100.0$  & $100.0$  &    &
                 & $8^{+}_{g}$      &   $5.1$  &   $3.57$ &   $3.5$  \\[2pt]
$5^{+}_{\gamma}$ & $4^{+}_{g}$      &   $2.91$ &   $4.15$ &   $4.3$  &    &
                 & $6^{+}_{\gamma}$ & $100.0$  & $100.0$  & $100.0$  \\[2pt]
                 & $6^{+}_{g}$      &   $3.6$  &   $3.31$ &   $3.1$  &    &
                 & $7^{+}_{\gamma}$ & $135.0$  &  $28.64$ &  $29.0$  \\[2pt]
                 & $3^{+}_{\gamma}$ & $100.0$  & $100.0$  & $100.0$  &    &
                 &                  &          &          &          \\[2pt]
                 & $4^{+}_{\gamma}$ & $122.0$  &  $98.22$ &  $98.5$  &    &
                 &                  &          &          &          \\[4pt]
\hline
\end{tabular}
\end{center}
\label{Tabbe2su3}
\end{table}

Electromagnetic transitions are a more sensitive probe to the structure
of states, hence are an important indicator for testing the predictions 
of SU(3)-PDS. Based on the analytic expressions of Eq.~(\ref{E2tran}), 
the parameters $\alpha$ and $\theta$ of the E2 operator 
can be extracted from the experimental values of
$B(E2;0^{+}_{g}\rightarrow 2^{+}_{g})$ and 
$B(E2;0^{+}_{g}\rightarrow 2^{+}_{\gamma})$.
The corresponding ratio for $^{168}$Er is $\theta/\alpha=4.261$.
As shown in Table~\ref{Tab3Er168}, the resulting SU(3) PDS 
E2 rates for transitions originating within the $\gamma$ band 
are found to be in excellent agreement with 
experiment and are similar to the WCD broken-SU(3) calculation~\cite{WCD}. 
In particular, the SU(3) PDS calculation reproduces correctly the
$(\gamma\rightarrow\gamma)/(\gamma\rightarrow g)$ strengths. 
The only significant discrepancy is that for the $8^{+}_{\gamma}\rightarrow
7^{+}_{\gamma}$ transition which is very weak experimentally, with an
intensity error of $50\%$ and an unknown M1 component~\cite{WCD}. 
The expressions in Eq.~(\ref{E2tran}) imply that 
$\gamma\rightarrow g$ $B(E2)$ ratios are independent of both E2 parameters 
$\alpha$ and $\theta$. Furthermore, since the
ground and $\gamma$ bands have pure SU(3) character, $(2N,0)$ and
$(2N-4,2)$ respectively, the corresponding wave-functions do not depend
on parameters of the Hamiltonian and hence are determined solely by
symmetry. Consequently, the 
B(E2) ratios for $\gamma\rightarrow g$ transitions quoted in 
Table~\ref{Tab3Er168} are 
parameter-free predictions of SU(3) PDS. 
The agreement between these predictions and the data
confirms the relevance of SU(3)-PDS to the spectroscopy of $^{168}$Er.

PDS has important implications not only for the structure 
of the pure (solvable) states, but it also affects 
the mixing of the remaining (non-solvable) states. 
Of particular interest is the nature of the lowest $K=0^{+}_2$ excitation, 
for which the role of double-$\gamma$-phonon admixtures is still subject 
to controversy~\cite{garrett01}.
A closer look at the SU(3) decomposition 
in Fig.~\ref{figsu3decomp}, shows that in the SU(3)-PDS calculation, 
the $K=0_2$ band is mixed and has the structure~\cite{LevSin99}
\ba
\vert L,K=0_2\rangle &=& A_1\,\tilde\phi_{E}((2N-4,2)\tilde K=0,L)
+ A_2\,\tilde\phi_{E}((2N-8,4)\tilde K=0,L)\nonumber\\
&& + A_3\,\phi_{E}((2N-6,0)K=0,L) ~.
\label{pdsk0wf}
\ea
Here $\tilde\phi_{E}$ denote states orthogonal to the solvable 
$\gamma^k_{K=2k}$ Elliott's states. The notation $\tilde{K}=0$ signifies 
that $K=0$ is only the dominant component in these states.  
For example, 
\ba
&&\tilde\phi_{E}((2N-4,2)\tilde K=0,L) =
\nonumber\\
&&
\qquad\qquad
\Bigl[\phi_{V}((2N-4,2)\tilde{\chi}=0,L)
+x^{(L)}_{20}\phi_{V}((2N-4,2)\tilde{\chi}=2,L)\Bigr ]/x^{(L)}_{22} ~,
\qquad\quad
\label{phitil}
\ea
is the state orthogonal to the Elliott state in Eq.~(\ref{phigam}). 
For $^{168}$Er, the $K=0_2$ 
band was found~\cite{LevSin99} to contain $9.6\%$ $(26,0)$ and 
$2.9\%$ $(24,4)$ admixtures into the dominant $(28,2)$ irrep. 
Using the geometric analogs of the SU(3) bands~\cite{WC82}, 
$(2N-4,2)K=0 \sim \beta$, 
$(2N-8,4)K=0 \sim (\sqrt{2}\beta^2 + \gamma{^2}_{K=0})$, 
$(2N-6,0)K=0 \sim (\beta^2 - \sqrt{2}\gamma^{2}_{K=0})$, the wave function
of Eq.~(\ref{pdsk0wf}) can be expressed in terms of the probability amplitudes
for single- and double- phonon $K=0$ excitations
\ba
A_{\beta} &=& A_1 ~,\quad
A_{\gamma^2} = (A_2 -\sqrt{2}A_3)/\sqrt{3} ~,\quad
A_{\beta^2} = (\sqrt{2}A_2 + A_3)/\sqrt{3} ~.
\label{Ai}
\ea
It follows
that in the PDS calculation, the $K=0_2$ band in $^{168}$Er contains 
admixtures of $12.4\%$ $\gamma^{2}_{K=0}$ and $0.1\%$ $\beta^2$ into the 
$\beta$ mode, {\it i.e.} $12.5\%$ double-phonon admixtures into the dominant
single-phonon component. These findings support the conventional single-phonon 
interpretation for this band with small but significant 
double-$\gamma$-phonon admixture. 

General properties of the $K=0_2$ band have been studied~\cite{LevSin99} 
by examining the SU(3)-PDS Hamiltonian 
of Eq.~(\ref{HPSsu3}). 
The empirical value of the ratio of $K=0_2$ and $\gamma$ 
bandhead energies  
$E(0^{+}_{2})/[E(2^{+}_{\gamma})-E(2^{+}_{g})] = 0.8-1.8$, 
in the rare-earth region~\cite{casten94,casten94b} constrains the 
parameters of $\hat{H}(h_0,h_2)$ to be in the range 
$0.7 \leq h_0/h_2 \leq 2.4$. 
In general, one finds that the $K=0_2$ wave function retains the form as in 
Eq.~(\ref{pdsk0wf}) and, therefore, a three-band mixing calculation is 
sufficient to describe its structure.
The SU(3) breaking and double-phonon admixture is more pronounced 
when the $K=0_2$ band is above the $\gamma$ band. 
For most of the relevant range of $h_0/h_2$, 
corresponding to bandhead ratio in the range $0.8-1.65$, 
the double-phonon admixture is at most $\sim 15\%$. Only for higher 
values of the bandhead ratio can one obtain larger admixtures and even 
dominance of the $\gamma^{2}_{K=0}$ component in the $K=0_2$ wave function.
\begin{table}
\begin{center}
\caption{\label{Tabbe2K0}
\protect\small
Comparison of 
calculated (Calc) and experimental (Exp)~\cite{lehmann98} 
absolute B(E2) values [W.u.] for transitions from the $K=0_2$ band in 
$^{168}$Er. PDS is the SU(3) partial dynamical symmetry 
calculation~\cite{LevSin99}, while WCD~\cite{WCD} and CQF~\cite{CQF} 
are broken SU(3) calculations. Adapted from~\cite{LevSin99}.}
\vspace{1mm}
\begin{tabular}{lcc|ccc}
\hline
& & & & &\\[-3mm]
\multicolumn{3}{c|}{Exp} &
\multicolumn{3}{c}{Calc}\\
Transition & B(E2) & Range & PDS & 
WCD & CQF \\
& & & & &\\[-3mm]
\hline
& & & & &\\[-2mm]
$2^{+}_{K=0_2}\rightarrow 0^{+}_g$ & 0.4 & 0.06--0.94 &
0.65  & 0.15  & 0.03  \\[2pt]
$2^{+}_{K=0_2}\rightarrow 2^{+}_g$ & 0.5 & 0.07--1.27 &
1.02  & 0.24  & 0.03  \\[2pt]
$2^{+}_{K=0_2}\rightarrow 4^{+}_g$ & 2.2 & 0.4--5.1 &
2.27 & 0.50 & 0.10 \\[2pt]
$2^{+}_{K=0_2}\rightarrow 2^{+}_{\gamma}$ $^{a)}$ & 6.2 (3.1) 
& 1--15 (0.5--7.5) &
4.08 & 4.2 & 4.53 \\[2pt]
$2^{+}_{K=0_2}\rightarrow 3^{+}_{\gamma}$ $^{a)}$ & 7.2 (3.6) 
& 1--19 (0.5--9.5) &
7.52 & 7.9 & 12.64 \\[4pt]
\hline
\multicolumn{6}{l}{}\\[-3mm]
\multicolumn{6}{l}{\small $^{a)}$ The two numbers in each entry
correspond to an assumption of pure E2}\\
\multicolumn{6}{l}{\small 
and (in parenthesis) 
50\% E2 multipolarity.}\\
\end{tabular}
\end{center}
\end{table}

An important clue to the structure of $K=0_2$ collective excitations
comes from 
E2 transitions connecting the $K=0_2$ and $K=0_1$ bands. 
If we recall that only the ground
band has the SU(3) component $(\lambda,\mu)=(2N,0)$, 
that $Q^{(2)}$, as a generator, cannot connect different SU(3) irreps 
and that $\Pi^{(2)}$, 
as a $(2,2)$ tensor under SU(3), can connect the $(2N,0)$ irrep only with 
the $(2N-4,2)$ irrep, we obtain the following expression for the 
B(E2) values of $K=0_2\rightarrow g$ transitions
\ba
B(E2;K=0_2,L\rightarrow g,L') =
\qquad\qquad\qquad\qquad\qquad
\qquad\qquad\qquad\qquad\qquad\qquad
\nonumber\\
A_{\beta}^2\,\theta^2\,
\frac{\vert\langle\phi_{E}((2N,0)K=0,L')||\Pi^{(2)}||
\tilde\phi_{E}((2N-4,2)\tilde K=0,L)\rangle\vert^{2}}{(2L+1)}
~. \qquad
\label{be2K0}
\ea
Employing Eq.~(\ref{phitil}), 
this B(E2) value can be expressed in terms of known B(E2) 
values in the Vergados basis~\cite{arimaiac78,ISA}.
Using Eq.~(\ref{gamtog}), the E2 parameter $\theta$ 
can be determined from the 
known $2^{+}_{\gamma}\rightarrow 0^{+}_{g}$ E2 rates, and for
$^{168}$Er is found to be $\theta^2=2.175$ W.u. 
As seen from Eq.~(\ref{be2K0}), 
the B(E2) values for $K=0_2\rightarrow g$ transitions
are proportional to $(A_{\beta})^2$, hence, 
they provide a direct way for extracting the amount of SU(3) breaking and 
the admixture of double-phonon excitations in the $K=0_2$ wave function.
In Table~\ref{Tabbe2K0} we compare the predictions of the PDS and broken-SU(3) 
calculations with the B(E2) values deduced from
a lifetime measurement of the $2^{+}_{K=0_2}$ level in $^{168}$Er 
\cite{lehmann98} (the indicated range for the B(E2) values 
correspond to different assumptions on the feeding of the level). 
The PDS and WCD calculations are seen to agree well with the 
empirical values but the CQF calculation under predicts the measured 
$K=0_2\rightarrow g$ data. 

The SU(3)-PDS discussed above, is relevant to 
rotational-vibrational states of a prolate deformed shape, 
with equilibrium deformations ($\beta=\sqrt{2},\gamma=0)$ and a 
symmetry $z$-axis. It is also possible to identify an  
$\overline{{\rm SU(3)}}$-PDS corresponding to an oblate shape with 
equilibrium deformations ($\beta=\sqrt{2},\gamma=\pi/3)$ and a symmetry 
$y$-axis. The Hamiltonian with $\overline{{\rm SU(3)}}$-PDS has the same 
form as in Eq.~(\ref{hPDSsu3}) but the $L=0,2$ boson-pairs are obtained from 
Eq.~(\ref{PL}) by a change of phase: $s^{\dag}\to -s^{\dag}$, $s\to-s$. 
The generators and quadratic Casimir operator of $\overline{{\rm SU(3)}}$ 
are listed in the Appendix. The relevant $\overline{{\rm SU(3)}}$ irreps, 
$(\bar{\lambda},\bar{\mu})=(2k,2N-4k-6m)$, 
are conjugate to the SU(3) irreps, $(\lambda,\mu)$, encountered in the 
SU(3) chain, Eq.~(\ref{chainsu3}).

\subsection{O(6) PDS (type I)}
\label{subsec:o6PDStypeI}

The O(6) DS chain of the IBM and related quantum numbers are given 
by~\cite{arimaiac79}
\ba
\begin{array}{ccccccc}
{\rm U}(6)&\supset&{\rm O}(6)&\supset&{\rm O}(5)&
\supset&{\rm O}(3)\\
\downarrow&&\downarrow&&\downarrow&&\downarrow\\[0mm]
[N]&&\langle\sigma\rangle&&(\tau)&n_\Delta& L
\end{array} ~,
\label{chaino6}
\ea
where the generators of the above groups are listed in Table~\ref{TabIBMcas} 
of the Appendix. 
For a given U(6) irrep $[N]$, the allowed O(6) and O(5) irreps are 
$\sigma=N,\,N-2,\dots 0$ or $1$, and  
$\tau=0,\,1,\,\ldots \sigma$, respectively. 
The ${\rm O(5)}\supset {\rm O(3)}$ reduction is the same as in the 
U(5) chain. 
The eigenstates $|[N]\langle\sigma\rangle(\tau)n_\Delta LM\rangle$
are obtained with a Hamiltonian
with O(6) DS which, for one- and two-body interactions, can be
transcribed in the form
\ba
\hat{H}_{\rm DS} &=& h_{0}\left [-\hat C_{{\rm O(6)}} 
+ \hat N (\hat{N} +4)\right ]
+ B\, \hat{C}_{{\rm O(5)}} + C\,\hat{C}_{{\rm O(3)}} ~.
\label{hDSo6}
\ea
The quadratic Casimir operators, $\hat{C}_{G}$, are defined 
in the Appendix. 
The spectrum of $\hat{H}_{\rm DS}$ is completely solvable with eigenenergies
\ba
E_{\rm DS} &=& 
h_{0}\,(N-\sigma)(N+\sigma + 4) + B\,\tau(\tau+3) +\, C\,L(L+1) ~.
\label{eDSo6}
\ea
The spectrum resembles that of a $\gamma$-unstable deformed rotovibrator, 
where states are arranged in O(6) multiplets with quantum number $\sigma$.
The ground band corresponds to the O(6) irrep with $\sigma=N$. 
The splitting of states in a given O(6) multiplet is governed  
by the O(5) and O(3) terms in 
$\hat{H}_{\rm DS}$~(\ref{hDSo6}). 
The lowest members in each band have quantum numbers 
$(\tau=0,\, L=0)$, $(\tau=1,\, L=2)$ and $(\tau=2,\, L=2,4)$. 
\begin{table}
\begin{center}
\caption{\label{Tabo6tensII}
\protect\small
Normalized one- and two-boson O(6) tensors.}
\vspace{1mm}
\begin{tabular}{cccccl}
\hline
& & & & &\\[-3mm]
$n$&$\sigma$&$\tau$&$n_{\Delta}$&$\ell$&
$\hat B^\dag_{[n]\langle\sigma\rangle(\tau)n_{\Delta};\ell m}$\\
& & & & &\\[-3mm]
\hline
& & & & &\\[-2mm]
1& 1& 0& 0 & 0& $s^{\dag}$\\[2pt]
1& 1& 1& 0 & 2& $d^{\dag}_{m}$\\[2pt]
2& 2& 0& 0 & 0& $\sqrt{\frac{1}{12}}(d^\dag d^\dag)^{(0)}_0
                +\sqrt{\frac{5}{12}}(s^\dag )^2$\\[2pt]
2& 2& 1& 0 & 2& $s^\dag d^{\dag}_m$\\[2pt]
2& 2& 2& 0 & 2& $\sqrt{\frac{1}{2}}(d^\dag d^\dag)^{(2)}_m$\\[2pt]
2& 2& 2& 0 & 4& $\sqrt{\frac{1}{2}}(d^\dag d^\dag)^{(4)}_m$\\[2pt]
2& 0& 0& 0 & 0& $\sqrt{\frac{5}{12}}(d^\dag d^\dag)^{(0)}_0
-\sqrt{\frac{1}{12}}(s^\dag)^2$\\
& & & & &\\[-3mm]
\hline
\end{tabular}
\end{center}
\end{table}

The construction of Hamiltonians with O(6)-PDS is based on identification 
of $n$-boson operators which annihilate all states in a given 
O(6) irrep, $\langle\sigma\rangle$, chosen here to be the ground band 
irrep $\langle\sigma\rangle=\langle N\rangle$. 
For that purpose, a relevant operator to consider is
\ba
P^{\dagger}_{0} &=& d^{\dagger}\cdot d^{\dagger} - (s^{\dagger})^2 ~.
\label{Pdag0o6}
\ea
As seen from Table~\ref{Tabo6tensII}, the above operator 
is proportional to a two-boson O(6) tensor, 
$\hat B^\dag_{[n]\langle\sigma\rangle(\tau)n_{\Delta}\ell m}$,
with $n=2$ and $\sigma=\tau=L=0$
\ba
\hat B^\dag_{[2]\langle0\rangle(0)0;00} &=& 
\frac{1}{2\sqrt{3}}\,P^{\dagger}_{0} ~.
\label{Pdag0}
\ea
The corresponding Hermitian conjugate boson-pair annihilation operator, 
$P_0$, transforms also as $\langle\sigma\rangle=\langle 0\rangle$ 
under O(6) and satisfies 
\ba
P_{0}\,\vert [N]\langle N \rangle (\tau)n_{\Delta}LM\rangle = 0 ~.
\label{P0o6}
\ea
Equivalently, this operator satisfies
\ba
P_{0}\vert c;\, N\rangle &=& 0
\label{P0cond}
\ea
where 
\ba 
\vert c; \, N\rangle &=& 
(N!)^{-1/2} (b^{\dag}_{c})^{N}\vert 0\rangle
\nonumber\\
b^{\dag}_{c} &=& 
[\,\cos\gamma\,d^{\dag}_{0} + \sin\gamma\, 
(d^{\dag}_{2}+d^{\dag}_{-2})/\sqrt{2} + s^{\dag}\,]/\sqrt{2} ~.
\label{condso6}
\ea
The state $\vert c; \, N\rangle$ 
is obtained by substituting the O(6) equilibrium deformations in the 
coherent state of Eq.~(\ref{condgen}), 
$\vert c; \, N\rangle = \vert\beta=1,\gamma ; N \rangle$.  
It is the lowest-weight state in the O(6) irrep 
$\langle\sigma\rangle = \langle N\rangle$ 
and serves as an intrinsic state 
for the O(6) ground band~\cite{lev84}. 
The rotational members of the band, 
$\vert [N]\langle N\rangle (\tau)n_{\Delta}LM\rangle$, Eq.~(\ref{P0o6}), 
are obtained by O(5) projection from $\vert c;\, N\rangle$ 
and span the entire O(6) irrep $\langle\sigma\rangle=\langle N\rangle$.
The relations in Eqs.~(\ref{P0o6})-(\ref{P0cond}) follow from the 
fact that the action of the operator $P_{0}$ leads to a state with 
$N-2$ bosons in the U(6) irrep $[N-2]$, 
which does not contain the O(6) irrep $\langle N\rangle$  obtained from the 
product of $\langle 0\rangle\times \langle N\rangle$.

Since both $P^{\dag}_{0}$ and $P_0$~(\ref{Pdag0o6}) 
are O(6) scalars, they give rise 
to the following O(6)-invariant interaction
\ba
P^{\dag}_{0}P_{0} &=& 
\left [-\hat C_{{\rm O(6)}} 
+ \hat N (\hat{N} +4)\right ] ~, 
\label{HPSo6}
\ea
which is simply the O(6) term in 
$\hat{H}_{\rm DS}$, Eq.~(\ref{hDSo6}), 
with an exact O(6) symmetry. Thus, in this case, 
unlike the situation encountered with SU(3)-PDS, 
the algorithm does not yield an O(6)-PDS of type I with two-body 
interactions. In the IBM framework, an Hamiltonian with a genuine 
O(6)-PDS of this class, requires higher-order terms. A construction of 
three-body Hamiltonians 
with such property will be presented in Subsection~\ref{subsec:o6PDS3bod}.

\section{PDS type II}
\label{sec:PDStypeII}

PDS of type II corresponds to a situation for which {\it all} the
states of the system preserve {\it part} of the dynamical symmetry, 
\ba
G_0 \supset G_1 \supset G_2 \supset \ldots \supset  G_{n} ~.
\label{DSchain}
\ea
In this case, there are no analytic solutions,
yet selected quantum numbers (of the conserved symmetries) are retained.
This occurs, for example, 
when the Hamiltonian contains interaction
terms from two different chains with
a common symmetry subalgebra, {\it e.g.}, 
\ba
G_0 \supset 
\left \{
\begin{array}{c}
G_1 \\
G_{1}'
\end{array}
\right \}\supset 
G_2 \supset \ldots \supset G_n
\label{G0chains}
\ea
If $G_{1}$ and $G_{1}'$ are incompatible, {\it i.e.}, do not commute, 
then their irreps are mixed in the eigenstates of the Hamiltonian. 
On the other hand, since $G_2$ and its subalgebras are common to both 
chains, then the labels of their irreps remain as good quantum numbers.  
An Hamiltonian based on a spectrum generating algebra $G_0$ and 
a symmetry algebra $G_n$ common to all chains of $G_0$, has by definition, 
a $G_n$-PDS of type II, albeit a trivial one [{\it e.g.}, 
$G_n={\rm O(3)}$]. Therefore, 
the notion of PDS type II is physically relevant 
when the common segment of the two DS chains, contains 
subalgebras which are different from the symmetry algebra, {\it i.e.}, 
$G_{2}\neq G_n$ in Eq.~(\ref{G0chains}).

An alternative situation where PDS of type II occurs is when the 
Hamiltonian preserves only some of the symmetries $G_i$ in the 
chain~(\ref{DSchain}) and only their irreps are unmixed. 
A~systematic procedure for identifying interactions with such property 
was proposed in~\cite{isa99}. 
Let $G_1\supset G_2\supset G_3$ be a set of nested algebras which 
may occur anywhere in the reduction~(\ref{DSchain}), 
in-between the spectrum generating algebra $G_0$ and the invariant symmetry 
algebra $G_n$. The procedure is based on 
writing the Hamiltonian in terms of generators, $g_i$, of $G_1$, 
which do not belong to its subalgebra $G_2$. 
By construction, such Hamiltonian preserves the 
$G_1$ symmetry but, in general, not the $G_2$ symmetry, and hence will have 
the $G_1$ labels as good quantum numbers but will mix different 
irreps of $G_2$. The Hamiltonians can still conserve the $G_3$ labels 
{\it e.g.}, by choosing it to be a scalar of $G_3$. 
The procedure for constructing Hamiltonians with the above properties 
involves the identification of the tensor character under $G_2$ and $G_3$ 
of the operators $g_i$ and their products, $g_i g_{j}\ldots g_{k}$. 
The Hamiltonians obtained in this manner
belong to the integrity basis of $G_3$-scalar operators in 
the enveloping algebra of $G_1$ and, hence, their 
existence is correlated with their order. 
In the discussion below we exemplify the two scenarios for constructing 
Hamiltonians with PDS of type II within the IBM framework.

\subsection{O(5) PDS (type II)}
\label{subsec:o5PDStypeII}

An example of mixing two incompatible chains with a common symmetry 
subalgebra is the U(5) and O(6) chains in the IBM~\cite{levnov86} 
\ba
{\rm U(6)} \supset 
\left \{
\begin{array}{c}
{\rm U(5)} \\
{\rm O(6)}
\end{array}
\right \}\supset 
{\rm O(5)} \supset {\rm O(3)} ~.
\label{u5o6}
\ea
The corresponding quantum numbers were discussed in 
Subsection~\ref{subsec:u5PDStypeI} and~\ref{subsec:o6PDStypeI}. 
The most general Hamiltonian which conserves the O(5) symmetry, involves 
a combination of terms from these two chains, and for one- and two-body 
interactions reads
\ba
\hat{H}_{O(5)} &=& \epsilon\,\hat{n}_d + \alpha\,\hat{n}_d(\hat{n}_d+4)
+ A\,P^{\dag}_{0}P_0 + B\, \hat{C}_{{\rm O(5)}} + 
C\,\hat{C}_{{\rm O(3)}} ~.
\label{hPDSo5}
\ea
The d-boson number operator, $\hat{n}_d$, is the linear Casimir of 
U(5), the O(6)-pairing term, $P^{\dag}_{0}P_0$, is related to the 
Casimir operator of O(6), Eq.~(\ref{HPSo6}), 
and the Casimir operators of O(5) and O(3) can be replaced by their 
eigenvalues $B\,\tau(\tau+1) +C\,L(L+1)$. 
In this case, 
all eigenstates of $\hat{H}_{{\rm O(5)}}$ have good O(5) symmetry 
but, with a few exceptions, none of them have good U(5) symmetry nor 
good O(6) symmetry, and hence only part of the dynamical symmetry of 
each chain in Eq.~(\ref{u5o6}) is observed. 
These are precisely the defining features of 
O(5)-PDS of type II. 
In general, $\hat{H}_{{\rm O(5)}}$~(\ref{hPDSo5}) mixes U(5) irreps, 
characterized by $n_d$, as well as mixes 
irreps of O(6) characterized by $\sigma$, but retains the 
${\rm O(5)}\supset {\rm O(3)}$ 
labels, $(\tau,L)$ as good quantum numbers. 
The conserved O(5) symmetry has important 
consequences which will be discussed briefly below~\cite{levnov86}.

The first four terms in~(\ref{hPDSo5}) 
are O(5) scalars, hence level-spacings 
within an O(5) irrep ($\tau$-multiplet) are the same throughout the 
U(5)-O(6) transition region. They are given only by $CL(L+1)$ and are 
independent of the values of the other parameters in 
$\hat{H}_{{\rm O(5)}}$. 
The various multiplets with the same value of $\tau$ are distinguished by 
another label, $\nu=1,\ldots$ which indicates its relative position in the 
spectrum. The actual positions of the various $\tau$-multiplets, as well as 
the wave functions of their states, are determined by diagonalization of 
the Hamiltonian~(\ref{hPDSo5}). The wave function of any state 
in the $\nu$th $\tau$-multiplet can be expressed in the U(5) basis as
\ba
\vert \nu; N,\tau, n_\Delta, L,M\rangle 
&=&
\sum_{n_d}\, \xi_{n_d}^{(\nu,\tau)}\,
\vert [N]\langle n_d\rangle (\tau)n_\Delta L M\rangle
\label{wfu5basis}
\ea
with $n_d=\tau,\tau+2,\ldots, N-1$ or $N$. Similarly, the same wave functions 
can be expressed in the O(6) basis as
\ba
\vert \nu; N,\tau, n_\Delta, L,M\rangle &=&
\sum_{\sigma}\, \eta_{\sigma}^{(\nu,\tau)}\,
\vert [N]\langle \sigma\rangle (\tau)n_\Delta L M\rangle
\label{wfo6basis}
\ea
with $\sigma=N, N-2,\ldots, (\tau+1)$ or $\tau$.
The amplitudes in Eqs.~(\ref{wfu5basis})-(\ref{wfo6basis}) depend, 
in general, on the actual values of $\epsilon,\,\alpha,\,A$ 
as well as on $N$. 

The O(5) symmetry of~(\ref{hPDSo5}) can be further used to derive special 
properties of electromagnetic transitions. 
The  $\Pi^{(2)}$ part of the E2 operator, Eq.~(\ref{Te2}), 
is a $\tau=1$ 
tensor under O(5) and connects states with $\Delta\tau=\pm 1$. 
The $U^{(2)}$ part is a $\tau=2$ 
tensor under O(5) and connects states with $\Delta\tau=0, \pm2$. 
The B(E2) values of transitions from a state with $\tau',n_{\Delta}',L'$ 
in the $\nu'$ multiplet to a state with $\tau,n_{\Delta},L$ in the $\nu$ 
multiplet can be written in the form
\bsub
\ba
&&B(E2; \nu',\tau',n_{\Delta}',L'\to \nu,\tau,n_{\Delta},L)
\nonumber\\
&&\qquad\qquad 
= e_{B}^2\,{\cal F}_{N}(\nu,\nu',\tau)
\left \langle
\begin{array}{cc|c}
\tau' & 1 & \tau\\
n_{\Delta}'L'    & 2 & n_{\Delta}L
\end{array}
\right \rangle^2
\quad\quad\;\;\; \tau'=\tau\pm 1 ~,\qquad\quad
\label{be2o5a}\\
&&\qquad\qquad 
= (e_{B}\chi)^2\,{\cal G}_{N}(\nu,\nu',\tau)
\left \langle
\begin{array}{cc|c}
\tau' & 2 & \tau
\\
n_{\Delta}'L'    & 2 & n_{\Delta}L
\end{array}
\right \rangle^2
\quad \tau'=\tau\, , \,\tau\pm 2~.\qquad\quad
\label{be2o5b}
\ea
\label{be2o5}
\esub
Each B(E2) is a product of two factors. The first factor which 
is determined by the Hamiltonian depends on $N$ and on $\nu,\tau$ of the 
initial and final states. The second factor is the 
${\rm O(5)}\supset {\rm O(3)}$ isoscalar factor (ISF). 
It is the same for every Hamiltonian~(\ref{hPDSo5}), 
is completely determined by O(5) symmetry and depends only on the 
$\tau,\, n_{\Delta},\, L$ of the intial and final states. 
Analytic expressions for some of these ISF are 
available~\cite{ibm,ibfm,piet87}, 
as well as a computer code for their evaluation~\cite{caprio09}.

The factorization observed in Eq.~(\ref{be2o5}) is a manifestation of the 
Wigner-Eckart theorem for the O(5) group and has important implications 
with respect to E2 rates. First, consider transitions between states in 
any $\nu,\tau$-multiplet to states in any $\nu'\tau'$-multiplet 
(including $\nu=\nu'$). A direct consequence of~(\ref{be2o5}) is  
that B(E2) ratios of such transitions are equal to the ratios of ISFs, 
hence, are independent of $\nu,\,\nu'$ throughout the O(6)-U(5) region. 
Such ratios are also independent of $N$ and the actual parameters 
in~(\ref{hPDSo5}), 
{\it i.e.}, they should be the same in all O(5) nuclei considered. 
Second, consider E2 transitions between a state with given 
$\tau,\,n_{\Delta},\,L$ and a  state with given $\tau',\,n_{\Delta}',\,L'$ 
either with $\nu=\nu'$ or $\nu\neq\nu'$, and even in different nuclei. 
In B(E2) ratios of such transitions the ISFs in Eq.~(\ref{be2o5}) cancel, 
and the ratio of the other factors [{\it e.g.}, 
$e_{B}^2{\cal F}_{N}(\nu,\nu',\tau)/e_{B}'^{2}
{\cal F}_{N'}(\nu'',\nu''',\tau'')$] 
is independent of 
$n_{\Delta},L,n_{\Delta}',L'$. Thus, such ratios should be the same for 
all transitions between the states of the $\tau,\,\tau'$-multiplets 
considered.

The very definite statements made above about E2 transitions follow directly 
from O(5) symmetry. 
They hold exactly for any values of $N$ and of 
parameters in the Hamiltonian~(\ref{hPDSo5}) and, in particular, at the O(6) 
and U(5) limits. They played an instrumental role in identifying empirical 
signatures of O(5) symmetry which are common to all IBM Hamiltonians 
(\ref{hPDSo5}) in the O(6)-U(5) region and which should be clearly 
distinguished from features, {\it e.g.}, absolute B(E2) values, 
that can yield crucial evidence for O(6) or U(5) symmetries or 
for deviations from these limits~\cite{casten87,brentano88,rain10}. 

Hamiltonians with O(5)-PDS of type II have been used 
extensively for studying transitional nuclei in the Ru-Pd~\cite{Stachel82} 
and Xe-Ba~\cite{Casten85} regions, whose structure 
varies from spherical [U(5)] to $\gamma$-unstable deformed [O(6)]. 
Such O(5)-PDS is also relevant to the coexistence of
normal and intruder levels in $^{112}$Cd~\cite{Jolie95}. 
The particular Hamiltonian $\hat{H}_{{\rm O(5)}}$ of Eq.~(\ref{hPDSo5}), 
with O(5)-PDS of type II, has also selected U(5) basis states as 
eigenstates and hence exhibits also U(5)-PDS of type I.  
This follows from the fact that $AP^{\dag}_{0}P_0$, 
which is the only term in $\hat{H}_{\rm{O(5)}}$ that breaks the U(5) DS, 
involves the operator $P_0 = \tilde{d}\cdot \tilde{d} -s^2$. 
The latter operator annihilates the U(5) basis states, 
$\vert [N],n_d=\tau=N,L\,\rangle$ and 
$\vert [N],n_d=\tau=N-1,L\,\rangle$, for reasons given after 
Eq.~(\ref{V0ndN}).

\subsection{O(6) PDS (type II)}
\label{subsec:o6PDStypeII}

An alternative situation where PDS of type II can occur is 
when the entire eigenspectrum retains 
only some of the symmetries $G_i$ in a given dynamical symmetry 
chain~(\ref{DSchain}).
Such a scenario was considered in~\cite{isa99} in relation to the 
O(6) chain 
\begin{equation}
\begin{array}{ccccccc}
{\rm U(6)} &\supset& {\rm O(6)} &\supset& {\rm O(5)} &
\supset& {\rm O(3)} \\
\downarrow&&\downarrow&&\downarrow&&\downarrow\\[0mm]
\left [N\right ] && \langle0,\sigma,0\rangle && (\tau,0) & n_{\Delta}& L
\end{array}.
\label{DSo6}
\end{equation} 
The following Hamiltonian, with O(6)-PDS of type II, has been proposed
\ba
\hat{H}_1 \;=\; \kappa_{0}P^{\dagger}_{0}P_{0}
+ \kappa_2 \Bigl (\Pi^{(2)}\times \Pi^{(2)}\Bigr )^{(2)}\cdot\Pi^{(2)} ~.
\label{h1}
\ea
The $\kappa_0$ term is the O(6) pairing term of Eq.~(\ref{HPSo6}). 
It is diagonal in the dynamical-symmetry basis 
$\vert [N]\langle\sigma\rangle (\tau)n_{\Delta}LM\rangle$ 
of Eq.~(\ref{DSo6}) with eigenvalues
$\kappa_0(N - \sigma)(N +\sigma +4)$.
The $\kappa_2$ term is constructed only from the O(6) generator,
$\Pi^{(2)}=d^{\dagger}s+s^{\dagger}\tilde{d}$, 
which is not a generator of O(5). 
Therefore, it cannot connect states in different O(6) irreps
but can induce O(5) mixing subject to $\Delta\tau=\pm 1,\pm 3$. 
Consequently, all eigenstates of $\hat{H}_1$ 
have good O(6) quantum number $\sigma$ 
but do not possess O(5) symmetry $\tau$.
These are the necessary ingredients of an O(6) PDS of type II 
associated with the chain of Eq.~(\ref{DSo6}). 

As shown in Fig.~\ref{figo6energy}, 
a typical spectra of $\hat{H}_1$ displays rotational bands of 
an axially-deformed nucleus. 
All bands of $\hat{H}_1$ are pure with respect to O(6). 
This is demonstrated in the left panel of Fig.~\ref{figo6decomp} for 
the $K=0_1,2_1,2_3$ bands 
which have $\sigma=N$, and for the $K=0_2$ band which has $\sigma=N-2$.
In this case, the diagonal $\kappa_0$-term in Eq.~(\ref{h1}) simply
shifts each band as a whole in accord with its $\sigma$ assignment.
On the other hand, the $\kappa_2$-term in Eq.~(\ref{h1}) 
is an O(5) tensor with $(\tau,0)=(3,0)$ and, therefore, 
all eigenstates of $\hat{H}_1$ are mixed with respect to O(5). 
This mixing is demonstrated in the right panel of 
Fig.~\ref{figo6decomp} for the $L=0,2$ members of the ground band.

A key element in the above procedure for 
constructing Hamiltonians with O(6)-PDS of type II, 
is the tensorial character of the generators contained in O(6) but not 
in O(5)~\cite{isa99}. 
In the present case, the tensor character of the operator 
$\Pi^{(2)}$ under O(5) is $(\tau,0)=(1,0)$ and under O(3), $L=2$. 
A quadratic interaction $\Pi^{(2)}\cdot \Pi^{(2)}$ corresponds to the 
O(5) multiplication $(1,0)\times (1,0) = (2,0)\oplus (1,1)\oplus (0,0)$. 
Since only the irrep $(0,0)$ contains an $L=0$, it follows that 
the quadratic terms must be an O(5) scalar. Indeed, 
from Table~\ref{TabIBMcas} of the Appendix, we find 
$\Pi^{(2)}\cdot \Pi^{(2)}= \hat{C}_{{\rm O(6)}} - \hat{C}_{{\rm O(5)}}$.
In the next cubic order, the interaction 
$(\Pi^{(2)}\times \Pi^{(2)})^{(2)}\cdot\Pi^{(2)}$ corresponds to 
$(1,0)\times (1,0)\times (1,0)$; O(5) multiplication show that 
there is only one O(3) scalar and it has O(5) character $(3,0)$. 
Consequently, $(\Pi^{(2)}\times \Pi^{(2)})^{(2)}\cdot\Pi^{(2)}$
is an example of a $\sigma$-conserving, $\tau$-violating interaction; 
it mixes $(\tau,0)$ with $(\tau\pm 1,0)$ and $(\tau\pm 3,0)$. 
This discussion highlights the fact that the existence of Hamiltonians 
with PDS of type~II, constructed in this manner, may necessitate 
higher-order terms. 

A similar procedure can been applied to the 
${\rm U(6)}\supset {\rm U(5)}\supset {\rm O(5)}\supset {\rm O(3)}$ 
chain of the IBM~\cite{isa99}. 
The generators contained in U(5) but not in O(5) are 
$U^{(L)}_{\mu}\equiv (d^{\dag}\tilde{d})^{(L)}_{\mu}$ with $L=0,2,4$. 
$U^{(0)}$ is a scalar in O(5) and hence does not mix O(5) irreps. 
The operators $U^{(2)}_{\mu}$ and $U^{(4)}_{\mu}$, on the other hand, 
have O(5) character $(2,0)$. O(3)-scalar interactions obtained from 
quadratic combinations of such tensors involve terms of the U(5) DS 
Hamiltonian, Eq.~(\ref{hDSu5}), hence do not induce O(5) mixing 
among symmetric, $(\tau,0)$ irreps. 
On the other hand, 
cubic O(3)-scalar combinations 
of $U^{(2)}_{\mu}$ and $U^{(4)}_{\mu}$ can lead to two independent 
$d$-boson interaction terms that can induce 
O(5) mixing but conserve the U(5) quantum 
number, $n_d$. By definition, such $n_d$-conserving but $\tau$-violating 
cubic terms exemplify a U(5) PDS of type II.

\section{PDS type III}
\label{sec:PDStypeIII}

PDS of type III
combines properties of both PDS of type I and II. Such a generalized 
PDS~\cite{levisa02} has a hybrid character, for which {\it part} of the 
states of the system under study preserve {\it part} 
of the dynamical symmetry. 
In relation to the dynamical symmetry chain of Eq.~(\ref{chain}), 
$G_{\rm dyn}\supset G\supset\cdots\supset G_{\rm sym}$, 
with associated basis, $\vert [h_N]\langle\Sigma\rangle\Lambda\rangle$, 
this can be accomplished by relaxing the condition of Eq.~(\ref{anni}), 
$\hat T_{[h_n]\langle\sigma\rangle\lambda}
|[h_N]\langle\Sigma_0\rangle\Lambda\rangle=0$, 
so that it holds only for {\it selected} 
states $\Lambda$ contained in a given irrep $\langle\Sigma_0\rangle$ 
of $G$ and/or selected (combinations of) components $\lambda$ of the tensor 
$\hat T_{[h_n]\langle\sigma\rangle\lambda}$. Under such circumstances, 
let $G'\neq G_{sym}$ be a subalgebra of $G$ in the aforementioned 
chain, $G\supset G'$. 
In general, the Hamiltonians, constructed from 
these tensors, in the manner shown in Eq.~(\ref{PS}), 
are not invariant under $G$ nor $G'$. 
Nevertheless, they do posses the subset of solvable states, 
$|[h_N]\langle\Sigma_0\rangle\Lambda\rangle$, 
with good $G$-symmetry, $\langle\Sigma_0\rangle$, 
while other states 
are mixed. At the same time, the symmetry associated with the subalgebra 
$G'$, is broken in all states (including the solvable ones).
Thus, part of the eigenstates preserve part of the symmetry. 
These are precisely the requirements of PDS of type III.
In what follows we explicitly construct Hamiltonians with such properties 
within the IBM framework.

\subsection{O(6) PDS (type III)}
\label{subsec:o6PDStypeIII}

PDS of type III associated with the O(6) chain, Eq.~(\ref{DSo6}), 
can be realized  in terms of Hamiltonians which have a subset of 
solvable states with good O(6) symmetry but broken O(5) symmetry. 
Hamiltonians with such property can be constructed~\cite{levisa02} 
by means of the following boson-pair operators with angular 
momentum $L =0,\,2$
\bsub
\ba
P^{\dagger}_{0} &=& d^{\dagger}\cdot d^{\dagger} - (s^{\dagger})^2 ~,\\
P^{\dagger}_{2\mu} &=& \sqrt{2}d^{\dagger}_{\mu}s^{\dagger} + 
\sqrt{7}\, (d^{\dagger}\,d^{\dagger})^{(2)}_{\mu} ~.
\ea
\label{PLo6}
\esub
\begin{figure}[t]
\begin{center}
\includegraphics[height=4in,angle=-90]{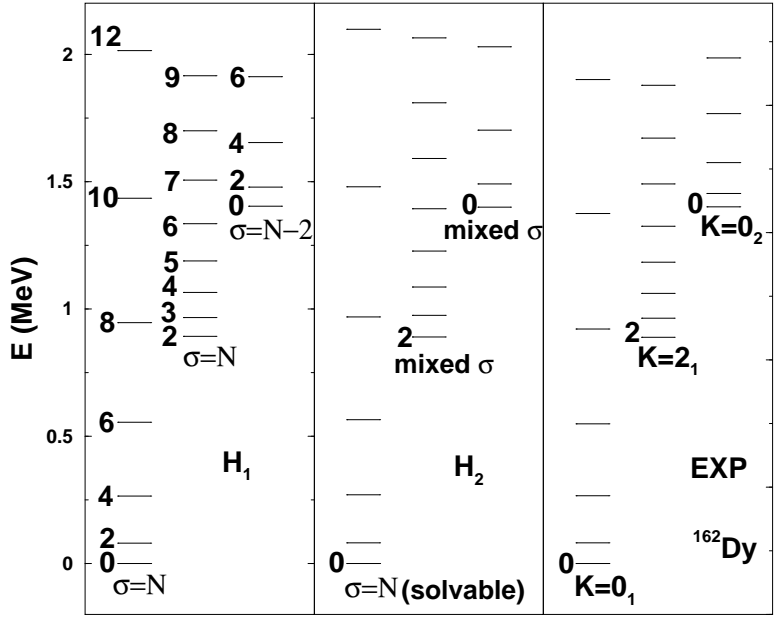}
\end{center}
\caption{
\small
Experimental spectra
(EXP) of $^{162}$Dy 
compared with calculated spectra 
of $\hat{H}_1+ C_1\,\hat{C}_{{\rm O(3)}}$, Eq.~(\ref{h1}), 
with O(6)-PDS type II 
and of $\hat{H}_2+ C_2\,\hat{C}_{{\rm O(3)}}$, Eq.~(\ref{h2}), 
with O(6)-PDS of type III. 
The parameters (in keV) are
$\kappa_0=8$, $\kappa_2=1.364$, $C_1=8$ and
$h_0=28.5$, $h_2=6.3$, $C_2=13.45$, and boson number $N=15$. 
Adapted from~\cite{levisa02}. 
\label{figo6energy}}
\end{figure}
\begin{figure}[t]
\begin{center}
\includegraphics*[width=2.49in,angle=-90]{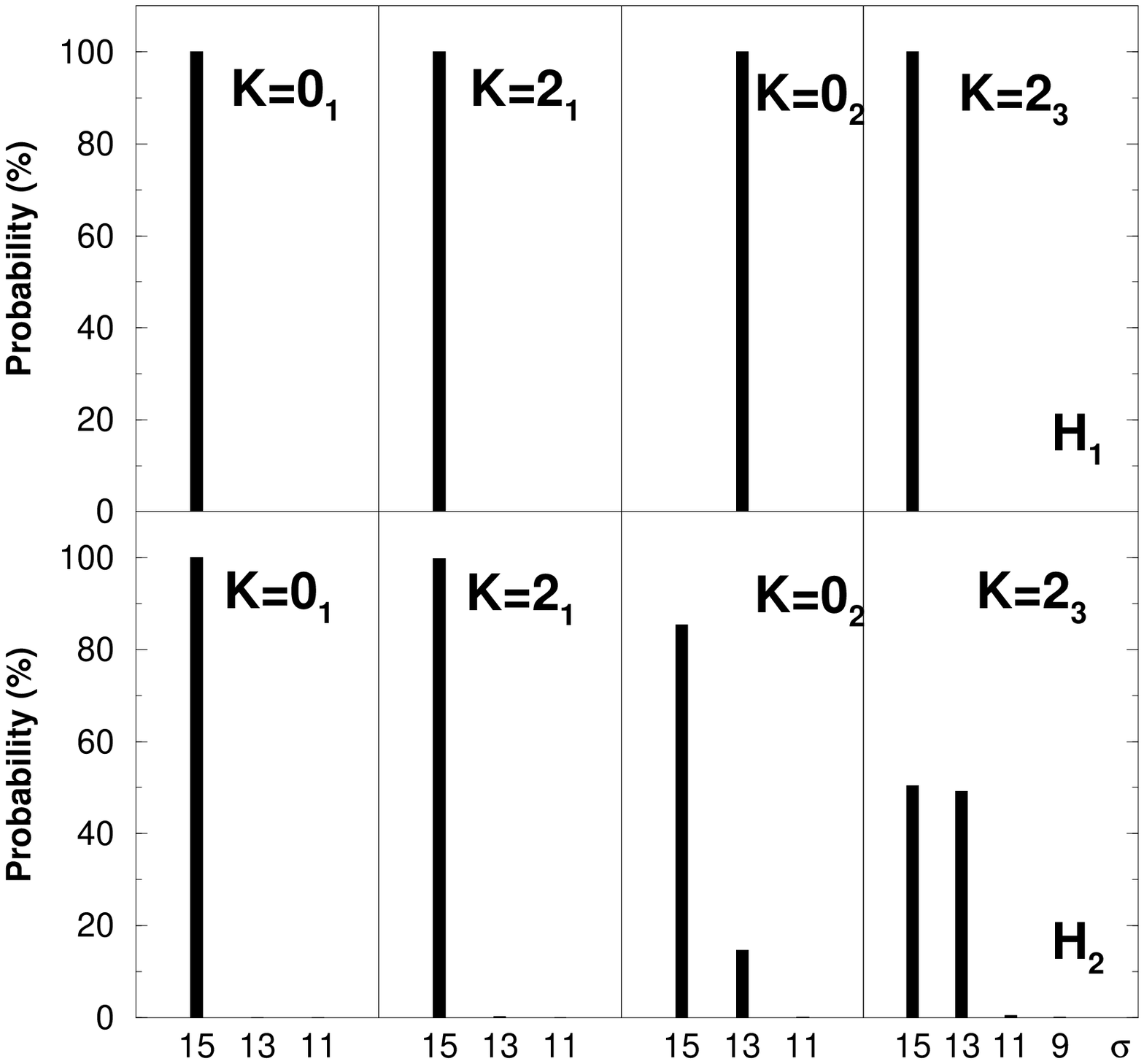}
\hspace*{0.2in}
\includegraphics*[width=2.49in,angle=-90]{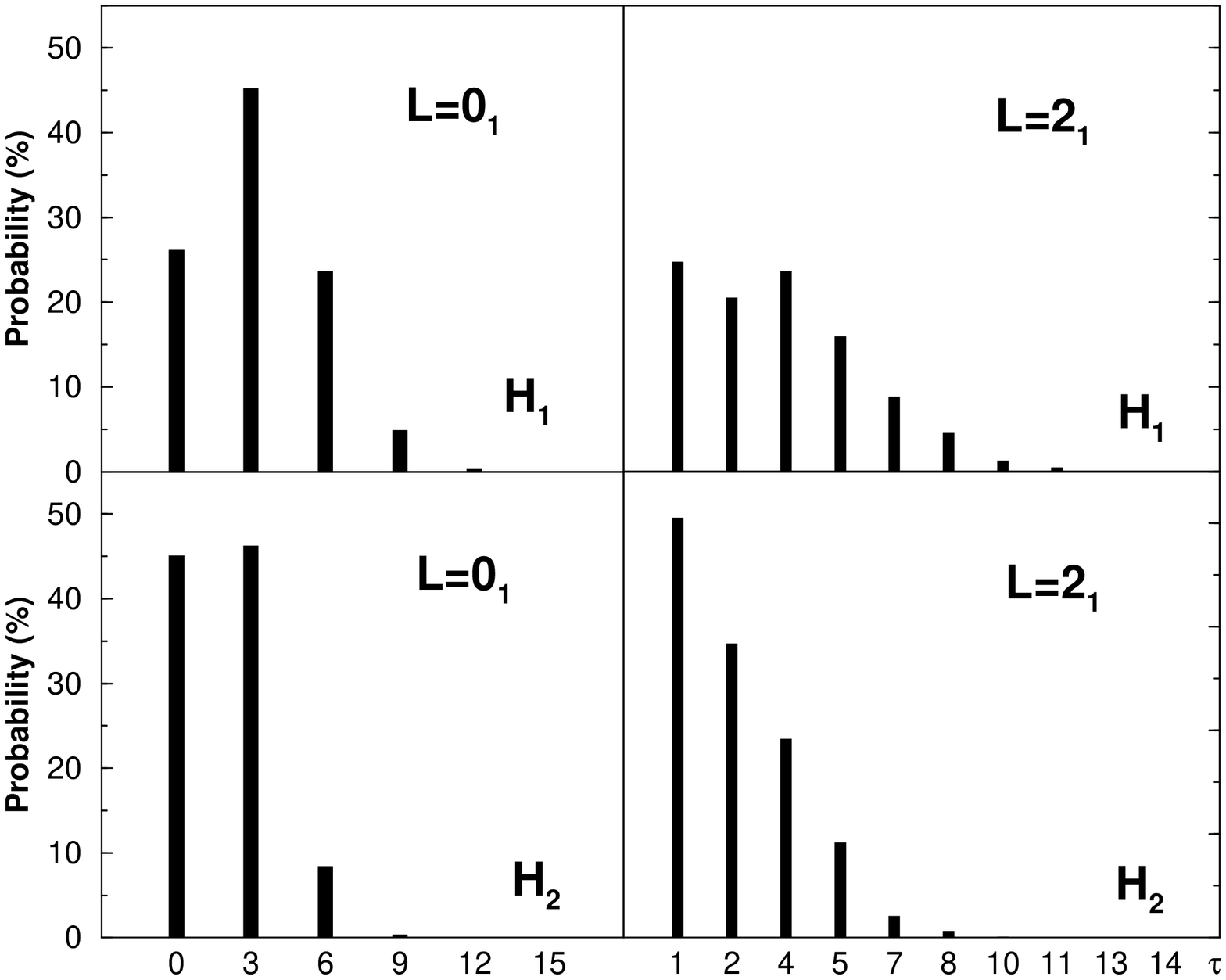}
\end{center}
\caption{
\small
Left: ~O(6) decomposition of wave functions of states in the bands 
$K=0_1,\,2_1,\,0_2,\,(L=K^{+})$, and $K=2_3,\,(L=3^{+})$, for 
$\hat{H}_1$~(\ref{h1}) with O(6)-PDS type II (upper portion) and 
$\hat{H}_2$~(\ref{h2}) with O(6)-PDS type III (lower portion). 
Right: O(5) decomposition of wave functions of $L=0,\,2$ states 
in the $\sigma=N$ ground bands ($K=0_1$) of $\hat{H}_1$ (upper portion) 
and $\hat{H}_2$ (lower portion). Adapted from~\protect\cite{levisa02}.
\label{figo6decomp}}
\end{figure}
From Table~\ref{Tabo6tensII} one sees that the 
$P^{\dag}_{0}$ pair 
is an O(6) tensor with $(\sigma=0,\tau=0,L=0)$, while $P^{\dag}_{2\mu}$ 
involves a combination 
of tensors with $(\sigma=2,\tau=1,L=2)$ and $(\sigma=2,\tau=2,L=2)$ 
\bsub
\ba
P^{\dagger}_{0} &=& 2\sqrt{3}\,
\hat B^\dag_{[2]\langle0\rangle(0)0;00} ~,\\
P^{\dagger}_{2\mu} &=& \sqrt{2}\,
\hat B^\dag_{[2]\langle2\rangle(1)0;2\mu}
+ \sqrt{14}\,\hat B^\dag_{[2]\langle2\rangle(2)0;2\mu} ~.
\label{Pdag2}
\ea
\label{Pdag02}
\esub
These operators satisfy 
\bsub
\ba
P_{0}\vert c;\, N\rangle &=& 0
\label{P0condo6}\\
P_{2\mu}\vert c;\, N\rangle &=& 0
\label{P2condo6}
\ea
\label{PLcondo6}
\esub
where 
\ba 
\vert c; \, N\rangle &=& 
(N!)^{-1/2} (b^{\dag}_{c})^{N}\vert 0\rangle \;\;\; , \;\;\;
b^{\dag}_{c}= (\,d^{\dag}_{0} + s^{\dag})/\sqrt{2} ~.
\label{condo6}
\ea
The state $\vert c; \, N\rangle$
is obtained by substituting the O(6) deformation, $\beta=1$, as well as 
$\gamma=0$ in the coherent state of Eq.~(\ref{condgen}), 
$\vert c; \, N\rangle = \vert\beta=1,\gamma=0 ; N \rangle$.
It has good O(6) character, $\langle\sigma\rangle=\langle N \rangle$, 
and serves as an intrinsic state for a prolate-deformed ground band.
Rotational members of the band with $\sigma=N$ and even values of $L$, 
are obtained by angular momentum projection,
$\vert [N]\langle N\rangle LM\rangle
\propto \hat{\cal P}_{LM}\vert c; \, N\rangle$. 
The projection operator, $\hat{\cal P}_{LM}$, involves 
an O(3) rotation which commutes with $P_{0}$ and transforms 
$\tilde{P}_{2\mu}$ among its various components. Consequently, 
$P_{0}$ and $P_{2\mu}$ annihilate also the projected states
\bsub
\ba
P_{0}\,\vert [N]\langle N\rangle LM\rangle &=& 0 ~,
\label{P0Lo6}\\
P_{2\mu}\,\vert [N]\langle N\rangle LM\rangle &=& 0 ~,
\qquad L=0,2,4,\ldots, 2N ~.
\label{P2Lo6}
\ea
\label{P0P2o6}
\esub
It should be noted that $P^{\dag}_{2\mu}$
and $P_{2\mu}$, Eq.~(\ref{Pdag2}), span only part of the 
$\sigma=2$ irrep. 
Consequently, the projected states 
$\vert [N]\langle N\rangle LM\rangle$ of Eq.~(\ref{P0P2o6}) 
span only part of the O(6) irrep, 
$\langle \sigma\rangle =\langle N\rangle$. 
The corresponding wave functions 
contain a mixture of components with different O(5) symmetry $\tau$, 
and their expansion in the O(6) basis 
$\vert [N]\langle\sigma\rangle (\tau)n_{\Delta} LM\rangle$ reads 
\bsub
\ba
\vert [N]\langle N\rangle LM\rangle &=& 
{\cal N}_{N}^{(L)}
\sum_{\tau,n_{\Delta}}\,a_{\tau,n_{\Delta}}^{(N,L)}\,
\vert [N]\langle N\rangle (\tau) n_{\Delta} 
L M\rangle ~,\\
a_{\tau, n_{\Delta}}^{(N,L)} &=& 
\left [(N-\tau)!(N+\tau+3)!\right ]^{-1/2}\,f_{\tau, n_{\Delta}}^{(L)} ~.
\ea
\label{solo6pds}
\esub
Here ${\cal N}_{N}^{(L)}$ is a normalization coefficient and 
explicit expressions of the factors $f_{\tau, n_{\Delta}}^{(L)}$ 
for $L=0,2,4$ are given in Table~\ref{Tabftau}. 
\noindent

Following the general algorithm, a two-body Hamiltonian with O(6) partial 
symmetry can now be constructed~\cite{levisa02} 
from the boson-pairs operators of Eq.~(\ref{PLo6}) as
\begin{equation}
\hat{H}_2 = h_{0}\, P^{\dagger}_{0}P_{0} + h_{2}\, P^{\dagger}_{2}
\cdot\tilde P_{2} ~.
\label{h2}
\end{equation}
The $h_0$ term is the O(6)-scalar interaction of Eq.~(\ref{HPSo6}). 
The multipole form of the $h_2$ term 
involves the Casimir operators of O(5) and O(3) 
which are diagonal in $\sigma$ and $\tau$, 
terms involving $\hat{n}_d$ which is a scalar under O(5) 
but can connect states differing by $\Delta\sigma=0,\pm 2$ and a 
$\Pi^{(2)}\cdot U^{(2)}$ term 
which induces both O(6) and O(5) mixing
subject to $\Delta\sigma=0,\pm 2$ and $\Delta\tau=\pm 1,\pm 3$.
Although $\hat{H}_2$ is not an O(6)-scalar, 
relations~(\ref{PLcondo6}) and~(\ref{P0P2o6})
ensure 
that it has an exactly solvable ground band with good O(6) 
symmetry, $\langle\sigma\rangle =\langle N\rangle$, but broken O(5) 
symmetry. The Casimir operator of O(3) can be added 
to $\hat{H}_2$ to obtain 
\ba
\hat{H}_{PDS} = \hat{H}_2 + C\, \hat{C}_{{\rm O(3)}} ~.
\label{h2PDS}
\ea
The solvable states of $\hat{H}_{PDS}$ form 
an axially--deformed ground band
\bsub
\ba
&&\vert [N]\langle N\rangle LM\rangle\qquad L=0,2,4,\ldots, 2N\\
&& E_{PDS} = C\,L(L+1) ~.
\ea
\label{ePDSo6III}
\esub
Thus, $\hat{H}_{PDS}$~(\ref{h2PDS}) 
has a subset of solvable states with good O(6) symmetry, 
which is not preserved by other states.
All eigenstates of $\hat{H}_{PDS}$ break the O(5) symmetry but preserve the
O(3) symmetry. These are precisely the required features of 
O(6)-PDS of type III. 
\begin{table}[t]
\begin{center}
\caption{\label{Tabbe2Dy162}
\protect\small
Calculated and observed (Exp) B(E2) values 
(in $10^{-2}e^2b^2)$ for $g\to g$ and $\gamma\to g$ transitions in 
$^{162}$Dy. The parameters of the E2 operator, Eq.~(\ref{Te2}), are 
$e_B=0.138$ $[0.127]$ $eb$ and $\chi=-0.235$ $[-0.557]$ 
for $\hat{H}_1$~(\ref{h1}) [$\hat{H}_2$~(\ref{h2})].
The Hamiltonian $\hat{H}_1$ ($\hat{H}_2$) has O(6)-PDS of type II 
(type III). Adapted from ~\protect\cite{levisa02}.} 
\vspace{1mm}
\begin{tabular}{llll|llll}
\hline
& & & & & & &\\[-3mm]
Transition & $H_{1}$ & $H_{2}$ & Exp & 
Transition & $H_{1}$ & $H_{2}$ & Exp \\
\hline
& & & & & & &\\[-2mm]
$2^{+}_{K=0_1}\rightarrow 0^{+}_{K=0_1}$  & 107   & 107  & 107(2) &
$2^{+}_{K=2_1}\rightarrow 0^{+}_{K=0_1}$  & 2.4   & 2.4  &   2.4(1)  \\[2pt]
$4^{+}_{K=0_1}\rightarrow 2^{+}_{K=0_1}$  & 151   & 152  & 151(6) &
$2^{+}_{K=2_1}\rightarrow 2^{+}_{K=0_1}$  & 3.8   & 4.0  & 4.2(2)  \\[2pt]
$6^{+}_{K=0_1}\rightarrow 4^{+}_{K=0_1}$  & 163   & 165   & 157(9) &
$2^{+}_{K=2_1}\rightarrow 4^{+}_{K=0_1}$  & 0.24  & 0.26 & 0.30(2) \\[2pt]
$8^{+}_{K=0_1}\rightarrow 6^{+}_{K=0_1}$  & 166   & 168 & 182(9)   &
$3^{+}_{K=2_1}\rightarrow 2^{+}_{K=0_1}$  & 4.2 & 4.3 &           \\[2pt]
$10^{+}_{K=0_1}\rightarrow 8^{+}_{K=0_1}$ & 164   & 167 & 183(12)  &
$3^{+}_{K=2_1}\rightarrow 4^{+}_{K=0_1}$  & 2.2  & 2.3  &           \\[2pt]
$12^{+}_{K=0_1}\rightarrow 10^{+}_{K=0_1}$& 159  & 163 & 168(21)  &
$4^{+}_{K=2_1}\rightarrow 2^{+}_{K=0_1}$ & 1.21 & 1.14 & 0.91(5)  \\[2pt]
 & & & &
$4^{+}_{K=2_1}\rightarrow 4^{+}_{K=0_1}$ & 4.5  & 4.7  & 4.4(3)   \\[2pt]
 & & & &
$4^{+}_{K=2_1}\rightarrow 6^{+}_{K=0_1}$ & 0.59 & 0.61 & 0.63(4)  \\[2pt]
 & & & &
$5^{+}_{K=2_1}\rightarrow 4^{+}_{K=0_1}$ & 3.4 & 3.3  & 3.3(2)   \\[2pt]
 & & & &
$5^{+}_{K=2_1}\rightarrow 6^{+}_{K=0_1}$ & 2.9  & 3.1  & 4.0(2)   \\[2pt]
 & & & &
$6^{+}_{K=2_1}\rightarrow 4^{+}_{K=0_1}$ & 0.84 & 0.72 & 0.63(4)  \\[2pt]
 & & & &
$6^{+}_{K=2_1}\rightarrow 6^{+}_{K=0_1}$ & 4.5  & 4.7  & 5.0(4)   \\[4pt]
\hline
\end{tabular}
\end{center}
\end{table}
\begin{table}
\begin{center}
\caption{\label{Tab2be2Dy162}
\protect\small
Calculated and observed (Exp)~\cite{apra06} 
B(E2) values (in $e^2b^2$) 
for transitions from the $K=0_2$ band in $^{162}$Dy. 
The calculations involve the Hamiltonian $\hat{H}_2$~(\ref{h2}) with 
O(6)-PDS of type III and the CQF Hamiltonian with 
broken O(6) symmetry. Adapted from~\cite{apra06}.} 
\vspace{1mm}
\begin{tabular}{llll|llll}
\hline
& & & & & & &\\[-3mm]
Transition & $H_{2}$ & CQF & Exp & 
$\;$Transition & $H_{2}$ & CQF & Exp \\
& & & & & & &\\[-3mm]
\hline
& & & & & & &\\[-2mm]
$0^{+}_{K=0_2}\rightarrow 2^{+}_{K=0_1}$ & 0.0023 & 0.0011 &      & 
$4^{+}_{K=0_2}\rightarrow 2^{+}_{K=0_1}$ & 0.0005 & 0.0002 &        \\[2pt]
$0^{+}_{K=0_2}\rightarrow 2^{+}_{K=2_1}$ & 0.1723 & 0.151  &      &
$4^{+}_{K=0_2}\rightarrow 4^{+}_{K=0_1}$ & 0.0004 & 0.0001 &        \\[2pt]
$2^{+}_{K=0_2}\rightarrow 0^{+}_{K=0_1}$ & 0.0004 & 0.0002 &       &
$4^{+}_{K=0_2}\rightarrow 6^{+}_{K=0_1}$ & 0.0015 & 0.0006 & 0.0034(7) \\[2pt]
$2^{+}_{K=0_2}\rightarrow 2^{+}_{K=0_1}$ & 0.0005 & 0.0002 &      &
$4^{+}_{K=0_2}\rightarrow 2^{+}_{K=2_1}$ & 0.0005 & 0.0001 & 0.0015(5)  \\[2pt]
$2^{+}_{K=0_2}\rightarrow 4^{+}_{K=0_1}$ & 0.0014 & 0.0006 & 0.013(2)  &
$4^{+}_{K=0_2}\rightarrow 3^{+}_{K=2_1}$ & 0.0085 & 0.0030 & 0.0011(3)  \\[2pt]
$2^{+}_{K=0_2}\rightarrow 2^{+}_{K=2_1}$ & 0.0369 & 0.0242 & 0.016(3)  &
$4^{+}_{K=0_2}\rightarrow 4^{+}_{K=2_1}$ & 0.0446 & 0.0283 & 0.011(2) \\[2pt]
$2^{+}_{K=0_2}\rightarrow 3^{+}_{K=2_1}$ & 0.0849 $\;\;\;$ & 0.0716 & 
0.052(5)  & 
$4^{+}_{K=0_2}\rightarrow 5^{+}_{K=2_1}$ & 0.0737 & 0.0631 & 0.018(4)  \\[2pt]
$2^{+}_{K=0_2}\rightarrow 4^{+}_{K=2_1}$ & 0.0481 & 0.0474 & $\equiv$0.048 &  
$4^{+}_{K=0_2}\rightarrow 6^{+}_{K=2_1}$ & 0.0373 & 0.0361 &   \\[4pt]
\hline
\end{tabular}
\end{center}
\end{table}

The calculated spectra of $\hat{H}_2$~(\ref{h2}) 
and $\hat{H}_1$~(\ref{h1}), supplemented with an O(3) term, 
are compared with the experimental spectrum of $^{162}$Dy 
in Fig.~\ref{figo6energy}. 
The spectra display rotational bands of an axially-deformed nucleus,
in particular, a ground band $(K=0_1)$ and excited $K=2_1$ and $K=0_2$ 
bands. The O(6) and O(5) decomposition of selected bands are shown in 
Fig.~\ref{figo6decomp}. 
For $\hat{H}_1$, characteristic features of the results were discussed in 
Subsection~\ref{subsec:o6PDStypeII}. 
For $\hat{H}_2$, the solvable $K=0_1$ ground band has $\sigma=N$ 
and all eigenstates are mixed with respect to O(5). 
However, in contrast to $\hat{H}_1$, 
excited bands of $\hat{H}_2$ can have 
components with different O(6) character. For example, 
the $K=0_2$ band of $\hat{H}_2$ has components with $\sigma=N$ $(85.50\%)$,
$\sigma=N-2$ $(14.45\%)$, and $\sigma=N-4$ $(0.05\%)$.
These $\sigma$-admixtures can, in turn, be interpreted in terms of 
multi-phonon excitations. 
Specifically, the $K=0_2$ band is composed of 
$36.29\%$ $\beta$, $63.68\%$ $\gamma^2_{K=0}$,
and $0.03\%$ $\beta^2$ modes,
{\it i.e.}, it is dominantly a double-gamma phonon excitation
with significant single-$\beta$ phonon admixture. The $K=2_1$ band 
has only a small O(6) impurity and is 
an almost pure single-gamma phonon band. 
The results of Fig.~\ref{figo6decomp} illustrate 
that $\hat{H}_2$~(\ref{h2}) possesses O(6)-PDS of type III 
which is distinct from the O(6)-PDS of type II exhibited by 
$\hat{H}_1$~(\ref{h1}).

In Table~\ref{Tabbe2Dy162} the experimental B(E2) values for 
E2 transitions in $^{162}$Dy, are compared with PDS calculations. 
The B(E2) values predicted by $\hat{H}_1$~(\ref{h1}) and 
$\hat{H}_2$~(\ref{h2}) for $K=0_1\rightarrow K=0_1$ and 
$K=2_1\rightarrow K=0_1$ transitions
are very similar and agree well with the measured values.
On the other hand, their predictions for interband transitions
from the $K=0_2$ band are very different~\cite{levisa02}. 
For $\hat{H}_1$, the $K=0_2\rightarrow K=0_1$ and $K=0_2\rightarrow K=2_1$
transitions are comparable and weaker than $K=2_1\rightarrow K=0_1$.
In contrast, for $\hat{H}_2$,
$K=0_2\rightarrow K=2_1$ and $K=2_1\rightarrow K=0_1$ transitions
are comparable and stronger than $K=0_2\rightarrow K=0_1$. 
The results of a recent detailed measurement~\cite{apra06} of $^{162}$Dy, 
shown in Table~\ref{Tab2be2Dy162}, indicate that characteristic features 
of the $K=0_2$ band in this nucleus are reproduced by 
both $\hat{H}_2$ with O(6)-PDS of type III, and the CQF Hamiltonian 
with broken O(6) symmetry, but refinements are necessary.

\section{Partial Solvability}
\label{sec:PartialSolv}

The PDS of type I and III, discussed so far,  
involve subsets of solvable states with good symmetry character, with 
respect to algebras in a given dynamical symmetry chain. 
A~further extension of this concept is possible, for which the selected 
solvable states are not associated with 
any underlying symmetry. Such a situation can be referred to as 
partial solvability. In the PDS examples considered within the IBM framework, 
the solvable states were obtained by choosing specific deformations and 
projecting from an intrinsic state, Eq.~(\ref{condgen}), 
representing the ground band 
\ba
\vert\,\beta,\gamma ; N \rangle &\propto&
\left [\,\beta\cos\gamma\,
d^{\dagger}_{0} + \beta\sin{\gamma}\,
( d^{\dagger}_{2} + d^{\dagger}_{-2})/\sqrt{2} + s^{\dagger}\,
\right ]^N\vert 0\rangle ~.
\label{condbet}
\ea
Specifically, for SU(3)-PDS of type I, the solvable ground band was 
associated with deformations $(\beta=\sqrt{2},\gamma=0)$, while for 
O(6)-PDS of type III, it was associated with $(\beta=1,\gamma=0)$. 
More generally, a natural candidate for a solvable ground band would be 
the set states of good O(3) symmetry $L$, projected from 
the prolate-deformed intrinsic state, $\vert\,\beta,\gamma=0 ; N \rangle$, 
with arbitrary deformation $\beta>0$~\cite{lev06}
\ba
\vert\, \beta; N, L M\rangle &\propto&
\left [\Gamma_{N}^{(L)}(\beta)\right ]^{-1/2} 
\hat{\cal{P}}_{LM}\vert \beta,\gamma=0; N\rangle 
\qquad L=0,2,4,\ldots, 2N
\nonumber\\
\Gamma_{N}^{(L)}(\beta) &=& 
\frac{1}{N!}\int_{0}^{1}dx 
\left [ 1 + \beta^2\,P_{2}(x)\right ]^N P_{L}(x) ~. \qquad
\label{wfqpt1}
\ea
Here $P_{L}(x)$ is a Legendre polynomial with $L$ even 
and $\Gamma_{N}^{(L)}(\beta)$ is a normalization factor.
In general, these $L$-projected states do not have good symmetry 
properties with respect to any of the IBM dynamical 
symmetry chains~(\ref{IBMds}). 
Their wave functions have the following expansion in the U(5) 
basis 
\ba
\vert\, \beta; N, L M\rangle &=&
\sum_{n_d,\tau,n_{\Delta}}\frac{1}{2}
\left [1 + (-1)^{n_d-\tau}\right ]\, \xi_{n_d,\tau,n_{\Delta}}^{(N,L)}
\vert\, [N]\langle n_d\rangle (\tau) n_{\Delta} L M\rangle ~,
\label{projndt}
\ea
where $(\tau,n_{\Delta})$ take the values compatible with the 
${\rm O(5)}\supset {\rm O(3)}$ reduction and the $n_d$ summation 
covers the range $\tau\leq n_d\leq N$. 
The coefficients $\xi_{n_d,\tau,n_{\Delta}}^{(N,L)}$ are of the 
form~\cite{lev05} 
\ba
\xi_{n_d,\tau,n_{\Delta}}^{(N,L)} &=& 
\left [\Gamma_{N}^{(L)}(\beta)\right ]^{-1/2} 
\frac{\beta^{n_d}}
{\left [
(N-n_d)!(n_d-\tau)!!(n_d +\tau +3)!!\right ]^{1/2}}\,
f_{\tau,n_{\Delta}}^{(L)} ~.
\label{xindt}
\ea
Explicit expressions~\cite{lev05} for some of the factors 
$f_{\tau,n_{\Delta}}^{(L)}$ are given in Table~\ref{Tabftau}. 
\begin{table}[t]
\begin{center}
\caption{\label{Tabftau}
\protect\small
The factors $f_{\tau,n_{\Delta}}^{(L)}$, Eq.~(\ref{xindt}), for 
the states $\vert\beta; N,LM\rangle$, Eq.~(\ref{projndt}), with $L=0,2,4$. 
The label $n_{\Delta}$ is not required, since 
these $L$-states are multiplicity-free.} 
\vspace{1mm}
\begin{tabular}{llll}
\hline
& & &\\[-3mm]
   & $f_{\tau=0,3,6,\ldots}^{(L)}$ & $f_{\tau=1,4,7,\ldots}^{(L)}$ &
$f_{\tau=2,5,8,\ldots}^{(L)}$\\[4pt]
\hline
& & &\\[-2mm]
$L=0$ & $(-)^{\tau}\sqrt{2\tau+3}$ & &  \\[4pt]
$L=2$ &   &  $(-)^{\tau+1}\sqrt{\tau+2}$ & 
$(-)^{\tau+1}\sqrt{\tau+1}$ \\[6pt]
$L=4\quad$ & 
$(-1)^{\tau}\sqrt{\frac{7(2\tau+3)\tau(\tau+3)}{3(2\tau+5)(2\tau+1)}}\quad$ &
$(-1)^{\tau}\sqrt{\frac{5(\tau+2)(\tau-1)}{6(2\tau+5)}}\quad$ &
$(-1)^{\tau}\sqrt{\frac{5(\tau+1)(\tau+4)}{6(2\tau+1)}}$\\[6pt]
& & &\\[-3mm]
\hline
\end{tabular}
\end{center}
\end{table}

The construction of partially-solvable Hamiltonians which have the 
set of states~(\ref{wfqpt1}) as eigenstates, can be 
accomplished~\cite{lev85} by means of the following boson-pair operators 
with angular momentum $L=0,\,2$
\bsub
\ba
P^{\dagger}_{0}(\beta_0) &=& 
d^{\dagger}\cdot d^{\dagger} - \beta_{0}^2(s^{\dagger})^2 ~,
\label{P0b0}\\
P^{\dagger}_{2\mu}(\beta_0) &=& 
\beta_{0}\sqrt{2}d^{\dagger}_{\mu}s^{\dagger} + 
\sqrt{7}\, (d^{\dagger}\,d^{\dagger})^{(2)}_{\mu} ~.
\label{P2b0}
\ea
\label{PLb0}
\esub
These operators satisfy
\bsub
\ba
P_{0}(\beta_0)\,\vert\,\beta_0,\gamma=0 ; N \rangle &=& 0 ~,\\
P_{2\mu}(\beta_0)\,\vert\,\beta_0,\gamma=0 ; N \rangle &=& 0 ~,
\label{PLcondb0}
\ea
\esub  
or equivalently, 
 \ba
P_{0}(\beta_0)\,\vert\, \beta_0; N, L M\rangle &=& 0 ~,
\nonumber\\
P_{2\mu}(\beta_0)\,\vert\, \beta_0 ; N, L M\rangle &=& 0 ~.
\label{P0P2b0}
\ea
The following Hamiltonian~\cite{lev85,kirlev85,lev87}
\ba
\hat{H}(h_0,h_2,\beta_0) &=& h_{0}\, 
P^{\dagger}_{0}(\beta_{0})P_{0}(\beta_0) 
+ h_{2}\,P^{\dagger}_{2}(\beta_0)\cdot \tilde{P}_{2}(\beta_0) ~,
\label{hPS}
\ea
has a solvable zero-energy prolate-deformed ground band, 
composed of the states in 
Eq.~(\ref{wfqpt1}). The Casimir operator of O(3) can be added to it to 
form a partially solvable (PSolv) Hamiltonian
\ba
\hat{H}_{PSolv} &=& \hat{H}(h_0,h_2,\beta_0) + C\,\hat{C}_{{\rm O(3)}} ~.
\label{hPSolv}
\ea
The solvable states and energies are
\bsub
\ba
&&
\vert\, \beta_0; N, L M\rangle
\;\;\;\; L=0,2,4,\ldots, 2N\\
&& E_{PSolv}= CL(L+1) ~.
\ea
\label{ePSolv}
\esub
Since the wave functions of these 
states are known, it is possible to obtain closed form expressions for 
related observables. For example, for the E2 operator of Eq.~(\ref{Te2}), 
the B(E2) values for transitions between 
members of the solvable ground band read~\cite{lev06,lev07}
\ba
&&B(E2; L+2\to L) = e_{B}^2\,(L+2,0;2,0\vert L,0)^2\,\beta^2\,
\frac{
[\,a_1\,\Gamma_{N-1}^{(L)}(\beta) 
+ a_2\, \Gamma_{N-1}^{(L+2)}(\beta)\,]^2}
{\Gamma_{N}^{(L)}(\beta)\,\Gamma_{N}^{(L+2)}(\beta)} ~,
\qquad\quad
\nonumber\\
&&
a_1= 1-\beta\sqrt{\frac{2}{7}}\chi L/(2L+3) \;\;\; , \;\;\; 
a_2= 1-\beta\sqrt{\frac{2}{7}}\chi (L+3)/(2L+3) ~, 
\label{be2beta}
\ea
where the symbol $(...\vert ..)$ denotes an O(3) Clebsch Gordan coefficient.

The Hamiltonian $\hat{H}_{PSolv}$ of Eq.~(\ref{hPSolv}) is partially solvable 
for any value of $\beta_0>0$. 
For $\beta_0=\sqrt{2}$, it reduces to the Hamiltonian of Eq.~(\ref{hPDSsu3}) 
with SU(3)-PDS of type I. In this case, the solvable states span 
the SU(3) irrep $(2N,0)$ and the normalization factor 
in Eq.~(\ref{wfqpt1}) is given by 
\ba
\Gamma_{N}^{(L)}(\beta=\sqrt{2}) &=& 
\frac{3^{N}(2N)!}{(2N-L)!!(2N+L+1)!!N!} ~.
\ea
Relation~(\ref{projndt}) then provides transformation brackets 
between these SU(3) states and the U(5) basis and 
Eq.~(\ref{be2beta}) reduces to a 
well-known expression for E2 transitions among states in the 
SU(3) ground band~\cite{arimaiac78,ISA}. 
When $\beta_0 =1$, the Hamiltonian $\hat{H}_{PSolv}$~(\ref{hPSolv}) 
coincides with the Hamiltonian of Eq.~(\ref{h2PDS}) 
with O(6)-PDS of type III.

When $h_2=0$, the Hamiltonian of Eq.~(\ref{hPS}) takes the form 
\ba
\hat{H}(h_0,\beta_0) &=& h_{0}\, 
P^{\dagger}_{0}(\beta_{0})P_{0}(\beta_0) ~. 
\label{hPS0}
\ea 
Both $P^{\dag}(\beta_0)$ and $P_0(\beta_0)$, Eq.~(\ref{P0b0}), 
are O(5)-scalars. Futhermore, $P_0(\beta_0)$ annihilates the intrinsic 
state, Eq.~(\ref{condbet}), with $\beta=\beta_0$ and 
arbitrary $\gamma$ 
\ba
P_{0}(\beta_0)\,\vert\,\beta_0,\gamma ; N \rangle &=& 0 ~.
\label{P0condb0}
\ea
Equivalently, 
\ba
P_{0}(\beta_0)\,\vert\beta_0 ; N,\tau, n_{\Delta},LM\rangle &=& 0 ~,
\label{P0b0L}
\ea
where the indicated states, with good $\tau$ and $L$ quantum numbers, 
are obtained by O(5) projection
from the deformed intrinsic state 
$\vert\,\beta_0,\gamma ; N \rangle$~(\ref{condbet})
\ba
\vert\beta; N,\tau, n_{\Delta},LM\rangle &\propto& 
\left [F_{N}^{(\tau)}(\beta)\right ]^{-1/2} 
\hat{\cal{P}}_{\tau,n_{\Delta},LM}\vert \beta,\gamma; N\rangle
\nonumber\\
F_{N}^{(\tau)}(\beta) &=&  
\sum_{n_d}\frac{1}{2}\left [1 + (-1)^{n_d-\tau}\right ]
\frac{\beta^{2n_d}}{(N-n_d)!\,(n_d-\tau)!!\,(n_d+\tau+3)!!} ~.
\qquad
\label{wfqpt2}
\ea
Here $F_{N}^{(\tau)}(\beta)$ is a normalization factor and the $n_d$ 
summation covers the range $\tau\leq n_d \leq N$. 
The corresponding wave 
functions have the following expansion in the U(5) basis~\cite{levgin03}
\bsub
\ba
\vert\, \beta; N, \tau, n_{\Delta},L M\rangle &=&
\sum_{n_d}\frac{1}{2}
\left [1 + (-1)^{n_d-\tau}\right ]\, \theta_{n_d}^{(N,\tau)}
\vert\, [N]\langle n_d\rangle(\tau)n_{\Delta} L M\rangle ~,\\
\theta_{n_d}^{(N,\tau)} &=& 
\left [ F_{N}^{(\tau)}(\beta)\right ]^{-1/2}
\frac{\beta^{n_d}}{[(N-n_d)!(n_d -\tau)!!(n_d+\tau+3)!!]^{1/2}}~.
\qquad
\ea
\label{projnd}
\esub
The Hamiltonian~(\ref{hPS0}) mixes the U(5) and O(6) chains 
but preserves the common O(5) subalgebra. 
This is explicitly seen from its multipole form
\ba
\hat{H}(h_0,\beta_0) &=& 
h_{0}\beta_{0}^2\left [\, 
-\hat{C}_{{\rm O(6)}} + 5\hat{N}
+\beta_{0}^2\hat{N}(\hat{N}-1)
\,\right ]
\nonumber\\
&&
+h_{0}(1-\beta_{0}^2)\left [ 
4\hat{n}_d + 2\beta_{0}^2(\hat{N}-1)\hat{n}_d + 
(1-\beta_{0}^2)\hat{n}_{d}(\hat{n}_d-1)
-\hat{C}_{{\rm O(5)}} \right ] ~.
\qquad\;
\ea
It has a solvable zero-energy $\gamma$-unstable deformed ground band, 
composed of the states in 
Eq.~(\ref{wfqpt2}). 
The Casimir operators of O(5) and O(3) can be added to it to form 
a partially solvable (PSolv) Hamiltonian 
\ba
\hat{H}_{PSolv} &=& \hat{H}(h_0,\beta_0) + 
B\, \hat{C}_{{\rm O(5)}} + C\,\hat{C}_{{\rm O(3)}} ~. 
\label{hPSolvt}
\ea
$\hat{H}_{PSolv}$~(\ref{hPSolvt}) has also 
an O(5)-PDS of type II in the sense discussed in 
Subsection~\ref{subsec:o5PDStypeII}. 
The solvable states and energies are
\bsub
\ba
&&
\vert\, \beta_0; N, \tau, n_{\Delta}, L M\rangle ~,\\
&& E_{PSolv}= B\, \tau(\tau+3) + C\, L(L+1) ~,
\ea
\label{ePSolvt}
\esub
where the $(\tau,n_{\Delta},L)$ assignments are the same as for 
states in the O(6) irrep with $\sigma=N$. 
Closed form expressions can be derived for observables in these states. 
For example, for the general E2 operator of Eq.~(\ref{Te2}), 
we find~\cite{levgin03}
\ba
&&B(E2; \tau+1,n_{\Delta}',L'\to \tau,n_{\Delta},L)
\nonumber\\
&&\qquad\qquad 
= e_{B}^2
\frac{\tau+1}{2\tau+5}
\beta^2\,
\frac{[\,F^{(\tau)}_{N-1}(\beta) 
+ F^{(\tau+1)}_{N-1}(\beta)\,]^2}
{F^{(\tau)}_{N}(\beta)F^{(\tau+1)}_{N}(\beta)}
\left \langle
\begin{array}{cc|c}
\tau+1 & 1 & \tau\\
n_{\Delta}'L'    & 2 & n_{\Delta}L
\end{array}
\right \rangle^2 ~. 
\label{be2beta0}
\ea
This expression is similar in form to that encountered in Eq.~(\ref{be2o5a}), 
but now the factor in front of the O(5) isoscalar factor 
is explicitly known. The Hamiltonian of Eq.~(\ref{hPSolvt}) is partially 
solvable for any value of $\beta_0>0$. 
For $\beta_0=1$, it reduces to the Hamiltonian of Eq.~(\ref{hDSo6}) 
with O(6) dynamical symmetry. The solvable states~(\ref{ePSolvt}) 
then span the O(6) irrep $\langle\sigma\rangle = \langle N\rangle$ 
and the normalization factor~(\ref{wfqpt2}) becomes 
\ba
F_{N}^{(\tau)}(\beta=1) &=& \frac{2^{N+1}(N+1)!}{(N-\tau)!(N+\tau+3)!} ~.
\label{Fo6}
\ea  
In this case, relation~(\ref{projnd}) corresponds to known 
transformation brackets between these O(6) states and the U(5) basis, 
and one recovers from Eq.~(\ref{be2beta0}) a familiar expression for the 
indicated $B(E2)$ in the O(6) limit of the IBM~\cite{arimaiac79}. 

In addition to the states shown in Eq.~(\ref{P0b0L}), the operator 
$P_{0}(\beta_0)$ (\ref{P0b0}) annihilates also the following 
U(5) basis states
\bsub
\ba
P_{0}\,\vert [N],n_d=\tau=N,n_{\Delta},L M\,\rangle &=& 0 ~,\\
P_{0}\,\vert [N],n_d=\tau=N-1,n_{\Delta},L M\,\rangle &=& 0 ~,
\label{P0ndN}
\ea
\esub
for reasons explained after Eq.~(\ref{V0ndN}). Consequently, 
$\hat{H}_{PSolv}$ of Eq.~(\ref{hPSolvt}) has also 
a U(5)-PDS of type I, in the sense discussed in 
Subsection~\ref{subsec:u5PDStypeI}. 
The additional solvable eigenstates and energies are
\bsub
\ba
&&\vert [N], n_d=\tau=N,n_{\Delta},L M\,\rangle \;\;\qquad\quad 
E_{Psolv} = B\,N(N+3) + C\,L(L+1) ~,\qquad\qquad\\
&&\vert [N], n_d=\tau=N-1,n_{\Delta},L M\,\rangle \;\;\;\quad 
E_{Psolv} = B\,(N-1)(N+2) + C\,L(L+1) ~, \qquad\qquad
\ea
\label{ePSolvndt}
\esub
where the $(\tau,n_{\Delta},L)$ assignments are those of the 
${\rm O(5)}\supset{\rm O(3)}$ reduction. 

The Hamiltonian $\hat{H}(h_0,h_2,\beta_0)$ of Eq.~(\ref{hPS}) 
is a prototype of an intrinsic 
Hamiltonian which generate band-structure~\cite{lev85,kirlev85,lev87}. 
Its energy surface, defined as in Eq.~(\ref{enesurf}),  
\ba
E_{N}(\beta,\gamma) &=& 
N(N-1)(1+\beta^2)^{-2}
\left [ h_{0}(\beta^2-\beta_{0}^2)^2 + 
2h_{2}\beta^2(\beta^2 - 2\beta_{0}\beta\cos 3\gamma + \beta_{0}^2)\right ]
\qquad\;
\label{esurfint}
\ea
has a global minimum at $(\beta_0>0,\gamma_0=0)$, corresponding to 
a prolate-deformed shape. $\hat{H}(h_0,h_2,\beta_0)$ 
is O(3)-invariant, but has the 
deformed equilibrium intrinsic state, 
$\vert\beta_0,\gamma=0; N\rangle$~(\ref{condbet}), 
as a zero-energy eigenstate. The O(3) symmetry is thus spontaneously broken. 
The two Goldstone modes are associated with rotations about directions 
perpendicular to the symmetry axis. 
The intrinsic modes involve the one-dimensional 
$\beta$ mode and two-dimensional $\gamma$ modes of vibrations. 
For large N, the spectrum of $\hat{H}(h_0,h_2,\beta_0)$~(\ref{hPS}) 
is harmonic, involving $\beta$ and $\gamma$ vibrations about the 
deformed minimum with frequencies given by~\cite{lev85,lev87} 
\ba
\epsilon_{\beta} &=& 2N \beta_{0}^2(2h_{0}+h_{2}) \;\; , \;\;
\epsilon_{\gamma}= 18N \beta_{0}^2(1+\beta_{0}^2)^{-1}h_{2} ~.
\label{emodes}
\ea
The importance of $\hat{H}(h_0,h_2,\beta_0)$ 
lies in the fact that the most general one- and two-body IBM Hamiltonian 
with equilibrium deformations $(\beta_0>0,\gamma_0=0)$, can be resolved 
into intrinsic and collective parts~\cite{kirlev85,lev87}
\ba
\hat{H}_{IBM} &=& \hat{H}(h_0,h_2,\beta_0) + \hat{H}_c ~.
\label{resol}
\ea
The intrinsic part is the partially-solvable Hamiltonian of 
Eq.~(\ref{hPS}). The collective part, $\hat{H}_c$, involves kinetic 
rotational terms which do not affect the shape of the energy surface
\ba
\hat{H}_{c} &=& 
c_3 \left [\, \hat{C}_{{\rm O(3)}} - 6\hat{n}_d \,\right ]
+ c_5 \left [\, \hat{C}_{{\rm O(5)}} - 4\hat{n}_d \,\right ]
+\, c_6 \left [\, \hat{C}_{\overline{{\rm O(6)}}} 
- 5\hat{N}\,\right ] + E_0~.
\label{hcol}
\ea
The various Casimir operators in Eq.~(\ref{hcol}) are defined in the 
Appendix. 
The $L$-projected states, $\vert\, \beta; N, L M\rangle$, 
of Eq.~(\ref{wfqpt1}) can now be used to 
construct an $L$-projected energy surface, $E^{(N)}_{L}(\beta) = 
\langle\beta;N,LM\vert \hat{H}_{IBM} \vert\beta;N,LM\rangle$, 
for the IBM Hamiltonian~(\ref{resol})
\ba
E^{(N)}_{L}(\beta) &=& 
h_0\,(\beta^2-\beta^{2}_{0})^2\, S_{2,L}^{(N)} +
2h_2\,(\beta-\beta_0)^2\,\Sigma_{2,L}^{(N)} + 
c_3\left [ L(L+1) - 6D_{1,L}^{(N)}\right ]
\nonumber\\
&&
+\, c_5\left [ D_{2,L}^{(N)} - \beta^4\,S_{2,L}^{(N)}\right ]
+\, c_6\left [ N(N-1) -(1+\beta^2)^2\,S_{2,L}^{(N)}\right ] + E_0 ~.
\qquad
\label{eneL}
\ea
Here $D_{1,L}^{(N)}$, $S_{2,L}^{(N)}$, $D_{2,L}^{(N)}$ and 
$\Sigma_{2,L}^{(N)}$ denote the expectation values in 
the states $\vert \beta;N,L M\rangle$ 
of $\hat{n}_d$, $\hat{n}_s(\hat{n}_s-1)$, $\hat{n}_d(\hat{n}_d-1)$ 
and $\hat{n}_s\hat{n}_d$ 
respectively. All these quantities are expressed in terms of 
the expectation value of $\hat{n}_s$, denoted by $S^{(N)}_{1,L}$. 
Specifically, 
$D_{1,L}^{(N)}= N - S_{1,L}^{(N)}$, 
$S^{(N)}_{2,L} = S^{(N)}_{1,L}S^{(N-1)}_{1,L}$, 
$\Sigma^{(N)}_{2,L} = (N-1)S^{(N)}_{1,L} - S^{(N)}_{2,L}$, 
$D_{2,L}^{(N)} = N(N-1) - 2(N-1)S^{(N)}_{1,L} + S^{(N)}_{2,L}$.
The quantity $S^{(N)}_{1,L}$ itself is determined by the 
normalization factors of Eq.~(\ref{wfqpt1}) 
\ba
S^{(N)}_{1,L} = 
\langle\beta;N,LM\vert\hat{n}_s\vert\beta;N,LM\rangle =
\Gamma^{(L)}_{N-1}(\beta)/\Gamma^{(L)}_{N}(\beta) ~.
\label{S1L}
\ea 
It also satisfies the following recursion relation~\cite{lev06}
\ba
S^{(N)}_{1,L}
 = \frac{(N-L/2)(2N+L+1)}
{(\beta^2+4)(N-1)+3 
+ (\beta^2-2)(1+\beta^2)S^{(N-1)}_{1,L}} ~.
\label{S1Lrecur}
\ea

For $h_2=0$, the intrinsic part of the IBM Hamiltonian in Eq.~(\ref{resol}) 
reduces to the Hamiltonian $\hat{H}(h_0,\beta_0)$ of Eq.~(\ref{hPS0}), 
which is O(5)-invariant. The unprojected energy surface~(\ref{esurfint}) 
is now independent of $\gamma$ and the equilibrium shape is deformed 
$(\beta_0>0)$ and $\gamma$-unstable. 
The O(5) symmetry is spontaneously broken in 
the intrinsic state, $\vert\beta_0,\gamma;N\rangle$~(\ref{condbet}), 
which is a zero-energy eigenstate of $\hat{H}(h_0,\beta_0)$. 
As a result, the $\gamma$ and three rotational modes are Goldstone modes, 
and only the $\beta$ vibration in Eq.~(\ref{emodes}) survives as a genuine 
mode~\cite{lev85,lev87}. 
In this case, $\hat{H}_{IBM}(h_2=0)$ in Eq.~(\ref{resol}) preserves 
the O(5) symmetry, $\tau$, and the O(5)-projected states of 
Eq.~(\ref{wfqpt2}) can be used to construct its $\tau$-projected 
energy surface, $E^{(N)}_{\tau,L}(\beta) = 
\langle\beta;N,\tau,n_{\Delta},LM
\vert \hat{H}_{IBM}(h_2=0) \vert\beta;N,\tau,n_{\Delta},LM\rangle$, 
\ba
E_{\tau,L}^{(N)}(\beta) &=&
h_0\,(\beta^2-\beta^{2}_{0})^2\, S_{2,\tau}^{(N)} +
c_3\left [ L(L+1) - 6D_{1,\tau}^{(N)}\right ]
\nonumber\\
&&
+\, c_5\left [ \tau(\tau+3) -4D_{1,\tau}^{(N)}\right ]
+\, c_6\left [ N(N-1) -(1+\beta^2)^2\,S_{2,\tau}^{(N)}\right ] + E_0 ~.
\qquad
\label{enetau}
\ea
Here $D_{1,\tau}^{(N)}$ and $S_{2,\tau}^{(N)}$ 
denote the expectation values in 
the states $\vert \beta;N,\tau,n_{\Delta},L M\rangle$ 
of $\hat{n}_d$ and $\hat{n}_s(\hat{n}_s-1)$ respectively. 
All these quantities are expressed in terms of 
the expectation value of $\hat{n}_s$, denoted by $S^{(N)}_{1,\tau}$. 
Specifically, 
$D_{1,\tau}^{(N)}= N - S_{1,\tau}^{(N)}$, 
$S^{(N)}_{2,\tau} = S^{(N)}_{1,\tau}S^{(N-1)}_{1,\tau}$. 
The quantity $S^{(N)}_{1,\tau}$ itself is determined by the 
normalization factors of Eq.~(\ref{wfqpt2}) 
\ba
S^{(N)}_{1,\tau} = 
\langle\beta;N,\tau,n_{\Delta},L M\vert\hat{n}_{s} 
\vert\beta;N,\tau,n_{\Delta},L M\rangle = 
F^{(\tau)}_{N-1}(\beta)/F^{(\tau)}_{N}(\beta) ~.
\label{S1t}
\ea 
It also satisfies the following recursion relation
\ba
S^{(N)}_{1,\tau}
 = \frac{(N-\tau)(N+\tau+3)}
{2(N+1) + (\beta^4-1)S^{(N-1)}_{1,\tau}} ~.
\label{S1trecur}
\ea

\section{PDS and Quantum Phase Transitions}
\label{sec:PDSQPT}

Symmetry plays a profound role in quantum phase transitions (QPT). 
The latter occur at zero temperature as a function of a 
coupling constant in the Hamiltonian. Such ground-state energy phase 
transitions~\cite{gilmore79} are a pervasive 
phenomenon observed in many branches of physics, and are realized 
empirically in nuclei as transitions between different shapes. 
QPTs occur as a result of a competition between terms in the Hamiltonian 
with different symmetry character, which lead to considerable mixing in the 
eigenfunctions, especially at the critical-point where the structure 
changes most rapidly. An interesting question to address is 
whether there are any symmetries (or traces of) still present at the 
critical points of QPT. As shown below, unexpectedly, partial dynamical 
symmetries (PDS) can survive at the critical point in spite of the 
strong mixing~\cite{lev07}. The feasibility of such persisting symmetries 
gains support from the recently proposed~\cite{iac0001} and empirically  
confirmed~\cite{caszam0001} analytic descriptions of critical-point nuclei,  
and the emergence of quasi-dynamical symmetries~\cite{rowe0405} 
in the vicinity of such critical-points.

A convenient framework to study symmetry-aspects of QPT in nuclei 
is the IBM~\cite{ibm}, whose dynamical symmetries~(\ref{IBMds}) 
correspond to possible phases of the system. The starting point 
is the energy surface of the Hamiltonian, Eq.~(\ref{enesurf}), which 
for one- and two- body interactions has the form 
\ba
E_{N}(\beta,\gamma) &=& E_0 + N(N-1)f(\beta,\gamma) ~,
\nonumber\\
f(\beta,\gamma) &=& 
(1+\beta^2)^{-2}\beta^2\left [ a - b\beta\cos 3\gamma + c\beta^2 \right ] ~.
\label{eLan}
\ea 
The coefficients $E_0,a,b,c$ involve particular linear 
combinations of the Hamiltonian's parameters~\cite{lev87}. 
\begin{figure}[t]
\begin{center}
\includegraphics[height=0.3\textheight,width=7cm,angle=270]{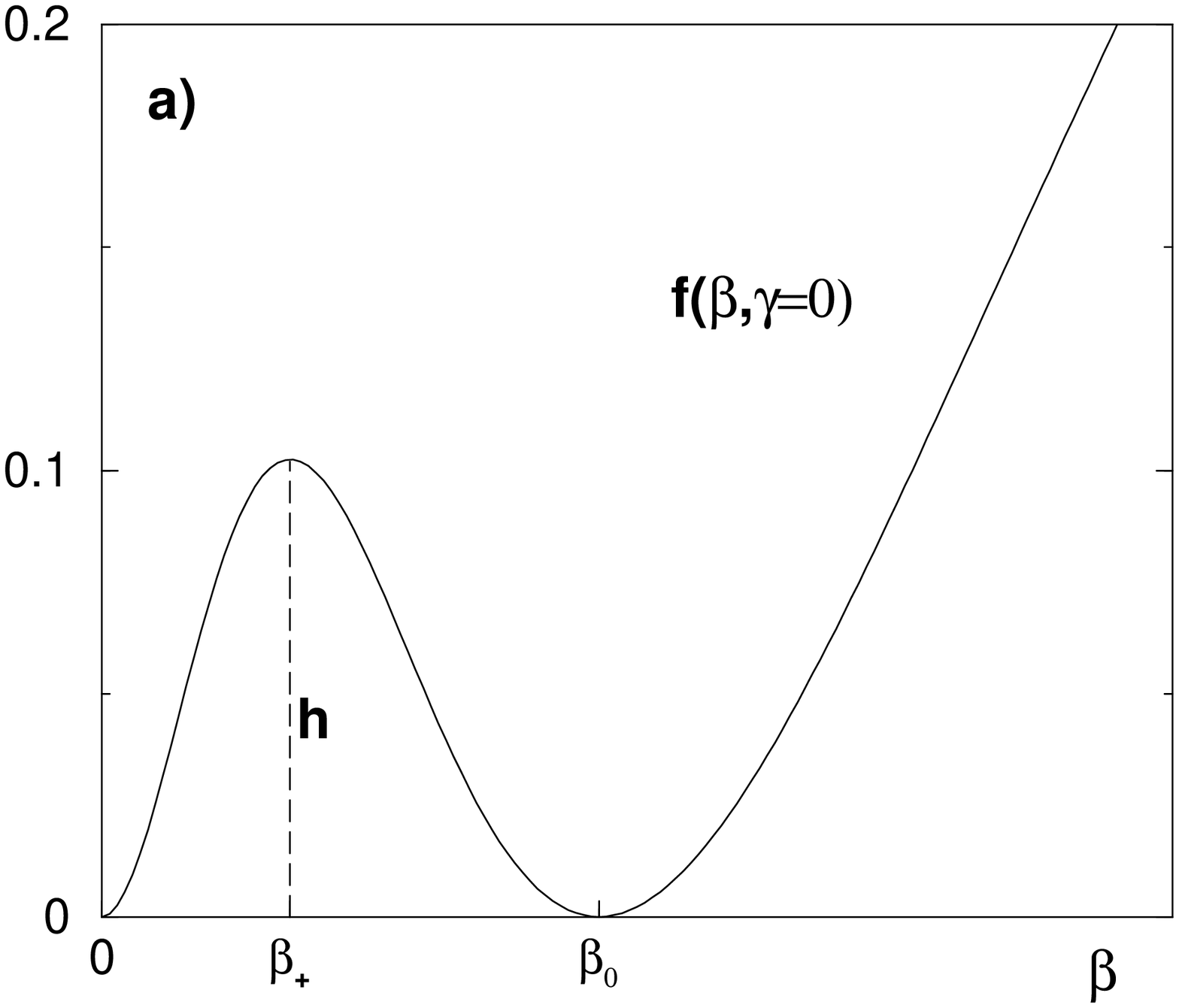}
\hspace{0.5cm}
\includegraphics[height=0.3\textheight,width=7cm,angle=270]{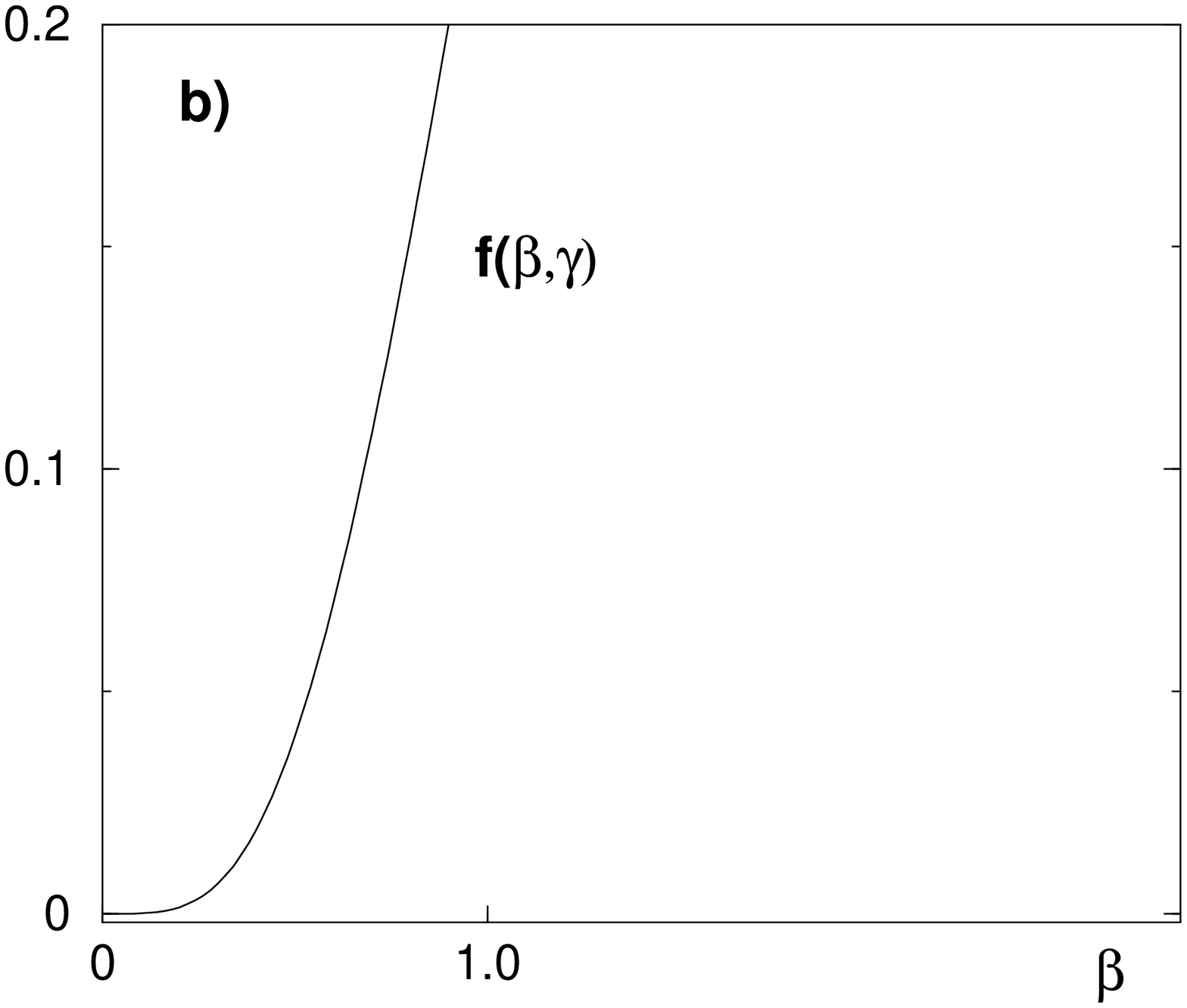}
\caption{
\small
Energy surfaces at the critical points, Eq.~(\ref{1st2nd}). 
(a)~First-order transition. The position and height of the barrier 
are $\beta= \beta_{+} = (-1 + \sqrt{1+\beta_{0}^2}\;)/\beta_0\,$ 
and $h = f(\beta_{+},\gamma=0) = 
( -1 + \sqrt{1+\beta_{0}^2}\,)^2/4 \,$ respectively. 
(b)~Second-order transition. In this case 
$f(\beta,\gamma)$ is independent of $\gamma$. 
Asymptotically, $f(\beta\to\infty,\gamma)=1$. 
\label{figcrisurf}}
\end{center}
\end{figure} 
Phase transitions can be studied 
by IBM Hamiltonians of the form, 
$\hat{H}(\alpha) = (1-\alpha)\,\hat{H}_{1} + \alpha\, \hat{H}_{2}$, involving 
terms from different dynamical symmetry chains~\cite{diep80}. 
The nature of the phase transition is 
governed by the topology of the corresponding surface~(\ref{eLan}), 
which serves as a Landau's potential 
with the equilibrium deformations as order parameters. 
The conditions on the parameters 
and resulting surfaces at the critical-points 
of first- and second-order transitions are given by
\bsub
\ba
1^{st}\, {\rm order}\quad 
b^{2}=4ac,\;a>0,\; b >0
&& f(\beta,\gamma=0) = 
c(1+\beta^2)^{-2}\beta^2\left ( \beta-\beta_0\right )^2 ~,
\qquad\;\;
\label{1st}
\\
2^{nd}\, {\rm order} \quad
a=0,\; b=0,\; c>0 \;\;\;\;
&& f(\beta,\gamma) =  c(1+\beta^2)^{-2}\beta^4 ~.
\label{2nd}
\ea
\label{1st2nd}
\esub 
As shown in Fig.~\ref{figcrisurf}, 
the first-order critical-surface has degenerate spherical and 
deformed minima at $\beta=0$ and $(\beta=\beta_0>0,\gamma=0)$, 
where $\beta_0 =2a/b$. The position ($\beta_{+}$) and height ($h$) of 
the barrier are indicated in the caption. 
The second-order critical-surface is independent of $\gamma$ 
and is flat bottomed $(\sim \beta^4)$ for small $\beta$. 
The conditions on $a,\,b,\,c$ in Eq.~(\ref{1st2nd}) 
fix the critical value of the control 
parameter $(\alpha=\alpha_c)$ which, in turn, determines the critical-point 
Hamiltonian, $\hat{H}_{cri}=\hat{H}(\alpha=\alpha_c)$. 
IBM Hamiltonians of this type have been used extensively for studying 
shape-phase transitions in 
nuclei~\cite{diep80,iaczam04,rowe0405,lev07,levgin03,gilmore79,
iac0001,caszam0001,lev05,lev06,QPTrev}. 
We now show that a large class of such critical-point Hamiltonians 
exhibit PDS~\cite{lev07}.

The spherical to deformed $\gamma$-unstable shape-phase transition 
is modeled in the IBM by the Hamiltonian
\ba
\hat{H}_{cri} &=& \epsilon\,\hat{n}_d + 
A \left[\, d^{\dagger}\cdot d^{\dagger} -  (s^{\dagger})^2\,\right ]
\left[\, H.c.\,\right]
\nonumber\\
\epsilon &=& 4(N-1)A ~.
\label{hcri2nd}
\ea
The $A$-term is the O(6) pairing term of Eq.~(\ref{HPSo6}). 
$\hat{H}_{cri}$ satisfies 
condition~(\ref{2nd}) with $c=4A$, hence 
qualifies as a second-order critical Hamiltonian. 
It involves a particular combination of the U(5) and O(6) Casimir 
operators, hence is recognized to be a special case of the Hamiltonian 
$\hat{H}_{{\rm O(5)}}$ of Eq.~(\ref{hPDSo5}) with O(5)-PDS of type II. 
In fact, since O(5) is a good symmetry 
common to both the U(5) and O(6) chains~(\ref{u5o6}), 
the O(5) PDS is valid 
throughout the U(5)-O(6) transition region. 
As mentioned at the end of Subsection~\ref{subsec:o5PDStypeII}, 
$\hat{H}_{{\rm O(5)}}$ and, therefore, $\hat{H}_{cri}$~(\ref{hcri2nd}), 
has also U(5)-PDS of type I, with the following solvable U(5) basis states
\bsub
\ba
&&\vert [N], n_d=\tau=N,L\,\rangle \;\;\qquad\quad 
E = \epsilon\,N ~,\qquad\qquad\\
&&\vert [N], n_d=\tau=N-1,L\,\rangle \;\;\;\quad 
E = \epsilon\,\,(N-1) ~, \qquad\qquad
\ea
\label{ecriu5o6}
\esub
where $L$ takes the values compatible with the 
${\rm O(5)}\supset{\rm O(3)}$ reduction. 

The dynamics at the critical point of a spherical to prolate-deformed 
shape-phase transition can be modeled in the IBM by the 
following Hamiltonian~\cite{lev06}
\begin{figure}[t]
\begin{center}
\rotatebox{270}{\includegraphics[width=2.6in,clip=]{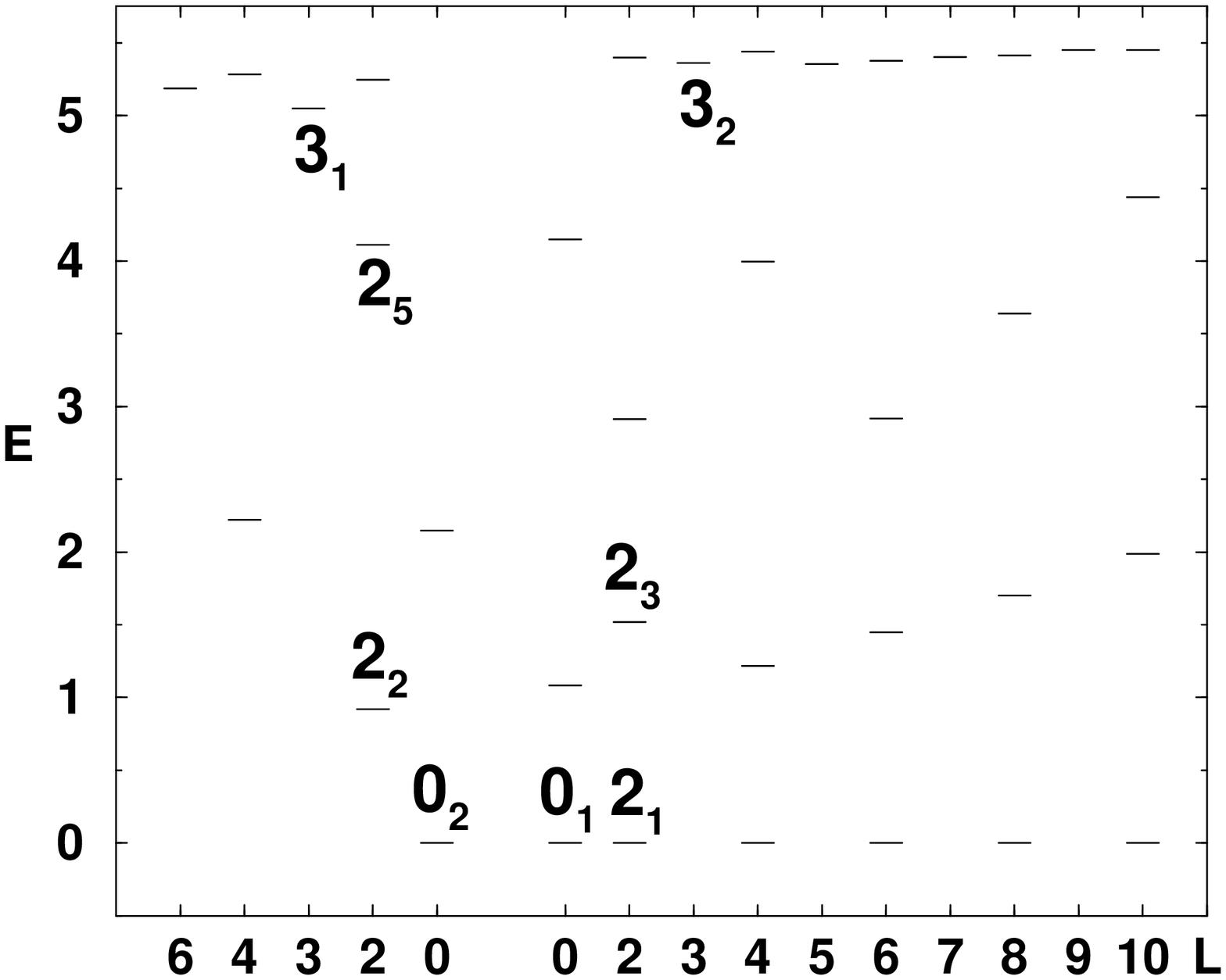}} 
\hspace{0.1cm}
\rotatebox{270}{\includegraphics[width=2.6in,clip=]{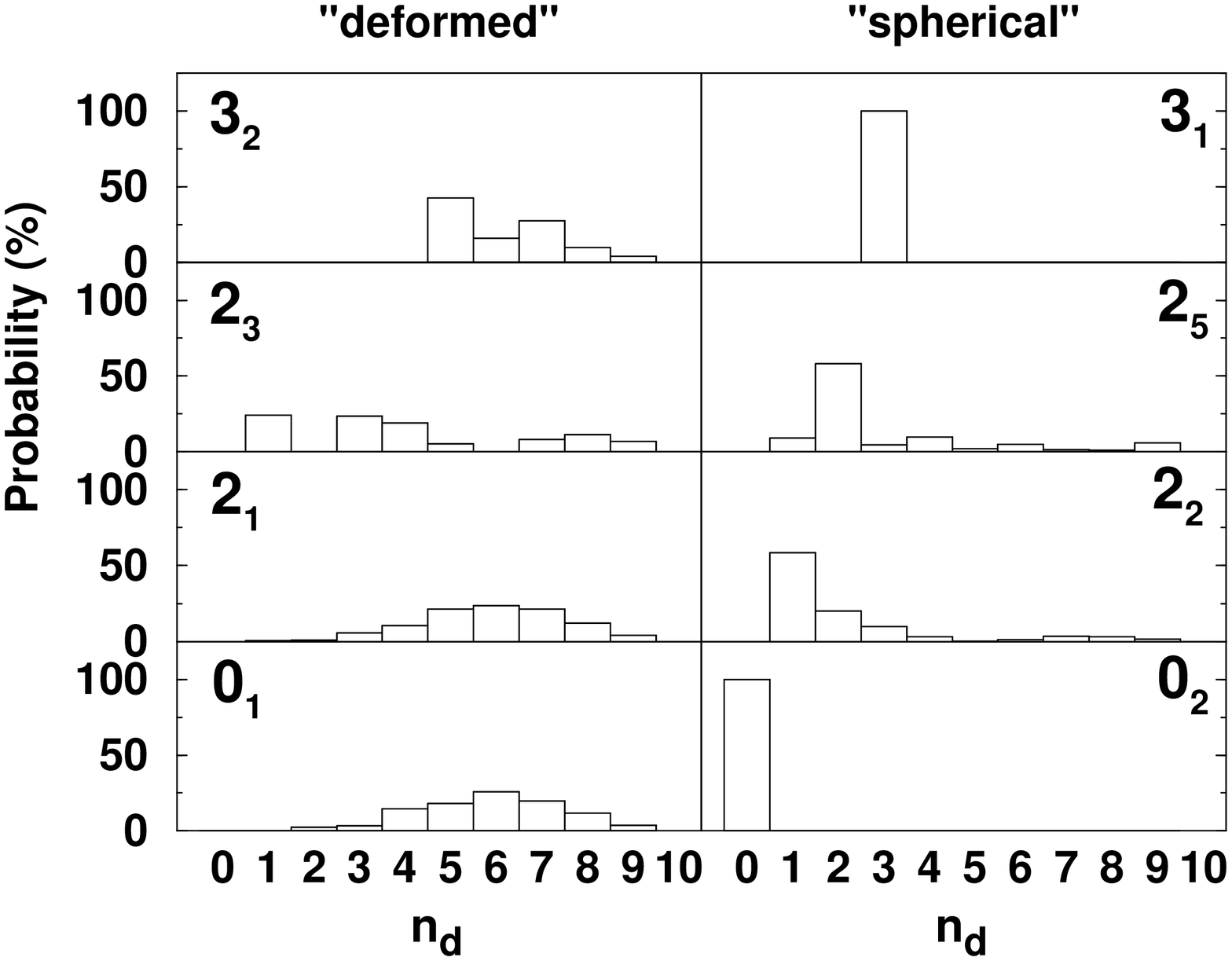}}
\hspace{0.1cm}
\end{center}
\vspace{-0.5cm}
\caption{
\small
Left: spectrum of a first-order critical Hamiltonian 
$\hat{H}_{cri}(\beta_0)$, Eq.~(\ref{hcri1st}), 
with $h_2=0.05$, $\beta_0 =1.3$ and $N=10$. 
The solvable eigenstates are 
the deformed states, Eq.~(\ref{deform}), forming a zero-energy 
$K=0_1$ ground band and the spherical states, 
$L=0_2,\,3_1$, Eq.~(\ref{spher}), with good U(5) symmetry. 
Right: U(5) ($n_d$) decomposition for 
selected eigenstates of $\hat{H}_{cri}(\beta_0)$. 
Adapted from~\cite{lev06}. 
\label{fighcri1stspec}}
\end{figure}
\ba
\hat{H}_{cri}(\beta_0) &=& h_{2}\, 
P^{\dagger}_{2}(\beta_0)\cdot\tilde{P}_{2}(\beta_0) ~,
\label{hcri1st}
\ea 
where $P^{\dagger}_{2\mu}(\beta_0)$ 
is the $L=2$ boson-pair of Eq.~(\ref{P2b0}) and $h_2,\,\beta_0>0$. 
The corresponding surface in Eq.~(\ref{eLan}) has coefficients  
$a=2h_2\beta_{0}^2, b=4h_{2}\beta_{0},c=2h_2$, which satisfy 
condition~(\ref{1st}). This qualifies $\hat{H}_{cri}(\beta_0)$ 
as a first-order critical Hamiltonian whose potential accommodates 
two degenerate minima at $\beta=0$ and $(\beta,\gamma)=(\beta_0,0)$. 
$\hat{H}_{cri}(\beta_0)$ is recognized to be a special case of the 
partially-solvable Hamiltonian, $\hat{H}_{PSolv}$ of Eq.~(\ref{hPSolv}). 
As such, it has a solvable prolate-deformed ground band, composed 
of the states of Eq.~(\ref{ePSolv}) 
\ba
\vert\beta_0;N,L\rangle \quad E=0\qquad\;\; L=0,2,4,\ldots, 2N~.
\qquad\quad
\label{deform}
\ea
On the other hand, the following multipole form of 
$\hat{H}_{cri}(\beta_0)$
\ba
&&\hat{H}_{cri}(\beta_0) =
\nonumber\\
&&\qquad\;\;\;
h_{2}\left [2(\beta_{0}^2\hat{N} -2)\hat{n}_d 
+ 2(1-\beta_{0}^2)\hat{n}_{d}^2 
+ 2 \hat{C}_{{\rm O(5)}} - \hat{C}_{{\rm O(3)}}
+ \sqrt{14}\,\beta_{0}\,\Pi^{(2)}\cdot U^{(2)}\right ]
\qquad\quad
\label{hcri1stmult}
\ea
identifies it as the Hamiltonian of Eq.~(\ref{hPDSu5}) 
with U(5)-PDS of type I. As such, it has also 
the solvable spherical eigenstates of Eq.~(\ref{ePDSu5}), 
with good U(5) symmetry 
\bsub
\ba 
\vert N,n_d=\tau=L=0 \rangle  \;\; &&E = 0
\label{nd0b0}\\
\vert N,n_d=\tau=L=3 \rangle \;\;
&&E = 6 h_2[\beta_{0}^2 (N-3) + 5]~.
\qquad\quad
\label{nd3b0}
\ea
\label{spher}
\esub
The spectrum of $\hat{H}_{cri}(\beta_0)$~(\ref{hcri1st}) and the 
U(5) ($n_d$) decomposition of selected eigenstates is shown in 
Fig.~\ref{fighcri1stspec}. The spectrum displays a coexistence of 
spherical states (some of which solvable with 
good U(5) symmetry) and deformed states (some of which solvable), 
signaling a first-order transition. 
The remaining non-solvable states in the spectrum are either predominantly 
spherical (with characteristic dominance of single $n_d$ 
components) or deformed states (with a broad $n_d$ distribution) 
arranged in several excited bands~\cite{lev06}. 

\begin{figure}[t]
\begin{center}
\rotatebox{270}{\includegraphics[width=2.5in,clip=]
{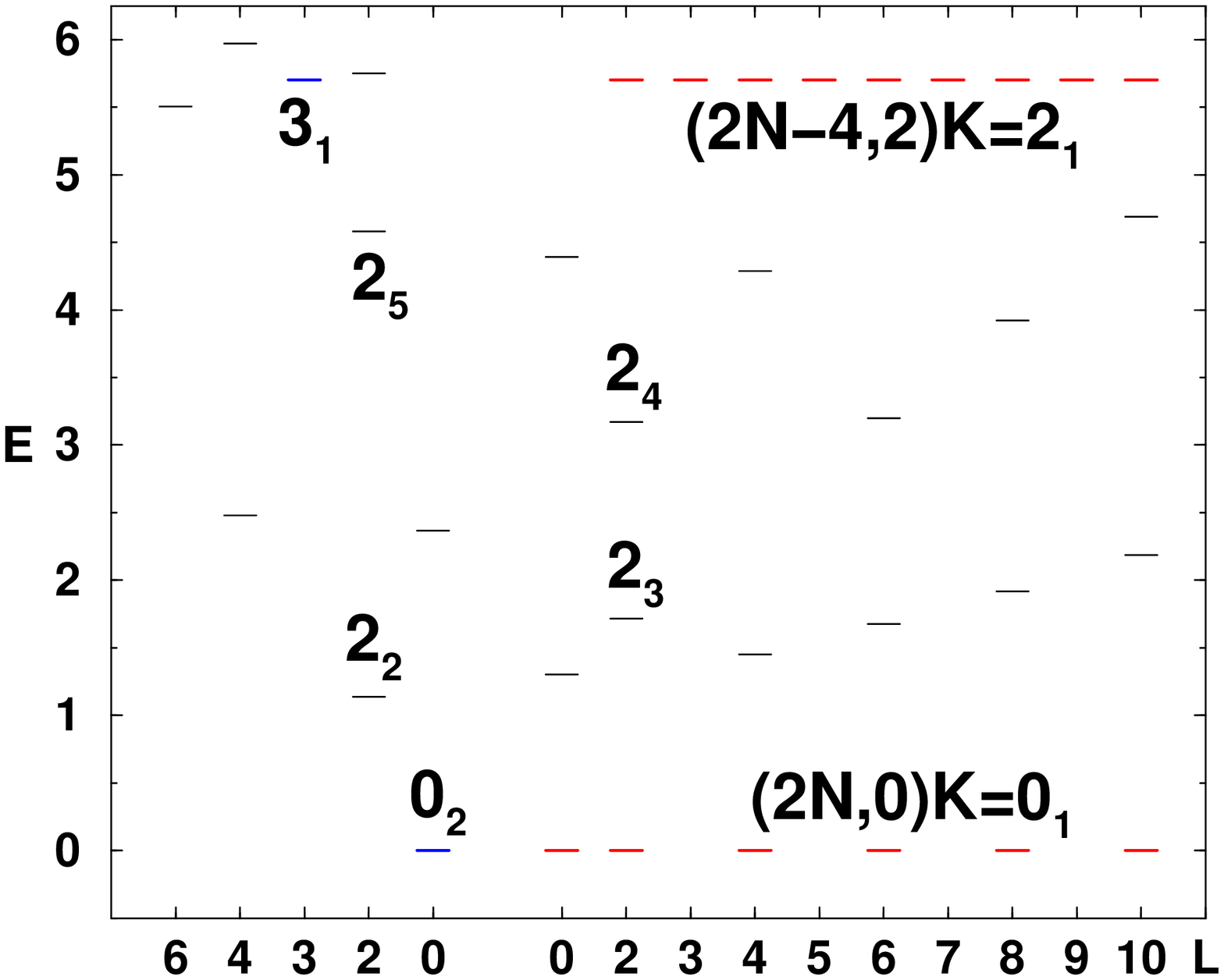}}\hspace{0.2cm}
\hspace{0.1cm}
\rotatebox{270}{\includegraphics[width=2.5in,clip=]
{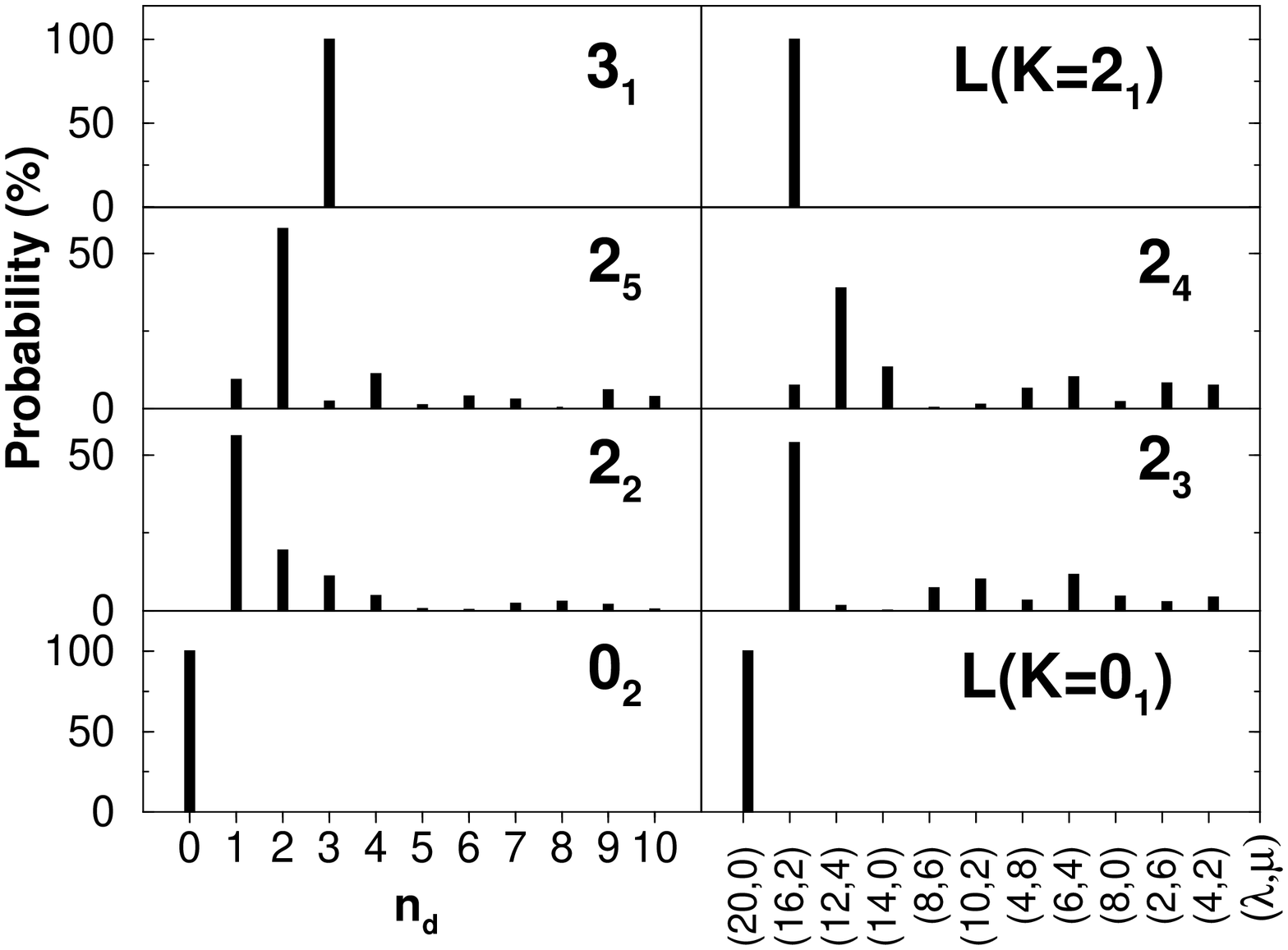}}\hspace{0.2cm}
\caption{
\small
Left: spectrum of $\hat{H}_{cri}(\beta_0=\sqrt{2})$, 
Eq.~(\ref{hcri1st}), with $h_2=0.05$ and $N=10$. 
$L(K=0_1)$ and $L(K=2_1)$ are the solvable SU(3) states 
of Eq.~(\ref{solsu3g}) and Eq.~(\ref{solsu3gam}) with $k=1$, respectively. 
$L=0_2,3_1$ are the solvable U(5) states of Eq.~(\ref{spher}). 
Right: U(5) ($n_d$) and SU(3) $[(\lambda,\mu)]$ decomposition for 
selected eigenstates of $\hat{H}_{cri}(\beta_0=\sqrt{2})$. 
Adapted from~\cite{lev07}.
\label{fighcrisu3}}
\end{center}
\end{figure}

The critical Hamiltonian of Eq.~(\ref{hcri1st}) with $\beta_0=\sqrt{2}$ 
is a special case of the Hamiltonian of Eq.~(\ref{hPDSsu3}), 
shown to have SU(3)-PDS of type I. 
As such, it has a subset of solvable states, 
Eqs.~(\ref{gband})-(\ref{gamband}), 
which are members of the ground  $g(K=0)$ and $\gamma^{k}(K=2k)$ bands, 
with good SU(3) symmetry, $(\lambda,\mu)=(2N-4k,2k)$
\bsub
\ba
&&\vert N,(2N,0)K=0,L\rangle \;\;\;\; E = 0
\qquad 
L=0,2,4,\ldots, 2N
\label{solsu3g}
\\
&&\vert N,(2N-4k,2k)K=2k,L\rangle
\;\;\;\;\;\;\;\; 
E =  h_{2}\,6k \left (2N - 2k+1 \right )
\qquad\qquad
\nonumber\\
&&
L=K,K+1,\ldots, (2N-2k) \qquad\; k>0 ~. 
\label{solsu3gam}
\ea
\label{solsu3}
\esub
In addition, $\hat{H}_{cri}(\beta_0=\sqrt{2})$ has the spherical states of 
Eq.~(\ref{spher}), with good U(5) symmetry, as eigenstates. 
The spherical $L=0$ state, Eq.~(\ref{nd0b0}), is 
exactly degenerate with the SU(3) ground band, Eq.~(\ref{solsu3g}), 
and the spherical $L=3$ state, Eq.~(\ref{nd3b0}), 
is degenerate with the SU(3) $\gamma$-band, Eq.~(\ref{solsu3gam}) 
with $k=1$. 
The remaining levels of $\hat{H}_{cri}(\beta_0=\sqrt{2})$, shown 
in Fig.~\ref{fighcrisu3} are calculated numerically and 
their wave functions are spread over many 
U(5) and SU(3) irreps. This situation, 
where some states are solvable with good U(5) symmetry, 
some are solvable with good SU(3) symmetry and all other 
states are mixed with respect to both U(5) and SU(3), 
defines a U(5) PDS of type I 
coexisting with a SU(3) PDS of type I.

\begin{figure}[t]
\begin{center}
\rotatebox{270}{\includegraphics[width=2.5in,clip=]
{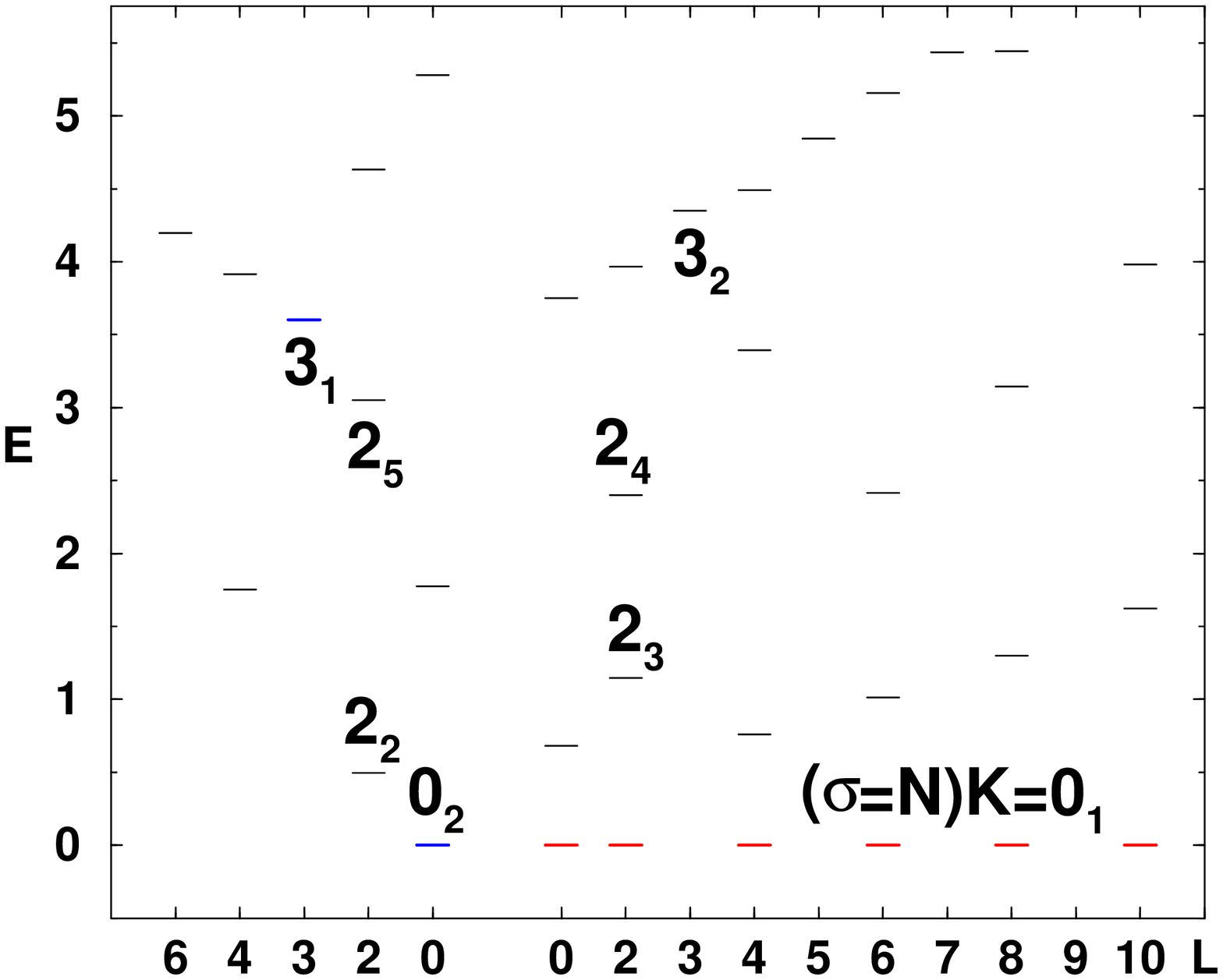}}\hspace{0.2cm}
\end{center}
\begin{center}
\rotatebox{270}{\includegraphics[width=2.5in,clip=]
{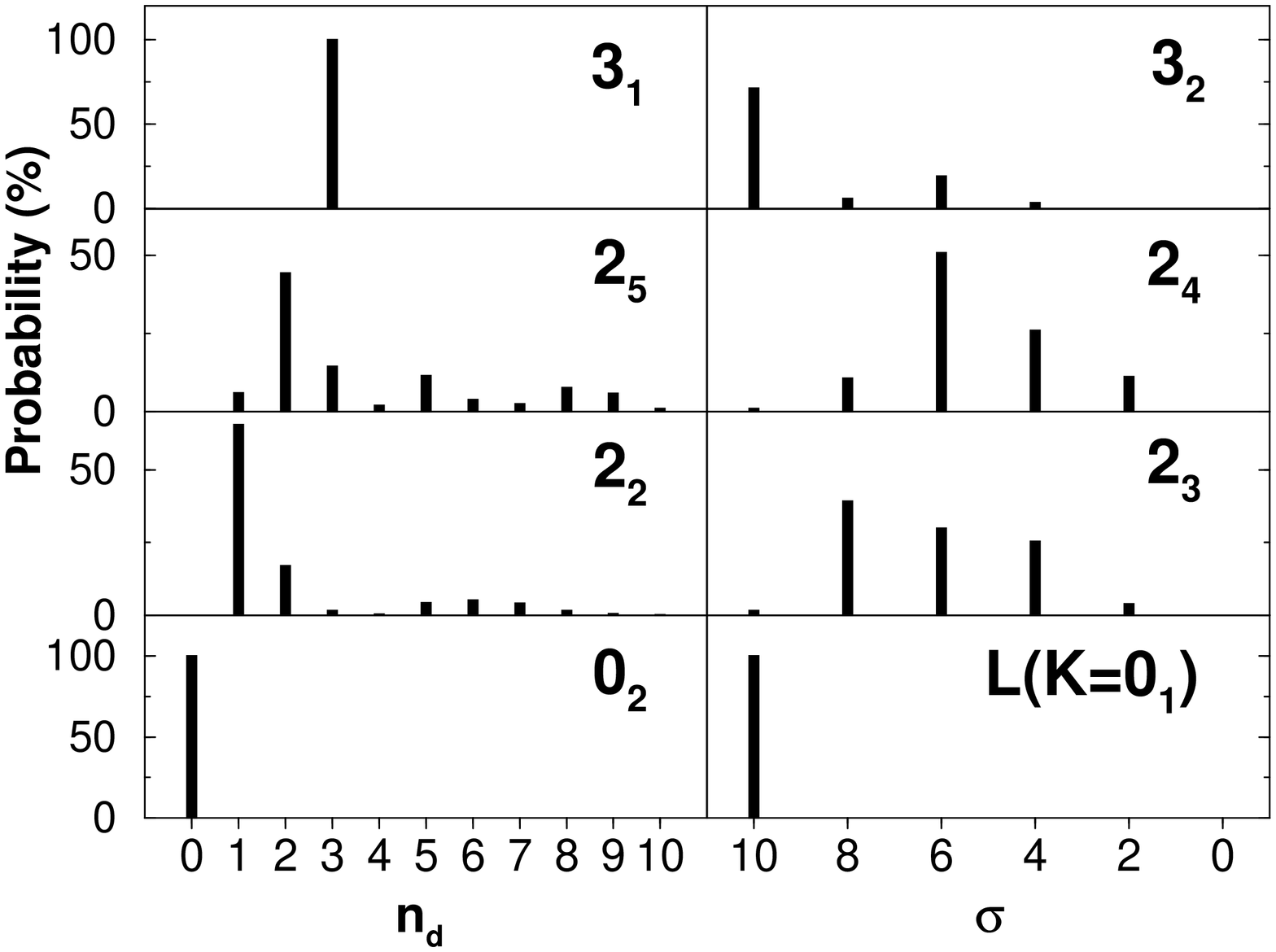}}
\hspace*{0.1in}
\rotatebox{270}{\includegraphics[width=2.5in,clip=]
{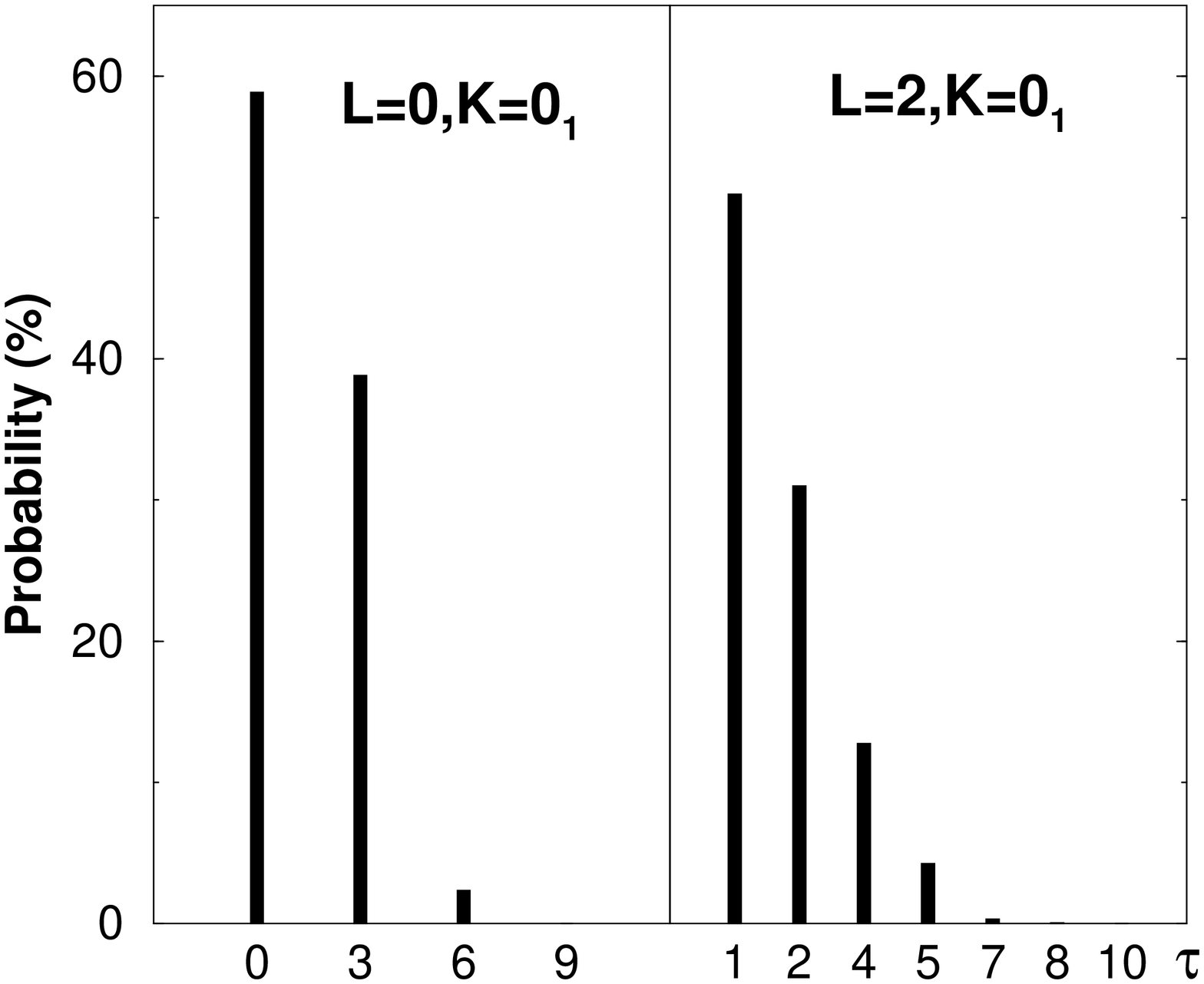}}
  \caption{
\footnotesize 
Upper panel: spectrum of $\hat{H}_{cri}(\beta_0=1)$, Eq.~(\ref{hcri1st}), 
with $h_2=0.05$ and $N=10$. 
$L(K=0_1)$ are the solvable states of Eq.~(\ref{solo6}) with good 
O(6) but broken O(5) symmetry.
$L=0_2,3_1$ are the solvable U(5) states of Eq.~(\ref{spher}). 
Bottom left panel: U(5) ($n_d$) and O(6) $(\sigma)$ decomposition for 
selected spherical and deformed eigenstates of $\hat{H}_{cri}(\beta_0=1)$. 
Bottom right panel: O(5) ($\tau$) decomposition for the $L=0,\,2$ states, 
Eq.~(\ref{solo6}), members 
of the ground band ($K=0_1$) of $\hat{H}_{cri}(\beta_0=1)$.
Both states have O(6) symmetry $\sigma=N$.
Adapted from~\cite{lev07}.
\label{fighcrio6}}
\end{center}
\end{figure}

The critical Hamiltonian of Eq.~(\ref{hcri1st}) with $\beta_0=1$ 
is a special case of the Hamiltonian of Eq.~(\ref{h2PDS}), shown 
to have O(6)-PDS of type III. 
As such, it has a subset of solvable states
Eq.~(\ref{ePDSo6III}), which are members of a prolate-deformed 
ground band, with good O(6) symmetry, 
$\langle\sigma\rangle = \langle N \rangle$, but broken O(5) symmetry
\ba
\vert N,\sigma=N,L\rangle \qquad E=0
\qquad L=0,2,4,\ldots, 2N ~.
\quad
\label{solo6}
\ea
In addition, $\hat{H}_{cri}(\beta_0=1)$ has the spherical states of 
Eq.~(\ref{spher}), with good U(5) symmetry, as eigenstates. 
The remaining eigenstates of $\hat{H}_{cri}(\beta_0=1)$ shown 
in Fig.~\ref{fighcrio6} are mixed with respect to both U(5) and O(6). 
Apart from the solvable U(5) states 
of Eq.~(\ref{spher}), all eigenstates of $\hat{H}_{cri}(\beta_0=1)$ 
are mixed with respect to O(5) [including the solvable 
O(6) states of Eq.~(\ref{solo6}), 
as shown in the bottom right panel of Fig.~\ref{fighcrio6}]. 
It follows that the Hamiltonian has a subset of states 
with good U(5) symmetry and a subset of states with good O(6) 
but broken O(5) symmetry, and all other states are mixed with respect 
to both U(5) and O(6). These are precisely the required features of 
U(5) PDS of type I coexisting with O(6) PDS of type III.

In conclusion, the above results demonstrate the relevance of the PDS notion 
to critical-points of QPT, with phases characterized by Lie-algebraic 
symmetries. In the example considered, second-order critical Hamiltonians 
mix incompatible symmetries 
but preserve a common lower symmetry, resulting in a single PDS 
with selected quantum numbers conserved. 
First-order critical Hamiltonians 
exhibit distinct subsets of solvable states with good symmetries, 
giving rise to a coexistence of different PDS. 
The ingredients of an algebraic description 
of QPT is a spectrum generating algebra and an associated geometric 
space, formulated in terms of coherent (intrinsic) states. 
The same ingredients are used in the construction of Hamiltonians 
with PDS. These, in accord with the present discussion, 
can be used as tools to explore the role of partial symmetries in 
governing the critical behaviour of dynamical systems undergoing QPT.

\section{PDS and Mixed Regular and Chaotic Dynamics}
\label{sec:PDSChaos}

Partial dynamical symmetries can play a role not only for discrete 
spectroscopy but also for analyzing statistical aspects of 
nonintegrable systems~\cite{walev93,levwhe96}. 
Hamiltonians with dynamical symmetry are always completely 
integrable~\cite{zhang88}. 
The Casimir invariants of the algebras in the chain provide a set 
of constants of the motion in involution. The classical motion is purely 
regular. A dynamical symmetry-breaking is connected 
to nonintegrability and may give rise to chaotic 
motion~\cite{zhang88,zhang89,zhang90}. 
Hamiltonians with PDS are not completely integrable, 
hence can exhibit stochastic behavior, nor are they completely chaotic, 
since some eigenstates preserve the symmetry exactly. 
Consequently, such Hamiltonians are optimally suitable to the study
of mixed systems with coexisting regularity and chaos. 

The dynamics of a generic classical Hamiltonian system is 
mixed~\cite{bohig93};  
KAM islands of regular motion and chaotic regions coexist
in phase space. 
In the associated quantum system, if no separation between regular and 
irregular states is done, the statistical properties of the spectrum 
are usually intermediate between the Poisson and the Gaussian orthogonal 
ensemble (GOE) statistics. 
In a PDS of type I, the symmetry of the subset of solvable states 
is exact, yet does not arise from invariance properties of the 
Hamiltonian. This offers an important 
opportunity to study how the existence of partial (but exact) symmetries 
affects the dynamics of the system. If the fraction of solvable states 
remains finite in the classical limit, one might expect that a corresponding 
fraction of the phase space would consist of KAM tori and 
exhibit regular motion. It turns out that 
PDS has an even greater effect on the dynamics. 
It is strongly 
correlated with suppression ({\it i.e.}, reduction) of chaos even though the 
fraction of solvable states approaches zero in the classical 
limit~\cite{walev93,levwhe96}. 

We consider the IBM Hamiltonian of Eq.~(\ref{hPS})
\ba
\hat{H}(\beta_0) &=& h_{0}\, 
P^{\dagger}_{0}(\beta_{0})P_{0}(\beta_0) 
+ h_{2}\,P^{\dagger}_{2}(\beta_0)\cdot \tilde{P}_{2}(\beta_0) ~.
\label{Hchaos}
\ea 
As discussed in Section~\ref{sec:PartialSolv}, 
when $\beta_0=\sqrt{2}$, the 
Hamiltonian~(\ref{Hchaos}) has an SU(3)-PDS of type I. 
In this case, the solvable states 
are those of Eqs.~(\ref{gband})-(\ref{gamband}).
At a given spin per boson $l=L/N$, and 
to leading order in $1/N$, the fraction $f$ of solvable states 
decreases like $1/N^2$ with boson number.
However, at a given boson number $N$, this fraction increases with $l$, 
a feature which is valid also for finite $N$~\cite{walev93}.
The classical limit of~(\ref{Hchaos}) is 
obtained~\cite{hatch82,alnovo90,alw91} 
through the use of coherent states parametrized by the six complex numbers 
$\{\alpha_s,\alpha_{\mu};\mu=-2,\ldots,2\}$ and taking $N\to\infty$. 
The classical Hamiltonian is then obtained from~(\ref{Hchaos}) 
by the substitution 
$s^{\dag},d^{\dag}_{\mu}\to\alpha_{s}^{*},\alpha_{\mu}^{*}$ and 
$s,d_{\mu}\to\alpha_{s},\alpha_{\mu}$ and rescaling the parameters 
$h_{i}\to Nh_{i}$ $(i=0,2)$. Here $1/N$ plays the role of $\hbar$. 
\begin{figure}[t]
\begin{center} 
\includegraphics[height=5in]{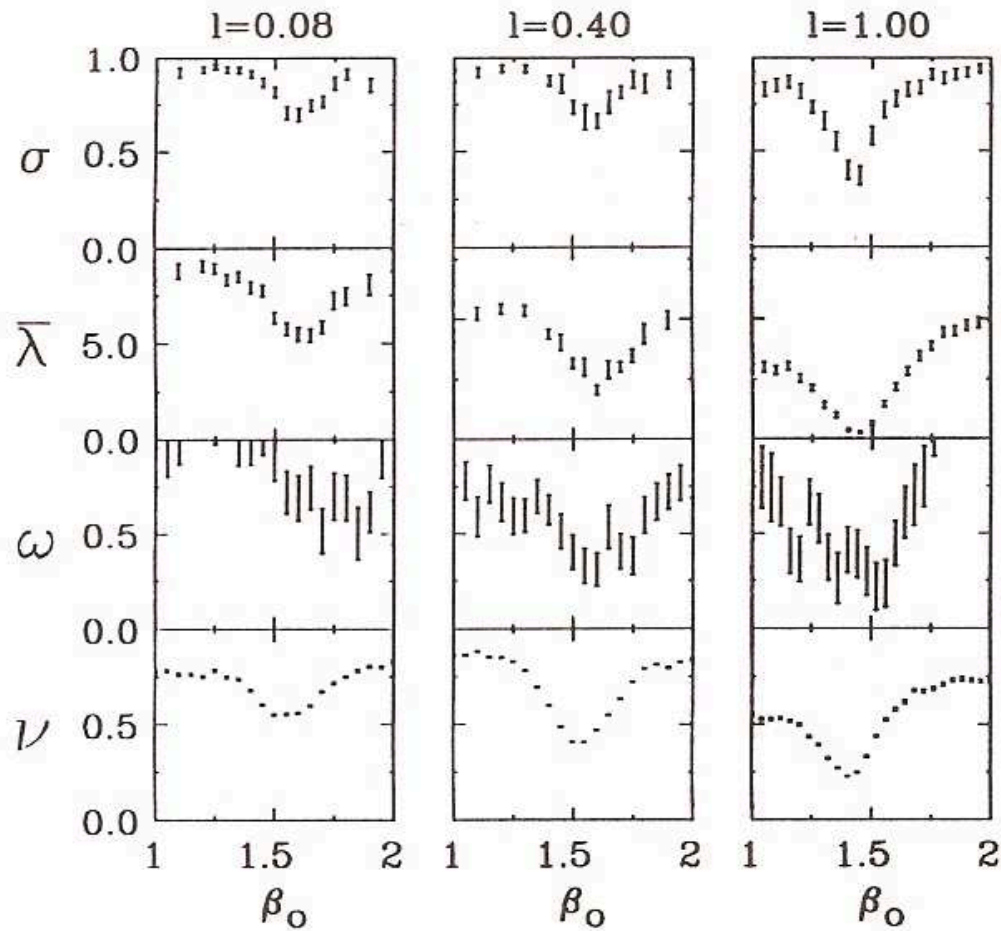}
\caption{
\small
Classical $(\sigma,\bar{\lambda})$ and quantal $(\omega,\nu)$ measures of 
chaos versus $\beta_0$ for the Hamiltonian~(\ref{Hchaos}) with $h_2/h_0=7.5$. 
Shown are three cases with classical spins $l=0.08,\, 0.4$, and $1$. 
The quantal calculations $(\omega,\nu)$ are done for $N=25$ bosons and spins 
$L=2,10$, and $25$, respectively. Notice that with increasing spin the 
minimum gets deeper and closer to $\beta_0=\sqrt{2}$. The suppression of 
chaos near $\beta_0=\sqrt{2}$ is seen both for finite $N$ through the 
measures $\omega,\,\nu$ and in the classical limit $N\to\infty$ through the 
measures $\sigma,\,\bar{\lambda}$.
Adapted from~\cite{walev93}.
\label{figchaossu3}}
\end{center}
\end{figure}

To study the effect of the SU(3) PDS on the dynamics, 
we fix the ratio $h_{2}/h_{0}$ at a value far from the exact SU(3) symmetry 
(for which $h_0/h_2 =1)$. 
We then change $\beta_0$ in the range $1\leq\beta_0\leq 2$. 
Classically, we determine the fraction $\sigma$ of chaotic volume and the 
average largest Lyapunov exponent $\bar{\lambda}$. To analyze 
the quantum Hamiltonian, we study spectral and transition intensity 
distributions. The nearest neighbors level spacing distribution is 
fitted by a Brody distribution, 
$P_{\omega}(S) = AS^{\omega}\exp(-\alpha S^{1+\omega})$, 
where $A$ and $\alpha$ are determined by 
the conditions that $P_{\omega}(S)$ is normalized to $1$ and 
$\langle S\rangle =1$. For the Poisson statistics $\omega=0$ and for 
GOE $\omega=1$, corresponding to integrable and fully chaotic classical 
motion~\cite{berry77, bohig84}, respectively. The intensity distribution 
of the SU(3) E2 operator, $Q^{(2)}$ of Eq.~(\ref{Te2su3}), 
is fitted by a $\chi^2$ distribution in $\nu$ degrees of 
freedom~\cite{levine86},
$P_{\nu}(y) = 
[(\nu/2\langle y\rangle)^{\nu/2}/\Gamma(\nu/2)]y^{\nu/2-1}
\exp(-\nu y/2\langle y\rangle)$. 
For the GOE, $\nu=1$ and $\nu$ decreases as the dynamics become more 
regular.

Fig.~\ref{figchaossu3} 
shows the two classical measures $\sigma$, $\bar{\lambda}$ and 
the two quantum measures $\omega$, $\nu$ for the Hamiltonian~(\ref{Hchaos}) 
as a function of $\beta_0$. The parameters of the Hamiltonian are taken to be 
$h_2/h_0=7.5$ and the number of bosons is $N=25$. Shown are three 
classical spins $l=0.08,\, 0.4$ and $1$, which correspond in the 
quantum case to $L=2,\, 10$ and $25$. All measures show a pronounced minimum 
which gets deeper and closer to $\beta_0=\sqrt{2}$ [where the partial SU(3) 
symmetry occurs] as the classical spin increases. This behaviour is 
correlated with the fraction of solvable states (at a constant $N$) being 
larger at higher $l$. 
We remark that the classical measures show a clear enhancement of the 
regular motion near $\beta_0=\sqrt{2}$ even though the fraction of solvable 
states vanishes as $1/N^2$ in the classical limit $N\to\infty$.  
\begin{figure}[t]
\begin{center} 
\includegraphics[height=5in]{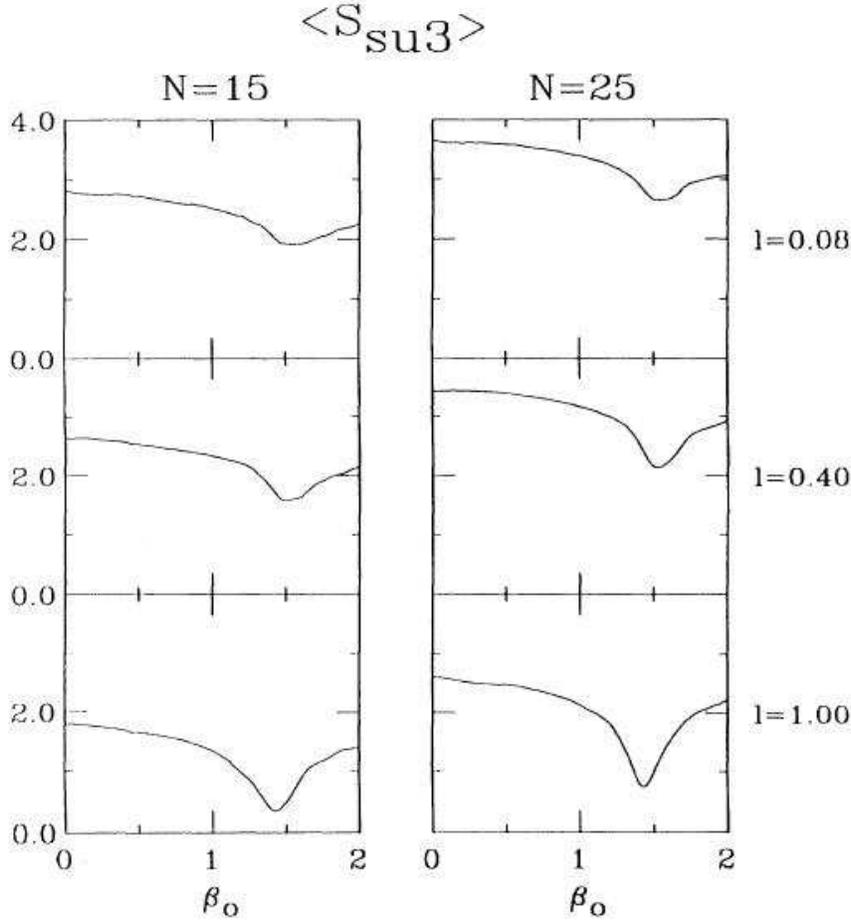}
\caption{
\small
The average SU(3) entropy of the eigenstates of the 
Hamiltonian~(\ref{Hchaos}) (for $h_2/h_0=7.5$) versus $\beta_0$, for 
three values of the spin (per boson), $l=0.08,\,0.4$, and $1$. Left: 
$N=15$ bosons; right: $N=25$ bosons. Adapted from~\cite{walev93}.
\label{figentropysu3}}
\end{center}
\end{figure}

To confirm that the observed suppression of chaos is related to the SU(3) 
PDS, we employ the concept of an entropy~\cite{iaclevine87,izra89} 
associated with a given symmetry. To determine the SU(3) entropy, we expand 
any eigenstate $\vert\alpha LM\rangle$ in an SU(3) basis, 
$\vert\alpha LM\rangle = 
\sum_{(\lambda\mu),K} c^{(\alpha)}_{(\lambda,\mu)K}
\vert(\lambda,\mu)KLM\rangle$. 
Denoting by $p^{(\alpha)}_{\lambda\mu}$ the probability to be in the SU(3) 
irrep $(\lambda,\mu)$, $p^{(\alpha)}_{\lambda\mu} = 
\sum_{K}\vert c^{(\alpha)}_{(\lambda,\mu)K}\vert^2$, 
the SU(3) entropy of the state $\vert\alpha L M\rangle $ is defined as
$S^{(\alpha)}_{SU(3)} = - \sum_{\lambda,\mu} p^{(\alpha)}_{\lambda\mu} 
\ln p^{(\alpha)}_{\lambda\mu}$. 
The entropy vanishes when the state has a good SU(3) symmetry. The averaged 
entropy $\langle S_{{\rm SU(3)}}\rangle$ over all eigenstates is 
then a measure of the global SU(3) symmetry. This quantity is plotted in 
Fig.~\ref{figentropysu3}, versus 
$\beta_0$ for $N=15$ and 25 and for the same spin values (per boson) 
$l$ as in Fig.~\ref{figchaossu3}. 
We observe a minimum which is well correlated with the 
minimum in Fig.~\ref{figchaossu3}. 
The maximum SU(3) entropy is the logarithm of the number 
of allowed SU(3) irreps for the given $N$ and $l$. The average SU(3) entropy 
therefore increases with $N$. The depth of the minimum increases with 
$N$ and $l$ though the fraction of solvable states is smaller at $N=25$ 
than at $N=15$ by a factor of about 3. The existence of an SU(3) PDS 
seems to have an effect of increasing the SU(3) symmetry 
of all states, not just those with an exact SU(3) symmetry~\cite{walev93}. 

In order to better understand the strong suppression of classical chaos 
induced by PDS, we consider a simpler model and use its PDS 
to infer relationships between the classical and quantum dynamics 
of a Hamiltonian in a mixed KAM r\'egime~\cite{levwhe96}. 
The model is based on a U(3) spectrum generating algebra and its 
building blocks are three types of bosons
$a^{\dagger}$, $b^{\dagger}$, $c^{\dagger}$ satisfying the usual
commutation relations.
The nine number-conserving bilinear products of creation and destruction
operators comprise the U(3) algebra. The conservation
of the total boson-number $\hat{N}=\hat{n}_a+\hat{n}_{b}+\hat{n}_c$
($\hat{n}_a =a^{\dagger}a$ with eigenvalue $n_a$ etc.) 
ensures that the model describes a system
with only two independent degrees of freedom. All states of the model are
assigned to the totally symmetric representation [N] of U(3).
One of the dynamical symmetries of the model is associated with the
following chain of algebras
\begin{equation} \label{chainu3}
{\rm U(3)} \supset {\rm U(2)} \supset {\rm U(1)}
\end{equation}
Here ${\rm U}(2)\equiv {\rm SU}(2)\times {\rm U}_{ab}(1)$ with a 
linear Casimir $\hat{n}_{ab}=\hat{n}_a+\hat{n}_b$ [which is also the 
generator of ${\rm U}_{ab}(1)$]. 
The generators of SU(2) are $\hat{J}_+ = b^{\dagger}a$,
$\hat{J}_-= a^{\dagger}b$, $\hat{J}_z=(\hat{n}_b-\hat{n}_a)/2$ and its 
Casimir $\mbox{\boldmath $J^2$}=\hat{n}_{ab}(\hat{n}_{ab}+2)/4$. 
The subalgebra U(1) 
in Eq.~(\ref{chainu3}) is composed of the operator $\hat{J}_z$.
A choice of Hamiltonian with a U(2) dynamical symmetry is
\begin{eqnarray}
\hat{H}_0 & = &
\omega_a a^{\dagger}a + \omega_b b^{\dagger}b
\;\; = \;\;
\hat{n}_{ab} - 2A\hat{J}_z
\label{h0}
\end{eqnarray}
where $\omega_{a,b} = 1 \pm A$, and $A$ is introduced to break degeneracies.
Diagonalization of this Hamiltonian is trivial and leads to eigenenergies
$E_{n_a,n_b}= \omega_a n_a + \omega_b n_b$ and eigenstates
$\vert n_a,n_b,n_c\rangle$ or equivalently $\vert N,J,J_z\rangle$
where the label $J=n_{ab}/2$ identifies the SU(2) irrep.
These are states with well defined $n_a$, $n_b$ and $n_c=N-n_a-n_b$.
To create a PDS 
we add the term
\begin{equation} 
\hat{H}_1 = b^{\dagger}(b^\dagger a + b^\dagger c + a^\dagger b
+ c^\dagger b)b ~,
\label{h1chaos}
\end{equation}
which preserves the total boson number but not the individual boson
numbers, so it breaks the dynamical symmetry.
However, states of the form $\vert n_a,n_b=0,n_c\rangle$ (or
equivalently $\vert N,J=n_a/2,J_z=-J\rangle$ ) with
$n_a=0,1,2,\ldots N$ are annihilated by $\hat{H}_1$ and therefore remain
eigenstates of $\hat{H}_0+B \hat{H}_1$. The latter Hamiltonian
is not an SU(2) scalar yet has a subset of $(N+1)$ ``special''
solvable states with SU(2) symmetry, and therefore has PDS.
There is one special state per SU(2) irrep
$J=n_{a}/2$ (the lowest weight state in each case) with energy
$\omega_an_a$ independent of the parameter B.
Other eigenstates are mixed. Although $\hat{H}_0$ and $\hat{H}_1$ 
do not commute,
when acting on the ``special'' states they satisfy
\begin{equation} \label{comm}
\Bigl [\hat{H}_0\, ,\, \hat{H}_1\Bigr ]\vert n_a,n_b=0,n_c\rangle\; = 0 ~.
\end{equation}
To break the PDS 
we introduce a third interaction
\begin{equation} 
\hat{H}_2 = a^{\dagger}c + c^{\dagger}a + b^{\dagger}c + c^{\dagger}b ~.
\label{h2chaos}
\end{equation}
The complete Hamiltonian is then
\begin{equation} 
\hat{H} = \hat{H}_0 + B\,\hat{H}_1 + C\,\hat{H}_2 ~.
\label{htot}
\end{equation}
For $B=C=0$ we have the full dynamical symmetry; for
$B\neq 0,\,C=0$ we have partial dynamical symmetry
and for $C\neq 0$ we have neither.

The classical Hamiltonian ${\cal H}_{cl}$ is obtained from~(\ref{htot})
by replacing $(a^\dagger,b^\dagger,c^\dagger)$ by complex c-numbers
$(\alpha^*,\beta^*,\gamma^*)$ and taking $N\rightarrow\infty$.
The latter limit is obtained 
by rescaling ${\bar B}=NB$, $\alpha\rightarrow \alpha/\sqrt{N}$ etc.
and considering the classical Hamiltonian per boson
${\cal H} = {\cal H}_{cl}/N$.
In the present model the latter has the form
\ba 
{\cal H} &=&
{\cal H}_0 + \bar{B}\,{\cal H}_1 + C\,{\cal H}_2 ~.
\label{cham}
\ea
Number conservation imposes a constraint
$\alpha^*\alpha+\beta^*\beta+\gamma^*\gamma=1$,
so that the phase space is compact and four-dimensional with a
volume $2\pi^2$. The total number of quantum states is $(N+1)(N+2)/2$.
Assigning, to leading order in $N$,
one state per $(2\pi\hbar)^2$ volume of phase space,
we identify
$\hbar=1/N$, so that the classical limit is
$N\rightarrow\infty$.
\begin{figure}[t]
\begin{center}
\includegraphics[height=3in]{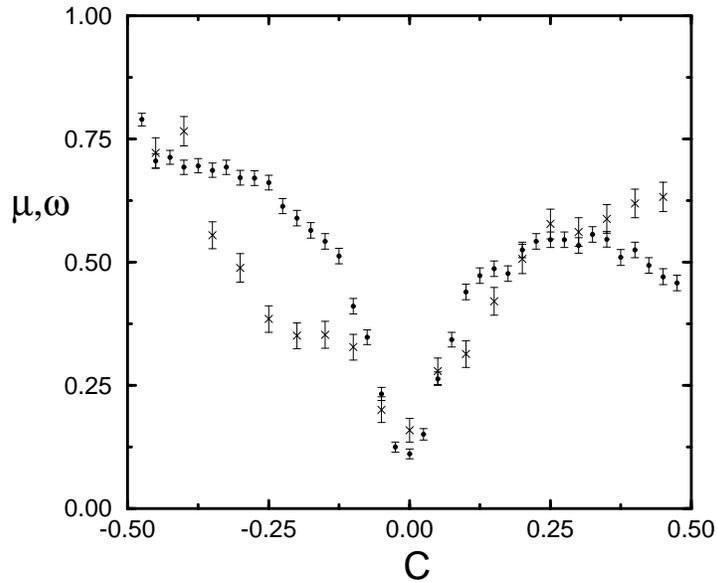}
\caption{
\small
Classical ($\mu$) and quantum ($\omega$) measures of chaos
[denoted by ($\bullet$) and ($\times$), respectively] versus $C$
for the Hamiltonian~(\ref{htot}) with ${\bar B}=0.5$.
Adapted from~\cite{levwhe96}.
\label{figchaoswl}}
\end{center}
\end{figure}

\begin{figure}[t]
\begin{center}
\includegraphics[height=12cm]{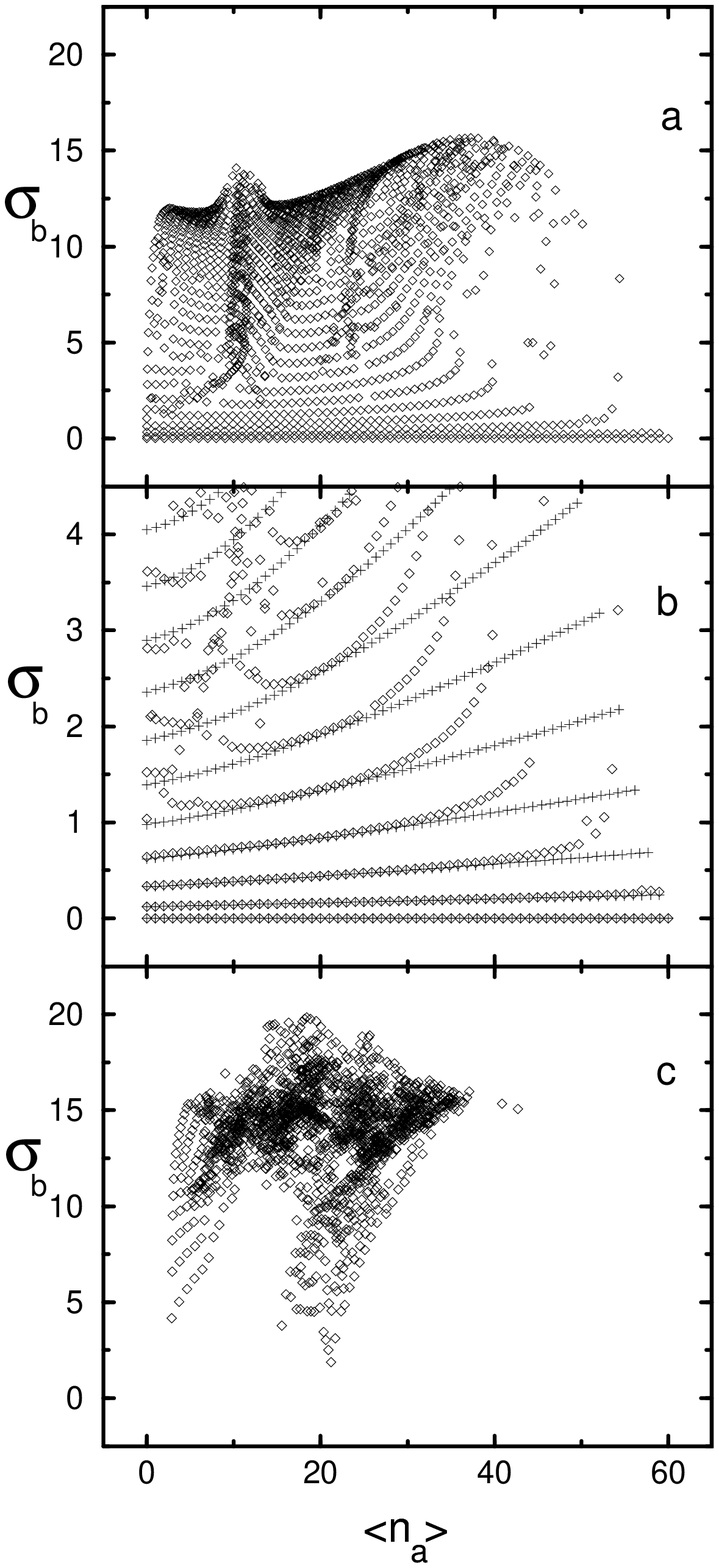}
\hspace{1cm}
\includegraphics[height=12cm]{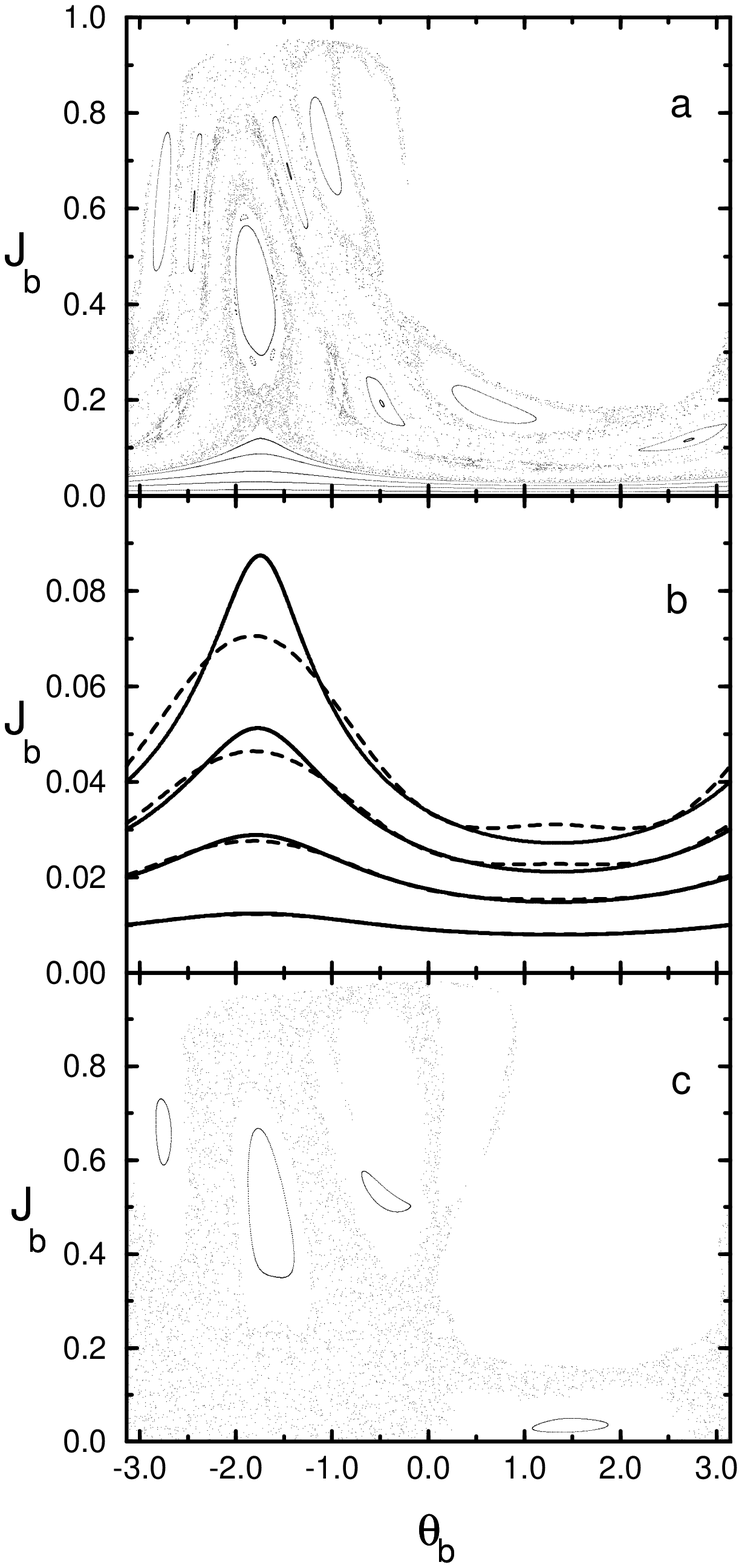}
\caption{\footnotesize 
(Left panel). 
The values of $\langle n_a\rangle$ and of the variance $\sigma_b$
(denoted by $\diamond$) of each eigenstate of the Hamiltonian~(\ref{htot}).
(a)~${\bar B}=0.5$, $C=0$ (partial dynamical symmetry).
(b)~a blow up of (a) with superimposed results [denoted by ($+$)] 
of quantum perturbation theory.
The families of states with low $\sigma_b$
have small values of $\langle n_b\rangle$.
(c)~${\bar B}=0.3$, $C=0.5$ (broken symmetry). 
Adapted from~\cite{levwhe96}.
\vspace{2pt}
\hfil\break 
{\normalsize {\rm Figure 14:}}
(Right panel). 
Poincar\'{e} sections $J_b$ versus $\theta_b$ at energy $E=1.0$.
(a)~${\bar B}=0.5$, $C=0$ (partial dynamical symmetry).
(b)~a blow up of (a) with superimposed results (dashed curves)
of classical perturbation theory.
(c)~${\bar B}=0.3$, $C=0.5$ (broken symmetry).
Adapted from~\cite{levwhe96}.
\label{figvariance}}
\end{center}
\end{figure}

In all calculations reported below~\cite{levwhe96} 
we take $A=0.8642$ and $N=60$.
As a first step, we fix ${\bar B}=0.5$ and vary $C$. 
As previously done, 
for the classical
analysis we randomly sample the phase space and determine the fraction
$\mu$ of chaotic volume (same as $\sigma$ in Fig.~\ref{figchaossu3}). 
For the quantum analysis we evaluate the energy
levels, calculated the nearest neighbors level spacing distribution of the
unfolded spectrum and fitted it to a Brody distribution. The latter is 
specified by the fit parameter $\omega$, mentioned above. 
As shown in Fig.~\ref{figchaoswl}, 
both of these measures indicate a suppression of chaos near $C=0$
similar to the results of Fig.~\ref{figchaossu3}. 
To appreciate the strong effect of the PDS (at $C=0$) 
on the underlying dynamics, it should be noted that the fraction of the
solvable states $\vert n_a,n_b=0,n_c\rangle$ is $2/(N+2)$, which approaches
zero in the classical limit.
To measure the extent to which each
eigenstate $|\Psi\rangle$ has SU(2) symmetry,
we define variances $\sigma_{i}^2 =
\langle \Psi|\hat{n}_{i}^2| \Psi\rangle -
\langle \Psi|\hat{n}_{i}|\Psi\rangle^2$ $(i=a,b)$.
A state which belongs to just one irrep of SU(2) (with well defined
$J,J_z$) has zero variances, while a mixed state
has large variances. These variances have the same physical content
as the entropies considered before. 
It is instructive to display the average $\langle\hat{n}_a\rangle$ and
variance of each state, as done in Fig.~\ref{figvariance}. 
SU(2) PDS is present in Fig.~\ref{figvariance}(a) ($B\neq0$, $C=0$), 
Fig.~\ref{figvariance}(b) is a blow up of Fig~\ref{figvariance}(a), and 
in Fig.~\ref{figvariance}(c) the symmetry is completely broken ($C\neq 0$).
In Figs.~\ref{figvariance}(a) and \ref{figvariance}(b) 
we see states with zero variance. These
are just the special $N+1$ states ($n_b=0$) discussed before, which preserve
the SU(2) symmetry. In addition, we see families of states with small
variance and small $\langle n_b\rangle$
which suggests that the presence of PDS 
increases the purity of states other than the special ones. By contrast, in
Fig.~\ref{figvariance}(c) we see no particular structure because of the 
destruction of the PDS for $C\neq 0$.

Considerable insight is gained by examining the classical phase space
structure in terms of action-angle variables
$\alpha =\sqrt{J_a}\exp(-i\theta_a)$,
$\beta =\sqrt{J_b}\exp(-i\theta_b)$ and
$\gamma = \sqrt{J_c} = \sqrt{1-J_a-J_a}$.
The $\theta_a=-\pi/2$
Poincar\'e section is shown in Fig.~14 for energy $E=1.0$.
When SU(2) PDS is present (${\bar B}\neq 0,\, C=0$), 
we see in Figs.~14(a) and 14(b) a torus
with $J_b=0$, and additional perturbed tori in its neighborhood
(small $J_b$). This structure
is absent when the symmetry is completely broken ($C\neq 0$), as shown in
Fig.~14(c). The features in Fig.~14 
persist also at other energies. To understand
them, we recall that for ${\bar B}=C=0$, the Hamiltonian~(\ref{cham}) 
is integrable and all trajectories wind around invariant tori.
By standard torus quantization (without turning points)
the actions are quantized as
$J_{i}=n_{i}\hbar=n_{i}/N$ $(i=a,b)$.
In the integrable limit, quantum states are associated with toroidal
manifolds in phase space. In case of a partial symmetry 
(${\bar B}\neq 0,\,C=0$) we are led by analogy with Eq.~(\ref{comm}) 
to seek manifolds ${\cal M}$ in phase space on which
\begin{equation} \label{poiss}
\Bigl.\Bigl \{{\cal H}_0\, ,\,{\cal H}_1\Bigr \}\Bigr|_{\cal M}\; =0
\end{equation}
vanishes even though the Poisson bracket is not zero everywhere.
In addition, we demand that 
$\{\{{\cal H}_0\,,\,{\cal H}_1\},{\cal H}_0+{\cal H}_1\}|_{\cal M}=0$
(in analogy to the quantum relation
$[[\,H_0,H_1]\,,\,H_0+H_1]|n_a,n_b=0,n_c\rangle =0$)
so that a trajectory starting on
${\cal M}$ remains on ${\cal M}$. The solution to these conditions
is the manifold $J_b=\beta^*\beta=0$, which may be
interpreted as a (degenerate) torus of the ${\cal H}_0$
Hamiltonian. It is also a stable isolated periodic orbit of
${\cal H}_0+\bar{B}{\cal H}_1$.
Quantization of the torus with $J_b=0$ proceeds exactly
as before, so we correctly predict no change in the quantum energies
associated with it.
The manifold ${\cal M}$ ($J_b=0$) is the direct classical
analogue of the special quantum states $\vert n_b=0\rangle$.
It refers, however, to a region of phase space of measure zero, and so
cannot by itself explain the observed (global) suppression of chaos.
However, as suggested by Fig.~14, the presence of PDS 
induces a quasi-regular region foliated by tori in the vicinity
of the special torus. 
The dynamics on a finite measure of phase space can be understood 
by performing a perturbative calculation in the neighbourhood 
of ${\cal M}$~\cite{levwhe96}.

For the classical perturbation calculation we set $C= 0$ in
Eq.~(\ref{cham}) and treat
$\bar{B}$ as an expansion parameter, assuming
$\bar{B}{\cal H}_1$ in Eq.~(\ref{cham}) is small in the neighbourhood
of the special periodic orbit. 
The second order correction reproduces well
the perturbed tori on the Poincar\'e sections as shown in Fig.~14(b). 
The variances can be calculated in quantum perturbation theory. 
In Fig.~\ref{figvariance}(b) we show the results 
[denoted by ($+$)] of the quantum perturbation theory (to order $B^5$).
We see that the first few families of states are reproduced.
It is these states which we can recover from perturbation
theory and whose approximate symmetry is induced by the
symmetry of the special states.

The following physical picture emerges from the foregoing analysis.
Near the special orbit, there are KAM tori,
some of which are quantized. The quantum eigenstates lie on these
tori, so knowing the classical variance of the actions of the tori tells us
the variances of the states themselves, in the semiclassical limit.
Large variances indicate the extent to which the corresponding states fail
to respect the symmetry. This provides a measure for a separation of regular
and irregular levels, as conceived in~\cite{perc}. 
In the present model, the
quantum states can be grouped into three classes:
i) the special states, which observe the symmetry;
ii) the ``almost special states'' which are accessible by perturbation
theory; iii) the rest of the states, which are mixed.
As in~\cite{bohig93}, the frontier between regular states 
(sets (i) and (ii) ) and irregular states (set (iii) ) is not sharp.

The above discussion illustrates the effect of PDS 
on the quantum and classical dynamics of a mixed system. At the
quantum level, PDS 
by definition implies the existence
of a ``special'' subset of  states, which observe the symmetry. The PDS 
affects the purity
of other states in the system; in particular, neighboring states,
accessible by perturbation theory,
possess approximately good symmetry. Analogously,
at the classical level, the region of phase space near the ``special''
torus also has toroidal structure. As a consequence
of having PDS, 
a finite region of phase space is
regular and a finite fraction of states is approximately ``special''.
This clarifies the observed suppression of chaos. Based on these
arguments and the above results, 
it is anticipated that the suppression of chaos will persist in higher 
dimensional systems with PDS.

\section{PDS and Higher-Order Terms}
\label{PDS3bod}

In applications of algebraic modeling to dynamical systems, there is 
occasionally a need, based on phenomenological and/or microscopic grounds, 
to include higher-order terms in the Hamiltonian. 
For example, in the IBM, accommodating rigid triaxial shapes and 
describing large anharmonicities in excited bands, 
requires at least cubic terms in the boson Hamiltonian.
From a microscopic point of view, many-body boson interactions in 
the IBM, are generated by the mapping of fermion pairs into 
bosons~\cite{ariyoshgin81,klein91} and the truncation to only monopole 
and quadrupole bosons, with the associated renormalization of the 
effective interaction in the truncated 
space~\cite{duval83,levkir84,otsuka85}. 
From this perspective, to confine to two-body interactions in the boson 
space, is only a convenient lowest-order approximation. 

The advantages of using higher-order terms with PDS are twofold. First, 
the algorithms for realizing such symmetry structures provide 
a systematic procedure for identifying and selecting interactions 
of a given order. Having at hand a selection criteria is highly desirable, 
since, if higher-order terms or new degrees of freedom are added, 
one is immediately faced with the problem of many possible interactions and 
a proliferation of free parameters. Second, Hamiltonians with PDS break the 
dynamical symmetry but retain selected subsets of solvable eigenstates 
with good symmetry. Such qualities are a virtue, since interactions with 
a PDS can be introduced, in a controlled manner, {\it without destroying} 
results previously obtained with a dynamical symmetry for a segment of 
the spectrum. 

In general, the existence of quantum Hamiltonians with PDS is closely related 
to the order of the interaction among the constituents. IBM Hamiltonians with 
higher-order terms, exhibiting PDS of type~II were already 
encountered in Subsection~\ref{subsec:o6PDStypeII}. In what follows, 
we present examples of three-body IBM Hamiltonians with U(5) and O(6) 
PDS of type~I. Work on three-body IBM Hamiltonians with SU(3)-PDS of type~I, 
is currently in progress and will be reported 
elsewhere~\cite{levramisa10}.

\subsection{U(5) PDS (type I) with three-body terms}
\label{subsec:u5PDS3bod}

The U(5) dynamical symmetry (DS) chain, 
${\rm U(6)}\supset {\rm U(5)} \supset {\rm O(5)}\supset {\rm O(3)}$, 
and its related basis states, 
$\vert[N]\langle n_d\rangle (\tau) n_\Delta LM\rangle$, 
were discussed in Section~\ref{subsec:u5PDStypeI}. 
In this case, new terms show up in the DS Hamiltonian 
at the level of three-body interactions. 
The DS Hamiltonian and related spectrum now read
\bsub
\ba
\hat{H}_{DS} &=&
t_1\,\hat{n}_d + t_2\,\hat{n}_{d}^2 + t_3\,\hat{n}_{d}^3 
+ t_4\,\hat{C}_{{\rm O(5)}} + t_5\,\hat{n}_d\hat{C}_{{\rm O(5)}} 
+ t_6\,\hat{C}_{{\rm O(3)}} + t_7\,\hat{n}_d\hat{C}_{{\rm O(3)}} ~,\qquad
\label{hDS3u5}\\
E_{DS} &=& 
t_1\, n_d + t_2\, n_{d}^2 + t_3\, n_{d}^3 
+ t_4\, \tau(\tau+3) + t_5\,n_{d}\tau(\tau+3) 
+ t_6\, L(L+1)
\nonumber\\ 
&& + t_7\, n_{d}L(L+1) ~.
\label{eDS3u5}
\ea
\label{ehDS3u5}
\esub
Terms of the form $\hat{N}\hat{C}_{G}$, with $\hat{C}_{G}$ a quadratic 
Casimir operator of $G={\rm U(5)},\,{\rm O(5)},\, {\rm O(3)}$, 
are included in $\hat{H}_{DS}$ by allowing the parameters 
$t_i$ to depend on $N$. 
\begin{table}[t]
\begin{center}
\caption{\label{Tabu5tens3}
\protect\small
Normalized three-boson U(5) tensors.}
\vspace{1mm}
\begin{tabular}{cccccl}
\hline
 & & & & &\\[-3mm]
$n$&$n_d$&$\tau$& $n_{\Delta}$ &$\ell$&
$\hat B^\dag_{[n]\langle n_d\rangle(\tau)n_{\Delta}\ell m}$\\
 & & & & &\\[-3mm]
\hline
 & & & & & \\[-2mm]
3& 0& 0& 0& 0& $\sqrt{\frac{1}{6}}(s^{\dag})^3$\\[2pt]
3& 1& 1& 0& 2& $\sqrt{\frac{1}{2}}(s^{\dag})^{2}d^{\dag}_{m}$\\[2pt]
3& 2& 0& 0& 0& $\sqrt{\frac{1}{2}}s^{\dag}(d^{\dag} d^{\dag})^{(0)}_0$\\[2pt]
3& 2& 2& 0& 2& $\sqrt{\frac{1}{2}}s^{\dag}(d^\dag d^\dag)^{(2)}_m$\\[2pt]
3& 2& 2& 0& 4& $\sqrt{\frac{1}{2}}s^{\dag}(d^\dag d^\dag)^{(4)}_m$\\[2pt]
3& 3& 1& 0& 2& $\sqrt{\frac{5}{14}}
((d^\dag d^\dag)^{(0)} d^\dag)^{(2)}_m$\\[2pt]
3& 3& 3& 1& 0& $\sqrt{\frac{1}{6}}
((d^\dag d^\dag)^{(2)} d^\dag)^{(0)}_m$\\[2pt]
3& 3& 3& 0& 3& $\sqrt{\frac{7}{30}}
((d^\dag d^\dag)^{(2)} d^\dag)^{(3)}_m$\\[2pt]
3& 3& 3& 0& 4& $\sqrt{\frac{7}{22}}
((d^\dag d^\dag)^{(2)} d^\dag)^{(4)}_m$\\[2pt]
3& 3& 3& 0& 6& $\sqrt{\frac{1}{6}}
((d^\dag d^\dag)^{(4)} d^\dag)^{(6)}_m$\\[4pt]
 & & & & & \\[-3mm]
\hline
\end{tabular} 
\end{center}
\end{table}

The construction of U(5)-PDS Hamiltonians with three-body terms follows 
the general algorithm by considering operators which annihilate, 
for example, the U(5) ground state, $\vert [N],n_d=\tau=L=0\rangle$. 
This can be accomplished by means of those U(5) tensors in 
Table~\ref{Tabu5tens3} with $n_d\neq 0$. Several families of 
U(5)-PDS Hamiltonians can be defined by identifying specific 
three-body terms which annihilate 
additional U(5) basis states. One such family involves the interaction 
\bsub
\ba
&&\hat{V}_{0} = r_0\,G^{\dag}_{0}G_{0} 
+ e_{0}\left (G^{\dag}_0 K_0 + K^{\dag}_{0}G_0 \right )
\label{V0a}\\
&&\hat{V}_{0}\vert [N], n_d=\tau, \tau, n_{\Delta}=0, L M\rangle = 0 
\qquad L=\tau,\tau+1,\ldots,2\tau-2,2\tau
\qquad\qquad
\label{V0b}
\ea
\label{V03bod}
\esub
where $G^{\dag}_{L,\mu} = [(d^\dag d^\dag)^{(\rho)} d^\dag]^{(L)}_{\mu}$ 
with $(\rho,L)=(2,0),\, (0,2),\, (2,3),\, (2,4),\, (4,6)$ and 
$K^{\dag}_{L,\mu} = s^{\dag}(d^{\dag} d^{\dag})^{(L)}$ 
with $L=0,\, 2,\, 4$. As shown in~\cite{talmi03}, 
the states of Eq.~(\ref{V0b}) may be projected from states created by acting 
on the vacuum by a product of $\tau$ operators $d^{\dag}_{m_i}$ with 
$m_i\geq 1$ for $i=1,\dots,\tau$. Hence, such states are guaranteed to 
contain no three $d$-boson states coupled to $L=0$ and, therefore, are 
annihilated by $G_0$. 
The same set of states are annihilated also by $K_0$, 
since all states with $n_d=\tau$ vanish under 
the action of  $\tilde{d}\cdot\tilde{d}$~\cite{arimaiac76}. 
The remaining eigenstates 
of $\hat{V}_0$~(\ref{V03bod}) are mixed with respect to both U(5) and 
O(5). Clearly, $\hat{H}_{DS} + \hat{V}_0$ exhibits a U(5)-PDS of type I. 

A second family of PDS Hamiltonians involves the interaction
\bsub
\ba
&&\hat{V}_{2} = a_{2}\, \Pi^{(2)}\cdot U^{(2)}
+ \sum_{L=0,2,4}e_{L}\left (\,G^{\dag}_L\cdot \tilde{K}_L
+ H.c.\,\right )
\qquad\qquad
\label{V2a}\\
&&\hat{V}_{2}\vert [N], n_d=\tau=L=3\rangle = 0 ~, 
\label{V2b}
\ea
\label{V23bod}
\esub 
where $\tilde{G}_{L,\mu} = (-1)^{\mu}G_{L,-\mu}$ and 
$\tilde{K}_{\mu} = (-1)^{\mu}K_{L,-\mu}$. 
The relation in Eq.~(\ref{V2b}) follows from arguments similar to those 
given after Eq.~(\ref{V2nd3}). Other eigenstates 
of $\hat{V}_2$~(\ref{V23bod}) are mixed in the U(5) basis. 
Clearly, $\hat{H}_{DS} + \hat{V}_2$ exhibits a U(5)-PDS of type I. 

A third family of PDS Hamiltonians involves the interaction
\ba
\hat{V}_{3} = r_3\,G^{\dag}_{3}\cdot{G}_{3} + r_0\,G^{\dag}_{0}G_{0} ~.
\label{V3}
\ea
Both terms in Eq.~(\ref{V3})
conserve the U(5) quantum number $n_d$, but are not O(5) scalars. They 
can induce O(5) mixing subject to $\Delta\tau=2,4,6$, and their multipole 
form involves U(5) generators, some of which are not contained in the O(5) 
subalgebra. As such, and in accord with the discussion at the end of 
Subsection~\ref{subsec:o6PDStypeII}, $\hat{H}_{DS} + \hat{V}_3$ 
exhibits U(5)-PDS of type II.  
Since both terms in $\hat{V}_3$ are rotational-scalars and 
are diagonal in $n_d$, it follows that in a given $n_d$ multiplet,  
those $L$-states which have a unique $\tau$-assignment, remain pure 
with respect to O(5), and hence are good U(5) eigenstates. 
For example, the states with $n_d=\tau, n_{\Delta}=0,\, 
L =2n_d,\, 2n_d-2, 2n_d-3, 2n_d-5$, or states with $L=0$ and $n_d\leq 5$, 
or states with $L=3$ and $n_d\leq 8$, are all eigenstates of $\hat{V}_3$, 
diagonal in the U(5) basis. 
In this sense, $\hat{H}_{DS} + \hat{V}_3$ exhibits also 
O(5)-PDS of type~I.

\subsection{O(6) PDS (type I) with three-body terms}
\label{subsec:o6PDS3bod}

The O(6) dynamical symmetry (DS) chain, 
${\rm U(6)}\supset {\rm O(6)}\supset {\rm O(5)}\supset {\rm O(3)}$, 
and its related 
basis states, $|[N]\langle\Sigma\rangle(\tau)n_\Delta LM\rangle$, 
were discussed in Subsection~\ref{subsec:o6PDStypeI}. 
The DS Hamiltonian is given in Eq.~(\ref{hDSo6}) and no new terms 
are added to it at the level of three-body interactions.

According to the general algorithm, the construction of interactions 
with O(6)-PDS of type I 
requires $n$-boson creation and annihilation operators
with definite tensor character in the O(6) basis: 
\begin{equation}
\hat B^\dag_{[n]\langle\sigma\rangle(\tau)n_{\Delta}\ell m},
\;\;
\tilde{B}_{[n^5]\langle\sigma\rangle(\tau)n_{\Delta}\ell m}
\equiv
(-1)^{\ell-m}
\left(\hat B^\dag_{[n]\langle\sigma\rangle(\tau)n_{\Delta}\ell,-m}
\right)^\dag.
\label{tenso6}
\end{equation}
Of particular interest are tensor operators with $\sigma<n$.
They have the property 
\begin{equation}
\tilde{B}_{[n^5]\langle\sigma\rangle(\tau)n_{\Delta}\ell m}
|[N]\langle N\rangle(\tau)n_{\Delta} LM\rangle=0,
\qquad
\sigma<n,
\label{anniso6}
\end{equation}
for all possible values of $\tau,n_{\Delta},L$ contained in the
O(6) irrep $\langle N\rangle$.
This is so because the action of
$\tilde{B}_{[n^5]\langle\sigma\rangle(\tau)n_{\Delta}\ell m}$
leads to an $(N-n)$-boson state
that contains the O(6) irreps
$\langle\Sigma\rangle=\langle N-n-2i\rangle,\,i=0,1,\dots$
which cannot be coupled with $\langle\sigma\rangle$
to yield $\langle\Sigma\rangle=\langle N\rangle$, since $\sigma<n$.
Number-conserving normal-ordered interactions that are constructed out
of such tensors with $\sigma<n$
(and their Hermitian conjugates)
thus have $|[N]\langle N\rangle(\tau)n_\Delta LM\rangle$
as eigenstates with zero eigenvalue~\cite{RamLevVan09}. 
\begin{table}[t]
\begin{center}
\caption{\label{Tabo6tens3}
\protect\small
Normalized three-boson O(6) tensors.}
\vspace{1mm}
\begin{tabular}{cccccl}
\hline
& & & & &\\[-3mm]
$n$&$\sigma$&$\tau$&$n_{\Delta}$&$\ell$&
$\hat B^\dag_{[n]\langle\sigma\rangle(\tau)n_{\Delta}\ell m}$\\[4pt]
& & & & &\\[-3mm]
\hline
 & & & & & \\[-2mm]
3& 3& 0& 0& 0& $\sqrt{\frac{3}{16}}s^\dag (d^\dag d^\dag)^{(0)}_0
             +\sqrt{\frac{5}{48}}(s^{\dag})^{3}$\\[2pt]
3& 3& 1& 0& 2& 
           $\sqrt{\frac{5}{112}}((d^\dag d^\dag)^{(0)} d^\dag)^{(2)}_m
         +\sqrt{\frac{7}{16}}(s^{\dag})^{2}d^{\dag}_m$\\[2pt]
3& 3& 2& 0& 2& $\sqrt{\frac{1}{2}}s^\dag (d^\dag
             d^\dag)^{(2)}_m$\\[2pt]
3& 3& 2& 0& 4& $\sqrt{\frac{1}{2}}s^\dag (d^\dag
             d^\dag)^{(4)}_m$\\[2pt]
3& 3& 3& 1& 0& $\sqrt{\frac{1}{6}}((d^\dag d^\dag)^{(2)}
             d^\dag)^{(0)}_0$\\[2pt]
3& 3& 3& 0& 3& $\sqrt{\frac{7}{30}}((d^\dag d^\dag)^{(2)}
             d^\dag)^{(3)}_m$\\[2pt]
3& 3& 3& 0& 4& $\sqrt{\frac{7}{22}}((d^\dag d^\dag)^{(2)}
             d^\dag)^{(4)}_m$\\[2pt]
3& 3& 3& 0& 6& $\sqrt{\frac{1}{6}}((d^\dag d^\dag)^{(4)}
             d^\dag)^{(6)}_m$\\[2pt]
3& 1& 0& 0& 0& $\sqrt{\frac{5}{16}}s^\dag (d^\dag d^\dag)^{(0)}_0
            -\sqrt{\frac{1}{16}}(s^{\dag})^{3}$\\[2pt]
3& 1& 1& 0& 2& $\sqrt{\frac{5}{16}}((d^\dag d^\dag)^{(0)} d^\dag)^{(2)}_m
            -\sqrt{\frac{1}{16}}(s^{\dag})^{2}d^{\dag}_m$\\[4pt]
& & & & &\\[-3mm]
\hline
\end{tabular}
\end{center}
\end{table}

As shown in Subsection~\ref{subsec:o6PDStypeI}, there is one 
two-boson operator
$P^{\dagger}_{0}=
d^{\dagger}\cdot d^{\dagger} - (s^{\dagger})^2$, Eq.~(\ref{Pdag0o6}), 
with $\sigma<n=2$, which 
gives rise to an O(6)-invariant interaction, $P^{\dag}_{0}P_{0}$, 
related to the completely solvable Casimir operator 
of O(6), Eq.~(\ref{HPSo6}).
On the other hand, from Table~\ref{Tabo6tens3}, one recognizes 
two three-boson O(6) tensors with $\sigma<n=3$ 
\begin{equation}
\hat B^\dag_{[3]\langle1\rangle(1)0;2m}=
\textstyle{\frac{1}{4}} P^{\dag}_{0}d^\dag_m,
\quad
\hat B^\dag_{[3]\langle1\rangle(0)0;00}=
\textstyle{\frac{1}{4}} P^{\dag}_{0}s^\dag,
\label{three}
\end{equation}
and from these one can construct
the interactions with an O(6) PDS.
The only three-body interactions that are partially solvable in O(6)
are thus $P^{\dag}_{0}\hat n_s P_{0}$
and $P^{\dag}_{0}\hat n_d P_{0}$. 
Since the combination $P^{\dag}_{0}(\hat n_s+\hat n_d)P_{0}
= (\hat{N} -2)P^{\dag}_{0}P_{0}$
is completely solvable in O(6),
there is only one genuine partially solvable three-body interaction
which can be chosen as $P^{\dag}_{0}\hat n_s P_{0}$,
with tensorial components $\sigma=0,\,2$. 
The O(6)-DS spectrum
\setcounter{figure}{14}
\begin{figure*}[t]
\begin{center}
\leavevmode
\includegraphics[width=0.90\linewidth]{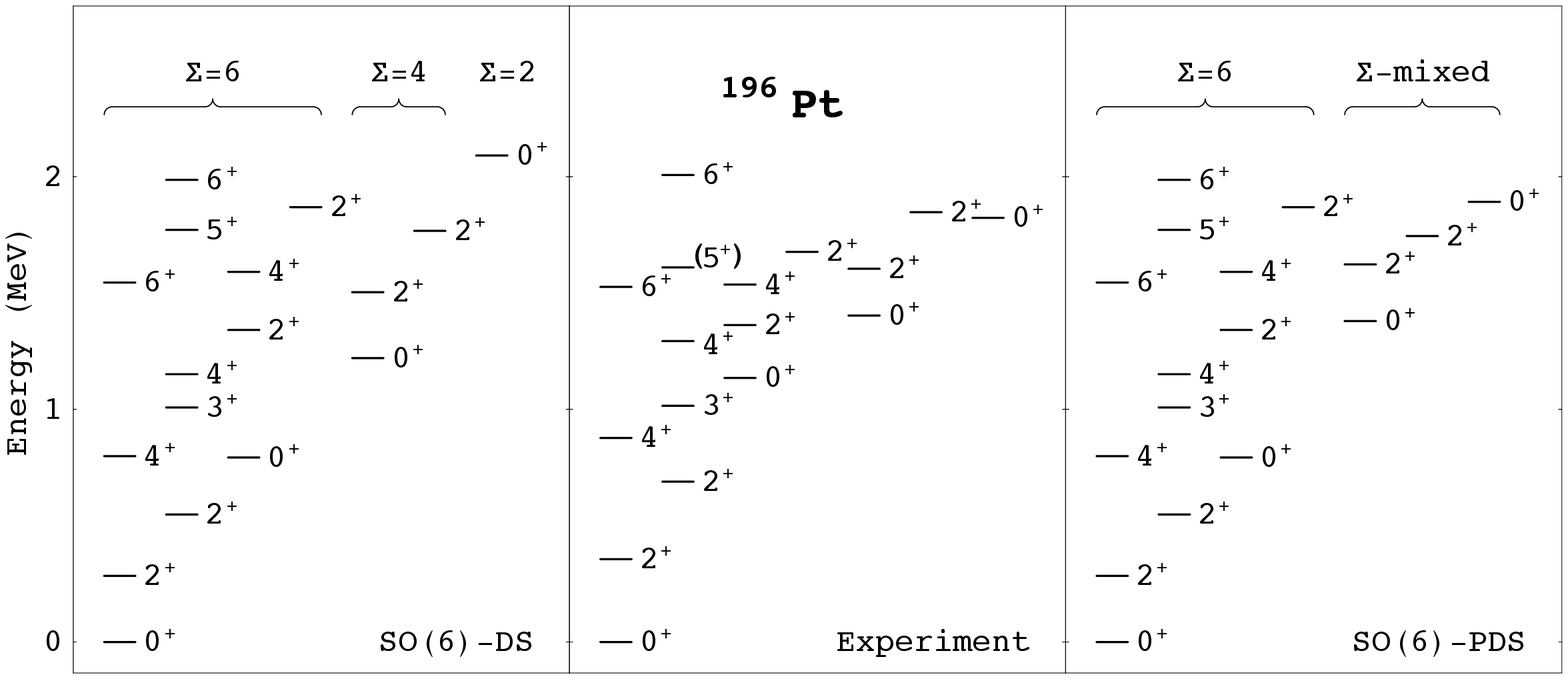}
\caption{
\small
Observed spectrum of $^{196}$Pt compared with the calculated spectra 
of $\hat H_{\rm DS}$~(\ref{hDSo6}), 
with O(6) dynamical symmetry (DS), 
and of $\hat H_{\rm PDS}$~(\ref{hPDSo63bod}) with 
O(6) partial dynamical symmetry (PDS). 
The parameters in $\hat H_{\rm DS}$ $(\hat H_{\rm PDS})$ are
$h_0=43.6\, (30.7)$, $B=44.0\, (44.0)$, $C=17.9\, (17.9)$, 
and $\eta=0\, (8.7)$ keV. The boson number is $N=6$ 
and $\Sigma$ is an O(6) label. 
Adapted from~\cite{RamLevVan09}.}
\label{figpt196}
\end{center}
\end{figure*}
\ba
E_{\rm DS} &=& 
4h_{0}\,(N-v +2)v + B\,\tau(\tau+3) +\, C\,L(L+1) ~,
\label{eDSo6v}
\ea
resembles that of a $\gamma$-unstable deformed rotovibrator, 
where states are arranged in bands with O(6) quantum number
$\Sigma=N-2v$, $(v=0,1,2,\ldots)$.
The O(5) and O(3) terms in the dynamical symmetry Hamiltonian, 
$\hat{H}_{\rm DS}$~(\ref{hDSo6}), 
govern the in-band rotational splitting. 
A comparison with the experimental spectrum and E2 rates of $^{196}$Pt
is shown in Fig.~\ref{figpt196} and Table~\ref{Tabbe2pt196}.
\begin{table}[t]
\begin{center}
\caption{\label{Tabbe2pt196}
\protect\small
Observed (EXP) and calculated B(E2) values 
(in $e^2{\rm b}^2$) for $^{196}$Pt. 
For both the exact (DS) and partial (PDS)
O(6) dynamical symmetry calculations, the E2 operator is
that of Eq.~(\ref{Te2}) with 
$e_{B}=0.151$ $e$b and $\chi=0.29$. Only the state $0^{+}_3$ has a mixed 
O(6) character. Adapted from~\cite{RamLevVan09}.} 
\vspace{1mm}
\begin{tabular}{clll|clll}
\hline
& & & & & & &\\[-3mm] 
Transition& EXP & DS &PDS &
Transition& EXP & DS &PDS\\
& & & & & & &\\[-3mm] 
\hline
& & & & & & &\\[-2mm] 
$2^+_1\rightarrow0^+_1$& 0.274~(1)  &  0.274&  0.274 &
$2^+_3\rightarrow0^+_2$& 0.034~(34) &  0.119&  0.119\\[2pt]
$2^+_2\rightarrow2^+_1$& 0.368~(9)  &  0.358&  0.358 &
$2^+_3\rightarrow4^+_1$& 0.0009~(8) &  0.0004& 0.0004\\[2pt]
$2^+_2\rightarrow0^+_1$& 3.10$^{-8}$(3) & 0.0018& 0.0018 &
$2^+_3\rightarrow2^+_2$& 0.0018~(16)& 0.0013& 0.0013\\[2pt]
$4^+_1\rightarrow2^+_1$& 0.405~(6)  &  0.358&  0.358 &
$2^+_3\rightarrow0^+_1$& 0.00002~(2)& 0        & 0         \\[2pt]
$0^+_2\rightarrow2^+_2$& 0.121~(67) &  0.365&  0.365 &
$6^+_2\rightarrow6^+_1$& 0.108~(34) & 0.103& 0.103\\[2pt]
$0^+_2\rightarrow2^+_1$& 0.019~(10) &  0.003&  0.003 &
$6^+_2\rightarrow4^+_2$& 0.331~(88) &  0.221&  0.221\\[2pt]
$4^+_2\rightarrow4^+_1$& 0.115~(40) &  0.174&  0.174 &
$6^+_2\rightarrow4^+_1$& 0.0032~(9) & 0.0008& 0.0008\\[2pt]
$4^+_2\rightarrow2^+_2$& 0.196~(42) &  0.191&  0.191 &
$0^+_3\rightarrow2^+_2$&$<0.0028$   & 0.0037& 0.0028\\[2pt]
$4^+_2\rightarrow2^+_1$& 0.004~(1)  &  0.001&  0.001 &
$0^+_3\rightarrow2^+_1$&$<0.034$    & 0          & 0        \\[2pt]
$6^+_1\rightarrow4^+_1$& 0.493~(32) &  0.365&  0.365 &
                       &            &            &  \\[4pt]
\hline
\end{tabular}
\end{center}
\end{table} 
The O(6)-DS limit is seen to provide a good description
for properties of states in the ground band $(\Sigma=N)$.
This observation was the basis of the claim~\cite{Cizewski78}
that the O(6)-DS is manifested empirically in $^{196}$Pt.
However, the resulting fit to energies of excited bands is quite poor.
The $0^+_1$, $0^+_3$, and $0^+_4$ levels of $^{196}$Pt
at excitation energies 0, 1403, 1823 keV, respectively,
are identified as the bandhead states
of the ground $(v=0)$, first- $(v=1)$
and second- $(v=2)$ excited vibrational bands~\cite{Cizewski78}.
Their empirical anharmonicity,
defined by the ratio $R=E(v=2)/E(v=1)-2$,
is found to be $R=-0.70$.
In the O(6)-DS limit these bandhead states
have $\tau=L=0$ and $\Sigma=N,N-2,N-4$, respectively.
The anharmonicity $R=-2/(N+1)$,
as calculated from Eq.~(\ref{eDSo6v}), is fixed by $N$.
For $N=6$, which is the appropriate boson number for $^{196}$Pt,
the O(6)-DS value is $R=-0.29$,
which is in marked disagreement with the empirical value.
A detailed study of double-phonon excitations within the IBM,
has concluded that large anharmonicities can be incorporated
only by the inclusion of at least cubic terms in the
Hamiltonian~\cite{ramos00b}.
In the IBM there are 17 possible three-body interactions~\cite{ibm}.
One is thus confronted with the need to select suitable higher-order terms
that can break the DS in excited bands but preserve it in the ground band. 
On the basis of the preceding discussion this can be accomplished by the 
following Hamiltonian with O(6)-PDS~\cite{RamLevVan09}
\begin{equation}
\hat{H}_{\rm PDS}=\hat{H}_{\rm DS}+ 
\eta\,P^{\dag}_{0}\hat{n}_s P_{0},
\label{hPDSo63bod}
\end{equation}
where the terms are defined in Eqs.~(\ref{hDSo6}) and~(\ref{three}).
The spectrum of $\hat{H}_{\rm PDS}$ is shown in Fig.~\ref{figpt196}.
The states belonging to the $\Sigma=N=6$ multiplet remain solvable
with energies given by the same DS expression, Eq.~(\ref{eDSo6v}).
States with $\Sigma < 6$ are generally admixed
but agree better with the data than in the DS calculation.
For example, the bandhead states of the first- (second-) excited bands
have the O(6) decomposition
$\Sigma=4$: $76.5\%\,(19.6\%)$,
$\Sigma=2$: $16.1\%\,(18.4\%)$,
and $\Sigma=0$: $7.4\%\,(62.0\%)$.
Thus, although the ground band is pure,
the excited bands exhibit strong O(6) breaking.
The calculated O(6)-PDS anharmonicity for these bands is $R=-0.63$,
much closer to the empirical value, $R=-0.70$. 
It should be emphasized that not only the energies 
but also the wave functions of the $\Sigma=N$ states remain unchanged
when the Hamiltonian is generalized from DS to PDS.
Consequently, the E2 rates for transitions among this class of states
are the same in the DS and PDS calculations. Thus, the additional 
three-body term in the Hamiltonian~(\ref{hPDSo63bod}), does not spoil 
the good O(6)-DS description for this segment of the spectrum. 
This is evident in Table~\ref{Tabbe2pt196}
where most of the E2 data concern transitions between $\Sigma=N=6$ states.
Only transitions involving states from excited bands ({\it e.g.},
the $0^+_3$ state in Table~\ref{Tabbe2pt196}) 
can distinguish between DS and PDS.
Unfortunately, such interband E2 rates
are presently poorly known experimentally.
Their measurement is highly desirable for further
testing the O(6)-PDS wave functions.

\section{PDS and Coupled Systems}
\label{PDSibm2}

So far the notion of partial dynamical symmetries was presented 
in the framework of algebraic models involving 
only one species of constituent particle. It is of great interest 
to extend this notion to the case of coupled systems involving 
two (or more) species of particles. In this case, the appropriate  
spectrum generating algebra, $G_1\times G_2$, contains the direct product 
of the two algebraic structures for systems 1 and 2. 

An example of such a coupled system is the proton-neutron version 
of the interacting boson model (IBM-2)~\cite{ibm,arima77,otsuka78}. 
The building blocks of the model are monopole and quadrupole bosons, 
$\{s^{\dag}_{\rho},\,d^{\dag}_{\rho\mu}\}$, 
of proton type ($\rho=\pi$) and of neutron type $(\rho=\nu$), 
representing pairs of identical valence nucleons. 
Number conserving bilinear combinations of operators in each set 
comprise the ${\rm U}_{\rho}(6)$ algebra as in the IBM-1, Eq.~(\ref{u6gen}), 
and bosons of different types commute. Since the separate proton- 
and neutron- boson numbers, 
$\hat{N}_{\pi}$ and $\hat{N}_{\nu}$, are conserved, 
the appropriate spectrum generating algebra of the model 
is ${\rm U}_{\pi}(6)\times {\rm U}_{\nu}(6)$. 
Subalgebras can be constructed with the aid of the individual subalgebras, 
${\rm U}_{\rho}(5),\, {\rm SU}_{\rho}(3),\, {\rm O}_{\rho}(6),\, 
{\rm O}_{\rho}(5),\,{\rm O}_{\rho}(3)$. 
For instance, for a given algebra $G_{\rho}$, 
with generators $g_{\rho}$, there is a combined algebra 
$G_{\pi+\nu}$, with generators $g_{\pi}+ g_{\nu}$. 
The dynamical symmetries of the IBM-2 are obtained by identifying the 
lattices of embedded algebras starting with 
${\rm U}_{\pi}(6)\times {\rm U}_{\nu}(6)$
and ending with the symmetry algebra ${\rm O}_{\pi+\nu}(3)$. 

A new aspect in coupled systems is the occurrence of states which 
are not symmetric with respect to interchange of the two constituents. 
This is clearly seen in the reduction 
\ba
\begin{array}{ccc}
{\rm U}_{\pi}(6)\times {\rm U}_{\nu}(6) & \supset & {\rm U}_{\pi+\nu}(6)\\
\downarrow&&\downarrow\\
\, [N_{\pi}]\times [N_{\nu}] && [N_1, N_2] 
\end{array} ~.
\label{u6pinu}
\ea
For a given irrep of ${\rm U}_{\pi}(6)\times {\rm U}_{\nu}(6)$, 
characterized by $N_{\pi}$ and $N_{\nu}$, 
the allowed irreps of ${\rm U}_{\pi+\nu}(6)$ are 
$[N_1,N_2] = [N_{\pi} + N_{\nu} -k, k]$, 
where $k=0,1,\ldots, {\rm min}\{N_{\pi},\, N_{\nu}\}$. 
States in the irreps with 
$N_2\neq 0$ ($k\neq 0$) are not symmetric with respect to $\pi$ and 
$\nu$ bosons. One of the successes of the IBM-2 has been the 
empirical discovery of such low-lying mixed symmetry states in nuclei,  
in which valence protons and neutrons move out of phase. 
A complete listing of all possible partial dynamical symmetries (PDS) 
of the IBM-2 is outside the scope of the present review. 
In what follows, we present a sample of such symmetry structures, 
illuminating new features of PDS in coupled systems.      

Coupled algebraic structure, $G_1\times G_2$, can involve also fermionic 
algebras, as well as Bose-Fermi algebras. An example of 
the latter is the interacting boson-fermion model (IBFM)~\cite{ibfm}, 
used for describing odd-mass nuclei and broken fermion-pairs 
in even-even nuclei. 
The model incorporates collective (bosonic) and quasi-particle (fermionic) 
degrees of freedom, and the associated spectrum generating algebra is 
${\rm U}_{B}(6)\times {\rm U}_{F}(m)$. 
Here ${\rm U}_{B}(6)$ and ${\rm U}_{F}(m)$ are the boson and fermion 
algebras respectively, and $m=\sum_{i}(2j_i+1)$ is the dimension of the 
single-particle space ($j_i$ are the angular momenta of the occupied 
shell-model orbits). 
Bose-Fermi symmetries correspond to dynamical symmetries of 
${\rm U}_{B}(6)\times {\rm U}_{F}(m)$. 
Supersymmetry corresponds to a further embedding of the Bose-Fermi 
symmetry into a graded Lie algebra 
${\rm U}(6/m)\supset {\rm U_{B}}(6)\times {\rm U_{F}}(m)$. 
Partial Bose-Fermi symmetries and partial supersymmetries 
have not been considered in detail so far. There are initial hints that 
such a structure can occur in the IBFM~\cite{jolos00}, 
however, an in-depth systematic study is called for.

\subsection{F-spin and selected PDS in the IBM-2}
\label{subsec:FspinPDS}

The proton-neutron degrees of freedom are naturally 
reflected in the IBM-2 via an ${\rm SU}_{F}(2)$ F-spin 
algebra~\cite{arima77} with generators 
\ba
\hat{F}_{+} = s^{\dagger}_{\pi} s_{\nu} + d^{\dagger}_{\pi}\cdot 
\tilde d_{\nu} \;\; , \;\;  
\hat{F}_{-} = (\hat{F}_{+})^{\dagger}\;\; , \;\; 
\hat{F}_{0} = (\hat{N}_{\pi} - \hat{N}_{\nu})/2 ~.
\label{su2F}
\ea
These generators commute with the total boson number operator, 
$\hat{N}=\hat{N}_{\pi}+\hat{N}_{\nu}$, which is a ${\rm U}_{N}(1)$ generator. 
The basic F-spin doublets are  $(s^{\dagger}_{\pi},s^{\dagger}_{\nu})$, 
and $(d^{\dagger}_{\pi \mu},d^{\dagger}_{\nu \mu})$, 
with F-spin projection +1/2 ($-$1/2) for proton (neutron) bosons.
The algebras ${\rm SU}_{F}(2)\times {\rm U}_{N}(1)$~(\ref{su2F}) 
and ${\rm U}_{\pi+\nu}(6)$~(\ref{u6pinu}) 
commute and obey a duality relationship, in the sense that their irreps 
are related by $F = (N_1-N_2)/2 = (N_{\pi} + N_{\nu})/2 -k$ and 
$N=N_1+N_2=N_{\pi}+ N_{\nu}$. 
In a given nucleus, with fixed $N_{\pi}$, $N_{\nu}$, all states have the 
same value of $F_{0}= (N_{\pi} - N_{\nu})/2$, while the allowed 
values of the F-spin quantum number F range from $|F_{0}|$ to 
$F_{max} \equiv (N_{\pi}+N_{\nu})/2 \equiv N/2$ in unit steps. 
F-spin characterizes the $\pi$-$\nu$ symmetry properties of IBM-2 states.
States with maximal F-spin, 
$F \equiv F_{max}$, are fully symmetric and correspond to the IBM-1 
states with only one type of bosons~\cite{ibm}. There 
are several arguments, {\it e.g.}, the empirical success of IBM-1, 
the identification of F-spin 
multiplets~\cite{harter85,brentano85,gupta89,zamfir92} 
(series of nuclei with constant $F$ and varying $F_{0}$ with nearly 
constant excitation energies), 
and weakness of M1 transitions, 
which lead to the belief that low lying collective states have predominantly 
$F=F_{max}$~\cite{lipas90}. States with $F<F_{max}$, correspond to 
`mixed-symmetry' states~\cite{iac84}, 
most notably, the orbital magnetic dipole scissors mode~\cite{boh84} 
has by now been established experimentally as a general 
phenomena in deformed even-even nuclei~\cite{rich95}. 
\begin{table}[t]
\begin{center}
\caption[]{\label{TabFspinmult}
\protect\small 
Energies (in MeV) of $2^{+}$ levels of the ground ($g$), 
$\gamma$ and $\beta$ bands in F-spin multiplets. The mass numbers are 
$A= 132 + 4F$. 
Adapted from~\cite{levgin00}.}
\vspace{1mm}
\begin{tabular}{lccccccc}
\hline
& & & & & & &\\[-3mm]
F & Energy & $^{A}$Dy & $^{A+4}$Er & $^{A+8}$Yb & $^{A+12}$Hf & 
 $^{A+16}$W & $^{A+20}$Os \\
& & & & & & &\\[-3mm]
\hline
& & & & & & &\\[-2mm]
6   & $E(2^{+}_{g})$      & 0.14 & 0.13 & 0.12 & 0.12 & 0.12 & 0.14 \\[2pt]
    & $E(2^{+}_{\gamma})$ & 0.89 & 0.85 & 0.86 & 0.88 &      & 0.86 \\[2pt]
    & $E(2^{+}_{\beta})$  & 0.83 & 1.01 & 1.07 & 1.06 &      & 0.74 \\
& & & & & & &\\[-3mm]
\hline
& & & & & & &\\[-2mm]
13/2 & $E(2^{+}_{g})$     & 0.10 & 0.10 & 0.10 & 0.10 & 0.11 & 0.13 \\[2pt] 
    & $E(2^{+}_{\gamma})$ & 0.95 & 0.90 & 0.93 & 0.96 &      &      \\[2pt]
    & $E(2^{+}_{\beta})$  & 1.09 & 1.17 & 1.14 & 0.99 &      &      \\
& & & & & & &\\[-3mm]
\hline
& & & & & & &\\[-2mm]
7   & $E(2^{+}_{g})$      & 0.09 & 0.09 & 0.09 & 0.10 & 0.11 & 0.13 \\[2pt]
    & $E(2^{+}_{\gamma})$ & 0.97 & 0.86 & 0.98 & 1.08 &      & 0.87 \\[2pt]
    & $E(2^{+}_{\beta})$  & 1.35 & 1.31 & 1.23 & 0.95 &      & 0.83 \\
& & & & & & &\\[-3mm]
\hline
& & & & & & &\\[-2mm]
15/2 & $E(2^{+}_{g})$     & 0.08 & 0.08 & 0.08 & 0.09 & 0.11 &      \\[2pt]
    & $E(2^{+}_{\gamma})$ & 0.89 & 0.79 & 1.15 & 1.23 & 1.11 &      \\[2pt]
    & $E(2^{+}_{\beta})$  & 1.45 & 1.53 & 1.14 & 0.90 & 1.08 &      \\
& & & & & & &\\[-3mm]
\hline
& & & & & & &\\[-2mm]
8   & $E(2^{+}_{g})$      & 0.07 & 0.08 & 0.08 & 0.09 &      &      \\ [2pt]
    & $E(2^{+}_{\gamma})$ & 0.76 & 0.82 & 1.47 & 1.34 &      &      \\[2pt]
    & $E(2^{+}_{\beta})$  &      & 1.28 & 1.12 & 1.23 &      &      \\
& & & & & & &\\[-3mm]
\hline
& & & & & & &\\[-2mm]
17/2 & $E(2^{+}_{g})$     & 0.08 & 0.08 & 0.08 &      &      &      \\[2pt]
    & $E(2^{+}_{\gamma})$ & 0.86 & 0.93 & 1.63 &      &      &      \\[2pt]
    & $E(2^{+}_{\beta})$  & 1.21 & 0.96 & 1.56 &      &      &      \\
& & & & & & &\\[-3mm]
\hline
\end{tabular}
\end{center}
\end{table}

\begin{figure}[t]
\begin{center} 
\includegraphics[height=11cm,angle=-90]{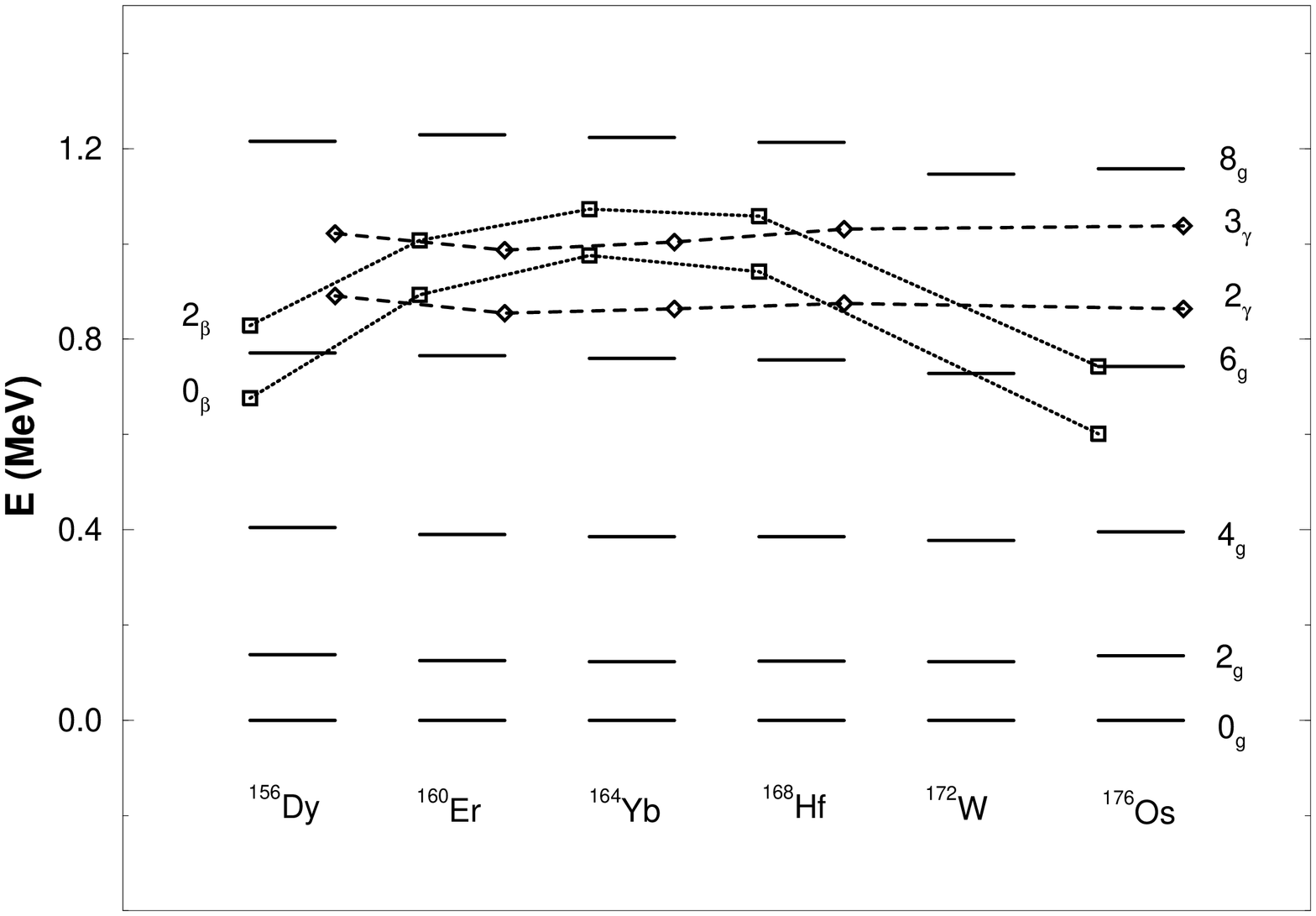}
\caption{
\small
Experimental levels of the ground $\gamma$ and $\beta$ bands 
in an F-spin multiplet $F=6$ of rare earth nuclei. 
Levels shown are up to $L=8^{+}_{g}$ for the 
ground band, $L=2^{+}_{\gamma},3^{+}_{\gamma}$ for the $\gamma$ band 
(diamonds connected by dashed lines)
and $L=0^{+}_{\beta},2^{+}_{\beta}$ for the $\beta$ band 
(squares connected by dotted lines). Adapted from~\cite{levgin00}.
\label{figFspinmult}}
\end{center}
\end{figure}

Various procedures have been proposed to estimate the F-spin purity of 
low lying states~\cite{lipas90}. In the majority of analyses, 
based on M1 transitions (which should vanish between pure $F=F_{max}$ 
states), magnetic moments and energy systematics of 
mixed-symmetry states, the F-spin admixtures in low 
lying states are found to be of a few percents ($<10\%$), typically 
$2\%-4\%$~\cite{lipas90}. 
In spite of its appeal, however, F-spin cannot be an exact symmetry of the 
Hamiltonian. The assumption of F-spin scalar Hamiltonians is at variance 
with the microscopic interpretation of the IBM-2, 
which necessitates different effective 
interactions between like and unlike nucleons~\cite{Talmi93}. 
Furthermore, if F-spin was a symmetry of the Hamiltonian, 
then {\it all} states 
would have good F-spin and would be arranged in F-spin multiplets. 
Experimentally this is not case. As noted in an 
analysis~\cite{brentano85,gupta89} of rare earth nuclei, 
the ground bands are in F-spin multiplets, whereas the 
vibrational $\beta$ bands and some $\gamma$ bands do not 
form good F-spin multiplets. The empirical situation in the deformed Dy-Os 
region is portrayed in Table~\ref{TabFspinmult} and Fig.~\ref{figFspinmult}. 
From Table~\ref{TabFspinmult} 
it is seen that, for $F>13/2$, the
energies of the $L=2^{+}$ members of the $\gamma$ bands vary fast 
in the multiplet and not always monotonically. The variation in the energies 
of the $\beta$ bands is large and irregular. Thus both microscopic and 
empirical arguments rule out F-spin invariance of the Hamiltonian. 
F-spin can at best be an approximate quantum number which is good only for 
a selected set of states while other states are mixed. We are thus 
confronted with a situation of having `special states' endowed with a good 
symmetry which does not arise from invariance of the Hamiltonian.
These are precisely the characteristics of a partial symmetry 
for which a non-scalar Hamiltonian produces a subset of special 
(at times solvable) states with good symmetry. 
In what follows we present IBM-2 Hamiltonians with F-spin as a partial 
symmetry~\cite{levgin00}. The construction process is similar to that 
employed in Section~\ref{sec:PartialSolv} 
for obtaining partially-solvable IBM-1 Hamiltonians.
Predictions of F-spin PDS are then confronted with empirical data. 

The ground band in the IBM-2 is represented by an intrinsic state which 
is a product of a proton condensate and a rotated neutron condensate with 
$N_{\pi}$ and $N_{\nu}$ bosons, respectively~\cite{levkir90}. 
It depends on the quadrupole 
deformations, $\beta_{\rho},\gamma_{\rho}$, ($\rho=\pi,\nu$) of the 
proton-neutron equilibrium shapes and on the relative orientation angles 
$\Omega$ between them. For $\beta_{\rho}>0$, the intrinsic state is deformed 
and members of the rotational ground-state band are obtained from it by 
projection. It has been shown in~\cite{lev90} that the intrinsic state will 
have a well defined F-spin, $F=F_{max}$, when the proton-neutron shapes are 
aligned and with equal deformations. The conditions ($\beta_{\pi}=
\beta_{\nu}$,$\gamma_{\pi}=\gamma_{\nu}$, $\Omega=0$) are weaker than the 
conditions for F-spin invariance, which makes it possible for a 
non-F-scalar IBM-2 Hamiltonian to have an equilibrium intrinsic state 
with pure F-spin. Focusing on the most likely situation, namely, 
aligned axially symmetric (prolate) deformed shapes ($\beta_{\rho}=\beta$, 
$\gamma_{\rho}=\Omega=0$), the equilibrium deformed 
intrinsic state for the ground band with $F=F_{max}$ has the form
\ba 
\vert c; K=0 \rangle &\equiv& \vert N_{\pi},N_{\nu} \rangle = 
(N_{\pi}!N_{\nu}!)^{- 1/2}(b^{\dagger}_{c,\pi})^{N_{\pi}} \,
(b^{\dagger}_{c,\nu})^{N_{\nu}} \vert 0 \rangle ~,
\nonumber\\
 b^{\dagger}_{c,\rho} &=& (1 + \beta^{2}\,)^{-1/2}
(\,s^{\dagger}_{\rho} + \beta\, d^{\dagger}_{\rho, 0} 
\, ) ~,
\label{condibm2}
\ea 
where $K$ denotes the angular momentum projection on the symmetry axis. 

The construction of partially-solvable IBM-2 Hamiltonian with F-spin 
partial symmetry, can be accomplished by means of the following boson-pair 
operators~\cite{levgin00}
\ba
\begin{array}{ll}
R^{\dagger}_{\rho,0} = 
d^{\dagger}_{\rho} \cdot d^{\dagger}_{\rho} -       
\beta ^{2}(s^{\dagger}_{\rho})^2,\;\;\; &
R^{\dagger}_{(\pi\nu),0} = 
\sqrt{2}\,(\,d^{\dagger}_{\pi}\cdot 
d^{\dagger}_{\nu} - \beta^{2}s^{\dagger}_{\pi}s^{\dagger}_{\nu}\,) \\[1mm]
R^{\dagger}_{\rho,2} = 
\sqrt{2}\,\beta\, s^{\dagger}_{\rho}d^{\dagger}_{\rho} 
+ \sqrt{7}(d^{\dagger}_{\rho}d^{\dagger}_{\rho})^{(2)}, \;\;\;&
R^{\dagger}_{(\pi\nu),2} = 
\beta (\,s^{\dagger}_{\pi}d^{\dagger}_{\nu} 
+ s^{\dagger}_{\nu}d^{\dagger}_{\pi}\,) +                            
\sqrt{14}(d^{\dagger}_{\pi} d^{\dagger}_{\nu})^{(2)} \\[1mm]
W^{\dagger}_{L} = (d^{\dagger}_{\pi} d^{\dagger}_{\nu})^{(L)}
\;\; (L=1,3),\;\;\; &
W^{\dagger}_{2} = 
s^{\dagger}_{\pi}d^{\dagger}_{\nu} - s^{\dagger}_{\nu}d^{\dagger}_{\pi}
\label{pairs}
\end{array}\qquad
\ea 
The $R^{\dagger}_{\rho,L}$ pairs ($\rho=\pi,\nu$) are the same $L=0,2$ pairs 
of Eq.~(\ref{PLb0}) and the $\pi$-$\nu$ pair, $R^{\dagger}_{(\pi\nu),L}$,  
completes the set to form an $F$-spin vector. 
Altogether, the $R^{\dagger}_{i,L}$ ($L=0,2$) 
are boson pairs with $F=1$ and 
$(F_0=1,0,-1)\leftrightarrow [i=\pi,(\pi\nu),\nu]$. 
The $W^{\dagger}_{L}$ 
$(L=1,2,3)$ are F-spin scalar ($F=0$) $\pi$-$\nu$ boson pairs. 
All these operators satisfy
\bsub
\ba
R_{i,L'\mu}\vert c; K=0 \rangle &=& 0 ~,\\
W_{L'\mu}\vert c; K=0 \rangle &=& 0 ~,
\ea
\label{pairscond}
\esub
or equivalently,
\bsub
\ba
R_{i,L'\mu}\vert c;\, [N_{\pi}],[N_{\nu}],\, 
F=F_{max}, LM\rangle &=& 0 ~,\\
W_{L'\mu}\vert c;\, [N_{\pi}],[N_{\nu}],\, 
F=F_{max}, LM\rangle &=& 0 ~.
\ea
\label{pairscondL}
\esub
The states in Eq.~(\ref{pairscondL}) are those obtained by 
${\rm O}_{\pi+\nu}(3)$ 
projection from the intrinsic state~(\ref{condibm2}). 
Since the angular momentum projection 
operator is an F-spin scalar, the projected states of good L will also have 
good $F=F_{max}$. Following the general algorithm, an 
IBM-2 Hamiltonian with partial F-spin symmetry can be constructed as 
\ba
\hat{H}  &=& 
\sum_{i}\sum_{L=0,2}
A^{(i)}_{L}R^{\dagger}_{i,L}\cdot\tilde{R}_{i,L}
+ \sum_{L=1,2,3}B_{L}W^{\dagger}_{L}\cdot\tilde W_{L}
+C_{2}\Bigl [
R^{\dagger}_{(\pi\nu),2}\cdot\tilde W_{2} 
+ H.c.
\Bigr ] ~,
\qquad\;\;\;
\label{hamilt}
\ea 
where $\tilde{R}_{i,L,\mu} = (-1)^{\mu}R_{i,L,-\mu}$, 
$\tilde{W}_{L,\mu} = (-1)^{\mu}W_{L,-\mu}$. 
The above Hamiltonian is an F-spin scalar only 
when $A^{(\pi)}_{L}=A^{(\nu)}_{L}=A^{(\pi\nu)}_{L}\, (L=0,2)$ 
and $C_{2}=0$. Nevertheless, 
the relations of Eqs.~(\ref{pairscond})-(\ref{pairscondL}) ensure that it 
has a solvable zero-energy band with good F-spin 
for {\it any} choice of parameters 
$A^{(i)}_{L},\,B_{L},\, C_{2}$ and {\it any} $N_{\pi},N_{\nu}$.
When $A^{(i)}_{L},\,B_{L},\,A^{(\pi\nu)}_{2}B_{2}-(C_{2})^{2}\geq 0$, 
$\hat{H}$~(\ref{hamilt}) becomes 
positive-definite and the solvable states form its ground band. 
We thus have a non-F-spin scalar Hamiltonian with a 
solvable (degenerate) ground band with $F=F_{max}$. 
The degeneracy can be lifted 
by adding to the Hamiltonian ${\rm O}_{\pi+\nu}(3)$ 
rotation terms which produce $L(L+1)$ type splitting but do not affect 
the wave functions. 
States in other bands can be mixed with respect to F-spin, 
hence the F-spin symmetry of $\hat{H}$ is partial. 
$\hat{H}$ trivially commutes with 
$\hat{F}_{0}$ but not with $\hat{F}_{\pm}$. However, 
$[\,\hat{H},\hat{F}_{\pm}\,]\vert c; K=0\rangle =0$ 
does hold and, therefore, $\hat{H}$ will yield 
F-spin multiplets for members of ground bands. On the other hand, states in 
other bands can have F-spin admixtures and are not compelled to form F-spin 
multiplets. These features which arise from the partial F-spin symmetry 
of the Hamiltonian are in line with the empirical situation as discussed 
above and as depicted in Table~\ref{TabFspinmult} and Fig.~\ref{figFspinmult}. 
It should be noted that the partial F-spin symmetry of $\hat{H}$ holds for 
any choice of parameters in Eq.~(\ref{hamilt}). In particular, one can 
incorporate realistic shell-model based constraints, by choosing
the $A^{(\rho)}_{2}$ ($\rho=\pi,\nu$) terms (representing 
seniority-changing interactions between like nucleons), to be small.
For the special choice $A^{(i)}_{2}=C_{2}=0$ 
and $B_{1}=B_{3}$, $\hat{H}$ of Eq.~(\ref{hamilt}) becomes 
${\rm O}_{\pi+\nu}(5)$-scalar which commutes, therefore, with the 
${\rm O}_{\pi+\nu}(5)$ projection operator and hence produces F-spin 
multiplets with good ${\rm O}_{\pi+\nu}(5)$ symmetry. 
Such multiplets were reported in the Yb-Os region of $\gamma$-soft 
nuclei~\cite{zamfir92}. 

The same conditions ($\beta_{\rho}=\beta$, $\gamma_{\rho}=\Omega=0$) which 
resulted in $F=F_{max}$  for the condensate of Eq.~(\ref{condibm2}), ensure 
also $F=F_{max}-1$ for the intrinsic state representing the scissors band
\ba
\vert sc ; K=1 \rangle &=& 
\Gamma^{\dagger}_{sc}\vert N_{\pi}-1,N_{\nu}-1 \rangle ~,
\nonumber\\
\Gamma^{\dagger}_{sc} &=& 
b^{\dagger}_{c,\pi}d^{\dagger}_{\nu,1} - 
d^{\dagger}_{\pi,1}b^{\dagger}_{c,\nu} ~.
\label{scissors}
\ea
Here $\Gamma^{\dagger}_{sc}$ is a $F=0$ deformed boson pair whose action 
on the condensate with $(N-2)$ bosons produces the scissors mode 
excitation. The states of good $L$ projected from this intrinsic state, 
$\vert sc;\, [N_{\pi}],[N_{\nu}],\, F=F_{max}-1, LM\rangle$ retain 
the F-spin quantum number, $F=F_{max}-1$. Futhermore, it can be shown that 
the operators $R^{\dag}_{i,L\mu}$ of Eq.~(\ref{pairs}) satisfy
\bsub
\ba
&&R_{i,L'\mu}\vert sc; K=1 \rangle = 0 ~,\\
&&R_{i,L'\mu}\vert sc;\, [N_{\pi}],[N_{\nu}],\, 
F=F_{max}-1, LM\rangle = 0 ~.
\ea
\label{Risc}
\esub
Consequently, the scissors intrinsic state~(\ref{scissors}) 
and corresponding $L$-projected states are 
exact eigenstates of the following Hamiltonian, obtained from 
Eq.~(\ref{hamilt}) for the special choice $C_{2} = 0$ and 
$B_{1}= B_{3} = 2B_{2} \equiv 2B$
\ba
\hat{H}' &=& 
\sum_{i}\sum_{L=0,2}
A^{(i)}_{L}R^{\dagger}_{i,L}\cdot\tilde{R}_{i,L} 
+ B\hat{{\cal M}}_{\pi\nu} ~.
\label{hprime}
\ea
The last term in Eq.~(\ref{hprime}) 
is the Majorana operator~\cite{ibm}, 
with eigenvalues $k(N-k+1)$ for states with $F=F_{max}-k$. 
It is related to the total F-spin operator 
and the quadratic Casimir operator of ${\rm U}_{\pi+\nu}(6)$ by 
${\hat{\cal M}}_{\pi\nu} 
= [\,\hat {N}(\hat{N}+2)/4 - \mbox{\boldmath $F^2$}\,]= 
[\,\hat{N}(\hat{N}+5) - \hat{C}_{2}[{\rm U}_{\pi+\nu}(6)]\,]/2$. 
Adding an ${\rm O}_{\pi+\nu}(3)$ rotation term we obtain the following 
Hamiltonian with F-spin PDS
\ba
\hat{H}_{PDS} = \hat{H}' + C\,\hat{C}_{2}[{\rm O}_{\pi+\nu}(3)] 
= \hat{H}_{DS} + \sum_{i}\sum_{L=0,2}
A^{(i)}_{L}R^{\dagger}_{i,L}\cdot\tilde{R}_{i,L} ~. 
\label{hPDSFspin}
\ea
Here $\hat{H}_{DS}$ contains the Majorana and 
${\rm O}_{\pi+\nu}(3)$ terms, associated with the dynamical symmetry chain 
${\rm U}_{\pi+\nu}(6)\supset {\rm O}_{\pi+\nu}(3)$. 
$\hat{H}_{PDS}$~(\ref{hPDSFspin}) has subsets of 
solvable states which form the $K=0$ ground band with $F=F_{max}$, 
\bsub
\ba
&&\vert c;\, [N_{\pi}],[N_{\nu}],\, F=F_{max}, LM\rangle
\;\;\;\;\; L=0,2,4,\ldots, 2N\\
&& E_{g}(L) = C\,L(L+1) ~,
\label{gbandibm2}
\ea
\esub
and the $K=1$ scissors band with $F=F_{max}-1$
\bsub
\ba
&&\vert sc;\, [N_{\pi}],[N_{\nu}],\, F=F_{max}-1, LM\rangle
\;\;\;\;\; L=1,2,3,\ldots, 2N-1\\
&& E_{sc}(L) = B\,N + C\,L(L+1) ~.
\label{scband}
\ea
\esub 
It follows that for such Hamiltonians, 
both the ground and scissors band have good F-spin and have the same 
moment of inertia. The latter derived property is in agreement with the 
conclusions of a comprehensive analysis of the scissors mode in 
heavy even-even nuclei~\cite{enders99}, which concluded that, 
within the experimental precisions ($\sim$ 10\%), the moment of inertia of 
the scissors mode are the same as that of the ground band. It is the 
partial F-spin symmetry of the Hamiltonian~(\ref{hPDSFspin}) which is 
responsible for the common signatures of collectivity in these two bands.
\begin{table}[t]
\begin{center} 
\caption[]{\label{TabM1Fspin}
\protect\small 
The ratio $R=\sum B(M1)\uparrow/(C_{F,F_0})^2$ 
for members of F-spin~multiplets. 
Here $\sum B(M1)\uparrow$ denotes the experimental 
summed M1 strength to the scissors mode~\cite{pietralla95,maser96} 
and $C_{F,F_0} = (F,F_0;1,0\vert F-1,F_0)$. 
Adapted from~\cite{levgin00}.}
\vspace{1mm}
\begin{tabular}{lccccc}
\hline
& & & & &\\[-3mm]
Nucleus & $F$ & $F_0$ & $\sum B(M1)\uparrow$ $[\mu_{N}^2]$ 
& $(C_{F,F_0})^2$ & $R$ \\
& & & & &\\[-3mm]
\hline
& & & & &\\[-2mm]
$^{148}$Nd & 4    & 1    & 0.78 (0.07) & 5/12     & 1.87 (0.17) \\
$^{148}$Sm &      & 2    & 0.43 (0.12) & 1/3      & 1.29 (0.36) \\
& & & & &\\[-3mm]
\hline
& & & & &\\[-2mm]
$^{150}$Nd & 9/2  & 1/2  & 1.61 (0.09) & 4/9      & 3.62 (0.20) \\[2pt]
$^{150}$Sm &      & 3/2  & 0.92 (0.06) & 2/5      & 2.30 (0.15) \\
& & & & &\\[-3mm]
\hline
& & & & &\\[-2mm]
$^{154}$Sm & 11/2 & 1/2  & 2.18 (0.12) & 5/11     & 4.80 (0.26) \\[2pt]
$^{154}$Gd &      & 3/2  & 2.60 (0.50) & 14/33    & 6.13 (1.18) \\
& & & & &\\[-3mm]
\hline
& & & & &\\[-2mm]
$^{160}$Gd & 7    & 0    & 2.97 (0.12) & 7/15     & 6.36 (0.26) \\[2pt]
$^{160}$Dy &      & 1    & 2.42 (0.18) & 16/35    & 5.29 (0.39) \\
& & & & &\\[-3mm]
\hline
& & & & &\\[-2mm]
$^{162}$Dy & 15/2 & 1/2  & 2.49 (0.13) & 7/15     & 5.34 (0.28) \\[2pt]
$^{166}$Er &      & $-1/2$ & 2.67 (0.19) & 7/15   & 5.72 (0.41) \\
& & & & &\\[-3mm]
\hline
& & & & &\\[-2mm]
$^{164}$Dy & 8    & 0    & 3.18 (0.15) & 8/17     & 6.76 (0.32) \\[2pt]
$^{168}$Er &      & $-1$   & 3.30 (0.12) & 63/136 & 7.12 (0.26) \\[2pt]
$^{172}$Yb &      & $-2$   & 1.94 (0.22)$^{a)}$ & 15/34 & 4.40 (0.50) \\
& & & & &\\[-3mm]
\hline
& & & & &\\[-2mm]
$^{170}$Er & 17/2 & $-3/2$ & 2.63 (0.16) & 70/153 & 5.75 (0.35) \\[2pt]
$^{174}$Yb &      & $-5/2$ & 2.70 (0.31) & 66/153 & 6.26 (0.72) \\
& & & & &\\[-3mm]
\hline
\end{tabular}\\[6pt]
{\small $^{a)}$ The low value of $\sum B(M1)\uparrow $ 
for $^{172}$Yb has been attributed to experimental deficiencies 
\cite{rich95}.
$\qquad\qquad$}\\
\end{center}
\end{table}

The Hamiltonian $\hat{H}'$ of Eq.~(\ref{hprime}) is not F-spin invariant, 
however, the following relations are satisfied, 
$[\,\hat{H}' , \vec{F} \,]\,\vert c; K=0 \rangle = 
[\,\hat{H}' , \vec{F} \,]\,\vert sc; K=1 \rangle = 0$. 
This implies that members of both the ground and scissors bands are 
expected to form F-spin multiplets. For ground bands such structures have 
been empirically established~\cite{harter85,brentano85,gupta89,zamfir92}. 
The prediction for F-spin multiplets of 
scissors states requires further elaboration. Although the mean energy of 
the scissors mode is at about 3 MeV~\cite{pietralla98}, 
the observed fragmentation of the 
M1 strength among several $1^{+}$ states prohibits, unlike ground bands, 
the use of nearly constant excitation energies as a criteria to 
identify F-spin multiplets of scissors states. 
Instead, a more sensitive test of this suggestion comes from the 
summed ground to scissors B(M1) strength. The IBM-2 
M1 operator $(\hat{L}_{\pi}- \hat{L}_{\nu})$ 
is an F-spin vector ($F=1,F_{0}=0$). Its matrix element between the  
ground state [$L=0^{+}_{g},\,(F = F_{max},F_{0})$] and scissors state 
[$L=1^{+}_{sc},\,(F' =F-1,F_{0}$)] is proportional to an F-spin 
Clebsch Gordan coefficient $C_{F,F_0} = (F,F_0;1,0\vert F-1,F_0)$ times 
a reduced matrix element. It follows that the ratio  
$B(M1;0^{+}_{g}\rightarrow 1^{+}_{sc})/(C_{F,F_0})^2$ does not 
depend on $F_{0}$ and should be a constant in a given F-spin multiplet. 
In Table~\ref{TabM1Fspin} we list 
{\it all} F-spin partners for which the summed B(M1) strength 
to the scissors mode has been measured~\cite{pietralla95,maser96}. 
It is seen that within the experimental errors, the above ratio is fairly 
constant. The most noticeable discrepancy for $^{172}$Yb (F=8), arises 
from its measured low value of summed B(M1) strength. 
The latter should be regarded as a lower limit due to experimental 
deficiencies (large background and strong fragmentation~\cite{rich95}).
These observations strengthen the contention of high F-spin purity and 
formation of F-spin multiplets of scissors states.

The Hamiltonian $\hat{H}'$~(\ref{hprime}) depends 
on $\beta$ through the operators $R^{\dag}_{i,L}$~(\ref{pairs}). 
It exhibits additional partial symmetries for specific choices of the 
deformation and/or parameters. 
Specifically, for $\beta=\sqrt{2}$, $\hat{H}'$ 
has both F-spin and ${\rm SU}_{\pi+\nu}(3)$ PDS of type I. 
In such circumstances, the ground (K=0) and scissors (K=1) bands 
have good F-spin and ${\rm SU}_{\pi+\nu}(3)$ symmetries: 
$[(\lambda,\mu),F] = [(2N,0),F_{max}]$ and $[(2N-2,1),F=F_{max}-1]$, 
respectively. 
If in addition, $A^{(\pi)}_{2}=A^{(\nu)}_{2}=A^{(\pi\nu)}_{2}$, 
then also the symmetric-$\gamma$ ($K=2$), and antisymmetric-$\gamma$ 
($K=2$) bands are solvable and have good SU(3) and F-spin symmetries: 
$[(2N-4,2),F=F_{max}]$ and $[(2N-4,2),F=F_{max}-1]$, respectively.
In this case, also states of the $\gamma$-bands will be arranged in 
F-spin multiplets.  At the same time, since the Hamiltonian is not F-spin 
scalar, the $\beta$ bands can have F-spin admixtures and need not form 
F-spin multiplets.  
As noted in~\cite{brentano85,gupta89} and shown in Table~\ref{TabFspinmult} 
and Fig.~\ref{figFspinmult}, 
such a behaviour is observed for nuclei with $F=6,\, 13/2$. 
For $\beta=1$, the ground (K=0) and scissors (K=1) bands 
have good F-spin and ${\rm O}_{\pi+\nu}(6)$ symmetries: 
$[\langle\sigma_1,\sigma_2\rangle,F] = [\langle N,0\rangle,F_{max}]$ 
and $[\langle N-1,1\rangle,F_{max}-1]$, respectively, 
but the projected states are mixed with respect to 
${\rm O}_{\pi+\nu}(5)$. Consequently, in this case, 
$\hat{H}'$~(\ref{hprime}) has ${\rm O}_{\pi+\nu}(6)$ PDS of type III. 
For $A_{2}^{(\pi)}=A_{2}^{(\nu)}=A_{2}^{(\pi\nu)}=0$, 
$\hat{H}'$ is ${\rm O}_{\pi+\nu}(5)$-invariant. 
It contains a mixture of terms from several chains:  
${\rm U}_{\pi}(5)\times {\rm U}_{\nu}(5)$, 
${\rm O}_{\pi}(6)\times {\rm O}_{\nu}(6)$ 
and ${\rm U}_{\pi+\nu}(6)$, all sharing a common  
${\rm O}_{\pi+\nu}(5)\supset {\rm O}_{\pi+\nu}(3)$ segment. 
In such circumstances, $\hat{H}'$ 
has a partially-solvable ${\rm O}_{\pi+\nu}(5)$ PDS of type II. 
Such a PDS was used in~\cite{smir02} to obtain an extended M1 sum rule for 
excited symmetric and mixed-symmetry states, and apply it to $^{94}$Mo.

A new aspect that can occur in an algebraic description of coupled systems, 
is the situation in which the set of Casimir operators in a given chain 
of subalgebras of $G$, may not be sufficient to express 
the most general Hamiltonian constructed from the generators of $G$. 
Such a scenario was considered in~\cite{Talmi97} in connection with the 
${\rm U}_{\pi+\nu}(5)$ chain of the IBM-2, and shown to be associated with 
a partial dynamical symmetry. 
The ${\rm U}_{\pi+\nu}(5)$ limit of the IBM-2 corresponds to the chain
\ba
\begin{array}{ccccccccc}
{\rm U}_{\pi}(6)\times {\rm U}_{\nu}(6) 
&\supset &{\rm U}_{\pi+\nu}(6) &\supset 
&{\rm U}_{\pi+\nu}(5)& \supset&{\rm O}_{\pi+\nu}(5) 
&\supset &{\rm O}_{\pi+\nu}(3)\\ 
\downarrow&&\downarrow&&\downarrow&&\downarrow&&\downarrow\\[0mm] 
[N_{\pi}]\times [N_{\nu}] && [N-k,k]&&\{n_1,n_2\} &&(\tau_1,\tau_2) 
&\alpha&L 
\end{array} ~.
\label{chainu5ibm2}
\ea
The total number of bosons is $N=N_{\pi}+N_{\nu}$ and their F-spin is 
$F= N/2 -k$. The states conserve also the total number of $d$-bosons, 
$n_d=n_1+n_2$, and their separate F-spin, $F_d = (n_1-n_2)/2$.
Apart from $\hat{N}_{\rho}$- and $\hat{N}$-dependent terms, the most general 
one- and two-body Hamiltonian which has a ${\rm U}_{\pi+\nu}(5)$  
dynamical symmetry (DS) is given by
\ba
\hat{H}_{DS} &=& 
\epsilon\,\hat{C}_{1}[{\rm U}_{\pi+\nu}(5)]
+ \eta\, (\hat{C}_{1}[{\rm U}_{\pi+\nu}(5)])^2
+ A\,\hat{C}_{2}[{\rm U}_{\pi+\nu}(5)]
\nonumber\\
&& + B\,\hat{C}_{2}[{\rm O}_{\pi+\nu}(5)] 
+ C\,\hat{C}_{2}[{\rm O}_{\pi+\nu}(3)] 
+ a\hat{{\cal M}}_{\pi\nu} ~,
\qquad\quad
\label{hDSu5ibm2}
\ea
where $\hat{C}_{p}[G]$ denoted the $p$-th order Casimir operator of $G$. 
However, it is not the most general Hamiltonian constructed from 
the generators of ${\rm U}_{\pi+\nu}(5)$. 
To obtain the latter, another independent 
operator must be added to $\hat{H}_{DS}$ 
which is not a Casimir operator of a subalgebra. 
A simple choice of such an operator can be~\cite{Talmi97}
\ba
\hat{V}_1 &=& \xi_1\,W^{\dag}_{1}\cdot \tilde{W}_{1} ~,
\label{V1}
\ea 
where $W^{\dag}_{1\mu}$ is the $(F=0,\,L=1)$ boson-pair defined 
in Eq.~(\ref{pairs}). 
The latter transforms as a $(\tau_1,\tau_2)=(1,1)$ tensor under 
${\rm O}_{\pi+\nu}(5)$. Consequently, $\hat{V}_1$ has components with 
$(\tau_1,\tau_2) = (2,2)\oplus (0,0)$, hence 
breaks the ${\rm O}_{\pi+\nu}(5)$ symmetry. 
The ${\rm U}_{\pi+\nu}(6)$ irrep $[N-k,k]$ with $k\neq 0$ contains 
${\rm O}_{\pi+\nu}(5)$ irreps $(\tau_1,\tau_2)$ with $\tau_2\neq 0$, 
which can be admixed by this term. 
Nevertheless, $\hat{V}_1$ has a subset of zero-energy solvable 
states with good ${\rm O}_{\pi+\nu}(5)$ symmetry. These are the 
${\rm U}_{\pi+\nu}(5)$ basis states with $F_d=n_d/2$, 
which are annihilated by $W_{1\mu}$~\cite{Talmi97}, 
\ba
W_{1\mu}\vert [N_{\pi}]\times [N_{\nu}];\,[N-k,k], \{n_d,0\}, 
(\tau,0), n_{\Delta}, LM\rangle &=& 0 ~.
\label{W1mu}
\ea
The interaction $\hat{V}_1$ can be added to $\hat{H}_{DS}$~(\ref{hDSu5ibm2}) 
to obtain the following Hamiltonian with ${\rm O}_{\pi+\nu}(5)$ PDS 
of type I
\ba
\hat{H}_{PDS} &=& \hat{H}_{DS} + \xi_1\, W^{\dag}_{1}\cdot \tilde{W}_{1} ~.
\label{hPDSu5ibm2}
\ea
$\hat{H}_{PDS}$ 
breaks the ${\rm U}_{\pi+\nu}(5)$ DS but retains a subset of 
solvable ${\rm U}_{\pi+\nu}(5)$ 
basis states with known eigenvalues 
\ba
\begin{array}{l}
\vert [N_{\pi}]\times [N_{\nu}];\,[N-k,k], \{n_d,0\}, 
(\tau,0), n_{\Delta}, LM\rangle\\
E_{PDS} = \epsilon n_d +\eta n_{d}^2
+ An_d(n_d+4) + B\tau(\tau+3) + CL(L+1)
+ a k(N-k+1) ~. 
\end{array} 
\label{ePDSu5ibm2}
\ea 
$\hat{H}_{PDS}$~(\ref{hPDSu5ibm2}) exhibits also a 
${\rm U}_{\pi+\nu}(6)\supset {\rm U}_{\pi+\nu}(5)$ PDS of type II, since 
the remaining eigenstates preserve the quantum numbers of 
${\rm U}_{\pi+\nu}(6),\,{\rm U}_{\pi+\nu}(5)$ and ${\rm O}_{\pi+\nu}(3)$ 
but not of ${\rm O}_{\pi+\nu}(5)$ in the chain~(\ref{chainu5ibm2}). 
$\hat{V}_1$~(\ref{V1}) 
has additional zero-energy eigenstates with 
$F_d< n_d/2$ which break, however, the ${\rm O}_{\pi+\nu}(5)$ 
symmetry~\cite{Talmi97}. 
These solvable states lead to additional PDS of the 
Hamiltonian~(\ref{hPDSu5ibm2}), provided $B=0$ in Eq.~(\ref{hDSu5ibm2}). 
In general, PDS associated with vanishing eigenvalues of 
$\hat{V}_1$ can 
explain the simple regularities in the spectra of the generalized 
Majorana operator, observed in an IBM-2 analysis of Pd and Ru 
nuclei~\cite{Gianatiempo98}.

\section{PDS in Fermion Systems}
\label{PDSfermion}

Partial symmetries are not confined to bosonic systems. 
The proposed algorithms for constructing Hamiltonians with PDS 
do not rely on the statistics of the constituents, hence can be 
implemented for both bosons and fermions. 
Identifying partial symmetries in fermion systems can 
proceed in two ways.
The first approach relies on a mapping of a bosonic Hamiltonian, 
which posses a partial symmetry, into its fermionic counterpart. 
If the bosonic generators of the spectrum generating algebra 
can be mapped into fermionic generators of the same algebra, 
then both Hamiltonians will exhibit the same type of partial symmetry. 
This approach was demonstrated in~\cite{Manana99} 
for schematic ${\rm U(2)}\times {\rm U(2)}$ Lipkin-type models. 
A second approach relies on a direct construction of fermion
Hamiltonians with partial symmetries. 
In what follows, we demonstrate this approach 
by identifying fermionic PDS related to properties of the 
quadrupole-quadrupole interaction in the framework of the symplectic 
shell-model~\cite{Escher00,Escher02} 
and to properties of seniority-conserving and non-conserving interactions 
in a single $j$ 
shell~\cite{rowerosen01,rosenrowe03,
escuder06,zamick07,isahein08,zamisa08,Talmi10}. 
Such findings constitute a first step towards 
understanding the microscopic origin of PDS in nuclei.

\subsection{PDS in the symplectic shell model}
\label{subsec:SymplecPDS}

The symplectic shell model (SSM)~\cite{Rowe85,SymplM} 
is an algebraic, fermionic, shell-model
scheme which includes multiple $2 \hbo$ one-particle one-hole excitations. 
The scheme is based on the symplectic algebra Sp(6,R) whose generators 
$\hat{A}^{(20)}_{\ell m},\,\hat{B}^{(02)}_{\ell m},\,
\hat{C}^{(11)}_{\ell m}$ and $\hat{H}_0$ have good
SU(3) [superscript $(\lambda,\mu)$]
and O(3) [subscript $\ell,m$] tensorial properties.
The $\hat{A}^{(20)}_{\ell m}$ [$\hat{B}^{(02)}_{\ell m} 
= (-1)^{\ell-m} (\hat{A}^{(20)}_{l,-m})^{\dagger}$], 
$\ell$ = 0 or 2,
create (annihilate) $2 \hbo$ excitations in the system.
The $\hat{C}^{(11)}_{\ell m}$, $\ell$ = 1 or 2, generate a SU(3) 
subgroup and act only within one harmonic oscillator (h.o.\/) shell
($\sqrt{3} \hat{C}^{(11)}_{2m}=$ $Q^E_{2m}$, the symmetrized quadrupole
operator of Elliott, which does not couple different
h.o.\/ shells~\cite{Elliott58}, and $\hat{C}^{(11)}_{1m}=\hat{L}_m$,
the orbital angular momentum operator).
The harmonic oscillator Hamiltonian, 
$\hat{H}_0$, is a SU(3) scalar and generates U(1) in 
U(3) $=$ SU(3) $\times$ U(1).
A fermion realization of these generators is given in~\cite{Escher98b}.
The model fully accommodates the action of the collective
quadrupole operator, 
$Q_{2m}=\sqrt{\frac{16\pi}{5}} \sum_s r^2_s Y_{2m} (\hat{r}_s)$,
which takes the form, $Q_{2m} = \sqrt{3} ( \hat{C}^{(11)}_{2m}
+ \hat{A}^{(20)}_{2m} + \hat{B}^{(02)}_{2m} )$. 

A basis for the symplectic model is generated by applying symmetrically
coupled products of the 2$\hbo$ raising operator $\hat{A}^{(20)}$ with
itself, to the usual $0 \hbo$ many-particle shell-model states.
Each $0 \hbo$ starting configuration
is characterized by the distribution of oscillator quanta into the three
cartesian directions, $\{ \s_1,\s_2,\s_3 \}$ ($\s_1 \geq \s_2 \geq \s_3$),
or, equivalently, by its U(1)$\times$SU(3) quantum numbers $N_{\s} \, \lms$.
Here $\la_{\s} = \s_1 - \s_2$,
$\mu_{\s} = \s_2 - \s_3$ are the Elliott SU(3) labels,
and $N_{\s} = \s_1 +\s_2 +\s_3$ is related to the eigenvalue of the
oscillator number operator. 
\begin{figure}[t]
\begin{center}
\includegraphics[height=5cm]{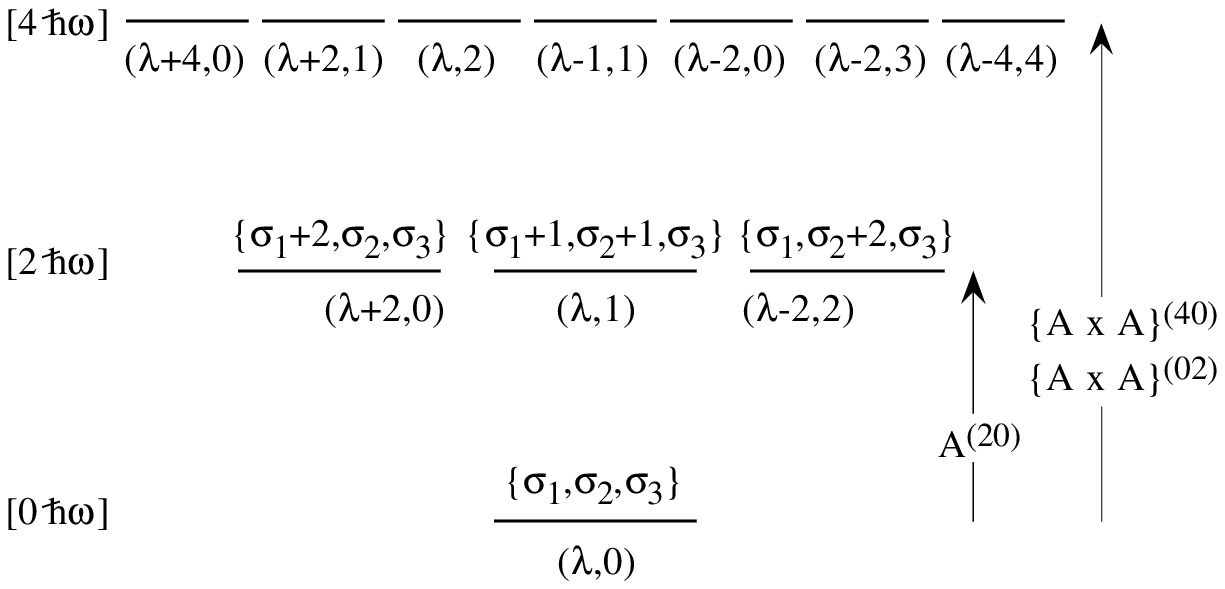}
\caption{
\small
Basis construction in the symplectic model.
SU(3)-coupled products of the raising operator $\hat{A}^{(20)}$ with
itself act on an Elliott starting state with $\lms = (\la,0)$
($\{\s_1,\s_2,\s_3=\s_2\}$) to generate symplectic $2\hbo$, $4\hbo$,
$\ldots$ excitations.
Also shown are the SU(3) labels $\lm$ and quanta distributions
$\{\om_1,\om_2,\om_3\}$ for some excited states. 
Adapted from~\cite{Escher00}.
\label{figSSMIrrep}}
\end{center}
\end{figure}
Each such set of U(3) quantum numbers uniquely determines an irrep 
of the symplectic group, since it characterizes a 
Sp(6,R) lowest weight state. 
The product of $N/2$, $N=0,2,4,\ldots$, 
raising operators $\hat{A}^{(20)}$ 
is multiplicity-free and 
generates $N\hbo$
excitations for each starting irrep $N_{\s} \, \lms$.
Each such product operator ${\cal P}^{N (\la_n,\mu_n)}$, labeled
according to its SU(3) content, $(\la_n,\mu_n)$, is coupled with
$| N_{\s} \, \lms \rangle$ to good SU(3) symmetry $\rho \lm$, with $\rho$
denoting the multiplicity of the coupling $(\la_n,\mu_n) \times \lms$.
The quanta distribution in the resulting state 
is given by
$\{ \om_1,\om_2,\om_3 \} $, with $N_{\s} + N = \om_1 + \om_2 + \om_3$,
$\om_1 \geq \om_2 \geq \om_3$, and $\la = \om_1 - \om_2$,
$\mu = \om_2 - \om_3$.
The basis state construction is schematically illustrated in
Fig.~\ref{figSSMIrrep} for a typical Elliott starting state with
$(\lambda_{\s},\mu_{\s}) = (\lambda, 0)$. 
To complete the basis state labeling, additional quantum numbers 
$\alpha = \kappa L M$ are required, where 
$\kappa$ is a multiplicity index, which enumerates
multiple occurrences of a particular $L$ value in the SU(3) irrep $\lm$
from 1 to
$\kappa^{max}_L \lm = [ (\la+\mu+2-L)/2 ]$ - $[ (\la+1-L)/2 ]$
- $[ (\mu+1-L)/2 ]$, where [$\ldots$] is the greatest non-negative integer
function~\cite{Lopez90}. 
The orthonormal SU(3) basis employed is that of Vergados~\cite{VER}, 
however, for convenience, the running index 
$\kappa = 1, 2, \ldots, \kappa^{max}_L$ is used instead of the 
usual Vergados label, $\tilde{\chi}$. 
The dynamical symmetry chain and the associated quantum labels of the above
scheme are given by~\cite{SymplM}:
\bea
\begin{array}{ccccc}
    {\rm Sp(6,R)} &\supset& {\rm U(3)} &\supset& {\rm SO(3)} \\
\downarrow & & \downarrow && \\
    N_{\sigma}(\lambda_{\sigma},\mu_{\sigma}) & N(\lambda_n,\mu_n) \rho &
    N_{\omega}(\lambda_{\omega},\mu_{\omega})& \kappa & L
\end{array} ~.
\label{eq:DSBasis}
\eea                                           

The following SSM Hamiltonians which has 
SU(3) partial symmetry have been proposed~\cite{Escher00,Escher02} 
\ba
\hat{H}(\beta_0,\beta_2) &=& \beta_0 \hat{A}_0 \hat{B}_0
+ \beta_2 \hat{A}_2 \cdot \hat{B}_2 
\nonumber\\
&=&  \frac{\beta_2}{18} ( 9\hat{C}_{{\rm SU(3)}} - 9\hat{C}_{{\rm Sp(6)}}
+ 3\hat{H}_0^2 - 36\hat{H}_0 )
   + ( \beta_0 - \beta_2) \hat{A}_0 \hat{B}_0 ~.  
\label{Eq:Hpds}
\ea
Here $\hat{A}_{\ell m}\equiv \hat{A}_{\ell m}^{(2,0)}$, 
$\hat{B}_{\ell m}\equiv \hat{B}_{\ell m}^{(0,2)}$ 
and the Casimir operators, $\hat{C}_{G}$, conform with the conventions 
used in~\cite{Escher00,Escher02}.  
For $\beta_0=\beta_2$, the Hamiltonian is an SU(3) scalar 
which is diagonal in the dynamical symmetry basis~(\ref{eq:DSBasis}). 
For $\beta_0=-5\beta_2$, the Hamiltonian transforms as a $(2,2)$ tensor 
under SU(3). Thus, in general, $\hat{H}(\beta_0,\beta_2)$ is not 
SU(3) invariant, however, it exhibits partial SU(3) symmetry. 
Specifically, among the eigenstates of $\hat{H}(\beta_0,\beta_2)$, 
there exists a subset of solvable pure-SU(3) states, 
$\vert\phi_{LM}(N)\rangle$, 
the SU(3)$\supset$O(3) classification of which depends on both the Elliott
labels $(\lambda_{\s},\mu_{\s})$ of the starting state and the symplectic
excitation $N$. In general, it is found that all L-states in the 
starting configuration ($N=0$) are solvable with good SU(3) 
symmetry~$\lms$. For excited configurations, with $N>0$ ($N$ even), 
one can distinguish between two possible cases:
\begin{figure}
\begin{center}
\includegraphics[height=5cm]{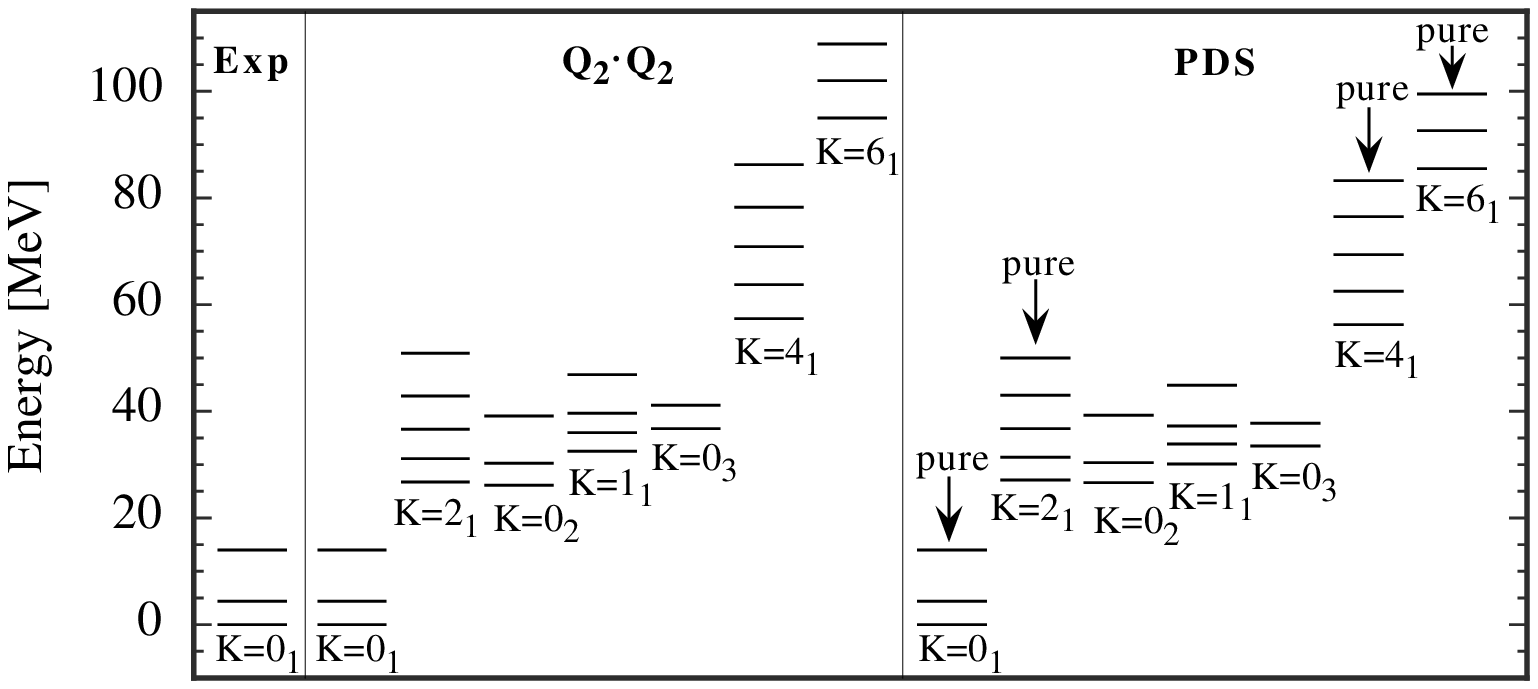}
\caption{\footnotesize 
Energy spectra for $^{12}$C.
Comparison between experimental values (left), results from a 
symplectic $8\hbo$ calculation (center) and a PDS calculation (right).
K=$0_1$ indicates the ground band in all three parts of the figure.
In addition, resonance bands dominated by 2$\hbo$ excitations
(K=$2_1,0_2,1_1,0_3$), 4$\hbo$ excitations (K=$4_1$), and 6$\hbo$
excitations (K=$6_1$) are shown for the Sp(6,R) and PDS calculations.
Additional mixed resonance bands (not shown), dominated by 4$\hbo$ and 6$\hbo$ 
excitations, exist for this nucleus.
The angular momenta of the positive parity states in the rotational bands are
$L$=0,2,4,$\ldots$ for K=0 and $L$=K,K+1,K+2, $\ldots$ otherwise.
Bands which consist of pure-SU(3) eigenstates of the 
PDS Hamiltonian~(\ref{hPDSsymp}) are indicated. 
Adapted from~\cite{Escher02}.
\label{figEnergies_C12}}
\end{center}
\end{figure}
\begin{table}
\begin{center}
\caption{\label{TabBE2_C12}
\protect\small
B(E2) values (in Weisskopf units) 
for ground band transitions in $^{12}$C.
Compared are several symplectic calculations, PDS results, 
and experimental data.
Q denotes the static quadrupole moment of the $L^{\pi}=2^+_1$ state 
in units of $eb$. 
PDS results are rescaled by an effective charge e$^*$=1.33 and
the symplectic calculations employ bare charges.
Adapted from~\cite{Escher02}.}
\vspace{1mm}
\begin{tabular}{ccccccccc}
\hline
& & & & & & & &\\[-3mm]
\multicolumn{1}{c}{Transition} & \multicolumn{5}{c}{Model B(E2) [W.u.]} &
B(E2) [W.u.] \\
 \cline{2-6}
 $J_i \rightarrow J_f$ & \multicolumn{1}{c}{$2\hbar\om$} &
 \multicolumn{1}{c}{$4\hbar\om$} & \multicolumn{1}{c}{$6\hbar\om$} &
 \multicolumn{1}{c}{$8\hbar\om$} & \multicolumn{1}{c}{PDS} & Exp \\
& & & & & & & &\\[-3mm]
\hline
& & & & & & & &\\[-2mm]
2 $\rightarrow$ 0 & 4.65 & 4.65 & 4.65 & 4.65 & 4.65 &  4.65 $\pm$ 0.26\\[2pt]
4 $\rightarrow$ 2 & 4.35 & 4.27 & 4.24 & 4.23 & 4.28 &    n/a          \\
& & & & & & & &\\[-3mm]
\hline
& & & & & & & &\\[-2mm]
\multicolumn{1}{c}{
     $Q$ [eb]}      & 0.059& 0.060& 0.060& 0.060& 0.058& 0.06$\pm$ 0.03\\
& & & & & & & &\\[-3mm]
\hline
\end{tabular}
\end{center}
\end{table}
\begin{figure}[t]
\begin{center}
\includegraphics[height=5cm]{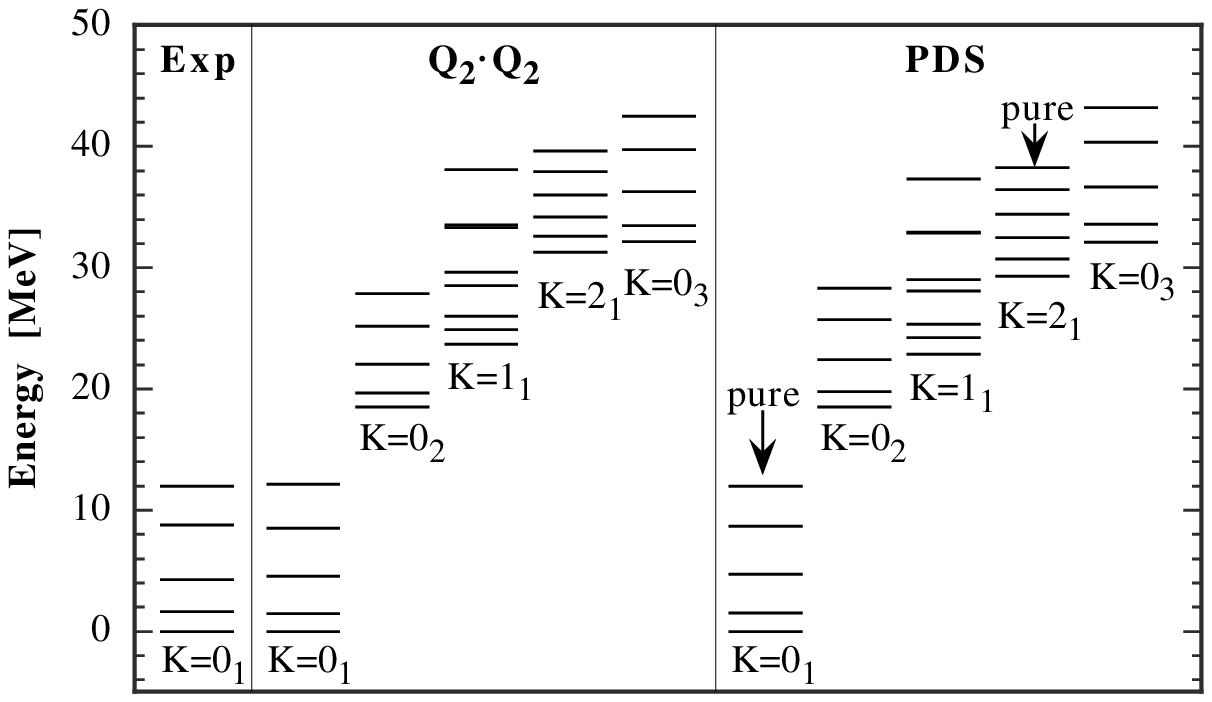}
\caption{
\small
Energy spectra for $^{20}$Ne. 
Experimental ground band (K=$0_1$) energies (left), compared to 
theoretical results for both the ground band 
and 2$\hbo$ resonances (K=$0_2,1_1,2_1,0_3$) 
for a symplectic $8\hbo$ calculation (center) 
and a PDS calculation (right). 
Rotational bands which consist of pure eigenstates of the PDS Hamiltonian
are indicated. Adapted from~\cite{Escher02}. 
\label{figNe20Energies}}
\end{center}
\end{figure}
\begin{figure}[t]
\hspace{3.5cm}$^{20}$Ne
\hspace{6.5cm}$^{12}$C
\begin{center}
\hspace{-0.5cm}
\includegraphics[height=12cm]{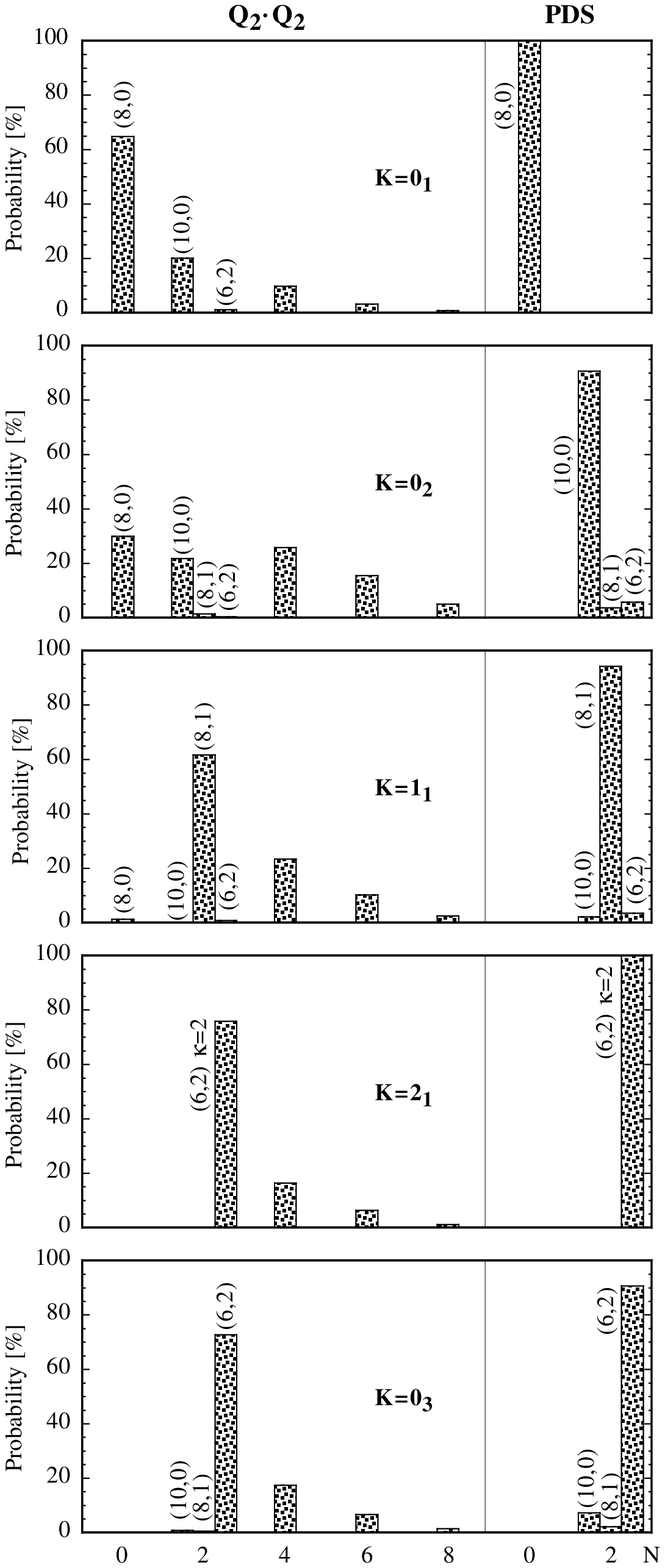}
\hspace*{-2cm}
\includegraphics[height=12cm]{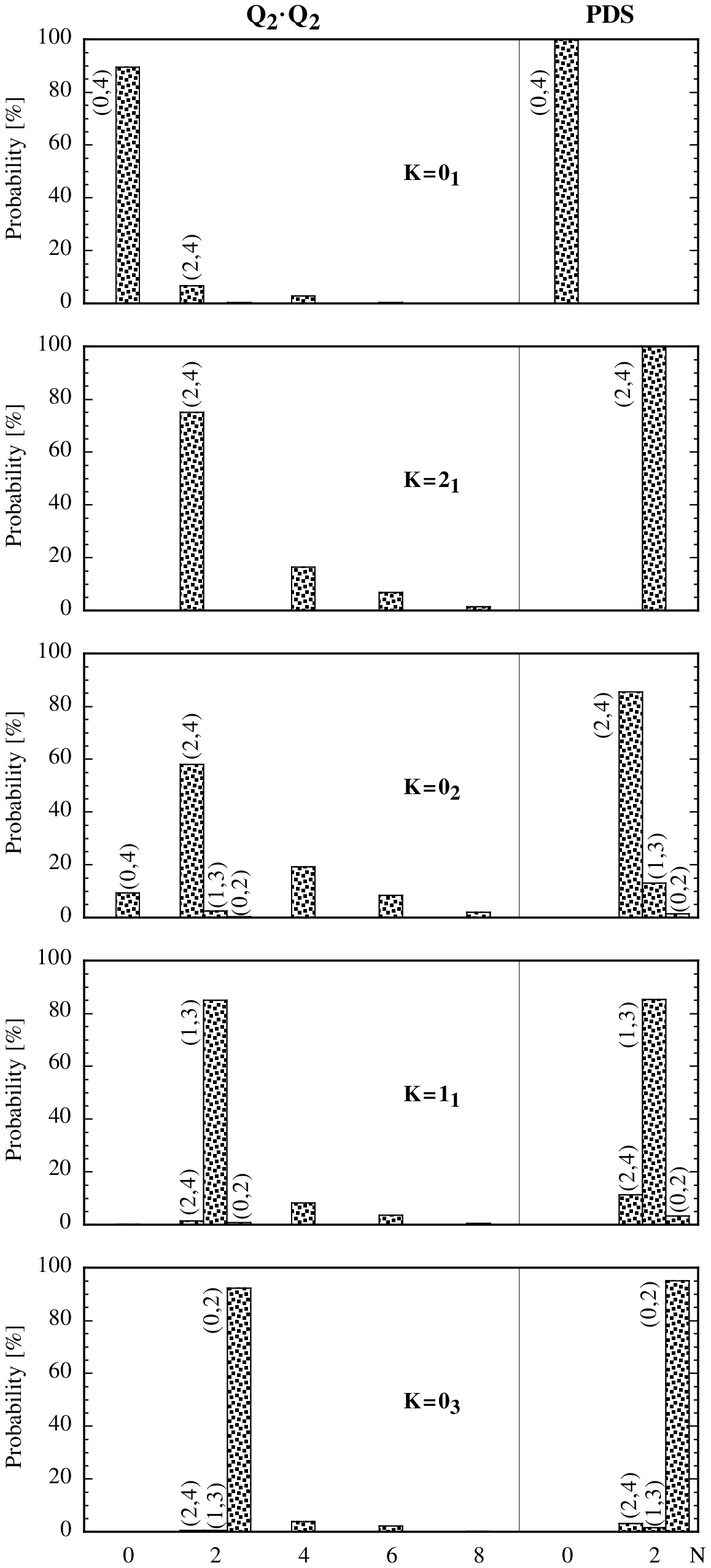}
\caption{\footnotesize 
Decompositions for calculated $2^+$ states of $^{20}$Ne (left) and 
$^{12}$C (right). 
Individual contributions from the relevant SU(3) irreps at the 0$\hbo$
and 2$\hbo$ levels are shown for both a symplectic $8\hbo$ calculation
(denoted $Q_{2}\cdot Q_{2}$) and a PDS calculation.
In addition, the total strengths contributed by the $N\hbo$ excitations
for $N>2$ are given for the symplectic case.
Adapted from~\cite{Escher02}.}
\label{figDecompC12Ne20}
\end{center}
\end{figure}
\begin{table}
\begin{center}
\caption{\label{Tab:Ne20Be2}
\protect\small
B(E2) values (in Weisskopf units)
for ground band transitions in $^{20}$Ne.
Compared are several symplectic calculations, PDS results,
and experimental data. 
The static quadrupole moment of the $2^+_1$ state is given in the last row.
PDS results are rescaled by an effective charge e$^*$=1.95 and
the symplectic calculations employ bare charges. 
Adapted from~\cite{Escher02}.} 
\vspace{1mm}
\begin{tabular}{ccccccccc}
\hline
& & & & & & & &\\[-3mm]
\multicolumn{1}{c}{Transition} & \multicolumn{5}{c}{Model B(E2) [W.u.]} &
B(E2) [W.u.] \\
 \cline{2-6}
 $J_i \rightarrow J_f$ & \multicolumn{1}{c}{$2\hbar\om$} &
 \multicolumn{1}{c}{$4\hbar\om$} & \multicolumn{1}{c}{$6\hbar\om$} &
 \multicolumn{1}{c}{$8\hbar\om$} & \multicolumn{1}{c}{PDS} & Exp \\
& & & & & & & &\\[-3mm]
\hline
& & & & & & & &\\[-2mm]
2 $\rightarrow$ 0 & 14.0 & 18.7 & 19.1 & 19.3 & 20.3 &  20.3 $\pm$ 1.0 \\[2pt]
4 $\rightarrow$ 2 & 18.4 & 24.5 & 24.6 & 24.5 & 25.7 &  22.0 $\pm$ 2.0 \\[2pt]
6 $\rightarrow$ 4 & 17.1 & 22.3 & 21.5 & 20.9 & 21.8 &  20.0 $\pm$ 3.0 \\[2pt]
8 $\rightarrow$ 6 & 12.4 & 15.2 & 13.3 & 12.4 & 12.9 &   9.0 $\pm$ 1.3 \\
& & & & & & & &\\[-3mm]
\hline
& & & & & & & &\\[-2mm]
\multicolumn{1}{c}{$Q$ [eb]}
    & -0.14 & -0.16 & -0.16 & -0.16 & -0.17 & -0.23 $\pm$ 0.03 \\
& & & & & & & &\\[-3mm]
\hline
\end{tabular}
\end{center}
\end{table}                                                
\begin{figure}[t]
\begin{center}
\vspace{-1cm}
\includegraphics[height=5in]{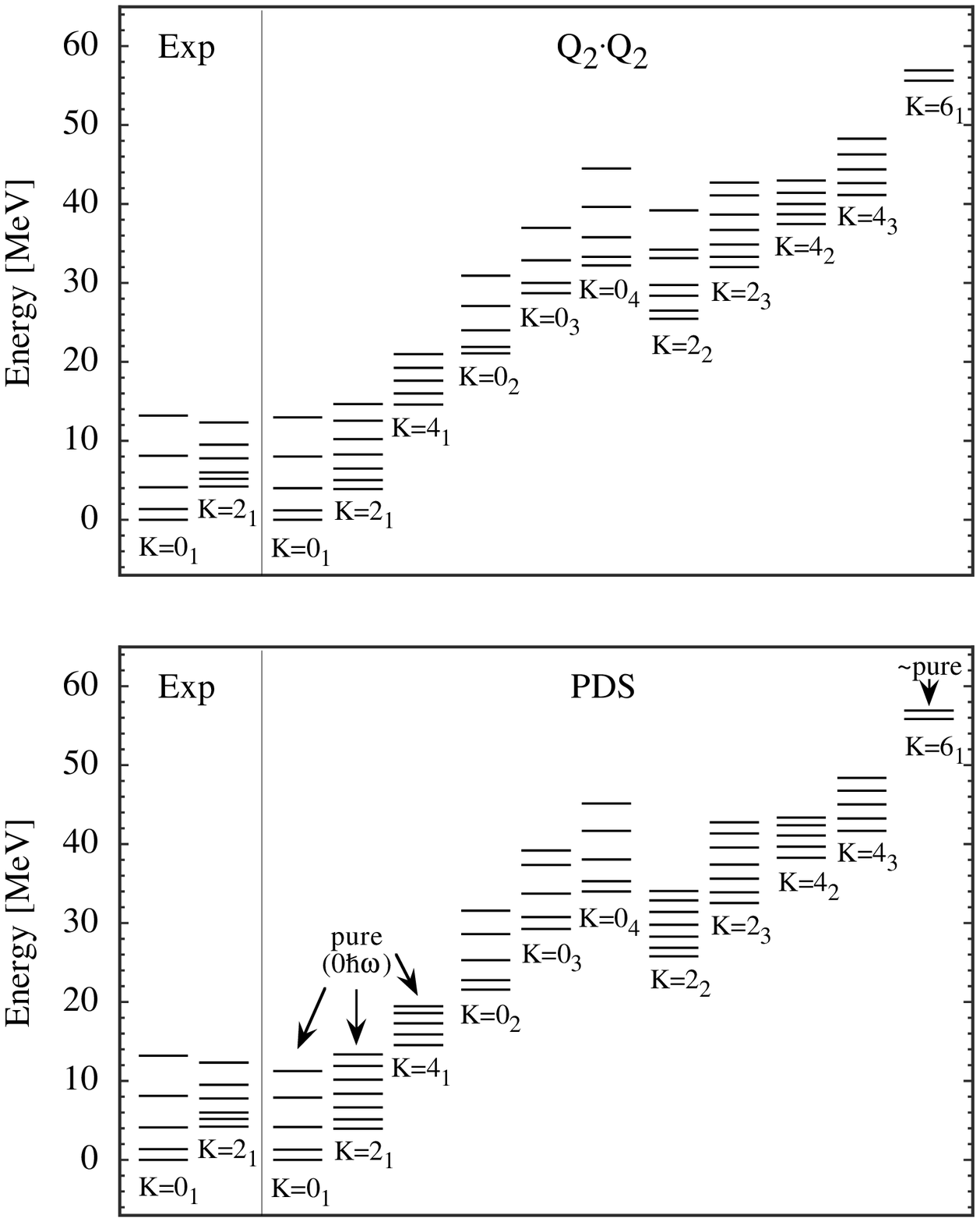}
\vspace{-2.5cm}
\caption{
\small
Energy spectra for $^{24}$Mg.
Energies from a PDS calculation (bottom) are compared to symplectic 
results (top). 
Both 0$\hbo$ dominated bands (K=$0_1,2_1,4_1$) and some 2$\hbo$ resonance
bands (K=$0_2,0_3,0_4,2_2,2_3,4_2,4_3,6_1$) are shown.
The K=$0_1,2_1,4_1$ ($6_1$) states are pure (approximately pure) 
in the PDS scheme. 
Experimental energies of the ground and $\gamma$ bands 
are shown on the left. 
Adapted from~\cite{Escher02}.
\label{figEnergies_Mg24}}
\end{center}
\end{figure}
\begin{figure}[t]
\begin{center}
\includegraphics[height=13cm]{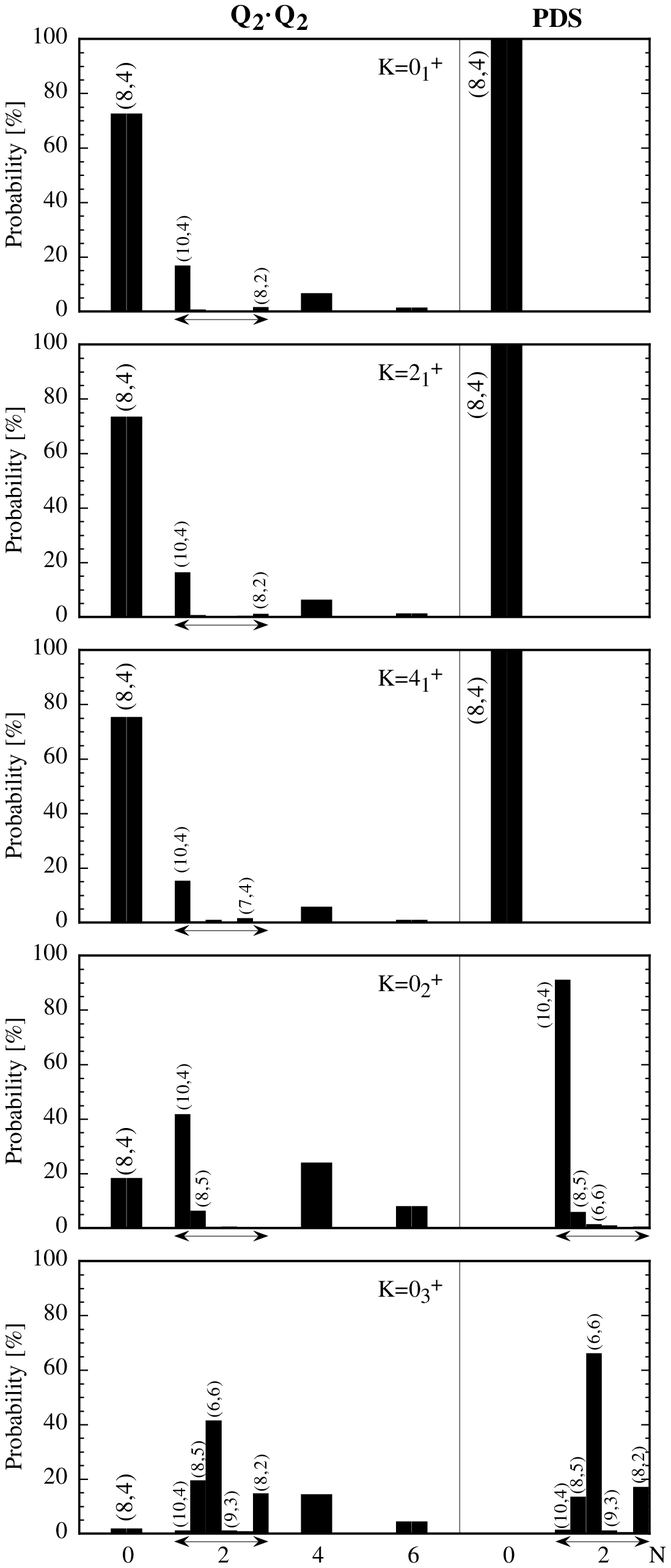}
\hspace*{-3cm}
\includegraphics[height=13cm]{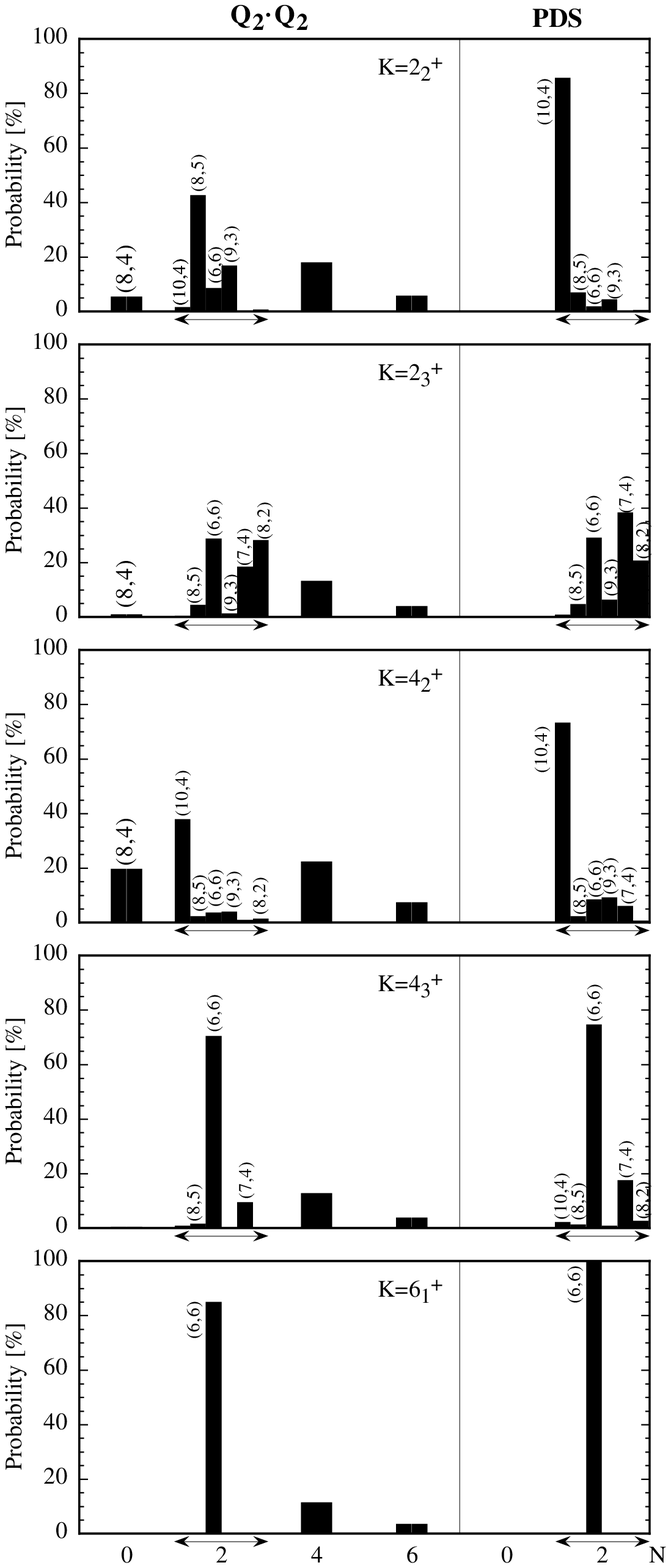}
\caption{\footnotesize 
Decompositions for calculated $L^{\pi}=6^+$ states of $^{24}$Mg.
Eigenstates resulting from the symplectic 6$\hbo$ calculation 
(denoted $Q_{2}\cdot Q_{2}$) are decomposed
into their 0$\hbo$, 2$\hbo$, 4$\hbo$, and 6$\hbo$ components. 
At the 0$\hbo$ and 2$\hbo$ levels, contributions from the individual SU(3)
irreps are shown, for higher excitations ($N>2$) only the summed strengths
are given.
Eigenstates of the PDS Hamiltonian belong entirely to one $N\hbo$ level of
excitation, here 0$\hbo$ or 2$\hbo$.
Contributions from the individual SU(3) irreps at these levels are shown.
Members of the K=$0_1,2_1,4_1$ bands are pure in the PDS scheme, and K=$6_1$
states are very nearly ($>99\%$) pure.
Adapted from~\cite{Escher02}.
\label{figDecomp_Mg24}}
\end{center}
\end{figure}
\begin{itemize}
\item[(a)] $\la_{\s} > \mu_{\s}$:
the pure states belong to $\lm = (\la_{\s}-N,\mu_{\s}+N)$ 
at the $N \hbar\omega$ level and have 
$L = \mu_{\s}+N, \mu_{\s}+N+1, \ldots , \la_{\s}-N+1$
with $N=2,4, \ldots$ subject to $2N \leq (\la_{\s} - \mu_{\s} + 1)$.
\item[(b)] $\la_{\s} \leq \mu_{\s}$:
the special states belong to $\lm =(\la_{\s}+N,\mu_{\s})$ 
at the $N \hbar\omega$ level and have 
$L = \la_{\s}+N, \la_{\s}+N+1, \ldots , \la_{\s}+N+\mu_{\s}$
with $N=2,4, \ldots$.
\end{itemize}
To prove the claim, it is sufficient to show that
$\hat{B}_0$ annihilates the states in question 
\ba
\hat{B}_0\,\vert \phi_{LM}(N)\rangle &=& 0 ~.
\label{B0phiL}
\ea
For $N=0$ this follows immediately from the fact that the $0 \hbo$
starting configuration is a Sp(6,R) lowest weight which, by definition,
is annihilated by the lowering operators of the Sp(6,R) algebra.
The latter include the generators $\hat{B}^{(02)}_{lm}$. For $N>0$,
let $\{ \sigma_1,\sigma_2,\sigma_3 \}$ be the quanta distribution for a
0$\hbo$ state with $\la_{\s} > \mu_{\s}$.
Adding $N$ quanta to the 2-direction yields a $N \hbo$ state
with quanta distribution $\{ \s_1, \s_2+N, \s_3 \}$, that is, 
$\lm = (\la_{\s}-N,\mu_{\s}+N)$.
Acting with the rotational invariant $\hat{B}_0$ on such a state does not
affect the angular momentum, but removes two quanta from
the 2-direction, giving a $(N-2) \hbo$ state with
$(\la',\mu') = (\la_{\s}-N+2,\mu_{\s}+N-2)$.
(The symplectic generator $\hat{B}_0$ cannot remove quanta from the other two
directions of this particular state, since this would yield a state belonging
to a different symplectic irrep.)
Comparing the number of $L$ occurrences in $\lm$ and $(\la',\mu')$, one finds
that as long as $\la_{\s} - N + 1 \geq \mu_{\s}+N$,
$\Delta_L(N) \equiv \kappa_L^{max}\lm - \kappa_L^{max}(\la',\mu') = 1$
for $L = \mu_{\s}+N, \mu_{\s}+N+1, \ldots, \la_{\s}-N+1$, and
$\Delta_L(N)=0$ otherwise.
Therefore, when $\Delta_L(N)$=1, a linear combination 
$|\phi_{L}(N)\rangle =
\sum_{\kappa} c_{\kappa} | N\hbo (\la_{\s}-N,\mu_{\s}+N) \kappa L M \rangle$
exists such that $\hat{B}_0 |\phi_L(N)\rangle = 0$, and thus 
the claim for family (a) holds. 
The proof for family (b) can be carried out analogously.
Here the special irrep $\lm =(\la_{\s}+N,\mu_{\s})$ is obtained by adding
$N$ quanta to the 1-direction of the starting configuration.
In this case there is no restriction on $N$, hence family (b) is infinite.

The special states have well defined symmetry with respect to the 
chain~(\ref{eq:DSBasis}) and the condition of Eq.~(\ref{B0phiL}) 
ensures that they are solvable eigenstates of 
$\hat{H}(\beta_0,\beta_2)$~(\ref{Eq:Hpds}) 
with eigenvalues $E(N=0) = 0$ for the 0$\hbo$ level, and
\bsub
\ba
E(N) &=& \beta_2 \frac{N}{3} ( N_{\s} - \la_{\s} + \mu_{\s} - 6 
+ \frac{3}{2} N )
     \;\;\;\; (\la_{\s} > \mu_{\s}) 
\label{familya}\\ 
E(N) &=& \beta_2 \frac{N}{3} ( N_{\s} + 2 \la_{\s} + \mu_{\s} - 3 
+ \frac{3}{2} N )
     \;\;\;\;  (\la_{\s} \leq \mu_{\s})
\label{familyb}
\ea
\label{Eq:PDSEnergies}
\esub
for $N > 0$. 
All 0$\hbo$ states are unmixed and span the entire $\lms$
irrep. In contrast, for the excited levels ($N > 0$),
the pure states span only part of the corresponding SU(3) irreps. 
There are other states at each excited level which do not preserve the
SU(3) symmetry and therefore contain a mixture of SU(3) irreps. 
The partial SU(3) symmetry of $\hat{H}(\beta_0,\beta_2)$ is converted 
into SU(3)-PDS of type I by adding to it O(3) rotation terms which lead
to L(L+1)-type splitting but do not affect the wave functions.
The solvable states then form rotational bands and since 
condition~(\ref{B0phiL}) determines their wave functions, 
one can obtain analytic expressions for 
the E2 rates between them~\cite{Escher02}. 

The Hamiltonian of Eq.~(\ref{Eq:Hpds}), with SU(3) partial symmetry, 
has a close relationship to the collective quadrupole- quadrupole interaction, 
\bea
\lefteqn{
Q_2 \cdot Q_2 = 3(\hat{C}_2 + \hat{A}_2 + \hat{B}_2 )
                  \cdot (\hat{C}_2 + \hat{A}_2 + \hat{B}_2) }
\nonumber\\
&&= \hat{H}(\beta_0=12,\beta_2=18)+ const - 3 \hat{L}^2 
+ \{ \mbox{terms coupling different h.o. shells} \}  ~.
\qquad\;\;\,
\label{q2q2}
\eea
Here $const = 6\hat{C}_{{\rm Sp(6)}} - 2\hat{H}_0^2 + 34\hat{H}_0$ is 
fixed for a given symplectic irrep $N_{\s}\lms$ and $N\hbo$ excitation.  
Although $\hat{H}(\beta_0,\beta_2)$ does not couple different 
harmonic oscillator shells, it breaks the SU(3)-symmetry 
and, due to the above relation, exhibits in-shell behavior similar to that 
of $Q_2 \cdot Q_2$. To study this connection further, and 
to illustrate that the PDS Hamiltonians of Eq.~(\ref{Eq:Hpds}) are physically
relevant, the formalism was applied to oblate prolate and triaxially 
deformed nuclei, by comparing the properties of the following 
SU(3)-PDS Hamiltonian 
\bea
\hat{H}_{PDS} &=& h(N) + \xi \hat{H}(\beta_{0}=12,\beta_{2}=18) 
+ \gamma_2 \hat{L}^2 + \gamma_4 \hat{L}^4
\label{hPDSsymp}
\eea
to those of the symplectic Hamiltonian
\bea
\hat{H}_{{\rm Sp(6)}} &=& \hat{H}_0 - \eta Q_2 \cdot Q_2 + d_2 \hat{L}^2 
+ d_4 \hat{L}^4 \; .
\label{hSp6}
\eea  
Here the function $h(N)$, which contains the harmonic oscillator term 
$\hat{H}_0$, is simply a constant for a given $N\hbo$ excitation. 
Details of the calculations can be found in~\cite{Escher02}.

The leading Sp(6,R) irrep for the oblate nucleus $^{12}$C 
is $N_{\s} \lms$ $=24.5(0,4)$.
In this case, $\hat{H}_{PDS}$~(\ref{hPDSsymp}) 
has the pure-SU(3) states of family~(b) 
as solvable eigenstates. In particular, 
all 0$\hbo$ states are pure $\lms = (0,4)$ 
states, and at 2$\hbo$ a rotational band with good SU(3) symmetry 
$\lm=(2,4)$ and $L=2,3,4,5,6$ exists. 
Similarly, at 4$\hbo$ there are pure-SU(3) bands with $\lm=(4,4)$ and
$L=4,5,6,7,8$, at 6$\hbo$ with $\lm=(6,4)$ and $L=6,7,8,9,10$, etc. 
On the other hand, 
the leading Sp(6,R) irrep for the prolate nucleus $^{20}$Ne is 
$N_{\s}\lms$ $=48.5(8,0)$. In this case,  
the solvable pure-SU(3) eigenstates of $\hat{H}_{PDS}$ 
are those of family~(a). 
They form a K=$0_1$ $L=0,2,4,6,8$ rotational band
with $\lm=$ (8,0) at 0$\hbo$, a  K=$2_1$ $L=2,3,4,5,6,7$ band with $\lm=$
(6,2) at 2$\hbo$, and a K=$4_1$ $L=4,5$ `band' with $\lm=$ (4,4) at 4$\hbo$. 
There are no other pure PDS states at higher levels of excitation. 
As shown in Fig.~\ref{figEnergies_C12} and Table~\ref{TabBE2_C12} for 
$^{12}$C, and in Fig.~\ref{figNe20Energies} and Table~\ref{Tab:Ne20Be2} 
for $^{20}$Ne, the resulting energies and transition rates in the PDS and 
symplectic approaches are similar and converge to
values which agree with the data. 
The PDS calculations of B(E2) values, however, require an effective charge. 
A better measure for the level of agreement between the PDS and symplectic
results is given by a comparison of the eigenstates. From 
Fig.~\ref{figDecompC12Ne20} we observe that, although 
$\hat{H}_{{\rm Sp(6)}}$~(\ref{hSp6}) 
[$\hat{H}_{PDS}$~(\ref{hPDSsymp})] does (does not) 
mix different major oscillator shells, the $N\hbo$ level 
to which a particular PDS band belongs is also dominant in the 
corresponding symplectic band. Furthermore, within this dominant 
excitation, eigenstates of $\hat{H}_{{\rm Sp(6)}}$ 
and $\hat{H}_{PDS}$ have very similar SU(3) structure, that is, 
the relative strengths of the various SU(3) irreps in the symplectic 
states are approximately reproduced in the PDS case. 
This holds for the ground and excited bands in both nuclei, with the 
exception of the K=$0_2$ resonance band in $^{20}$Ne, where 
significant differences appear in the structure of the wave functions. 
Thus the SU(3)-PDS Hamiltonian captures much of the physics of 
the $Q_{2}\cdot Q_{2}$ interaction. 

Calculations were also performed~\cite{Escher02} 
for the triaxially-deformed nucleus, $^{24}$Mg, whose 
leading Sp(6,R) irrep is $N_{\s}\lms$ $=62.5(8,4)$. 
Since now both $\la_{\s} \neq 0$ and $\mu_{\s} \neq 0$, the symplectic
Hilbert space has a very rich structure. 
In this case, the solvable pure-SU(3) states are those of family~(a) 
and include three rotational K=0,2,4, bands at 0$\hbo$ 
with $(\lambda,\mu)=(8,4)$, 
and at 2$\hbo$ a (short) rotational K=6 band with $L$=6,7, which belongs 
to the irrep $\lm=(6,6)$. 
Figure~\ref{figEnergies_Mg24} compares the experimental spectrum of 
$^{24}$Mg with energies obtained with 6$\hbo$ symplectic and 
PDS calculations. 
Both the PDS and symplectic Hamiltonians 
were supplemented with small cubic and quartic SU(3)-conserving interactions 
to account for $K$-band splitting. These extra terms break 
the partial symmetry slightly ($<1\%$) in the $K=6_1$ band, 
as can be inferred from the eigenstate decompositions plotted 
in Fig.~\ref{figDecomp_Mg24}. As in the previous examples, the eigenstates 
of the PDS and symplectic Hamiltonians are 
found to have very similar structures. 
The structural differences that do exist are reflected in the very
sensitive interband transition rates~\cite{Escher02}. 

The boson Hamiltonian, $\hat{H}(h_0,h_2)$ of Eq.~(\ref{HPSsu3}), 
and the fermion Hamiltonian, $\hat{H}(\beta_0,\beta_2)$ 
of Eq.~(\ref{Eq:Hpds}), have several features in common. Both display 
partial SU(3) symmetry, they are constructed to be rotationally 
invariant functions of $\lm=(2,0)$ and $\lm=(0,2)$ SU(3) tensor operators, 
and both contain components with $\lm=(0,0)$ and $(2,2)$. 
Both Hamiltonian have solvable pure-SU(3)
eigenstates, which can be organized into rotational bands. 
The ground bands are pure in both cases, and higher-energy pure bands 
coexist with mixed-symmetry states. 
There are, however, several significant differences between the bosonic 
and fermionic PDS Hamiltonians. For example, the ground band of the bosonic 
Hamiltonian $\hat{H}(h_0,h_2)$, 
Eq.~(\ref{HPSsu3}), is characterized by $\lm = (2N,0)$, i.e., it describes 
an axially-symmetric prolate nucleus. 
As mentioned at the end of Subsection~\ref{subsec:su3PDStypeI}, 
it is also possible to find an IBM Hamiltonian with partial SU(3) symmetry
for an oblate nucleus.
It can be shown that these two cases exhaust all possibilities for partial
SU(3) symmetry with a two-body Hamiltonian in the IBM with one type of 
monopole and quadrupole bosons.
In contrast, the fermionic Hamiltonian 
$\hat{H}(\beta_0,\beta_2)$~(\ref{Eq:Hpds}) 
can accommodate ground
bands of prolate [$(\la_{\s},0)$], oblate [$(0,\mu_{\s})$], and triaxial
[$(\la_{\s},\mu_{\s})$ with $\la_{\s} \neq 0$, $\mu_{\s} \neq 0$] shapes. 
Another difference 
lies in the physical interpretation of the excited solvable bands.
While these bands represent $\gamma^k$ 
excitations in the IBM, they correspond to giant monopole and quadrupole 
resonances in the fermion case. 
Furthermore, whereas the pure eigenstates of the IBM 
Hamiltonian $\hat{H}(h_0,h_2)$ can be
generated by angular momentum projection from intrinsic states~(\ref{k}), 
a similar
straightforward construction process for the special eigenstates of 
the symplectic Hamiltonian $\hat{H}(\beta_0,\beta_2)$ has not been 
identified yet. 
The situation seems to be more complicated in the fermion case, which is
also reflected in the fact that $\hat{H}(\beta_0,\beta_2)$ has two 
possible families of pure eigenstates, one finite, the other infinite.
The association of the special states to one or the other family depends on
the 0$\hbo$ symplectic starting configuration.
Thus, in spite of similar algebraic structures, the bosonic and fermionic 
PDS Hamiltonians involve different mechanisms for generating the 
partial symmetries in question.

\subsection{PDS and seniority}
\label{subsec:PDSseniority}

The notions of pairing and seniority for $n$ identical nucleons occupying 
a single $j$ shell, 
are conveniently encoded in the quasi-spin formalism~\cite{Kerman61}. 
The latter is based on an ${\rm SU}_{S}(2)$ algebra whose generators are
\ba
\hat{S}^{+}_{j} = 
\sqrt{\frac{\Omega}{2}}\,
(a^{\dag}_{j}\, a^{\dag}_{j})^{(0)}_{0}  
\;\;\; , \;\;\;
\hat{S}^{-}_{j} = (\hat{S}^{+}_{j})^{\dag}  
\;\;\; , \;\;\;
\hat{S}^{0}_{j} = \frac{1}{2}(\hat{n} - \Omega) ~.
\label{SU2Q}
\ea
Here $\hat{S}^{+}_{j}$ creates a pair of fermions with angular momentum 
$J=0$, $\hat{n}=\sum_{m}a^{\dag}_{jm}a_{jm}$ is the number-operator 
and $\Omega = (2j+1)/2$. 
The basic quasi-spin doublet is $(a^{\dag}_{jm},\,-\tilde{a}_{jm})$, 
with $\tilde{a}_{jm} = (-1)^{j+m}a_{j,-m}$. 
The seniority quantum number, $v$, refers to the number of nucleons 
which are not in zero-coupled $J=0$ pairs. 
The ${\rm SU}_{S}(2)$ quantum numbers, $(S,S_0)$ 
are related to $n$ and $v$ by 
\ba
S=(\Omega-v)/2\;\;\; , \;\;\; S_0 = (n-\Omega)/2 ~. 
\label{Sv}
\ea 
The operator $\hat{S}^{-}_{j}$ annihilates states with $(S,S_0=-S)$ 
corresponding to states in the $j^v$ configuration with seniority $v$ 
and $n=v$. 

The relevant chain of nested algebras for $n$ fermions in a single 
$j$ shell is~\cite{flowers52,helmers61}
\ba
\begin{array}{ccccc}
{\rm U(2j+1)} & \supset & {\rm Sp(2j+1)} & \supset & {\rm SU_{J}(2)}\\
\downarrow && \downarrow && \downarrow\\
\left [ 1^n \right ] &         &  v       & \rho    & J 
\end{array} ~,
\label{Senchain}
\ea 
where $\rho$ is a multiplicity index. 
The generators of the indicated algebras are 
\ba
&&{\rm U(2j+1)}=\{\hat{T}^{(L)}_{m}\}\;\; , \;\; 
{\rm Sp(2j+1)}=\{\hat{T}^{(L)}_{m}\; L\, {\rm odd}\} ~,
\nonumber\\
&&{\rm SU_{J}(2)}=\{\hat{J}_{m} = c_j\,\hat{T}^{(1)}_{m}\}\;\; , \;\;
c_j= \sqrt{j(j+1)(2j+1)/3}~,\qquad
\label{Unalgeb}
\ea
where 
\ba
\hat{T}^{(L)}_{m} = \left( a^{\dag}_{j}\,\tilde{a}_{j}\right )^{(L)}_{m} 
\qquad L=0,1,2,\ldots, 2j~.
\label{TLmS}
\ea
The unitary symplectic algebra, Sp(2j+1), and quasi-spin algebra, 
${\rm SU}_{S}(2)$, commute. The duality relationship between 
their respective irreps, $v$ and $S$, is given in Eq.~(\ref{Sv}) 
and their quadratic Casimir operators are related by 
$\hat{C}_{2}[{\rm Sp(2j+1)}] = 
\Omega(\Omega+2) - 4\mbox{\boldmath $\hat{S}^2$}$. 
The dynamical symmetry Hamiltonian associated with the 
chain~(\ref{Senchain}) is given by
\bsub
\ba
\hat{H}_{DS} &=& \beta\,\hat{n} 
+ \alpha \, \hat{n}(\hat{n}-1) 
+ a\,\hat{C}_{2}[{\rm Sp(2j+1)}] 
+ b\,\hat{C}_{2}[{\rm SU_{J}(2)}] ~,
\label{hDSsen}\\
E_{DS} &=& \beta\, n + \alpha\, n(n-1) + a\,v(2\Omega+2 -v) + b\,J(J+1) ~,
\label{eDSsen}
\ea
\label{heDSsen}
\esub
and $E_{DS}$ are the corresponding eigenvalues. 

The $\hat{T}^{(L)}_{m}$ operators of Eq.~(\ref{TLmS}), with $L$ odd, 
are quasi-spin scalars. Accordingly, the most general number-conserving 
rotational invariant Hamiltonian with seniority-conserving one- and 
two-body interactions, acting in a single-$j$ shell, 
can be transcribed in the form~\cite{Talmi93} 
\bsub
\ba
\hat{H} &=& \beta\,\hat{n} 
+ \alpha \, \hat{n}(\hat{n}-1) 
+ \sum_{L\, {\rm odd}}\lambda_{L}\,\hat{T}^{(L)}\cdot 
\hat{T}^{(L)}
\label{hSenCons}\\
E_{n,v,\rho, J} &=& \beta\,n +\alpha\, n(n-1) + Z_{v,\rho,J} ~.
\label{eSenCons}
\ea
\label{heSenCons}
\esub
For the special choice, $\lambda_{L}=2a + b\,c_{j}^2\,\delta_{L,1}$, 
the above Hamiltonian reduces to the dynamical symmetry Hamiltonian of 
Eq.~(\ref{hDSsen}). In general, 
eigenstates of $\hat{H}$~(\ref{heSenCons}) are basis states, 
$\vert n, v, \rho, J  M \rangle$, of the chain~(\ref{Senchain}) with 
eigenvalues $E_{n,v,\rho,J}$ and wave functions of the form 
\ba
\vert n, v, \rho, J  M \rangle \propto 
\left (\hat{S}_{j}^{+}\right )^{(n-v)/2}\vert n=v, j, v,\rho, J M\rangle ~.
\label{nvwf}
\ea
Since the last term in Eq.~(\ref{hSenCons}) is a quasi-spin scalar, 
its eigenvalues, $Z_{v,\rho,J}$, are independent of $n$. 
Consequently, for a given $n$, energy differences 
$E_{n, j, v, \rho, J}- E_{n, j, v',\rho',J'}$,  
are independent of $n$. Conservation of seniority does not, however, 
imply solvability. In general, eigenstates and eigenvalues 
of $\hat{H}$, Eq.~(\ref{heSenCons}), 
must be obtained from a numerical calculation.  
Nevertheless, it has been shown that 
some multiplicity-free eigenstates of $\hat{H}$, 
{\it i.e.}, with unique $n,v,J$ assignments, are completely solvable 
and closed algebraic expressions for their energies 
have been derived~\cite{rowerosen01,rosenrowe03}. 
These include states with $(v=2,J)$, $(v,J_{max})$ and 
$(v,J_{max}-2)$, where $J_{max}= v(2j+1-v)/2$. 
As an example, for $n$ even, 
the ground state of $\hat{H}$~(\ref{hSenCons}), 
with $v=J=0$, is analytically solvable with energy 
\ba
E_{n, j, v=0, J=0} &=& \beta\, n +\alpha\, n(n-1) ~. 
\label{ev0J0}
\ea
The above expression is identical to the ground-state energy 
of the dynamical symmetry Hamiltonian, Eq.~(\ref{heDSsen}). 
Since all eigenstates carry the seniority quantum number $v$, 
the Hamiltonian $\hat{H}$ of Eq.~(\ref{heSenCons}) 
exhibits PDS of type II with an added feature that some multiplicity-free 
states are analytically solvable. 

As is well known~\cite{Talmi93}, 
seniority remains a good quantum number for any 
two-body interaction acting within a $j$ shell when $j\leq 7/2$, but 
it need not be conserved for $j\geq 9/2$. 
Recently, it has been 
shown that it is possible to construct interactions that in general 
do not conserve seniority but which have {\it some} solvable 
eigenstates with good seniority~\cite{escuder06,isahein08}. 
Specifically, for $j=9/2,\,n=4$, 
there are two independent states ($\rho=1,2$) with seniority $v=4$ 
and $J=4,\,6$. For each such $J$ value, 
there exists one particular combination, which is completely solvable 
with good seniority, $v=4$, for {\it any} two-body interaction 
in the $j=9/2$ shell. The energies of the solvable states are 
given by~\cite{isahein08}
\bsub
\ba
E[(9/2)^4, v=4, J=4] &=& 
\frac{68}{33}\,\nu_2
+\nu_4
+ \frac{13}{15}\,\nu_6
+ \frac{114}{55}\,\nu_8 ~,\\[4mm]
E[(9/2)^4, v=4, J=6] &=& 
\frac{19}{11}\,\nu_2
+ \frac{12}{13}\,\nu_4
+\nu_6
+ \frac{336}{143}\,\nu_8~,
\ea
\esub
where $\nu_{\lambda} \equiv 
\langle (9/2)^2;\lambda\vert \hat{V}\vert (9/2)^2;\lambda\rangle$ are 
the matrix elements of an arbitrary rotational-invariant two-body 
interaction, $\hat{V}$. 
The indicated states retain their structure and are completely solvable, 
independent of whether the interaction conserves seniority or not. 
It follows that the 
most general one- and two-body rotational-invariant Hamiltonian in the 
$j=9/2$ shell, exhibits PDS of type I. 
The E2 matrix element between the two solvable states 
is interaction-independent, and the corresponding 
B(E2) value is given by~\cite{isahein08}
\ba
&&B(E2; (9/2)^4, v=4, J=6 \to (9/2)^4, v=4, J=4 ) =
\nonumber\\
&&
\qquad\qquad\qquad\qquad\qquad\qquad
\frac{209475}{176468}\,
B(E2; (9/2)^2, J=2 \to (9/2)^2, J=0 ) ~.
\label{be2sen}
\ea
This again defines a parameter-independent relation between a property 
of the two- and four-particle system. 
Some properties of these solvable states can be attributed to properties 
of certain coefficients of fractional 
parentage~\cite{zamick07,isahein08,zamisa08,Talmi10}. 
It has been suggested that 
this partial seniority conservation 
may shed some new light on the existence of isomers in nuclei 
with valence neutrons or protons predominantly confined 
to the $g_{9/2}$ or $h_{9/2}$ shell~\cite{isahein08}. 

In most medium-mass and heavy nuclei, the valence nucleons are distributed 
over several non-degenerate $j$ orbits in a major shell. This situation 
can be treated within the generalized seniority scheme~\cite{talmi71}, 
based on more general pair-operators with $J=0,2$ 
\bsub
\ba
\hat{S}^{\dag} &=& 
\sum_{j}\alpha_{j}\,(a^{\dag}_{j}\, a^{\dag}_{j})^{(0)}_{0} ~,
\label{Sgen}
\\
D^{\dag}_{\mu} &=& \sum_{j,j'}\beta_{jj'}\,
(a^{\dag}_{j}\, a^{\dag}_{j'})^{(2)}_{\mu} ~.
\ea
\label{vg0vg2}
\esub
The generalized seniority conditions~\cite{talmi71,shlomtalmi72,talmi73} 
are
\bsub
\ba
\hat{H}\vert 0\rangle &=& 0 ~,\\ 
\left [ \, \hat{H}\, , \,S^{\dag}\,\right ]
\vert 0 \rangle &=& V_{0}\,S^{\dag}\vert 0 \rangle \;\; , \;\;
\left [ \left [\, \hat{H}\, , \,S^{\dag}\,\right ]\, , \,S^{\dag}\,\right ]
= \Delta\,(S^{\dag})^2 ~,\\
\left [ \, \hat{H}\, , \,D^{\dag}_{\mu}\,\right ]
\vert 0 \rangle &=& V_{2}\,D^{\dag}_{\mu}\vert 0 \rangle \;\; , \;\;
\left [ \left [\, \hat{H}\, , \,S^{\dag}\,\right ]\, , 
\,D^{\dag}_{\mu}\,\right ]
= \Delta\,S^{\dag}D^{\dag}_{\mu} ~, 
\ea
\label{genSen}
\esub
where $\vert 0\rangle$ is the doubly-magic core state. 
These relations imply that the Hamiltonian has a solvable ground state 
with $J=0$ and generalized seniority $v_g=0$
\bsub
\ba
\vert n=2N, v_g =0, J=0\rangle &\propto& 
\left (S^{\dag}\right )^N\vert 0\rangle ~,\\ 
E_{N,v_g=0,J=0} &=& N\,V_0 + \frac{1}{2}N(N-1)\,\Delta ~,  
\ea
\label{vg0}
\esub
and a solvable excited state with $J=2$ and generalized seniority $v_g=2$
\bsub
\ba
\vert n=2N, v_g =2, J=2\rangle &\propto& 
\left (S^{\dag}\right )^{N-1}D^{\dag}_{\mu}\vert 0\rangle ~,\\ 
E_{N,v_g=2,J=2} &=& E_{N,v_g=0,J=0} + V_2-V_0 ~.
\ea
\label{vg2}
\esub
From the point of view of symmetry, the monopole pair-operators 
$S^{\dag},\, S$, and $\left [S^{\dag}\, , \,S\right ]/2$, of 
Eq.~(\ref{Sgen}), with unequal $\alpha_j$, 
do not form an SU(2) algebra. 
Nevertheless, an Hamiltonian $\hat{H}$ satisfying relations~(\ref{genSen}) 
has selected solvable states, {\it i.e.}, it is partially solvable. 
It should be noted that the energy and wave function 
of the $v_g=0$ ground state~(\ref{vg0}) have 
the same form as in the single-$j$ dynamical symmetry expressions, 
Eqs.~(\ref{nvwf})-(\ref{ev0J0}) and, by construction, both spectra 
exhibit linear two-nucleon separation energies and constant $2^{+}$-$0^{+}$ 
spacings. Furthermore, the Sp(2j+1) generators~(\ref{Unalgeb}), 
$\hat{T}^{(L)}_{m}$ with L odd and {\it any}~$j$, annihilate the state of 
Eq.~(\ref{vg0}). Therefore, the $v_g=0$ ground state is invariant under 
$\prod_{j}\otimes {\rm Sp(2j+1)}$, even though the latter is not a symmetry 
of the Hamiltonian.

\section{Concluding Remarks}
\label{conclusion}

The notion of partial dynamical symmetry (PDS) extends and complements the
fundamental concepts of exact and dynamical symmetries. 
It addresses situations in which a prescribed symmetry is neither 
exact nor completely broken. When found, such intermediate symmetry structure 
can provide analytic solutions and quantum numbers for a portion of the 
spectra, thus offering considerable insight into complex dynamics. 
In other circumstances a PDS can serve as a convenient 
starting point for further improvements.

As discussed in the present review, 
PDSs of various types have concrete applications to nuclear and 
molecular spectroscopy. Their empirical manifestations 
illustrate their ability to serve as a practical tool for calculation 
of observables in realistic systems. 
Hamiltonians with partial symmetries are not 
completely integrable nor fully chaotic. As such, they are relevant 
to the study of mixed systems with coexisting regularity and chaos, 
which are the most generic. Quantum phase transitions are driven by 
competing symmetries 
in the Hamiltonian. They provide a natural arena for PDSs, which 
incorporate such incompatible symmetries, and manage to survive at 
the critical points, in spite of the strong mixing. 

PDSs appear to be a common feature in algebraic descriptions of dynamical 
systems. They are not restricted to a specific model but can be applied 
to any quantal systems of interacting particles, bosons and fermions. 
The existence of PDS is closely related to the order of the interaction 
among the particles. The partial symmetry in question can be continuous 
(Lie-type), or discrete (point-group) and the associated dynamical 
algebra can involve a single or coupled algebraic structure. 
The examples of partial dynamical symmetries surveyed in the present 
review,  involved purely bosonic or purely fermionic algebras. 
It is clearly desirable to extend the PDS notion to mixed systems of 
bosons and fermions, and explore the possible role of partial 
Bose-Fermi symmetries and partial supersymmetries. 

On phenomenological grounds, having at hand concrete algorithms for 
identifying and constructing Hamiltonians with PDS, is a valuable asset. 
It provides selection criteria for the a priori huge number of 
possible symmetry-breaking terms, accompanied by a rapid proliferation 
of free-parameters. This is particularly important in complicated 
environments when many degrees of freedom take part in the dynamics 
and upon inclusion of higher-order terms in the Hamiltonian. 
Futhermore, Hamiltonians with PDS break the dynamical symmetry (DS) 
but retain selected solvable eigenstates with good symmetry. The 
advantage of using interactions with a PDS is that they can be introduced, 
in a controlled manner, without destroying results previously obtained 
with a DS for a segment of the spectrum. These virtues 
greatly enhance the scope of applications of algebraic modeling 
of quantum many-body systems.

On a more fundamental level, it is important to recall 
that dynamical systems often display simple patterns amidst 
a complicated environment. A representative example is the occurrence of both 
collective and quasi-particle type of states in the same nucleus. 
It is natural to associate the ``simple'' states with a symmetry that 
protects their purity and special character. 
This symmetry, however, is shared by only a subset of 
states, and is broken in the remaining eigenstates of the 
same Hamiltonian. It thus appears that realistic quantum many-body 
Hamiltonians can accommodate simultaneously eigenstates with different 
symmetry character. These are precisely the defining ingredients 
of a partial symmetry. In this context, PDS can offer a possible clue to 
the deep question of how simple features emerge from complicated dynamics. 

Underlying the PDS notion, is the recognition that 
it is possible for a non-invariant Hamiltonian to have selected eigenstates 
with good symmetry and good quantum numbers. 
In such a case, the symmetry in question is preserved 
in some states but is broken in the Hamiltonian (an opposite situation to 
that encountered in a spontaneously-broken symmetry). 
The PDS concept is, therefore, 
relevant to situations where selected states fulfill the predictions of 
a symmetry, which is otherwise known to be broken. 
Familiar examples of such a scenario include flavor symmetry in 
hadrons and chiral symmetry in nuclei. PDS may thus shed a new light on 
the related question of why, occasionally, symmetries seem to be obeyed 
beyond their domain of validity.

The realistic attributes and fundamental aspects of partial symmetries, 
as portrayed in the present review, 
illuminate their potential useful role 
in dynamical systems and motivate their ongoing and future in-depth study. 
Much has been learnt but much more awaits to be explored and understood.

\section*{Acknowledgments}
\label{acknow}

The author is pleased to acknowledge a fruitful collaboration with 
Y.~Alhassid, J.~Escher, J.E.~Garc\'\i a-Ramos, J.N.~Ginocchio, 
I.~Sinai, M. Mamistvalov, P.~Van~Isacker and N.D.~Whelan. 
Valuable discussions with F.~Iachello, D.J.~Rowe and I.~Talmi 
on fundamental aspects of partial symmetries, 
and with R.F.~Casten and N.~Pietralla on their empirical manifestations, 
are much appreciated. To them and to many other colleagues 
who have shown interest in this avenue 
of research, my warm thanks. 
This work was supported in part by the Israel Science Foundation 
and in part by the U.S.-Israel Binational Science Foundation. 

\section*{Appendix}
\label{Appendix}

\begin{table}
\begin{center}
\caption{\label{TabIBMcas}
\protect\small
Generators and Casimir operators, $\hat{C}_{G}$, of algebras $G$ in the IBM. 
Here $\hat{n}_d = \sqrt{5}\,U^{(0)}$, 
$L^{(1)} = \sqrt{10}\,U^{(1)}$, 
$Q^{(2)} = \Pi^{(2)} -{\textstyle\frac{\sqrt{7}}{2}}\,U^{(2)}$, 
$\bar{Q}^{(2)} = \Pi^{(2)} +{\textstyle\frac{\sqrt{7}}{2}}\,U^{(2)}$. 
The operators $U^{(L)},\, \Pi^{(2)},\, \bar{\Pi}^{(2)},\,\hat{n}_s$, 
are defined in Eq.~(\ref{u6gen}).} 
\vspace{1mm}
\begin{tabular}{ccccc}
\hline
& & & &\\[-3mm]
Algebra & Generators & Irrep. & Casimir operator & Eigenvalues\\[4pt]
& & & &\\[-3mm]
\hline
& & & &\\[-2mm]
{\rm O(3)} & $U^{(1)}$ & L & $L^{(1)}\cdot L^{(1)}$ & L(L+1) \\[2pt]
{\rm O(5)} & $U^{(1)},U^{(3)}$ & $\tau$ & 
$2(U^{(1)}\cdot U^{(1)} +U^{(3)}\cdot U^{(3)})$ & $\tau(\tau+3)$ \\[2pt]
{\rm O(6)} & $U^{(1)},U^{(3)},\Pi^{(2)}$ & $\sigma$ & 
$\hat{C}_{{\rm O(5)}} 
+ \Pi^{(2)}\cdot\Pi^{(2)}$ & $\sigma(\sigma+4)$ \\[2pt]
$\overline{{\rm O(6)}}$ & $U^{(1)},U^{(3)},\bar{\Pi}^{(2)}$ & $\bar{\sigma}$ & 
$\hat{C}_{{\rm O(5)}} 
+ \bar{\Pi}^{(2)}\cdot\bar{\Pi}^{(2)}$ & 
$\bar{\sigma}(\bar{\sigma}+4)$ \\[2pt]
{\rm SU(3)} & $U^{(1)},Q^{(2)}$ & $(\lambda,\mu)$ & 
$2Q^{(2)}\cdot Q^{(2)} + {\textstyle\frac{3}{4}}L^{(1)}\cdot L^{(1)}$ &
$\lambda^2 + (\lambda+\mu)(\mu+3)$\\[2pt]
$\overline{{\rm SU(3)}}$ & $U^{(1)},\bar{Q}^{(2)}$ & 
$(\bar{\lambda},\bar{\mu})$ & 
$2\bar{Q}^{(2)}\cdot \bar{Q}^{(2)} 
+ {\textstyle\frac{3}{4}}L^{(1)}\cdot L^{(1)}$ &
$\bar{\lambda}^2 + (\bar{\lambda}+\bar{\mu})(\bar{\mu}+3)$\\[2pt]
{\rm U(5)} & $U^{(L)}$ $L=0,\ldots,4$ & $n_d$ & 
$\hat{n}_d,\,\hat{n}_d(\hat{n}_d+4)$ & $n_d,\, n_d(n_d+4)$ \\[2pt]
{\rm U(6)} & $U^{(L)}$ $L=0,\ldots,4$ & $N$ & 
$\hat{N},\,\hat{N}(\hat{N}+5)$ & $N,\, N(N+5)$ \\[2pt]
      & $\Pi^{(2)},\, \bar{\Pi}^{(2)},\, \hat{n}_s$ & & & \\[2pt]
& & & &\\[-3mm]
\hline
\end{tabular}
\end{center}
\end{table}

The normal-order form of the most general IBM Hamiltonian with one- and 
two-body interactions is given by
\ba
\hat{H}_{IBM} &=&
\epsilon_s\,s^{\dag}s 
+ \epsilon_{d}\,d^{\dag}\cdot \tilde{d}
+ u_{0}\,(s^{\dag})^2s^2 
+ u_{2}\,s^{\dag}d^{\dag}\cdot\tilde{d}s 
+ v_{0}\,\left [\, 
(s^{\dag})^2\tilde{d}\cdot \tilde{d} + H.c. \, \right ] 
\nonumber\\
&&
+ v_{2}\,\left [\, 
s^{\dag}d^{\dag}\cdot (\tilde{d} \tilde{d})^{(2)} + H.c. \, \right ] 
+ \sum_{L=0,2,4}c_{L}\,(d^{\dag}d^{\dag})^{(L)}\cdot
(\tilde{d}\tilde{d})^{(L)} ~.
\label{hIBMnormal}
\ea
The corresponding multipole form, written in terms of the Casimir operators 
listed in Table~\ref{TabIBMcas}, is given by
\ba
\hat{H}_{IBM} &=&
\epsilon_s\,\hat{N} + u_{0}\,\hat{N}(\hat{N}-1)
+ (\epsilon_{d}-\epsilon_{s})\,\hat{n}_d
- (2u_0 -u_2 + 2v_0)\,(\hat{N}-1)\hat{n}_d 
\nonumber\\
&&
+(u_0 - u_2 +2v_0 + {\textstyle\frac{1}{2\sqrt{7}}}v_2 
+ {\textstyle\frac{1}{5}}c_0 + {\textstyle\frac{2}{7}}c_2 
+ {\textstyle\frac{18}{35}}c_4)
\hat{n}_{d}(\hat{n}_d-1)
\nonumber\\
&&
-(v_0 + {\textstyle\frac{3}{2\sqrt{7}}}v_2 
+ {\textstyle\frac{1}{5}}c_0 - {\textstyle\frac{2}{7}}c_2 
+ {\textstyle\frac{3}{35}}c_4)
\left [\, \hat{C}_{{\rm O(5)}} - 4\hat{n}_d\,\right ]
-{\textstyle\frac{1}{2\sqrt{7}}}\,v_2\,
\left [\, \hat{C}_{{\rm SU(3)}} - 10\hat{N}\,\right ] 
\nonumber\\
&&
+ (v_0 + {\textstyle\frac{1}{\sqrt{7}}}v_2)
\left [\, \hat{C}_{{\rm O(6)}} - 5\hat{N}\,\right ]
+({\textstyle\frac{1}{2\sqrt{7}}}v_2 
+ {\textstyle\frac{1}{7}}c_4 - {\textstyle\frac{1}{7}}c_2)
\left [\, \hat{C}_{{\rm O(3)}} - 6\hat{n}_d\,\right ] ~.
\label{hIBMcas}
\ea

\def\Journal#1#2#3#4{{#1} {#2} (#4) #3}
\def\ANNP{\em Ann. Phys. (N.Y.)}
\def\MATH{{\em J. Math. Phys.}}
\def\CHEM{{\em J. Chem. Phys.}}
\def\INT{{\em Int. J. Mod. Phys.} A}
\def\JPA{{\em J. Phys.} A} 
\def\JPAold{{\em J. Phys. A: Math. Gen.}} 
\def\JPB{{\em J. Phys.} B} 
\def\JPG{{\em J. Phys.} G} 
\def\NCA{{\em Nuovo Cimento} A }
\def\RNC{\em Rivista Nuovo Cimento}
\def\NP{{\em Nucl. Phys.}}
\def\NPA{{\em Nucl. Phys.} A}
\def\NPB{{\em Nucl. Phys.} B\ }
\def\PHYS{{\em Physica}}
\def\PRO{{\em Prog. Theor. Phys.}}
\def\PLA{{\em Phys. Lett.} A\ }
\def\PLB{{\em Phys. Lett.} B}
\def\PL{{\em Phys. Lett.}}
\def\PRL{{\em Phys. Rev. Lett.\/}\ }
\def\PREV{\em Phys. Rev.}
\def\PREP{\em Phys. Rep.}
\def\PRA{{\em Phys. Rev.} A}
\def\PRD{{\em Phys. Rev.} D}
\def\PRC{{\em Phys. Rev.} C}
\def\PRE{{\em Phys. Rev.} E}
\def\PRB{{\em Phys. Rev.} B}
\def\PS{\em Phys. Scr.}
\def\PhilML{{\em Phil. Mag. Lett.}}
\def\RMF{{\em Rev. Mex. Fis.}}
\def\RMP{{\em Rev. Mod. Phys.}}
\def\RPP{{\em Rep. Prog. Phys.}}
\def\ZPA{{\em Z. Phys.} A}
\def\PRSL{{\em Proc. Roy. Soc. Lond.} A}
\def\PPNP{\em Prog. Part. Nucl. Phys.}


\end{document}